\documentclass[12pt]{article}
\usepackage{SUW_style}

\begin{document}
\singlespacing

\title{\textsc{\Large{Abadie's Kappa and Weighting Estimators of the\\Local Average Treatment Effect}}\thanks{This version: February 28, 2024. For helpful comments, we thank the Editor, an Associate Editor, two anonymous referees, Alberto Abadie, Josh Angrist, Bryan Graham, Phillip Heiler, Toru Kitagawa, Chris Muris, Tomasz Olma, Pedro Sant'Anna, Yuya Sasaki, Liyang Sun, seminar participants at Brandeis University, Goethe University Frankfurt, University of Bonn, and University of Tübingen, and conference participants at CFE, CRC Retreat, EEA, ESEM, Frankfurt Econometrics Workshop, IAAE, MEG, NY Camp Econometrics, SEA, Statistical Week, VfS, and the World Congress of the Econometric Society. We also thank Frances Hoffen and Qihui Lei for excellent research assistance. S\l{}oczy\'{n}ski acknowledges financial support from the Theodore and Jane Norman Fund. Uysal acknowledges financial support from the German Research Foundation through CRC TRR 190 (project no.~280092119). Our companion Stata package, \texttt{kappalate}, is available on the SSC\@. To download this package, type \texttt{ssc install kappalate} in Stata.}
}
\author{\textsc{\large{Tymon S\l{}oczy\'{n}ski}}\thanks{Brandeis University} \and \textsc{\large{S. Derya Uysal}}\thanks{Ludwig Maximilian University of Munich} \and \textsc{\large{Jeffrey M. Wooldridge}}\thanks{Michigan State University}
}
\date{}

\begin{titlepage}
\maketitle
\begin{abstract}
\begin{normalsize}
\noindent
Recent research has demonstrated the importance of flexibly controlling for covariates in instrumental variables estimation. In this paper we study the finite sample and asymptotic properties of various weighting estimators of the local average treatment effect (LATE), motivated by \citeauthor{Abadie2003}'s \citeyearpar{Abadie2003} kappa theorem and offering the requisite flexibility relative to standard practice. We argue that two of the estimators under consideration, which are weight normalized, are generally preferable. Several other estimators, which are unnormalized, do not satisfy the properties of scale invariance with respect to the natural logarithm and translation invariance, thereby exhibiting sensitivity to the units of measurement when estimating the LATE in logs and the centering of the outcome variable more generally. We also demonstrate that, when noncompliance is one sided, certain weighting estimators have the advantage of being based on a denominator that is strictly greater than zero by construction. This is the case for only one of the two normalized estimators, and we recommend this estimator for wider use. We illustrate our findings with a simulation study and three empirical applications, which clearly document the sensitivity of unnormalized estimators to how the outcome variable is coded. We implement the proposed estimators in the Stata package \texttt{kappalate}.
\end{normalsize}
\end{abstract}
\thispagestyle{empty}
\end{titlepage}

\setcounter{page}{2}
\doublespacing

\section{Introduction}
\label{sec:intro}

The validity of many instrumental variables, as applied in economics and related fields, requires conditioning on additional covariates. In such cases empirical researchers often approximate the causal effects of interest using additive linear models and two-stage least squares (2SLS) estimation. However, recent work by \cite{Sloczynski2018,Sloczynski2021} and \cite{BBMT2022} questions the general validity of this approach and, in particular, the ability of the 2SLS estimand to uncover the local average treatment effect (LATE), that is, the average effect of treatment for ``compliers,'' as defined by \cite{IA1994} and \cite{AIR1996}. One concern is that covariate specifications used by empirical researchers are insufficiently flexible \citep{BBMT2022}. Another concern is that even when they are flexible, the 2SLS estimand does not generally correspond to the LATE or any other parameter of interest \citep{Sloczynski2018,Sloczynski2021}.

In this paper we study a class of simple yet flexible weighting estimators of the LATE, which are robust to the aforementioned limitations of 2SLS\@. The estimators we consider can be motivated by the identification result in \cite{Abadie2003}, which applies to any parameter defined in terms of moments of the joint distribution of the data for compliers, including the LATE\@. The result in 
\cite{Abadie2003} is based on ``kappa weighting,'' with weights that depend on the instrument propensity score. Some of the estimators we consider can alternatively be motivated by the identification result in \cite{Frolich2007}, which suggests a simple approach to estimating the LATE using the ratio of two conventional weighting estimators. Although the recent literature in econometrics and statistics has adopted this approach, it focuses primarily on the ratio of two \textit{unnormalized} weighting estimators \citep{Tan2006,Frolich2007,MCH2011,DHL2014a,DHL2014b,AANP2017}, despite the fact that the lack of normalization leads to poor finite sample properties in related contexts \citep{Imbens2004,MT2009,BDM2014}. Here, normalization means rescaling the weights so that they sum to one in each sample.

In this paper we unify and provide a comprehensive treatment of the two approaches to constructing weighting estimators of the LATE\@. We begin with an observation that the existing identification results enable the construction of multiple consistent estimators of the LATE, only two of which are normalized. One normalized estimator is the sample analogue of a particular expression in \cite{AC2018}, based on \cite{Abadie2003}. However, it is also straightforward, as in \cite{UysalDiss}, to construct a normalized version of \citeauthor{Tan2006}'s \citeyearpar{Tan2006} and \citeauthor{Frolich2007}'s \citeyearpar{Frolich2007} estimator and to interpret it through the lens of ``kappa weighting.'' We argue that these two normalized estimators are likely to dominate the unnormalized weighting estimators of the LATE in many cases. Unlike most other papers that stress the importance of normalization, we also provide an objective and intuitively appealing criterion that differentiates the normalized from the unnormalized estimators; see also \cite{Tille1998} and \cite{AM2013}. Indeed, we demonstrate that the former class of estimators, unlike the latter, satisfies the properties of \emph{(i)} translation invariance and \emph{(ii)} scale invariance with respect to the natural logarithm. This ensures that the normalized estimators are not sensitive to the centering of the outcome variable or, when estimating the LATE in logs, to the units of measurement of the untransformed outcome \citep[cf.][]{CR2023}.

We also identify an important context, namely settings with one-sided noncompliance, in which certain estimators have an additional advantage: they are based on a denominator that is strictly greater than zero by construction. This is the case for \emph{(i)} \citeauthor{Tan2006}'s \citeyearpar{Tan2006} and \citeauthor{Frolich2007}'s \citeyearpar{Frolich2007} unnormalized estimator whenever there are no always-takers, that is, individuals who participate in the treatment regardless of the value of the instrument; \emph{(ii)} a different unnormalized estimator whenever there are no never-takers, that is, individuals who never participate in the treatment; and \emph{(iii)} the normalized estimator originally proposed by \cite{UysalDiss} in both of these cases. We recommend this last estimator for wider use in practice.

Our observations about translation and scale invariance as well as settings with one-sided noncompliance apply equally when the instrument propensity score is known and when it is estimated using standard methods. In practice, the instrument propensity score is rarely known, and its estimation can greatly influence the properties of the final estimator of the LATE\@. We consider maximum likelihood and covariate balancing estimation of the instrument propensity score, where the latter approach follows \cite{GPE2012,GPE2016}, \cite{IR2014}, \cite{Heiler2022}, and \cite{SASX2022}, among others. Either approach is compatible with the construction of the estimator in \cite{UysalDiss}, and when appropriate covariate balancing propensity scores are used, this estimator is also equivalent to \citeauthor{Heiler2022}'s \citeyearpar{Heiler2022}.

Aside from the finite sample properties of weighting estimators of the LATE, we also study their asymptotic properties in a unified framework of M-estimation. Under standard regularity conditions, our weighting estimators are asymptotically normal, and we derive their asymptotic variances. To illustrate our findings, we also use three empirical applications and a simulation study. The simulations confirm the very good relative performance of our preferred normalized estimator, especially with covariate balancing propensity scores, which appear to be more robust to misspecification than their maximum likelihood counterparts.

Our empirical applications focus on causal effects of military service \citep{Angrist1990}, college education \citep{Card1995}, and childbearing \citep{AE1998}. In each of these cases, we document what we regard as superiority of normalized weighting. The bottom line is that unnormalized estimators are very sensitive to how the outcome variable is coded. In each application, the estimates are sensitive to the units of measurement (cents, dollars, \$1,000s, \$100,000s) of the income variable prior to the log transformation. In our replication of \cite{AE1998}, we also consider labor force participation as a binary outcome, and we document that unnormalized estimators are highly sensitive to whether working for pay is coded as, say, 1 or 0.

Our application of weighting to estimate the LATE appears to be somewhat rare in practice, although \citeauthor{Abadie2003}'s \citeyearpar{Abadie2003} result is more commonly used to estimate mean characteristics of compliers, as also recommended by \cite{AP2009}. We analyze two samples of applications of instrumental variables to verify this claim. First, our reading of the 30 papers replicated by \cite{Young2022}, each of which uses 2SLS, suggests that none of these papers uses weighting estimators of the LATE or applies \citeauthor{Abadie2003}'s \citeyearpar{Abadie2003} result for any other purpose. Second, we have also examined whether any of the papers published in journals of the American Economic Association in 2019 and 2020 consider weighting estimators of the LATE\@. Our best assessment is that the answer is likewise negative. Still, \cite{MT2019}, \cite
{GGS2020}, \cite{LOL2020}, and \cite{LVRS2020} apply \citeauthor{Abadie2003}'s \citeyearpar{Abadie2003} result to estimate mean characteristics of compliers, while \cite{Cohodes2020} uses this result to estimate the control complier mean (CCM), a parameter introduced by \cite{KKL2001}. In this paper we argue that ``kappa weighting'' can also be used more widely as a flexible alternative to 2SLS, and we provide a practical guide to using this method to estimate the LATE\@.

The remainder of the paper is organized as follows. Section \ref{sec:framework} introduces our framework. Section \ref{sec:estandinf} provides our theoretical results on estimation and inference. Section \ref{sec:applications} illustrates our results with three empirical applications. Section \ref{sec:simulation} discusses our simulation study. Section \ref{sec:conclusion} concludes. Proofs and derivations are collected in the appendix unless noted otherwise. The estimators considered in this paper are also implemented in the companion Stata package \texttt{kappalate}.

\section{Framework}
\label{sec:framework}

Our framework broadly follows \cite{Abadie2003}. Let $Y$ denote the outcome variable of interest, $D$ the binary treatment, and $Z$ the binary instrument for $D$. We also introduce a vector of observed covariates, $X$, that predict $Z$\@. The instrument propensity score is written as $p(X) = \pr (Z=1 \mid X)$.

There are two potential outcomes, $Y_{1}$ and $Y_{0}$, only one of which is observed for a given individual, $Y = D \cdot Y_{1} + \left( 1-D \right) \cdot Y_{0}$. Similarly, there are two potential treatments, $D_{1}$ and $D_{0}$, and it is $Z$ that determines which of them is observed, $D = Z \cdot D_{1} + \left( 1-Z \right) \cdot D_{0}$. It will also be useful to include $Z$ in the definition of potential outcomes, letting $Y_{zd}$ denote the potential outcome that a given individual would obtain if $Z=z$ and $D=d$.

\cite{AIR1996} divide the population into four mutually exclusive subgroups based on the latent values of $D_{1}$ and $D_{0}$. Individuals with $D_{1} = D_{0} = 1$ are referred to as \emph{always-takers}, as they get treatment regardless of whether they are encouraged to do so or not; similarly, individuals with $D_{1} = D_{0} =0$ are referred to as \emph{never-takers}. Individuals with $D_{1} = 1$ and $D_{0} = 0$ are referred to as \emph{compliers}, as they comply with their instrument assignment; they get treatment if they are encouraged to do so but not otherwise. Analogously, individuals with $D_{1} = 0$ and $D_{0} = 1$ are referred to as \emph{defiers}, as they defy their instrument assignment.

As usual, we define the treatment effect as the difference in the outcomes with and without treatment, $Y_{1} - Y_{0}$. Following \cite{IA1994}, a large literature has focused on identification and estimation of the local average treatment effect (LATE), defined as
\begin{equation*}
\late = \e \left( Y_{1} - Y_{0} \mid D_{1} > D_{0} \right),
\end{equation*}
i.e.~as the average treatment effect for compliers or, in other words, for those individuals who would be induced to get treatment by the change in $Z$ from zero to one.

Next, we review a general identification result due to \cite{Abadie2003}, which we will use, in turn, to discuss identification of $\late$. We begin by restating \citeauthor{Abadie2003}'s \citeyearpar{Abadie2003} assumptions.

\begin{uassumption}{IV}
\label{ass:abadie}
\emph{(i)} \emph{Independence of the instrument:} $\left( Y_{00}, Y_{01}, Y_{10}, Y_{11}, D_{0}, D_{1} \right) \perp Z \mid X$. \\
\emph{(ii)} \emph{Exclusion of the instrument:} $\pr(Y_{1d}=Y_{0d} \mid X) = 1$ for $d \in \left\lbrace 0,1 \right\rbrace$ a.s. \\
\emph{(iii)} \emph{First stage:} $0 < \pr (Z=1 \mid X) < 1$ and $\pr (D_{1}=1 \mid X) > \pr (D_{0}=1 \mid X)$ a.s. \\
\emph{(iv)} \emph{Monotonicity:} $\pr (D_{1} \geq D_{0} \mid X)=1$ a.s.
\end{uassumption}

\noindent
These assumptions are standard in the recent literature. Assumption \ref{ass:abadie}(i) states that, conditional on covariates, the instrument is ``as good as randomly assigned.'' Assumption \ref{ass:abadie}(ii) implies that the instrument only affects the outcome through its effect on treatment status; it follows that $Y_{0} = Y_{10} = Y_{00}$ and $Y_{1} = Y_{11} = Y_{01}$. Assumption \ref{ass:abadie}(iii) combines an overlap condition with a requirement that the instrument affects the conditional probability of treatment. Finally, Assumption \ref{ass:abadie}(iv) rules out the existence of defiers, and implies that the population consists of always-takers, never-takers, and compliers. Under Assumption \ref{ass:abadie}, as demonstrated by \cite{Abadie2003}, any feature of the joint distribution of $\left( Y,D,X \right)$, $\left( Y_{0},X \right)$, or $\left( Y_{1},X \right)$ is identified for compliers.

\begin{lemma}[\citealp{Abadie2003}]
\label{lemma:abadie}
Let $g(\cdot)$, $g_0(\cdot)$, and $g_1(\cdot)$ be measurable functions of their arguments such that $\e | g(Y,D,X) | < \infty$, $\e | g_0(Y_0,X) | < \infty$, and $\e | g_1(Y_1,X) | < \infty$. Define
\begingroup
\allowdisplaybreaks
\begin{eqnarray}
\kappa_{0} &=& \left( 1 - D \right) \frac{\left( 1 - Z \right) - \left( 1 - p(X) \right)}{p(X) \left( 1 - p(X) \right)},
\nonumber\\
\kappa_{1} &=& D \frac{Z - p(X)}{p(X) \left( 1 - p(X) \right)},
\nonumber\\
\kappa = \kappa_{0} \left( 1 - p(X) \right) + \kappa_{1} p(X) &=& 1 - \frac{D \left( 1 - Z \right)}{1 - p(X)} - \frac{\left( 1 - D \right) Z}{p(X)}.
\nonumber
\end{eqnarray}
\endgroup
Under Assumption \ref{ass:abadie},
\begin{enumerate}
\item[\emph{(a)}] $\e \left[ g(Y,D,X) \mid D_{1} > D_{0} \right] = \frac{1}{\pr(D_{1} > D_{0})} \e \left[ \kappa \; g(Y,D,X) \right]$. Also,
\item[\emph{(b)}] $\e \left[ g_0(Y_{0},X) \mid D_{1} > D_{0} \right] = \frac{1}{\pr(D_{1} > D_{0})} \e \left[ \kappa_{0} \; g_0(Y,X) \right]$, and
\item[\emph{(c)}] $\e \left[ g_1(Y_{1},X) \mid D_{1} > D_{0} \right] = \frac{1}{\pr(D_{1} > D_{0})} \e \left[ \kappa_{1} \; g_1(Y,X) \right]$.
\end{enumerate}
Moreover, \emph{(a--c)} also hold conditional on $X$.
\end{lemma}

\noindent
Both \cite{Abadie2003} and the subsequent applied literature have focused on the implications of Lemma \ref{lemma:abadie}(a). On the other hand, Lemma \ref{lemma:abadie}(b) and (c) have been used in the econometrics literature to identify and estimate $\late$ and quantile treatment effects \citep{FM2013,AC2018,SASX2022,SS2024}.

To see how Lemma \ref{lemma:abadie}(b) and (c) identifies $\late$, take $g_0(Y_{0},X) = Y_{0}$ and $g_1(Y_{1},X) = Y_{1}$, and write:
\begin{equation}
\label{eq:late1}
\late = \frac{1}{\pr(D_{1} > D_{0})} \e \left( \kappa_{1} Y \right) - \frac{1}{\pr(D_{1} > D_{0})} \e \left( \kappa_{0} Y \right).
\end{equation}
We can also rewrite equation (\ref{eq:late1}) to obtain the following expression for $\late$:
\begin{equation}
\label{eq:late2}
\late = \frac{1}{\pr(D_{1} > D_{0})} \e \left[ \left( \kappa_{1} - \kappa_{0} \right) Y \right] = \frac{1}{\pr(D_{1} > D_{0})} \e \left[ Y \frac{Z - p(X)}{p(X) \left( 1 - p(X) \right)} \right].
\end{equation}
As we will see later, it is useful to treat equations (\ref{eq:late1}) and (\ref{eq:late2}) as distinct. In any case, it is clear that $\late$ is identified as long as $\pr(D_{1} > D_{0})$ is identified. As noted by \cite{Abadie2003}, Lemma \ref{lemma:abadie}(a) implies that $\pr(D_{1} > D_{0}) = \e(\kappa)$, which follows from taking $g(Y,D,X) = 1$. Similarly, however, we can use Lemma \ref{lemma:abadie}(b) and (c) to obtain $\pr(D_{1} > D_{0}) = \e(\kappa_{1})$ and $\pr(D_{1} > D_{0}) = \e(\kappa_{0})$. This is not a novel observation but we will provide a more comprehensive discussion of its consequences than has been done in previous work. We conclude this section with the following remark.

\begin{remark}
\label{remark:kappaeq}
$\e(\kappa) \; = \; \e(\kappa_{1}) - \e \left[ \frac{Z - p(X)}{p(X)} \right] \; = \; \e(\kappa_{1}) \; = \; \e(\kappa_{1}) - \e \left[ \frac{Z - p(X)}{p(X) \left( 1 - p(X) \right)} \right] \; = \; \e(\kappa_{0})$.
\end{remark}

\noindent
The proof of Remark \ref{remark:kappaeq} follows from simple algebra and is omitted. The facts that $\e \left[ \frac{Z - p(X)}{p(X)} \right] = 0$ and $\e \left[ \frac{Z - p(X)}{p(X) \left( 1 - p(X) \right)} \right] = 0$ hold by iterated expectations. It follows that $\e(\kappa) = \e(\kappa_{1}) = \e(\kappa_{0})$. Additionally, Lemma \ref{lemma:abadie} implies that each of these objects identifies $\pr(D_{1} > D_{0})$.

\section{Estimation and Inference}
\label{sec:estandinf}

In this section we study estimation and inference for $\late$. We begin by introducing our preferred weighting estimator of this parameter. Then, we develop the argument in favor of this estimator, beginning with the case where $p(X)$ is known and later explaining how $p(X)$ can be estimated when it is not known. While $p(X)$ is rarely known in practice, our novel insights in Sections \ref{sec:normalized} and \ref{sec:nearzero} apply equally in that case and when $p(X)$ is estimated using standard methods.

\subsection{Recommended Estimator}
\label{sec:recommended}

Given a random sample $\big\{ (D_{i},Z_{i},X_{i},Y_{i}):i=1,\ldots,N \big\}$, and assuming that the instrument propensity score is known, our recommended weighting estimator of $\late$ can be written as:
\begin{equation}
\label{eq:tauu}
\hat{\tau}_{u} = \frac{\left[ \sum_{i=1}^{N} \frac{Z_{i}}{p(X_{i})} \right]^{-1} \sum_{i=1}^{N} \frac{Y_i Z_{i}}{p(X_{i})} - \left[ \sum_{i=1}^{N} \frac{1 - Z_{i}}{1 - p(X_{i})} \right]^{-1} \sum_{i=1}^{N} \frac{Y_i \left( 1 - Z_{i} \right)}{1 - p(X_{i})}}{\left[ \sum_{i=1}^{N} \frac{Z_{i}}{p(X_{i})} \right]^{-1} \sum_{i=1}^{N} \frac{D_i Z_{i}}{p(X_{i})} - \left[ \sum_{i=1}^{N} \frac{1 - Z_{i}}{1 - p(X_{i})} \right]^{-1} \sum_{i=1}^{N} \frac{D_i \left( 1 - Z_{i} \right)}{1 - p(X_{i})}}.
\end{equation}
This estimator was proposed by \cite{UysalDiss}, and is easily implementable as a function of six sample means. It is also implementable as the coefficient on $D$ in a weighted IV regression of $Y$ on $D$, with $Z$ as the instrument and weights equal to $\frac{Z}{p(X)} + \frac{1-Z}{1-p(X)}$. When the instrument propensity score is not known, a possibility we consider explicitly in Sections \ref{sec:unknown} and \ref{sec:asymptotic}, we would adopt a parametric model for $p(X)$, $F(X,\alpha)$, estimate the unknown parameters by an appropriate method, and replace the instrument propensity scores in equation (\ref{eq:tauu}) with their estimates, $\hat{p}(X) = F(X,\hat{\alpha})$. The leading model for $p(X)$ is logit, $F(X,\alpha) = \exp(X\alpha)/[1+\exp(X\alpha)]$, and the natural estimation methods are maximum likelihood and covariate balancing. Appropriate covariate balancing approaches include those in \cite{GPE2012,GPE2016} and \cite{IR2014}, both of which would lead to simple method of moments estimators of $\alpha$. We defer further details on estimation of $\alpha$ to Section \ref{sec:unknown}. Note that $\hat{\tau}_{u}$ with covariate balancing propensity scores is also recommended by \cite{Heiler2022} but we are the first to determine its advantages given in the analysis below.

Recent software implements $\hat{\tau}_{u}$ in R and Stata. Specifically, \cite{BH2018} implement this estimator in their \texttt{causalweight} package in R, although covariate balancing estimation of $\alpha$ is not currently supported and inference is based on the bootstrap. Our companion Stata package \texttt{kappalate} implements $\hat{\tau}_{u}$ and other weighting estimators, and we allow both maximum likelihood and covariate balancing estimation of $\alpha$, as well as computation of analytical standard errors. The package is downloadable from the Statistical Software Components (SSC) Archive.

Two further comments about $\hat{\tau}_{u}$ are in order. First, this is our preferred member of the class of weighting estimators, but there are other classes of estimators one may be willing to consider. One such class is doubly robust estimators, which combine weighting and models for conditional expectations of $Y$ and $D$\@. Doubly robust estimators of $\late$ have been developed by \cite{Tan2006}, \cite{UysalDiss}, \cite{ORR2015}, \cite{BCFVH2017}, \cite{SUW2022}, \cite{MSSU2023}, and others. In this paper, however, we restrict our attention to the class of weighting estimators.

Second, a prototypical weighting or doubly robust estimator, such as $\hat{\tau}_{u}$, might be poorly behaved when some instrument propensity scores are close to 0 or 1 \citep[cf.][]{KT2010}, even if Assumption \ref{ass:abadie} is not violated. In this scenario, usually referred to as ``limited'' or ``weak'' overlap, it might be preferable to use estimators of $\late$ that were designed to alleviate this problem, such as those in \cite{Hongetal2020} and \cite{MSSU2023}. See also \cite{CH2016}, \cite{Rothe2017}, \cite{MW2020}, \cite{HK2021}, and \cite{SU2022} for settings with limited overlap and exogenous $D$, as well as \cite{LDDFS2021} and \cite{MSW2022} for formal statistical tests of limited overlap.

\subsection{Estimation When the Instrument Propensity Score Is Known}
\label{sec:known}

In this section we introduce several seemingly intuitive weighting estimators of $\late$, which we will later show to have some undesirable finite sample properties. For now, we continue to assume that the instrument propensity score is known. In this case, equation (\ref{eq:late2}) suggests that we can consistently estimate $\late$ as follows:
\begin{equation*}
\latehat = \frac{1}{\hat{\pr}(D_{1} > D_{0})}\left[ N^{-1}\sum_{i=1}^{N} Y_i \frac{Z_{i} - p(X_{i})}{p(X_{i}) \left( 1 - p(X_{i}) \right)} \right],
\end{equation*}
where $\hat{\pr}(D_{1} > D_{0}) \overset{p}{\rightarrow} \pr(D_{1} > D_{0}) > 0$. Our discussion in Section \ref{sec:framework} also implies that there are at least three candidate estimators for $\pr(D_{1} > D_{0})$, namely $N^{-1}\sum_{i=1}^{N} \kappa_{i}$, $N^{-1}\sum_{i=1}^{N} \kappa_{i1}$, and $N^{-1}\sum_{i=1}^{N} \kappa_{i0}$, where $\kappa_{i} = 1 - \frac{D_{i} \left( 1 - Z_{i} \right)}{1 - p(X_{i})} - \frac{\left( 1 - D_{i} \right) Z_{i}}{p(X_{i})}$, $\kappa_{i1} = D_{i} \frac{Z_{i} - p(X_{i})}{p(X_{i}) \left( 1 - p(X_{i}) \right)}$, and $\kappa_{i0} = \left( 1 - D_{i} \right) \frac{\left( 1 - Z_{i} \right) - \left( 1 - p(X_{i}) \right)}{p(X_{i}) \left( 1 - p(X_{i}) \right)}$. Consequently, we have the following consistent estimators of $\late$:
\begingroup
\allowdisplaybreaks
\begin{eqnarray}
\label{eq:taua}
\hat{\tau}_{a} &=& \left[ \sum_{i=1}^{N} \kappa_{i} \right]^{-1} \left[ \sum_{i=1}^{N} Y_i \frac{Z_{i} - p(X_{i})}{p(X_{i}) \left( 1 - p(X_{i}) \right)} \right], \\
\label{eq:taua1}
\hat{\tau}_{a,1} &=& \left[ \sum_{i=1}^{N} \kappa_{i1} \right]^{-1} \left[ \sum_{i=1}^{N} Y_i \frac{Z_{i} - p(X_{i})}{p(X_{i}) \left( 1 - p(X_{i}) \right)} \right], \\
\label{eq:taua0}
\hat{\tau}_{a,0} &=& \left[ \sum_{i=1}^{N} \kappa_{i0} \right]^{-1} \left[ \sum_{i=1}^{N} Y_i \frac{Z_{i} - p(X_{i})}{p(X_{i}) \left( 1 - p(X_{i}) \right)} \right].
\end{eqnarray}
\endgroup
One might mistakenly expect that the choice of the estimator for $\pr(D_{1} > D_{0})$ is largely inconsequential. We discuss this issue extensively in what follows. For now, it should suffice to note that $N^{-1}\sum_{i=1}^{N} \frac{Z_{i} - p(X_{i})}{p(X_{i})}$ and $N^{-1}\sum_{i=1}^{N} \frac{Z_{i} - p(X_{i})}{p(X_{i}) \left( 1 - p(X_{i}) \right)}$ are not generally equal to zero or to each other, and hence $N^{-1}\sum_{i=1}^{N} \kappa_{i}$, $N^{-1}\sum_{i=1}^{N} \kappa_{i1}$, and $N^{-1}\sum_{i=1}^{N} \kappa_{i0}$ will also generally be different, unlike their population counterparts (cf.~Remark \ref{remark:kappaeq}).

Lemma \ref{lemma:abadie} is not the only identification result that allows us to construct consistent estimators of the LATE\@. An alternative result is provided by \citet[Theorem 1]{Frolich2007}. An implication of this result is that the ratio of any consistent estimator of the average treatment effect (ATE) of $Z$ on $Y$ and any consistent estimator of the ATE of $Z$ on $D$ is consistent for the LATE\@. Given our interest in weighting estimators, a natural candidate estimator is
\begin{equation}
\label{eq:taut}
\hat{\tau}_{t} = \left[ \sum_{i=1}^{N} \frac{D_i Z_{i}}{p(X_{i})} - \sum_{i=1}^{N} \frac{D_i \left( 1 - Z_{i} \right)}{1 - p(X_{i})} \right]^{-1} \left[ \sum_{i=1}^{N} \frac{Y_i Z_{i}}{p(X_{i})} - \sum_{i=1}^{N} \frac{Y_i \left( 1 - Z_{i} \right)}{1 - p(X_{i})} \right],
\end{equation}
as suggested by \cite{Tan2006} and \cite{Frolich2007}. This estimator is equal to the ratio of two weighting estimators of the ATE of $Z$ (on $Y$ and $D$) under unconfoundedness \citep[cf.][]{HIR2003}. The following remark, which has not been precisely stated in previous work, clarifies the relationship between $\hat{\tau}_{t}$ and the other estimators introduced above.

\begin{remark}
\label{remark:equiv}
$\hat{\tau}_{t} = \hat{\tau}_{a,1}$.
\end{remark}

\noindent
Remark \ref{remark:equiv} states that $\hat{\tau}_{t}$ and $\hat{\tau}_{a,1}$ are numerically identical, which can be seen by plugging in the expression for $\kappa_{i1}$ into equation (\ref{eq:taua1}):
\begin{equation}
\label{eq:taua1bis}
\hat{\tau}_{a,1} = \left[ \sum_{i=1}^{N} D_i \frac{Z_{i} - p(X_{i})}{p(X_{i}) \left( 1 - p(X_{i}) \right)} \right]^{-1} \left[ \sum_{i=1}^{N} Y_i \frac{Z_{i} - p(X_{i})}{p(X_{i}) \left( 1 - p(X_{i}) \right)} \right].
\end{equation}
As is easy to see, expressions (\ref{eq:taut}) and (\ref{eq:taua1bis}) are equivalent. It is also important to note that $\hat{\tau}_{t}$ ($= \hat{\tau}_{a,1}$), or at least its variant where $p(X)$ is estimated, is by far the most popular weighting estimator of the LATE in the econometrics literature. It has been considered by \cite{Tan2006}, \cite{Frolich2007}, \cite{MCH2011}, \cite{DHL2014a,DHL2014b}, and \cite{AANP2017}, among others. As we will see in the next section, however, this estimator has a major drawback in practice.

\subsection{Unnormalized and Normalized Weights}
\label{sec:normalized}

Following \cite{Imbens2004}, \cite{MT2009}, and \cite{BDM2014}, it is widely understood that weighting estimators of the ATE under unconfoundedness should be normalized, i.e.~their weights should sum to unity, an idea that is often attributed to \cite{Hajek1971}. More recently, \cite{KU2023} provide a general treatment of normalization under unconfoundedness while \cite{SAZ2020} and \cite{CSA2021} stress the importance of normalization in difference-in-differences methods. It is natural to expect that normalization will also be important when estimating the LATE \citep[cf.][]{Heiler2022}.

It follows immediately that $\hat{\tau}_{t}$ is likely inferior to the ratio of two normalized, \citeauthor{Hajek1971}-type estimators of the ATE of $Z$ under unconfoundedness:
\begin{equation*}
\hat{\tau}_{u} = \frac{\left[ \sum_{i=1}^{N} \frac{Z_{i}}{p(X_{i})} \right]^{-1} \sum_{i=1}^{N} \frac{Y_i Z_{i}}{p(X_{i})} - \left[ \sum_{i=1}^{N} \frac{1 - Z_{i}}{1 - p(X_{i})} \right]^{-1} \sum_{i=1}^{N} \frac{Y_i \left( 1 - Z_{i} \right)}{1 - p(X_{i})}}{\left[ \sum_{i=1}^{N} \frac{Z_{i}}{p(X_{i})} \right]^{-1} \sum_{i=1}^{N} \frac{D_i Z_{i}}{p(X_{i})} - \left[ \sum_{i=1}^{N} \frac{1 - Z_{i}}{1 - p(X_{i})} \right]^{-1} \sum_{i=1}^{N} \frac{D_i \left( 1 - Z_{i} \right)}{1 - p(X_{i})}}.
\end{equation*}
This estimator, first proposed by \cite{UysalDiss}, was introduced in equation (\ref{eq:tauu}) as our preferred estimator. It might not be immediately obvious how the importance of normalization affects our understanding of $\hat{\tau}_{a}$, $\hat{\tau}_{a,1}$, and $\hat{\tau}_{a,0}$. To see this, note that these estimators can equivalently be represented as sample analogues of equation (\ref{eq:late1}):
\begingroup
\allowdisplaybreaks
\begin{eqnarray*}
\hat{\tau}_{a} &=& \left[ \sum_{i=1}^{N} \kappa_{i} \right]^{-1} \left[ \sum_{i=1}^{N} \kappa_{i1} Y_i \right] - \left[ \sum_{i=1}^{N} \kappa_{i} \right]^{-1} \left[ \sum_{i=1}^{N} \kappa_{i0} Y_i \right], \\
\hat{\tau}_{a,1} &=& \left[ \sum_{i=1}^{N} \kappa_{i1} \right]^{-1} \left[ \sum_{i=1}^{N} \kappa_{i1} Y_i \right] - \left[ \sum_{i=1}^{N} \kappa_{i1} \right]^{-1} \left[ \sum_{i=1}^{N} \kappa_{i0} Y_i \right], \\
\hat{\tau}_{a,0} &=& \left[ \sum_{i=1}^{N} \kappa_{i0} \right]^{-1} \left[ \sum_{i=1}^{N} \kappa_{i1} Y_i \right] - \left[ \sum_{i=1}^{N} \kappa_{i0} \right]^{-1} \left[ \sum_{i=1}^{N} \kappa_{i0} Y_i \right].
\end{eqnarray*}
\endgroup
None of these estimators is normalized. First, $\hat{\tau}_{a}$ uses weights of $\left[ \sum_{i=1}^{N} \kappa_{i} \right]^{-1} \kappa_{i1}$ and $\left[ \sum_{i=1}^{N} \kappa_{i} \right]^{-1} \kappa_{i0}$, which do not necessarily sum to unity across $i$. Second, $\hat{\tau}_{a,1}$ is based on weights of $\left[ \sum_{i=1}^{N} \kappa_{i1} \right]^{-1} \kappa_{i1}$, which are properly normalized, and $\left[ \sum_{i=1}^{N} \kappa_{i1} \right]^{-1} \kappa_{i0}$, which are not. Finally, $\hat{\tau}_{a,0}$ uses weights of $\left[ \sum_{i=1}^{N} \kappa_{i0} \right]^{-1} \kappa_{i1}$, which do not necessarily sum to unity across $i$, and $\left[ \sum_{i=1}^{N} \kappa_{i0} \right]^{-1} \kappa_{i0}$, which are properly normalized.

It is straightforward to construct a normalized estimator based on equation (\ref{eq:late1}). To do this, the two denominators need to be estimated separately, using different estimators of $\pr(D_{1} > D_{0})$, $N^{-1}\sum_{i=1}^{N} \kappa_{i1}$ and $N^{-1}\sum_{i=1}^{N} \kappa_{i0}$. The resulting estimator becomes
\begin{equation*}
\hat{\tau}_{a,10} = \left[ \sum_{i=1}^{N} \kappa_{i1} \right]^{-1} \left[ \sum_{i=1}^{N} \kappa_{i1} Y_i \right] - \left[ \sum_{i=1}^{N} \kappa_{i0} \right]^{-1} \left[ \sum_{i=1}^{N} \kappa_{i0} Y_i \right],
\end{equation*}
where both sets of weights, $\left[ \sum_{i=1}^{N} \kappa_{i1} \right]^{-1} \kappa_{i1}$ and $\left[ \sum_{i=1}^{N} \kappa_{i0} \right]^{-1} \kappa_{i0}$, are properly normalized. This estimator has been considered by \cite{AC2018} and \cite{SASX2022}. While the literature on quantile treatment effects studies normalized kappa weighting estimators somewhat more often \citep[see, e.g.,][]{FM2013}, the importance of normalization is not explicitly recognized. Interestingly, if the goal is to estimate $\e \left( X \mid D_{1} > D_{0} \right)$ rather than $\late$ or quantile treatment effects, as in \cite{MT2019}, \cite
{GGS2020}, \cite{LOL2020}, and \cite{LVRS2020}, among others, then three normalized estimators of this object can readily be constructed: $\left[ \sum_{i=1}^{N} \kappa_{i} \right]^{-1} \sum_{i=1}^{N} \kappa_{i} X_i$, $\left[ \sum_{i=1}^{N} \kappa_{i0} \right]^{-1} \sum_{i=1}^{N} \kappa_{i0} X_i$, and $\left[ \sum_{i=1}^{N} \kappa_{i1} \right]^{-1} \sum_{i=1}^{N} \kappa_{i1} X_i$.

It should also be noted that $\hat{\tau}_{u}$ can likewise be interpreted as a normalized ``Abadie'' or ``kappa weighting'' estimator. To see this, note that $N^{-1} \sum_{i=1}^{N} \frac{Z_{i}}{p(X_{i})} \overset{p}{\rightarrow} 1$ and $N^{-1} \sum_{i=1}^{N} \frac{1 - Z_{i}}{1 - p(X_{i})} \overset{p}{\rightarrow} 1$. This implies that $\hat{\tau}_{u} \overset{p}{\rightarrow} \frac{\e \left[ \frac{YZ}{p(X)} \right] - \e \left[ \frac{Y\left( 1 - Z \right)}{1 - p(X)} \right]}{\e \left[ \frac{DZ}{p(X)} \right] - \e \left[ \frac{D\left( 1 - Z \right)}{1 - p(X)} \right]} = \frac{\e \left[ Y \frac{Z - p(X)}{p(X) \left( 1 - p(X) \right)} \right]}{\e(\kappa_{1})}$, which is the same as the expression for $\late$ in equation (\ref{eq:late2}), subject to $\pr(D_{1} > D_{0}) = \e(\kappa_{1})$.

So far, we have made it seem obvious that weighting estimators should be normalized. Yet, it is natural to ask: \emph{Why} is it so important that weights sum to unity? Many of the recommendations to date are based on simulation results \citep[e.g.,][]{MT2009,BDM2014}, and it is not clear to what extent such evidence should guide estimator choice \citep[cf.][]{AKS2019}. In what follows, we provide an objective and intuitively appealing criterion that differentiates the normalized from the unnormalized estimators.

To present our criterion, we need to introduce some additional notation. Let $\mathbf{Y}$ be a column vector of observed data on outcomes and $\mathbf{W} = \left( \mathbf{D} \; \mathbf{Z} \; \mathbf{X} \right)$ be a matrix of observed data on the remaining variables, namely the treatment status, the instrument, and the covariates. We postulate that any reasonable estimator of $\late$ should be translation invariant.

\begin{udefinition}{TI}[Translation Invariance]
We say that an estimator $\hat{\tau} = \hat{\tau} \left( \mathbf{Y}, \mathbf{W} \right)$ is translation invariant if $\hat{\tau} \left( \mathbf{Y}, \mathbf{W} \right) = \hat{\tau} \left( \mathbf{Y} + k, \mathbf{W} \right)$ for all $\mathbf{Y}$, $\mathbf{W}$, and $k$.
\end{udefinition}

\noindent
The property of translation invariance is defined as the invariance of an estimator to an additive change of the outcome values for all units by a fixed amount. Put differently, estimators that are not translation invariant will generally depend on how the outcome variable is centered. If this variable is binary, the estimate may change when we relabel the zeros and ones, on top of the obvious sign change that is due to relabeling. If the outcome is a logarithm of some other variable, the estimator is also not invariant to scale transformations of that variable.

\begin{udefinition}{SE}[Scale Equivariance]
\label{def:scale}
We say that an estimator $\hat{\tau} = \hat{\tau} \left( \mathbf{Y}, \mathbf{W} \right)$ is scale equivariant if $\hat{\tau} \left( f(a\mathbf{Y}), \mathbf{W} \right) = a^{\alpha_1} \hat{\tau} \left( f(\mathbf{Y}), \mathbf{W} \right)$, $f(\mathbf{Y}) = \left( g(Y_1), \ldots, g(Y_N) \right)$, $g(Y) = \alpha_2 Y^{\alpha_1} - \alpha_3$, for all $\mathbf{Y}>0$, $\mathbf{W}$, $a>0$, and $\alpha_1,\alpha_2,\alpha_3 \in \mathbb{R}$.
\end{udefinition}

\noindent
The property of scale equivariance, if satisfied by a given estimator, gives a guarantee that a broad class of multiplicative, power, and additive transformations of the outcome data can only lead to specific, intuitively sensible changes in the final estimate. An important special case of scale equivariance is scale invariance with respect to the natural logarithm, which follows from setting $\alpha_1 \to 0$, $\alpha_2 = 1/\alpha_1$, and $\alpha_3 = \alpha_2$ in Definition \ref{def:scale}\@. To be clear, the idea here is as follows: the researcher transforms the outcome data prior to analysis, perhaps because they want to interpret the estimates as percentages, in which case they would use $g(Y) = \log(Y)$; however, if their estimator is not scale invariant with respect to the natural logarithm, the resulting estimates will depend on the units of $Y$, which directly contradicts the idea of interpreting them as percentages.

The following result demonstrates that the unnormalized weighting estimators discussed so far are not translation invariant and not scale equivariant. Thus, they are also not scale invariant with respect to the natural logarithm. On the other hand, the normalized estimators, $\hat{\tau}_{u}$ and $\hat{\tau}_{a,10}$, satisfy the properties of translation invariance and scale equivariance, which means that they are also scale invariant with respect to $g(Y) = \log(Y)$\@.

\begin{proposition}
\label{prop:invariance}
$\hat{\tau}_{u}$ and $\hat{\tau}_{a,10}$ are translation invariant and scale equivariant. $\hat{\tau}_{a}$, $\hat{\tau}_{t}$ ($= \hat{\tau}_{a,1}$), and $\hat{\tau}_{a,0}$ are not translation invariant and not scale equivariant.
\end{proposition}

\noindent
The properties of translation invariance and scale equivariance are very appealing, and it makes intuitive sense to only use estimators that satisfy them. To conclude this section, we make three final observations. First, the point of Proposition \ref{prop:invariance} is similar but distinct from that of \cite{CR2023}, who focus on the sensitivity to scaling of $\log(1+Y)$ and similar transformations, and do not restrict their attention to any specific estimators (including weighting). Unlike in \cite{CR2023}, the problem we describe disappears in large samples. On the other hand, the problem described by \cite{CR2023} disappears when the outcome only assumes strictly positive values, which is not the case in Proposition \ref{prop:invariance}. Second, it is useful to note that doubly robust estimators of $\late$, which we mentioned briefly in Section \ref{sec:recommended}, are generally translation invariant and scale equivariant, subject to mild conditions on the outcome model. Finally, several previous papers, including \cite{Tille1998} and \cite{AM2013}, note that the usual unnormalized weighting estimator is not translation invariant in settings with exogenous $D$\@. We extend this result to a class of weighting estimators of the LATE and additionally examine the more general property of scale equivariance.

\subsection{Near-Zero Denominators}
\label{sec:nearzero}

Weighting estimators of $\late$, like two-stage least squares and many other IV methods, are an example of ratio estimators. A common problem with such estimators is that they behave badly if their denominator is close to zero \citep[cf.][]{ASS2019}. In this section we document that in settings with one-sided noncompliance, i.e.~when units with $Z=1$ or units with $Z=0$ fully comply with their instrument assignment, there is a choice of weighting estimators that have an important advantage: they are based on a denominator that is strictly greater than zero by construction.

To see this, note that Table \ref{tab:kappa} provides simplified formulas for $\kappa$, $\kappa_{1}$, and $\kappa_{0}$ in each of the four subpopulations defined by their values of $Z$ and $D$. For example, $\kappa = 1$ if $Z=1$ and $D=1$ or $Z=0$ and $D=0$; moreover, $\kappa = - \frac{1 - p(X)}{p(X)}$ if $Z=1$ and $D=0$, and $\kappa = - \frac{p(X)}{1 - p(X)}$ if $Z=0$ and $D=1$. It follows that $N^{-1}\sum_{i=1}^{N} \kappa_{i}$ is the mean of a collection of positive and negative values, and hence it can be positive, negative, or zero. This is despite the fact that $N^{-1}\sum_{i=1}^{N} \kappa_{i}$ is also a consistent estimator of $\pr(D_{1} > D_{0})$, which is strictly positive under Assumption \ref{ass:abadie}\@. Similarly, $N^{-1}\sum_{i=1}^{N} \kappa_{i1}$ and $N^{-1}\sum_{i=1}^{N} \kappa_{i0}$ are also not guaranteed to be positive in general.

However, the situation is different in settings with one-sided noncompliance. If all individuals with $Z=1$ get treatment or, equivalently, there are no never-takers, the second row of Table \ref{tab:kappa} is empty and $\pr (\kappa_{0} \geq 0) = 1$. This is the case, for example, in studies that use twin births as an instrument for fertility \citep[e.g.,][]{AE1998}. Similarly, if there are no always-takers, then $\pr (\kappa_{1} \geq 0) = 1$. This is the case, for example, in randomized trials with noncompliance that make it impossible to access treatment if not offered. An implication of these observations is that in settings with one-sided noncompliance there exist estimators of $\pr(D_{1} > D_{0})$, and perhaps also the LATE, that have some desirable properties in finite samples.

\begin{proposition}
\label{prop:kappa1sided}
If there are no always-takers, $N^{-1}\sum_{i=1}^{N} \kappa_{i1} > 0$. If there are no never-takers, $N^{-1}\sum_{i=1}^{N} \kappa_{i0} > 0$.
\end{proposition}
\begin{proof}
To prove the first statement, note that $\frac{1}{p(X)}>1$ by Assumption \ref{ass:abadie}(iii). If there are no always-takers, then $\pr (Z=0, D=1) = 0$. Thus, $N^{-1}\sum_{i=1}^{N} \kappa_{i1} > N^{-1} \left( \underbrace{1+1+\cdots+1}_{N \cdot \hat{\pr} (D=1)} + \underbrace{0+0+\cdots+0}_{N \cdot \hat{\pr} (D=0)} \right) = \hat{\pr} (D=1) > 0$. The proof of the second statement is analogous.
\end{proof}

\noindent
Proposition \ref{prop:kappa1sided} demonstrates that settings with one-sided noncompliance offer a choice of estimators of $\pr(D_{1} > D_{0})$, based on $\kappa_{1}$ and $\kappa_{0}$, that are strictly greater than zero by construction. Interestingly, the denominator of $\hat{\tau}_{u}$ is also strictly greater than zero when noncompliance is one sided, and this is true regardless of whether there are no always-takers or no never-takers.

\begin{table}[!t]
\begin{adjustwidth}{-1in}{-1in}
\centering
\renewcommand*{\arraystretch}{1.75}
\begin{threeparttable}
\caption{Simplified Formulas for $\kappa$, $\kappa_{1}$, and $\kappa_{0}$ in Subpopulations Defined by $Z$ and $D$\label{tab:kappa}}
\begin{tabular}{>{\centering\arraybackslash}m{3cm} >{\centering\arraybackslash}m{2cm} >{\centering\arraybackslash}m{1cm} >{\centering\arraybackslash}m{2cm} >{\centering\arraybackslash}m{1cm} >{\centering\arraybackslash}m{2cm} >{\centering\arraybackslash}m{1cm}}
\hline\hline
 & $\kappa$ & $\sgn(\kappa)$ & $\kappa_{1}$ & $\sgn(\kappa_{1})$ & $\kappa_{0}$ & $\sgn(\kappa_{0})$ \\
\hline
$Z=1, D=1$ & $1$ & $+$ & $\frac{1}{p(X)}$ & $+$ & $0$ & $0$ \\
$Z=1, D=0$ & $- \frac{1 - p(X)}{p(X)}$ & $-$ & $0$ & $0$ & $- \frac{1}{p(X)}$ & $-$ \\
$Z=0, D=1$ & $- \frac{p(X)}{1 - p(X)}$ & $-$ & $- \frac{1}{1 - p(X)}$ & $-$ & $0$ & $0$ \\
$Z=0, D=0$ & $1$ & $+$ & $0$ & $0$ & $\frac{1}{1 - p(X)}$ & $+$ \\
\hline
\end{tabular}
\begin{footnotesize}
\begin{tablenotes}[flushleft]
\item
\end{tablenotes}
\end{footnotesize}
\end{threeparttable}
\end{adjustwidth}
\end{table}

\begin{proposition}
\label{prop:tauu1sided}
Suppose there are no always-takers or no never-takers. Then
\begin{equation*}
\left[ \sum_{i=1}^{N} \frac{Z_{i}}{p(X_{i})} \right]^{-1} \sum_{i=1}^{N} \frac{D_i Z_{i}}{p(X_{i})} \; - \; \left[ \sum_{i=1}^{N} \frac{1 - Z_{i}}{1 - p(X_{i})} \right]^{-1} \sum_{i=1}^{N} \frac{D_i \left( 1 - Z_{i} \right)}{1 - p(X_{i})} \; > \; 0.
\end{equation*}
\end{proposition}
\begin{proof}
Begin with the case of no always-takers. Then, $\pr [ D(1-Z)=1 ] = 0$, which implies that $\sum_{i=1}^{N} \frac{D_i \left( 1 - Z_{i} \right)}{1 - p(X_{i})} = 0$ and, as a result, $\left[ \sum_{i=1}^{N} \frac{Z_{i}}{p(X_{i})} \right]^{-1} \sum_{i=1}^{N} \frac{D_i Z_{i}}{p(X_{i})} - \left[ \sum_{i=1}^{N} \frac{1 - Z_{i}}{1 - p(X_{i})} \right]^{-1} \sum_{i=1}^{N} \frac{D_i \left( 1 - Z_{i} \right)}{1 - p(X_{i})} \; = \; \left[ \sum_{i=1}^{N} \frac{Z_{i}}{p(X_{i})} \right]^{-1} \sum_{i=1}^{N} \frac{D_i Z_{i}}{p(X_{i})} \; > \; 0$. Next, consider the case of no never-takers. Then, $Z=1$ implies $DZ=1$, which means that $\left[ \sum_{i=1}^{N} \frac{Z_{i}}{p(X_{i})} \right]^{-1} \sum_{i=1}^{N} \frac{D_i Z_{i}}{p(X_{i})} \; = \; 1$. At the same time, $\pr [ (1-D)(1-Z)=1 ] > 0$, which implies that $\sum_{i=1}^{N} \frac{1 - Z_{i}}{1 - p(X_{i})} \; > \; \sum_{i=1}^{N} \frac{D_i \left( 1 - Z_{i} \right)}{1 - p(X_{i})}$ and $0 \; < \; \left[ \sum_{i=1}^{N} \frac{1 - Z_{i}}{1 - p(X_{i})} \right]^{-1} \sum_{i=1}^{N} \frac{D_i \left( 1 - Z_{i} \right)}{1 - p(X_{i})} \; < \; 1$. Finally, $\left[ \sum_{i=1}^{N} \frac{Z_{i}}{p(X_{i})} \right]^{-1} \sum_{i=1}^{N} \frac{D_i Z_{i}}{p(X_{i})} \; - \; \left[ \sum_{i=1}^{N} \frac{1 - Z_{i}}{1 - p(X_{i})} \right]^{-1} \sum_{i=1}^{N} \frac{D_i \left( 1 - Z_{i} \right)}{1 - p(X_{i})} \; = \; 1 \; - \; \left[ \sum_{i=1}^{N} \frac{1 - Z_{i}}{1 - p(X_{i})} \right]^{-1} \sum_{i=1}^{N} \frac{D_i \left( 1 - Z_{i} \right)}{1 - p(X_{i})} \; > \; 0$.
\end{proof}

\noindent
An implication of Propositions \ref{prop:kappa1sided} and \ref{prop:tauu1sided} is that certain weighting estimators have the advantage of avoiding near-zero denominators when noncompliance is one sided. There are two unnormalized estimators that have this property, $\hat{\tau}_{a,1}$ when there are no always-takers and $\hat{\tau}_{a,0}$ when there are no never-takers, and one normalized estimator, $\hat{\tau}_{u}$, which retains this property in both cases. The other normalized estimator, $\hat{\tau}_{a,10}$, does not generally share this property with $\hat{\tau}_{u}$. Indeed, if $N^{-1}\sum_{i=1}^{N} \kappa_{i1}$ is away from zero but $N^{-1}\sum_{i=1}^{N} \kappa_{i0}$ is not, then this may affect the performance of not only $\hat{\tau}_{a,0}$ but also $\hat{\tau}_{a,10}$. Likewise, if $N^{-1}\sum_{i=1}^{N} \kappa_{i1}$ is close to zero, then both $\hat{\tau}_{a,1}$ and $\hat{\tau}_{a,10}$ are affected.

\subsection{Estimation When the Instrument Propensity Score Is Unknown}
\label{sec:unknown}

Our discussion in Sections \ref{sec:known}, \ref{sec:normalized}, and \ref{sec:nearzero} assumed that $p(X)$ is known, which is often unrealistic. In practice, researchers typically adopt a parametric model for $p(X)$, say the logit, $F(X,\alpha) = \exp(X\alpha)/[1+\exp(X\alpha)]$, and estimate $\alpha$ by maximum likelihood (cf.~Section \ref{sec:recommended}). Our observations above apply equally in this case. Indeed, the normalized estimators are translation invariant and scale equivariant while the unnormalized estimators are not. At the same time, two specific unnormalized estimators and one normalized estimator avoid near-zero denominators in settings with one-sided noncompliance. From now on, if we wish to specify that $\alpha$ is estimated using maximum likelihood, we use an ``ml'' subscript or superscript. Thus, $\hat{\alpha}_{ml}$ is the maximum likelihood estimator of $\alpha$, $\hat{p}_{ml}(X) = F(X,\hat{\alpha}_{ml})$ are the estimated propensity scores, and $\hat{\tau}_{u}^{ml}$, $\hat{\tau}_{a,10}^{ml}$, $\hat{\tau}_{a}^{ml}$, $\hat{\tau}_{t}^{ml}$ ($= \hat{\tau}_{a,1}^{ml}$), and $\hat{\tau}_{a,0}^{ml}$ are the analogues of the previously introduced estimators, with $\hat{p}_{ml}(X)$ replacing $p(X)$.

Alternatively, we can estimate $\alpha$ using covariate balancing methods, such as those studied by \cite{GPE2012,GPE2016}, \cite{IR2014}, \cite{Heiler2022}, and \cite{SASX2022}. Following \cite{Heiler2022}, we focus on the approach of \cite{IR2014}, which amounts to estimating $\alpha$ using a different set of moment conditions than maximum likelihood. Indeed, the population moment conditions in \cite{IR2014} are
\begin{equation*}
\label{eq:popmoments}
\e \left[ \frac{Z}{F(X,\alpha)} X \right] \; = \; \e \left[ \frac{1-Z}{1-F(X,\alpha)} X \right],
\end{equation*}
and the corresponding sample moment conditions can be written as
\begin{equation}
\label{eq:samplemoments}
N^{-1} \sum_{i=1}^{N} \frac{Z_{i}}{F(X_{i},\hat{\alpha}_{cb})} X_{i} \; = \; N^{-1} \sum_{i=1}^{N} \frac{1 - Z_{i}}{1 - F(X_{i},\hat{\alpha}_{cb})} X_{i},
\end{equation}
where 
$\hat{\alpha}_{cb}$ is the method of moments estimator of $\alpha$. We also use $\hat{p}_{cb}(X) = F(X,\hat{\alpha}_{cb})$ to denote the covariate balancing propensity scores, and $\hat{\tau}_{u}^{cb}$, $\hat{\tau}_{a,10}^{cb}$, $\hat{\tau}_{a}^{cb}$, $\hat{\tau}_{t}^{cb}$ ($= \hat{\tau}_{a,1}^{cb}$), and $\hat{\tau}_{a,0}^{cb}$ to denote the analogues of the previously introduced estimators, with $\hat{p}_{cb}(X)$ replacing $p(X)$.

In a recent paper, $\hat{\tau}_{u}^{cb}$ is also recommended by \cite{Heiler2022}, who shows that it is numerically identical to $\hat{\tau}_{t}^{cb}$, as long as $X$ includes a constant. We add to \citeauthor{Heiler2022}'s \citeyearpar{Heiler2022} observation and determine that, when $X$ includes a constant, $\hat{\tau}_{u}^{cb}$ is also identical to $\hat{\tau}_{a,10}^{cb}$ and $\hat{\tau}_{a,0}^{cb}$.

\begin{proposition}
\label{prop:cb}
If $X$ includes a constant, $\hat{\tau}_{u}^{cb} = \hat{\tau}_{t}^{cb} = \hat{\tau}_{a,1}^{cb} = \hat{\tau}_{a,0}^{cb} = \hat{\tau}_{a,10}^{cb}$.
\end{proposition}

\noindent
Proposition \ref{prop:cb} demonstrates that using covariate balancing propensity scores solves the problem of choosing an appropriate weighting estimator of $\late$, because all the estimators we previously determined to have some desirable finite sample properties are identical when $\hat{p}_{cb}(X)$ replaces $p(X)$.

\subsection{Inference}
\label{sec:asymptotic}

So far, we have focused on the finite sample properties of several weighting estimators of $\late$. To determine the asymptotic distribution of each estimator, we apply general results on M-estimation \citep{Wooldridge2010,BoosStefanski2013}, as all the weighting estimators considered in this paper can be represented as an M-estimator.

Weighting estimators are all functions of the instrument propensity score, $p(X)$\@.  As in Section \ref{sec:unknown}, we assume a parametric model, $F(X,\alpha)$, for $p(X)$\@. Thus, the LATE can be estimated by a two-step procedure where $\alpha$ is estimated in the first step and the unknown $F(X,\alpha)$ is replaced with its estimate in the second step. Alternatively, one could jointly estimate $\alpha$ and $\late$ within an M-estimation framework using moment functions related to both $\alpha$ and $\late$. The moment function related to the estimation of $\alpha$ is either the score from the maximum likelihood estimation or the covariate balancing condition from \cite{IR2014}. The moment functions related to $\late$ are derived from the identification results in Section \ref{sec:framework}. All moment functions are summarized in Table \ref{tab:moments}. For different weighting estimators, different combinations of moment functions will be necessary. Provided that the standard regularity conditions \citep{NeweyMcFadden1994} are satisfied and the relevant moments exist, all the estimators considered here are asymptotically normal. The derivation of the asymptotic variance for each of the estimators is presented in the appendix. These variances are also estimated in our companion Stata package \texttt{kappalate}.

Although it would be interesting to compare the asymptotic variances of the different weighting estimators considered in this paper, we leave this task to future research. At this time, we instead make three additional points. First, we conjecture, as in \cite{KM2016} and \cite{KU2023}, that normalization may help reduce the asymptotic variance of an estimator, in which case $\hat{\tau}_{u}^{ml}$ would be more efficient than $\hat{\tau}_{t}^{ml}$ ($= \hat{\tau}_{a,1}^{ml}$). Second, we note that $\hat{\tau}_{u}^{cb}$ attains the semiparametric efficiency bound in \cite{Frolich2007} and \cite{HN2010} as long as the number of balancing constraints grows appropriately with the sample size \citep[see][]{Heiler2022}. Third, we recognize that our asymptotic analysis implicitly requires a restriction stronger than Assumption \ref{ass:abadie}(iii), namely the ``strong overlap'' assumption of \cite{KT2010}.

\section{Empirical Applications}
\label{sec:applications}

In this section we use three empirical applications to illustrate our findings from Section \ref{sec:estandinf}. The bottom line is that the proportion of compliers is sufficiently large in every application (i.e.~the instruments are sufficiently strong) so that the phenomenon of dividing by ``near zero'' never occurs. Ultimately, the three normalized estimators that we consider, $\hat{\tau}_{u}^{cb}$, $\hat{\tau}_{u}^{ml}$, and $\hat{\tau}_{a,10}^{ml}$, are practically indistinguishable from one another in all applications. At the same time, we document the lack of translation invariance and scale equivariance of the unnormalized estimators. We also report the corresponding 2SLS estimates, which are obtained with the covariates appearing additively in the linear equation. Both in this context and in the case of parametric estimation of the instrument propensity score, the relevant model may be misspecified in the absence of sufficiently flexible covariate specifications.

\subsection{Causal Effects of Military Service \citep{Angrist1990}}
\label{sec:vietnam}

In our first application, we revisit \citeauthor{Angrist1990}'s \citeyearpar{Angrist1990} study of causal effects of military service using the draft eligibility instrument. In the early 1970s, priority for induction in the U.S. was determined in a sequence of lotteries. The instrument in \cite{Angrist1990} takes the value 1 for individuals with dates of birth that were randomly determined as draft eligible and 0 otherwise. Because the fraction of eligible dates of birth was cohort specific, it is essential to control for age in this application.

\begin{table}[!t]
	\begin{adjustwidth}{-1in}{-1in}
		\centering
		\begin{threeparttable}
			\caption{Causal Effects of Military Service on Log Wages\label{tab:military}}
			\begin{normalsize}
				\begin{tabular}{l >{\centering\arraybackslash}m{1.5cm} >{\centering\arraybackslash}m{1.5cm} >{\centering\arraybackslash}m{1.5cm} >{\centering\arraybackslash}m{1.5cm} >{\centering\arraybackslash}m{1.5cm} >{\centering\arraybackslash}m{1.5cm}}
					\hline\hline
					\multicolumn{1}{c}{} & (1) & (2) & (3) & (4) & (5) & (6) \\
					\hline
					\multicolumn{1}{l}{\underline{A. 2SLS}} & 0.233 & 0.233 & 0.227 & 0.227 & 0.254 & 0.254 \\
					\multicolumn{1}{c}{} & (0.212) & (0.212) & (0.229) & (0.229) & (0.227) & (0.227) \\
					\multicolumn{1}{l}{} & & & & & & \\
					\multicolumn{1}{l}{\underline{B. Normalized estimates:}} & & & & & & \\
					\multicolumn{1}{l}{$\hat{\tau}_{u}^{cb}$} & 0.229 & 0.229 & 0.208 & 0.208 & 0.241 & 0.241 \\
					\multicolumn{1}{c}{} & (0.213) & (0.213) & (0.232) & (0.232) & (0.229) & (0.229) \\
					\multicolumn{1}{l}{$\hat{\tau}_{u}^{ml}$} & 0.234 & 0.234 & 0.202 & 0.202 & 0.241 & 0.241 \\
					\multicolumn{1}{c}{} & (0.211) & (0.211) & (0.235) & (0.235) & (0.229) & (0.229) \\
					\multicolumn{1}{l}{$\hat{\tau}_{a,10}^{ml}$} & 0.227 & 0.227 & 0.204 & 0.204 & 0.241 & 0.241 \\
					\multicolumn{1}{c}{} & (0.204) & (0.204) & (0.239) & (0.239) & (0.229) & (0.229) \\
					\multicolumn{1}{l}{} & & & & & \\
					\multicolumn{1}{l}{\underline{C. Unnormalized estimates:}} & & & & & \\
					\multicolumn{1}{l}{$\hat{\tau}_{a}^{ml}$} & --0.429* & 0.015 & 0.537* & 0.314 & 0.241 & 0.241 \\
					\multicolumn{1}{c}{} & (0.258) & (0.207) & (0.322) & (0.252) & (0.229) & (0.229) \\
					\multicolumn{1}{l}{$\hat{\tau}_{t}^{ml} = \hat{\tau}_{a,1}^{ml}$} & --0.455 & 0.016 & 0.515* & 0.302 & 0.241 & 0.241 \\
					\multicolumn{1}{c}{} & (0.279) & (0.219) & (0.301) & (0.240) & (0.229) & (0.229) \\
					\multicolumn{1}{l}{$\hat{\tau}_{a,0}^{ml}$} & --0.413* & 0.014 & 0.540* & 0.317 & 0.241 & 0.241 \\
					\multicolumn{1}{c}{} & (0.246) & (0.199) & (0.326) & (0.255) & (0.229) & (0.229) \\
					\multicolumn{1}{l}{} & & & & & \\
					\multicolumn{1}{l}{\underline{Outcome measurement:}} & & & & & \\
					\multicolumn{1}{l}{Cents} & \checkmark & & \checkmark & & \checkmark & \\
					\multicolumn{1}{l}{Dollars} & & \checkmark & & \checkmark & & \checkmark \\
					\multicolumn{1}{l}{} & & & & & \\
					\multicolumn{1}{l}{\underline{Covariates:}} & & & & & \\
					\multicolumn{1}{l}{Age} & \checkmark & \checkmark & & & & \\
					\multicolumn{1}{l}{Cubic in age} & & & \checkmark & \checkmark & & \\
					\multicolumn{1}{l}{Saturated in age} & & & & & \checkmark & \checkmark \\
					\multicolumn{1}{l}{} & & & & & \\
					\multicolumn{1}{l}{Observations} & 3,027 & 3,027 & 3,027 & 3,027 & 3,027 & 3,027 \\
					\hline
				\end{tabular}
			\end{normalsize}
			\begin{footnotesize}
				\begin{tablenotes}[flushleft]
					\item \textit{Notes:} The data are \citeauthor{Angrist1990}'s \citeyearpar{Angrist1990} subsample of the 1984 Survey of Income and Program Participation (SIPP)\@. The outcome is log hourly wages, with wages measured either in cents or in dollars prior to the log transformation. The treatment is an indicator for whether an individual is a veteran. The instrument is an indicator for whether an individual had a lottery number below the draft eligibility ceiling. The logit model is used for the instrument propensity score, with the unknown parameters estimated using maximum likelihood or the moment conditions in equation (\ref{eq:samplemoments}). Standard errors are in parentheses. For 2SLS, we use robust standard errors. For the remaining estimators, we calculate the standard errors using the asymptotic variance formulas in the appendix.
					\item *Statistically significant at the 10\% level; **at the 5\% level; ***at the 1\% level.
				\end{tablenotes}
			\end{footnotesize}
		\end{threeparttable}
	\end{adjustwidth}
\end{table}

In what follows, we use a sample of 3,027 individuals from the 1984 Survey of Income and Program Participation (SIPP), which is also considered by \cite{MW2017}. Our outcome of interest is log wage. To illustrate the invariance properties in Proposition \ref{prop:invariance}, we consider the natural logarithm of hourly wages as measured in cents or dollars. We also consider three sets of covariates: age, a cubic in age, and a set of indicator variables for each value of age. Summary statistics for these data are reported in Table 6 of \cite{MW2017}.

Table \ref{tab:military} reports our estimates of causal effects of military service. Panels A and B, which report 2SLS and normalized weighting estimates, suggest that these effects were positive and economically meaningful in the period under study, with a narrow range of estimates from 20--25 log points. The differences between the 2SLS and weighting estimates (as well as their standard errors) are always very minor. Although the estimated effects are all positive, they are not statistically significant. The estimates do not depend on whether we measure wages in cents or dollars.

Panel C of Table \ref{tab:military} reports unnormalized weighting estimates. Unlike in panels A and B, these estimates are heavily dependent on the exact specification and, except in the case of the saturated specification, on whether we measure wages in cents or dollars prior to the log transformation. For example, in columns 1 and 2, we only control for age, and yet the estimates are negative and marginally significant when wages are measured in cents prior to the log transformation, while becoming marginally positive when wages are measured in dollars. When the covariate specification is saturated, as in columns 5 and 6, the unnormalized estimates do not depend on the units of measurement of the original outcome variable; they also become identical to each other and to the normalized estimates. This demonstrates the virtue of flexible covariate specifications.

\subsection{Causal Effects of College Education \citep{Card1995}}
\label{sec:college}

In our second application, we revisit \citeauthor{Card1995}'s \citeyearpar{Card1995} study of causal effects of education using the college proximity instrument. \cite{Card1995} uses data from the National Longitudinal Survey of Young Men (NLSYM) and restricts his attention to a subsample of 3,010 individuals who were interviewed in 1976 and reported valid information on wage and education. His endogenous variable of interest is years of schooling, which is instrumented by an indicator for the presence of a four-year college in the respondent's local labor market in 1966.

\begin{table}[!t]
	\begin{adjustwidth}{-1in}{-1in}
		\centering
		\begin{threeparttable}
			\caption{Causal Effects of College Education on Log Wages\label{tab:college}}
			\begin{normalsize}
				\begin{tabular}{l >{\centering\arraybackslash}m{1.5cm} >{\centering\arraybackslash}m{1.5cm} >{\centering\arraybackslash}m{1.5cm} >{\centering\arraybackslash}m{1.5cm} >{\centering\arraybackslash}m{0.05cm} >{\centering\arraybackslash}m{1.5cm} >{\centering\arraybackslash}m{1.5cm}
						>{\centering\arraybackslash}m{1.5cm}
						>{\centering\arraybackslash}m{1.5cm}}
					\hline\hline
					\multicolumn{1}{c}{} & \multicolumn{4}{c}{Some college attendance} &  & \multicolumn{4}{c}{College completion} \\
					\cline{2-5}
					\cline{7-10}
					\multicolumn{1}{c}{} & (1) & (2) & (3) & (4) &  & (5) & (6) & (7) & (8) \\
					\hline
					\multicolumn{1}{l}{\underline{A. 2SLS}} & 0.661** & 0.661** & 0.575* & 0.575* & & 1.392* & 1.392* & 0.991 & 0.991 \\
					\multicolumn{1}{c}{} & (0.294) & (0.294) & (0.308) & (0.308) & & (0.798) & (0.798) & (0.610) & (0.610) \\
					\multicolumn{1}{l}{} & \\
					\multicolumn{1}{l}{\underline{B. Normalized estimates:}} & & & & & & & & \\
					\multicolumn{1}{l}{$\hat{\tau}_{u}^{cb}$} & 0.376* & 0.376* & 0.331 & 0.331 &       & 0.853 & 0.853 & 0.588 & 0.588 \\
					\multicolumn{1}{c}{} & (0.223) & (0.223) & (0.236) & (0.236) &       & (0.549) & (0.549) & (0.433) & (0.433) \\
					\multicolumn{1}{l}{$\hat{\tau}_{u}^{ml}$} & 0.331 & 0.331 & 0.356 & 0.356 &       & 0.619 & 0.619 & 0.628 & 0.628 \\
					\multicolumn{1}{c}{} & (0.202) & (0.202) & (0.244) & (0.244) &       & (0.387) & (0.387) & (0.448) & (0.448) \\
					\multicolumn{1}{l}{$\hat{\tau}_{a,10}^{ml}$} & 0.346* & 0.346* & 0.293 & 0.293 &       & 0.586* & 0.586* & 0.836 & 0.836 \\
					\multicolumn{1}{c}{} & (0.200) & (0.200) & (0.252) & (0.252) &       & (0.356) & (0.356) & (0.821) & (0.821) \\
					\multicolumn{1}{l}{} & & & & & & & & \\
					\multicolumn{1}{l}{\underline{C. Unnormalized estimates:}} & & & & & & & & \\
					\multicolumn{1}{l}{$\hat{\tau}_{a}^{ml}$} & --0.319 & 0.170 & 2.248** & 0.842** &       & --0.594 & 0.315 & 4.317* & 1.617* \\
					\multicolumn{1}{c}{} & (1.182) & (0.370) & (0.971) & (0.362) &       & (2.184) & (0.696) & (2.485) & (0.891) \\
					\multicolumn{1}{l}{$\hat{\tau}_{t}^{ml} = \hat{\tau}_{a,1}^{ml}$} & --0.321 & 0.171 & 2.053** & 0.769** &       & --0.601 & 0.319 & 3.651** & 1.367** \\
					\multicolumn{1}{c}{} & (1.201) & (0.367) & (0.813) & (0.308) &       & (2.251) & (0.687) & (1.780) & (0.648) \\
					\multicolumn{1}{l}{$\hat{\tau}_{a,0}^{ml}$} & --0.290 & 0.154 & 2.846* & 1.066* &       & --0.501 & 0.266 & 7.241 & 2.712 \\
					\multicolumn{1}{c}{} & (1.036) & (0.354) & (1.592) & (0.574) &       & (1.728) & (0.639) & (7.246) & (2.577) \\
					\multicolumn{1}{l}{} & & & & & \\
					\multicolumn{1}{l}{\underline{Outcome measurement:}} & & & & & \\
					\multicolumn{1}{l}{Cents} & \checkmark & & \checkmark & & & \checkmark & & \checkmark & \\
					\multicolumn{1}{l}{Dollars} & & \checkmark & & \checkmark & & & \checkmark & & \checkmark \\
					\multicolumn{1}{l}{} & & & & & \\
					\multicolumn{1}{l}{\underline{Specification:}} & \citeauthor{Card1995} & \citeauthor{Card1995} & \citeauthor{Kitagawa2015} & \citeauthor{Kitagawa2015} & & \citeauthor{Card1995} & \citeauthor{Card1995} & \citeauthor{Kitagawa2015} & \citeauthor{Kitagawa2015} \\
					\multicolumn{1}{l}{} & & & & & & & & \\
					\multicolumn{1}{l}{Observations} & 3,010 & 3,010 & 3,010 & 3,010 & & 3,010 & 3,010 & 3,010 & 3,010 \\
					\hline
				\end{tabular}
			\end{normalsize}
			\begin{footnotesize}
				\begin{tablenotes}[flushleft]
					\item \textit{Notes:} The data are \citeauthor{Card1995}'s \citeyearpar{Card1995} subsample of the National Longitudinal Survey of Young Men (NLSYM)\@. The outcome is log hourly wages, with wages measured either in cents or in dollars prior to the log transformation. The treatment is an indicator for whether an individual has at least thirteen (``some college attendance'') or sixteen years of schooling (``college completion''). The instrument is an indicator for whether an individual grew up in the vicinity of a four-year college. The logit model is used for the instrument propensity score, with the unknown parameters estimated using maximum likelihood or the moment conditions in equation (\ref{eq:samplemoments}). The first specification (``Card'') follows \cite{Card1995} and includes experience, experience squared, nine regional indicators, and indicators for whether Black, whether lived in an SMSA in 1966 and 1976, and whether lived in the South in 1976. The second specification (``Kitagawa'') follows \cite{Kitagawa2015} and includes indicators for whether Black, whether lived in an SMSA in 1966 and 1976, and whether lived in the South in 1966 and 1976. Standard errors are in parentheses. For 2SLS, we use robust standard errors. For the remaining estimators, we calculate the standard errors using the asymptotic variance formulas in the appendix.
					\item *Statistically significant at the 10\% level; **at the 5\% level; ***at the 1\% level.
				\end{tablenotes}
			\end{footnotesize}
		\end{threeparttable}
	\end{adjustwidth}
\end{table}

This study has been revisited by numerous papers, many of which focus on binarized versions of \citeauthor{Card1995}'s \citeyearpar{Card1995} education variable. For example, \cite{Tan2006} and \cite{Sloczynski2021} study the effects of having at least thirteen years of schooling (``some college attendance'') while \cite{HM2015}, \cite{Kitagawa2015}, \cite{MW2017}, and \cite{AH2021} focus on having at least sixteen years of schooling (``college completion''). In what follows, we consider both binarizations. Our outcome of interest is log hourly wage, with wages measured either in cents or in dollars. We also consider two sets of covariates: a quadratic in experience, nine regional indicators, and indicators for whether Black, whether lived in an SMSA in 1966 and 1976, and whether lived in the South in 1976, as in \cite{Card1995}; and indicators for whether Black, whether lived in an SMSA in 1966 and 1976, and whether lived in the South in 1966 and 1976, as in \cite{Kitagawa2015}. Summary statistics for these data are reported in Table 1 of \cite{Card1995}.

Table \ref{tab:college} reports our estimates of causal effects of college education on log wages. Many of these estimates seem implausible, often because they are ``too large.'' This is unsurprising given the possible failures of the exclusion restriction and monotonicity in this application \citep[cf.][]{AH2021, Sloczynski2021}. From our perspective, these concerns are less relevant, however, because we use Table \ref{tab:college} as another illustration of Proposition \ref{prop:invariance}. The normalized estimates (as well as 2SLS) clearly do not depend on the units of measurement of the outcome variable prior to the log transformation. This is no longer the case for the unnormalized estimates, as reported in Panel C of Table \ref{tab:college}. For example, when focusing on the ``some college attendance'' treatment and using \citeauthor{Card1995}'s \citeyearpar{Card1995} specification, we obtain negative estimates when wages are measured in cents but positive when they are measured in dollars. Both sets of estimates are economically meaningful even if insignificant; regardless, the lack of invariance is disconcerting. When we use \citeauthor{Kitagawa2015}'s \citeyearpar{Kitagawa2015} specification instead, all estimates are positive and statistically different from zero, but more than twice as large when wages are originally measured in cents rather than dollars.

\subsection{Causal Effects of Childbearing \citep{AE1998}}
\label{sec:children}

In our third empirical application, we revisit \citeauthor{AE1998}'s \citeyearpar{AE1998} study of causal effects of childbearing using the sibling sex composition instrument. \cite{AE1998} use the incidence of a twin birth and the sex of the first two children as two alternative instruments for having at least three children in a sample of women with two or more children. In what follows, we restrict our attention to the sex composition instrument.

\begin{table}[!p]
	\begin{adjustwidth}{-1in}{-1in}
		\centering
		\begin{threeparttable}
			\caption{Causal Effects of Childbearing on Labor Force Participation and Log Income\label{tab:fertility}}
			\begin{normalsize}
				\begin{tabular}{l >{\centering\arraybackslash}m{1.825cm} >{\centering\arraybackslash}m{1.825cm} >{\centering\arraybackslash}m{1.825cm} >{\centering\arraybackslash}m{0.05cm} >{\centering\arraybackslash}m{1.6cm} >{\centering\arraybackslash}m{1.6cm} >{\centering\arraybackslash}m{1.6cm}
						>{\centering\arraybackslash}m{1.6cm}}
					\hline\hline
					\multicolumn{1}{c}{} & \multicolumn{3}{c}{Labor force participation} &  & \multicolumn{4}{c}{Log income} \\
					\cline{2-4}
					\cline{6-9}
					\multicolumn{1}{c}{} & (1) & (2) & (3) &  & (4) & (5) & (6) & (7) \\
					\hline
					\multicolumn{1}{l}{\underline{A. 2SLS}} & --0.117*** & --0.117*** & --0.117*** &       & --0.135 & --0.135 & --0.135 & --0.135 \\
					\multicolumn{1}{c}{} & (0.025) & (0.025) & (0.025) & & (0.092) & (0.092) & (0.092) & (0.092) \\
					\multicolumn{1}{l}{} & \\
					\multicolumn{1}{l}{\underline{B. Normalized estimates:}} & & & & & & & \\
					\multicolumn{1}{l}{$\hat{\tau}_{u}^{cb}$} & --0.117*** & --0.117*** & --0.117*** &       & --0.135 & --0.135 & --0.135 & --0.135 \\
					\multicolumn{1}{c}{} & (0.025) & (0.025) & (0.025) &       & (0.092) & (0.092) & (0.092) & (0.092) \\
					\multicolumn{1}{l}{$\hat{\tau}_{u}^{ml}$} & --0.117*** & --0.117*** & --0.117*** &       & --0.135 & --0.135 & --0.135 & --0.135 \\
					\multicolumn{1}{c}{} & (0.025) & (0.025) & (0.025) &       & (0.092) & (0.092) & (0.092) & (0.092) \\
					\multicolumn{1}{l}{$\hat{\tau}_{a,10}^{ml}$} & --0.117*** & --0.117*** & --0.117*** &       & --0.132 & --0.132 & --0.132 & --0.132 \\
					\multicolumn{1}{c}{} & (0.025) & (0.025) & (0.025) &       & (0.093) & (0.093) & (0.093) & (0.093) \\
					\multicolumn{1}{l}{} & & & & & & & \\
					\multicolumn{1}{l}{\underline{C. Unnormalized estimates:}} & & & & & & & \\
					\multicolumn{1}{l}{$\hat{\tau}_{a}^{ml}$} & --0.100*** & --0.070*** & --0.131*** &       & 0.286** & 0.143 & --0.073 & --0.216** \\
					\multicolumn{1}{c}{} & (0.025) & (0.026) & (0.025) &       & (0.113) & (0.102) & (0.093) & (0.093) \\
					\multicolumn{1}{l}{$\hat{\tau}_{t}^{ml} = \hat{\tau}_{a,1}^{ml}$} & --0.099*** & --0.069*** & --0.129*** &       & 0.282** & 0.140 & --0.072 & --0.213** \\
					\multicolumn{1}{c}{} & (0.025) & (0.025) & (0.025) &       & (0.111) & (0.100) & (0.092) & (0.091) \\
					\multicolumn{1}{l}{$\hat{\tau}_{a,0}^{ml}$} & --0.102*** & --0.071*** & --0.133*** &       & 0.291** & 0.145 & --0.074 & --0.220** \\
					\multicolumn{1}{c}{} & (0.026) & (0.026) & (0.026) &       & (0.115) & (0.104) & (0.094) & (0.094) \\
					\multicolumn{1}{l}{} & & & & & & & \\
					\multicolumn{1}{l}{\underline{Outcome measurement:}} & & & & & & & \\
					\multicolumn{1}{l}{Cents} & & & & & \checkmark \\
					\multicolumn{1}{l}{Dollars} & & & & & & \checkmark \\
					\multicolumn{1}{l}{\$1,000s} & & & & & & & \checkmark \\
					\multicolumn{1}{l}{\$100,000s} & & & & & & & & \checkmark \\
					\multicolumn{1}{l}{1 if worked, 0 otherwise} & \checkmark \\
					\multicolumn{1}{l}{2 if worked, 1 otherwise} & & \checkmark \\
					\multicolumn{1}{l}{1 if did not work, 0 otherwise} & & & \checkmark \\
					\multicolumn{1}{l}{} & & & & & & & \\
					\multicolumn{1}{l}{Observations} & 394,840 & 394,840 & 394,840 &       & 220,502 & 220,502 & 220,502 & 220,502 \\
					\hline
				\end{tabular}
			\end{normalsize}
			\begin{footnotesize}
				\begin{tablenotes}[flushleft]
					\item \textit{Notes:} The data are \citeauthor{FGV2018}'s \citeyearpar{FGV2018} subsample of the 1980 US Census, which is based on \cite{AE1998}. The outcome is an indicator for whether a woman worked for pay in the preceding year (``labor force participation'') or log annual income, with income measured in cents, dollars, \$1,000s, or \$100,000s prior to the log transformation. In the case of labor force participation, we also recode the outcome as 2 if worked for pay and 1 otherwise; and as 0 if worked for pay and 1 otherwise. In the latter case, we report the additive inverse of each estimate. The treatment is an indicator for whether a woman has at least three children. The instrument is an indicator for whether a woman's first two children are either two boys or two girls. The logit model is used for the instrument propensity score, with the unknown parameters estimated using maximum likelihood or the moment conditions in equation (\ref{eq:samplemoments}).  The set of covariates consists of age, age at first birth, sex of the first and second children, and indicators for whether Black, whether Hispanic, and whether another race. Standard errors are in parentheses. For 2SLS, we use robust standard errors. For the remaining estimators, we calculate the standard errors using the asymptotic variance formulas in the appendix.
					\item *Statistically significant at the 10\% level; **at the 5\% level; ***at the 1\% level.
				\end{tablenotes}
			\end{footnotesize}
		\end{threeparttable}
	\end{adjustwidth}
\end{table}

This study has been revisited in many papers, including \cite{FGV2018}. In what follows, we use \citeauthor{FGV2018}'s \citeyearpar{FGV2018} subsample of the 1980 US Census that consists of all women aged 21--35 with at least two children. The number of observations is 394,840, which is nearly identical to the sample size in \cite{AE1998}. Summary statistics for these data are reported in Table 2 of \cite{AE1998}. Our outcomes of interest are log annual income and an indicator for labor force participation. In the case of log income, we implicitly condition on reported income being greater than zero (as in Sections \ref{sec:vietnam} and \ref{sec:college}). The treatment is having more than two children. The set of covariates consists of age, age at first birth, sex of the first and second children, and indicators for whether Black, whether Hispanic, and whether another race. The instrument is an indicator for whether the first two children are of the same sex.

We consider a broader set of transformations of the outcome variables relative to the previous applications. In the case of labor force participation, we originally code working for pay as 1 and not working for pay as 0. Subsequently, however, we also recode working for pay as 2 and not working for pay as 1, as well as not working for pay as 1 and working for pay as 0. In the case of income, we consider four different units of measurement: cents, dollars, thousands of dollars, and hundreds of thousands of dollars. While the first and the last unit of measurement may appear impractical for annual income, our goal is to demonstrate the fragility of the unnormalized estimates with respect to such transformations.

Table \ref{tab:fertility} reports our estimates of causal effects of childbearing on labor market outcomes. Panels A and B, which report 2SLS and normalized weighting estimates, respectively, suggest that these effects are negative and economically meaningful, although the effects on log income are not statistically different from zero. As in our replication of \cite{Angrist1990}, the differences between the 2SLS and weighting estimates (as well as their standard errors) are always very minor. Transformations of the outcome variables do not influence any of the estimates.

Panel C of Table \ref{tab:fertility} reports the unnormalized estimates. The fragility of these estimates is immediately evident. In the case of income, the estimated effects of childbearing are positive and highly significant when income is measured in cents, positive and insignificant when in dollars, negative and insignificant when in thousands of dollars, and negative and highly significant when in hundreds of thousands of dollars. This is obviously very disconcerting. Likewise, in the case of labor force participation, the estimates are quite fragile, although less so than in the case of income, perhaps because of the binary nature of the outcome. Still, the estimates in column 3 are nearly twice larger than those in column 2, even though the only difference between these two columns is in a particular recoding of the binary outcome.

\section{Simulation Study}
\label{sec:simulation}

In this section we use a simulation study to illustrate our findings on the properties of weighting estimators of the LATE\@. To reduce the number of researcher degrees of freedom, we focus on data-generating processes from \cite{Heiler2022}, which leads to the following system of equations:
\begingroup
\allowdisplaybreaks
\begin{eqnarray*}
Z &=& 1 [ u < \pi(X) ], \\
\pi(X) &=& 1/\left( 1 + \exp \left( - \mu_z(X) \cdot \theta_0 \right) \right), \\
D_z &=& 1 [ \mu_d(X,z) > v ], \\
Y_1 &=& \mu_{y_1}(X) + \varepsilon_1, \\
Y_0 &=& \varepsilon_0,
\end{eqnarray*}
\endgroup
where $u$ and $X$ are i.i.d.~standard uniform, $\left( \begin{array}{c}
\varepsilon_1 \\
\varepsilon_0 \\
v
\end{array} \right) \sim \mathcal{N} \left( \left[ \begin{array}{c}
0 \\
0 \\
0
\end{array} \right], \left[ \begin{array}{ccc}
1 & 0 & 0.5 \\
0 & 1 & 0 \\
0.5 & 0 & 1
\end{array} \right] \right)$, $\theta_0 = \ln((1-\delta)/\delta)$, and $\delta \in \left\lbrace 0.01,0.02,0.05 \right\rbrace$. What remains to be specified is three functions, namely $\mu_d(x,z)$, $\mu_{y_1}(x)$, and $\mu_z(x)$. Our choices for these functions are listed in Table \ref{tab:DGPs}. It is useful to note that, given these choices and the fact that $X$ has a standard uniform distribution, $\delta$ is equal to the lowest possible value of the instrument propensity score and (symmetrically) one minus the instrument propensity score, that is, $\delta \leq \pr (Z=1 \mid X) \leq 1 - \delta$. Thus, $\delta$ controls the degree of overlap in the data.

Note that Designs A.1, B, C, and D in Table \ref{tab:DGPs} are identical to Designs A, B, C, and D, respectively, in \cite{Heiler2022}. It is easy to see that Design A.1 corresponds to a setting with (near) one-sided noncompliance, as $\pr(D=1 \mid Z=1) = \Phi(4) = 0.99997$, where $\Phi(\cdot)$ is the standard normal cdf. It follows that there are essentially no never-takers in Design A.1. To illustrate our findings from Section \ref{sec:nearzero} on near-zero denominators, we are also interested in a design with (nearly) no always-takers. This is accomplished by Design A.2, which is identical to Design A.1 except for a small change to $\mu_d(x,z)$ that reverses the direction of noncompliance. Indeed, in Design A.2, $\pr(D=1 \mid Z=0) = \Phi(-4) = 0.00003$, which means that there are essentially no always-takers.

It is also useful to note that Designs A.1 and A.2 correspond to the case of a fully independent instrument while in the remaining designs the instrument is conditionally independent. Additionally, in Designs A.1, A.2, and B, treatment effect heterogeneity is only due to the correlation between $\varepsilon_1$ and $v$; in Designs C and D, on the other hand, the dependence of $\mu_{y_1}(X)$ on $X$ constitutes another source of heterogeneity. In the end, the 2SLS estimator that controls for $X$ is expected to perform very well in Designs A.1, A.2, and B but not necessarily elsewhere \citep[cf.][]{Heiler2022}.

\begin{table}[!t]
\begin{adjustwidth}{-1in}{-1in}
\centering
\renewcommand*{\arraystretch}{1.75}
\begin{threeparttable}
\caption{Simulation Designs\label{tab:DGPs}}
\begin{tabular}{>{\centering\arraybackslash}m{2cm} >{\centering\arraybackslash}m{2.85cm} >{\centering\arraybackslash}m{2.85cm} >{\centering\arraybackslash}m{2.85cm} >{\centering\arraybackslash}m{2.85cm} >{\centering\arraybackslash}m{2.85cm}}
\hline\hline
 & Design A.1 & Design A.2 & Design B & Design C & Design D \\
\hline
$\mu_d(x,z)$ & $4z$ & $4 \left( z-1 \right)$ & $-1 + 2x + 2.122z$ & $-1 + 2x + 2.122z$ & $-1 + 2x + 2.122z$ \\
$\mu_{y_1}(x)$ & $0.3989$ & $0.3989$ & $0.3989$ & $9 \left( x+3 \right) ^2 $ & $9 \left( x+3 \right) ^2 $ \\
$\mu_z(x)$ & $2x-1$ & $2x-1$ & $2x-1$ & $2x-1$ & $x + x^2 - 1$ \\
\hline
\end{tabular}
\begin{footnotesize}
\begin{tablenotes}[flushleft]
\item
\end{tablenotes}
\end{footnotesize}
\end{threeparttable}
\end{adjustwidth}
\end{table}

In our simulations, similar to \cite{Heiler2022}, we thus use the 2SLS estimator as a benchmark that the weighting estimators will not be able to outperform in Designs A.1, A.2, and B while almost certainly being able to do so in Designs C and D\@. We also consider $\hat{\tau}_{u}^{cb}$, $\hat{\tau}_{u}^{ml}$, $\hat{\tau}_{a,10}^{ml}$, $\hat{\tau}_{a}^{ml}$, $\hat{\tau}_{a,1}^{ml}$ ($= \hat{\tau}_{t}^{ml}$), and $\hat{\tau}_{a,0}^{ml}$, also controlling for $X$\@. This leads to a misspecification in Design D, where $\mu_z(X)$ is quadratic in $X$ but we mistakenly omit the quadratic term. We consider three sample sizes, $N=500$, $N=1{,}000$, and $N=5{,}000$, and 10,000 replications for each combination of a design, a value of $\delta$, and a sample size.

Our main results are reported in Tables \ref{tab:simA1} to \ref{tab:simD}. For each estimator, we report the mean squared error (MSE), normalized by the MSE of the 2SLS estimator, the absolute bias, and the coverage rate for a nominal 95\% confidence interval.

In Design A.1, as expected, the 2SLS estimator outperforms all weighting estimators of the LATE, with MSEs of these estimators always at least 31\% larger, and sometimes orders of magnitude larger, than that of 2SLS\@. With better overlap and larger sample sizes, all estimators have small biases. When overlap is poor and/or samples small, 2SLS is better than the weighting estimators in terms of bias, too. Coverage rates are close to the nominal coverage rate for all estimators in all cases. At the same time, in a comparison of different weighting estimators, three of them, $\hat{\tau}_{t}^{ml}$, $\hat{\tau}_{a}^{ml}$, and $\hat{\tau}_{a,10}^{ml}$, are very unstable when overlap is sufficiently poor, $\delta \in \left\lbrace 0.01,0.02 \right\rbrace$, and samples are small, $N=500$. This is documented by very large MSEs in these cases. However, as predicted by Section \ref{sec:nearzero}, $\hat{\tau}_{a,0}^{ml}$, $\hat{\tau}_{u}^{ml}$, and $\hat{\tau}_{u}^{cb}$ do not suffer from instability, even in the most challenging case with $\delta = 0.01$ and $N=500$. This is because there are (nearly) no never-takers in Design A.1. More generally, $\hat{\tau}_{u}^{cb}$ and $\hat{\tau}_{u}^{ml}$ perform better than $\hat{\tau}_{a,0}^{ml}$, which is likely due to normalization.

Our results for Design A.2 are generally similar, except for the relative performance of 2SLS in terms of bias and, especially, the exact list of weighting estimators that suffer from instability. Unlike in Design A.1, when overlap is poor and/or samples small, the bias of 2SLS is not clearly smaller than that of (most of) the weighting estimators. Also, it is $\hat{\tau}_{a,0}^{ml}$, $\hat{\tau}_{a,10}^{ml}$, and perhaps $\hat{\tau}_{a}^{ml}$ that suffer from instability in such cases---but clearly not $\hat{\tau}_{t}^{ml}$. As discussed in Section \ref{sec:nearzero}, this is because there are (nearly) no always-takers in Design A.2. As before, $\hat{\tau}_{u}^{cb}$ and $\hat{\tau}_{u}^{ml}$ perform marginally better than the best unnormalized estimator (in this case, $\hat{\tau}_{t}^{ml}$).

In Design B, the instrument is no longer fully independent and noncompliance is no longer one sided. While 2SLS remains dominant in terms of MSE, it is always outperformed by most of the weighting estimators in terms of bias, often substantially and sometimes by all of them. In a comparison of different weighting estimators, $\hat{\tau}_{u}^{cb}$ and $\hat{\tau}_{u}^{ml}$ remain best overall while $\hat{\tau}_{t}^{ml}$, $\hat{\tau}_{a}^{ml}$, and $\hat{\tau}_{a,10}^{ml}$ clearly suffer from instability when overlap is sufficiently poor and samples sufficiently small. The case of $\hat{\tau}_{a,0}^{ml}$ is borderline, which is perhaps due to the fact that there are many more always-takers than never-takers in this design (although both groups clearly exist, unlike before).

Next, in Design C, we introduce another source of treatment effect heterogeneity through the dependence of $\mu_{y_1}(X)$ on $X$\@. The 2SLS estimator is no longer consistent for the LATE, which is illustrated by its large bias in all cases, including the least challenging case with $\delta = 0.05$ and $N=5{,}000$. Given that we define the coverage rate as the fraction of replications in which the LATE is contained in a nominal 95\% confidence interval, we also obtain very low coverage rates for 2SLS, never exceeding 66\% and approaching 0\% when the sample size is sufficiently large. Coverage rates for all the weighting estimators are close to the nominal level when overlap is good and samples large enough. The only weighting estimators that never suffer from instability are $\hat{\tau}_{u}^{cb}$ and $\hat{\tau}_{u}^{ml}$, although $\hat{\tau}_{u}^{cb}$ is now dominant, with substantial improvements in MSE in all cases.

Finally, in Design D, the instrument propensity score is misspecified, as we mistakenly omit the quadratic in $X$\@. The 2SLS estimator remains inconsistent, too, and its coverage rates are close to 0\% in all cases. While the weighting estimators clearly differ in performance, sometimes in unexpected ways, the most striking feature of the simulation results for Design D is the dominance of $\hat{\tau}_{u}^{cb}$, in terms of MSE, bias, and coverage. The relative efficiency of $\hat{\tau}_{u}^{cb}$, here and elsewhere, can be understood through the lens of a heuristic argument in \cite{Heiler2022}, who explained that covariate balancing implicitly regularizes the propensity score estimates away from the boundary and thereby decreases variance. It is also useful to note that, despite misspecification of the instrument propensity score, the coverage rate for $\hat{\tau}_{u}^{cb}$ approaches the nominal level when overlap is sufficiently good and samples sufficiently large, which is not the case for any other estimator.

It seems natural to interpret the instability of different weighting estimators of the LATE as a consequence of near-zero denominators, as we have done so far. To corroborate this interpretation, in Figures \ref{fig:simA1} to \ref{fig:simD}, we present box plots with simulation evidence on all estimators of the proportion of compliers that we consider: the first-stage coefficient on $Z$ in 2SLS; the denominator of $\hat{\tau}_{u}^{ml}$; $N^{-1}\sum_{i=1}^{N} \hat{\kappa}_{i1}$, $N^{-1}\sum_{i=1}^{N} \hat{\kappa}_{i0}$, and $N^{-1}\sum_{i=1}^{N} \hat{\kappa}_{i}$, with the maximum likelihood propensity scores; the denominator of $\hat{\tau}_{u}^{cb}$; and $N^{-1}\sum_{i=1}^{N} \hat{\kappa}_{i1} = N^{-1}\sum_{i=1}^{N} \hat{\kappa}_{i0}$, with the covariate balancing propensity scores. A straightforward comparison of Tables \ref{tab:simA1} to \ref{tab:simD} with Figures \ref{fig:simA1} to \ref{fig:simD} reveals that instability of weighting estimators of the LATE is indeed associated with situations in which the supports of their denominators, the estimators of the proportion of compliers, are crossing zero. In fact, it is not negative estimates of this proportion that are particularly problematic, even if they make no logical sense, but rather those estimates that are very close to zero, as this results in dividing by ``near zero'' to construct an estimate of the LATE, which leads to instability. Additional simulation evidence is also provided in Figures \ref{fig:hist_first} to \ref{fig:hist_last}, which present histograms for each combination of an estimator, a design, a value of $\delta$, and a sample size. In cases with instability, the normal approximation to the sampling distribution is clearly inappropriate.

\section{Conclusion}
\label{sec:conclusion}

In this paper we study the properties of several weighting estimators of the local average treatment effect (LATE), which are based on the identification results of \cite{Abadie2003} and \cite{Frolich2007}. We make several novel observations. First, we show that some of the most popular weighting estimators of the LATE are not translation invariant or scale invariant with respect to the natural logarithm, which translates to their sensitivity to the units of measurement when estimating the LATE in logs and the centering of the outcome variable more generally. In contrast, normalized weighting estimators generally have these important properties. Second, we demonstrate that certain weighting estimators of the LATE have an advantage of being based on a denominator that is strictly greater than zero in settings with one-sided noncompliance. There is only one estimator under consideration in this paper, originally proposed by \cite{UysalDiss}, that possesses both these advantages. When the instrument propensity score is estimated using an appropriate covariate balancing approach, this estimator is also equivalent to the one in \cite{Heiler2022}.

We illustrate our findings with three empirical applications and a simulation study. In simulations, our preferred estimator performs relatively well in every setting under consideration. In empirical applications, we clearly document the lack of translation invariance and scale equivariance of the unnormalized estimators. Our preferred estimator is fully robust to the underlying transformations of the outcome data.

\singlespacing

\setlength\bibsep{0pt}
\bibliographystyle{aer}
\bibliography{SUW_references}

\pagebreak
\begin{appendices}
\setcounter{page}{1}

\doublespacing

\begin{center}
\textsc{\Large{Online Appendix for \\ ``Abadie's Kappa and Weighting Estimators of the Local Average Treatment Effect''}} \\
\medskip
\textsc{\large{Tymon S\l{}oczy\'{n}ski, S. Derya Uysal, and Jeffrey M. Wooldridge}} \\
\end{center}

\paragraph{Review of Recent Empirical Applications.}

In Section \ref{sec:intro}, we included the following statement: ``Our application of weighting to estimate the LATE appears to be somewhat rare in practice, although \citeauthor{Abadie2003}'s \citeyearpar{Abadie2003} result is more commonly used to estimate mean characteristics of compliers, as also recommended by \cite{AP2009}. We analyze two samples of applications of instrumental variables to verify this claim. First, our reading of the 30 papers replicated by \cite{Young2022}, each of which uses 2SLS, suggests that none of these papers uses weighting estimators of the LATE or applies \citeauthor{Abadie2003}'s \citeyearpar{Abadie2003} result for any other purpose. Second, we have also examined whether any of the papers published in journals of the American Economic Association in 2019 and 2020 consider weighting estimators of the LATE\@. Our best assessment is that the answer is likewise negative. Still, \cite{MT2019}, \cite
{GGS2020}, \cite{LOL2020}, and \cite{LVRS2020} apply \citeauthor{Abadie2003}'s \citeyearpar{Abadie2003} result to estimate mean characteristics of compliers, while \cite{Cohodes2020} uses this result to estimate the control complier mean (CCM), a parameter introduced by \cite{KKL2001}.'' In what follows, we briefly explain how we reached these conclusions.

To examine whether any of the papers published in journals of the American Economic Association in 2019 and 2020 consider weighting estimators of the LATE, we first searched for the string ``instrument'' in the main text of each such paper. We retained every paper where this string appeared at least once and it was not immediately clear that the context in which it appeared had nothing to do with instrumental variables (e.g., financial instruments, Texas Instruments).

For every paper that was retained in the search described above and additionally for every paper replicated by \cite{Young2022}, we subsequently verified whether it cited any single-authored papers by Alberto Abadie, Markus Frölich, or Zhiqiang Tan, and whether any of the following strings appeared in its main text: ``propensity score,'' ``IPW,'' or ``weighting.'' In the case of any such citation and any appearance of any of these strings, we subsequently read the relevant part of the paper to determine whether \citeauthor{Abadie2003}'s \citeyearpar{Abadie2003} result and/or weighting estimators of the LATE may have been used. Our statement in Section \ref{sec:intro}, also restated above, summarizes our conclusions from this exercise.

\paragraph{Proof of Proposition \ref{prop:invariance}.}

We begin with the case of translation invariance. For $\hat{\tau}_{u}$, we can write
\begin{eqnarray}
\hat{\tau}_{u} \left( \mathbf{Y} + k, \mathbf{W} \right) &=& \frac{\left[ \sum_{i=1}^{N} \frac{Z_{i}}{p(X_{i})} \right]^{-1} \sum_{i=1}^{N} \frac{\left( Y_i + k \right) Z_{i}}{p(X_{i})} - \left[ \sum_{i=1}^{N} \frac{1 - Z_{i}}{1 - p(X_{i})} \right]^{-1} \sum_{i=1}^{N} \frac{\left( Y_i + k \right) \left( 1 - Z_{i} \right)}{1 - p(X_{i})}}{\left[ \sum_{i=1}^{N} \frac{Z_{i}}{p(X_{i})} \right]^{-1} \sum_{i=1}^{N} \frac{D_i Z_{i}}{p(X_{i})} - \left[ \sum_{i=1}^{N} \frac{1 - Z_{i}}{1 - p(X_{i})} \right]^{-1} \sum_{i=1}^{N} \frac{D_i \left( 1 - Z_{i} \right)}{1 - p(X_{i})}} \nonumber\\
&=& \hat{\tau}_{u} \left( \mathbf{Y}, \mathbf{W} \right) \; + \; \frac{\left[ \sum_{i=1}^{N} \frac{Z_{i}}{p(X_{i})} \right]^{-1} \sum_{i=1}^{N} \frac{k Z_{i}}{p(X_{i})} - \left[ \sum_{i=1}^{N} \frac{1 - Z_{i}}{1 - p(X_{i})} \right]^{-1} \sum_{i=1}^{N} \frac{k \left( 1 - Z_{i} \right)}{1 - p(X_{i})}}{\left[ \sum_{i=1}^{N} \frac{Z_{i}}{p(X_{i})} \right]^{-1} \sum_{i=1}^{N} \frac{D_i Z_{i}}{p(X_{i})} - \left[ \sum_{i=1}^{N} \frac{1 - Z_{i}}{1 - p(X_{i})} \right]^{-1} \sum_{i=1}^{N} \frac{D_i \left( 1 - Z_{i} \right)}{1 - p(X_{i})}} \nonumber\\
&=& \hat{\tau}_{u} \left( \mathbf{Y}, \mathbf{W} \right), \nonumber
\end{eqnarray}
which means that $\hat{\tau}_{u}$ is indeed translation invariant. Similarly,
\begin{eqnarray}
\hat{\tau}_{a,10} \left( \mathbf{Y} + k, \mathbf{W} \right) &=& \left[ \sum_{i=1}^{N} \kappa_{i1} \right]^{-1} \left[ \sum_{i=1}^{N} \kappa_{i1} \left( Y_i + k \right) \right] \; - \; \left[ \sum_{i=1}^{N} \kappa_{i0} \right]^{-1} \left[ \sum_{i=1}^{N} \kappa_{i0} \left( Y_i + k \right) \right] \nonumber\\
&=& \hat{\tau}_{a,10} \left( \mathbf{Y}, \mathbf{W} \right) \; + \; \left[ \sum_{i=1}^{N} \kappa_{i1} \right]^{-1} \left[ \sum_{i=1}^{N} \kappa_{i1} k \right] \; - \; \left[ \sum_{i=1}^{N} \kappa_{i0} \right]^{-1} \left[ \sum_{i=1}^{N} \kappa_{i0} k \right] \nonumber\\
&=& \hat{\tau}_{a,10} \left( \mathbf{Y}, \mathbf{W} \right), \nonumber
\end{eqnarray}
which means that $\hat{\tau}_{a,10}$ is translation invariant, too. On the other hand, we can write
\begin{eqnarray}
\hat{\tau}_{a} \left( \mathbf{Y} + k, \mathbf{W} \right) &=& \left[ \sum_{i=1}^{N} \kappa_{i} \right]^{-1} \left[ \sum_{i=1}^{N} \kappa_{i1} \left( Y_i + k \right) \right] \; - \; \left[ \sum_{i=1}^{N} \kappa_{i} \right]^{-1} \left[ \sum_{i=1}^{N} \kappa_{i0} \left( Y_i + k \right) \right] \nonumber\\
&=& \hat{\tau}_{a} \left( \mathbf{Y}, \mathbf{W} \right) \; + \; \left[ \sum_{i=1}^{N} \kappa_{i} \right]^{-1} \left[ \sum_{i=1}^{N} \kappa_{i1} k \right] \; - \; \left[ \sum_{i=1}^{N} \kappa_{i} \right]^{-1} \left[ \sum_{i=1}^{N} \kappa_{i0} k \right] \nonumber\\
&=& \hat{\tau}_{a} \left( \mathbf{Y}, \mathbf{W} \right) \; + \; k \left( \left[ \sum_{i=1}^{N} \kappa_{i} \right]^{-1} \left[ \sum_{i=1}^{N} \kappa_{i1} \right] \; - \; \left[ \sum_{i=1}^{N} \kappa_{i} \right]^{-1} \left[ \sum_{i=1}^{N} \kappa_{i0} \right] \right) \nonumber
\end{eqnarray}
and
\begingroup
\allowdisplaybreaks
\begin{eqnarray}
\hat{\tau}_{a,1} \left( \mathbf{Y} + k, \mathbf{W} \right) &=& \left[ \sum_{i=1}^{N} \kappa_{i1} \right]^{-1} \left[ \sum_{i=1}^{N} \kappa_{i1} \left( Y_i + k \right) \right] \; - \; \left[ \sum_{i=1}^{N} \kappa_{i1} \right]^{-1} \left[ \sum_{i=1}^{N} \kappa_{i0} \left( Y_i + k \right) \right] \nonumber\\
&=& \hat{\tau}_{a,1} \left( \mathbf{Y}, \mathbf{W} \right) \; + \; \left[ \sum_{i=1}^{N} \kappa_{i1} \right]^{-1} \left[ \sum_{i=1}^{N} \kappa_{i1} k \right] \; - \; \left[ \sum_{i=1}^{N} \kappa_{i1} \right]^{-1} \left[ \sum_{i=1}^{N} \kappa_{i0} k \right] \nonumber\\
&=& \hat{\tau}_{a,1} \left( \mathbf{Y}, \mathbf{W} \right) \; + \; k \left( 1 \; - \; \left[ \sum_{i=1}^{N} \kappa_{i1} \right]^{-1} \left[ \sum_{i=1}^{N} \kappa_{i0} \right] \right) \nonumber
\end{eqnarray}
\endgroup
and also
\begin{eqnarray}
\hat{\tau}_{a,0} \left( \mathbf{Y} + k, \mathbf{W} \right) &=& \left[ \sum_{i=1}^{N} \kappa_{i0} \right]^{-1} \left[ \sum_{i=1}^{N} \kappa_{i1} \left( Y_i + k \right) \right] \; - \; \left[ \sum_{i=1}^{N} \kappa_{i0} \right]^{-1} \left[ \sum_{i=1}^{N} \kappa_{i0} \left( Y_i + k \right) \right] \nonumber\\
&=& \hat{\tau}_{a,0} \left( \mathbf{Y}, \mathbf{W} \right) \; + \; \left[ \sum_{i=1}^{N} \kappa_{i0} \right]^{-1} \left[ \sum_{i=1}^{N} \kappa_{i1} k \right] \; - \; \left[ \sum_{i=1}^{N} \kappa_{i0} \right]^{-1} \left[ \sum_{i=1}^{N} \kappa_{i0} k \right] \nonumber\\
&=& \hat{\tau}_{a,0} \left( \mathbf{Y}, \mathbf{W} \right) \; + \; k \left( \left[ \sum_{i=1}^{N} \kappa_{i0} \right]^{-1} \left[ \sum_{i=1}^{N} \kappa_{i1} \right] \; - \; 1 \right). \nonumber
\end{eqnarray}
Even though $k \left( \left[ \sum_{i=1}^{N} \kappa_{i0} \right]^{-1} \left[ \sum_{i=1}^{N} \kappa_{i1} \right] \; - \; 1 \right) = o_p(1)$, $k \left( 1 \; - \; \left[ \sum_{i=1}^{N} \kappa_{i1} \right]^{-1} \left[ \sum_{i=1}^{N} \kappa_{i0} \right] \right) = o_p(1)$, and $k \left( \left[ \sum_{i=1}^{N} \kappa_{i} \right]^{-1} \left[ \sum_{i=1}^{N} \kappa_{i1} \right] \; - \; \left[ \sum_{i=1}^{N} \kappa_{i} \right]^{-1} \left[ \sum_{i=1}^{N} \kappa_{i0} \right] \right) = o_p(1)$, none of these objects is generally equal to zero in finite samples, which means that $\hat{\tau}_{a,0}$, $\hat{\tau}_{a,1}$, and $\hat{\tau}_{a}$, respectively, are not translation invariant.

We next turn to the case of scale equivariance. Begin by denoting $\frac{Z_i}{p(X_{i})} = \omega_{i1}$ and $\frac{1-Z_i}{1-p(X_{i})} = \omega_{i0}$. Then, for $\hat{\tau}_{u}$, we can write
\begingroup
\allowdisplaybreaks
\begin{eqnarray}
	\hat{\tau}_{u} \left( f(a\mathbf{Y}), \mathbf{W}\right) &=& \frac{\left[\sum_{i=1}^{N}\omega_{i1} \right]^{-1}\left[\sum_{i=1}^{N} (\alpha_2(aY_i)^{\alpha_1}-\alpha_3)\omega_{i1}\right]-\left[\sum_{i=1}^{N}\omega_{i0} \right]^{-1}\left[\sum_{i=1}^{N} (\alpha_2(aY_i)^{\alpha_1}-\alpha_3)\omega_{i0}\right]}{\left[\sum_{i=1}^{N}\omega_{i1} \right]^{-1}\left[\sum_{i=1}^{N} D_i\omega_{i1}\right]-\left[\sum_{i=1}^{N}\omega_{i0} \right]^{-1}\left[\sum_{i=1}^{N} D_i\omega_{i0}\right]} \nonumber\\
	&=& \frac{\left[\sum_{i=1}^{N}\omega_{i1} \right]^{-1}\left[\sum_{i=1}^{N} \alpha_2(aY_i)^{\alpha_1}\omega_{i1}\right]-\left[\sum_{i=1}^{N}\omega_{i0} \right]^{-1}\left[\sum_{i=1}^{N} \alpha_2(aY_i)^{\alpha_1}\omega_{i0}\right]}{\left[\sum_{i=1}^{N}\omega_{i1} \right]^{-1}\left[\sum_{i=1}^{N} D_i\omega_{i1}\right]-\left[\sum_{i=1}^{N}\omega_{i0} \right]^{-1}\left[\sum_{i=1}^{N} D_i\omega_{i0}\right]} \nonumber\\
	&&- \; \; \frac{\alpha_3\left(  \left[\sum_{i=1}^{N}\omega_{i1} \right]^{-1}\left[\sum_{i=1}^{N} \omega_{i1}\right]-\left[\sum_{i=1}^{N}\omega_{i0} \right]^{-1}\left[\sum_{i=1}^{N} \omega_{i0}\right]\right) } {\left[\sum_{i=1}^{N}\omega_{i1} \right]^{-1}\left[\sum_{i=1}^{N} D_i\omega_{i1}\right]-\left[\sum_{i=1}^{N}\omega_{i0} \right]^{-1}\left[\sum_{i=1}^{N} D_i\omega_{i0}\right]} \nonumber\\
		&=& \frac{a^{\alpha_1}\left(\left[\sum_{i=1}^{N}\omega_{i1} \right]^{-1}\left[\sum_{i=1}^{N} \alpha_2 Y_i^{\alpha_1}\omega_{i1}\right]-\left[\sum_{i=1}^{N}\omega_{i0} \right]^{-1}\left[\sum_{i=1}^{N} \alpha_2 Y_i^{\alpha_1}\omega_{i0}\right]\right)}{\left[\sum_{i=1}^{N}\omega_{i1} \right]^{-1}\left[\sum_{i=1}^{N} D_i\omega_{i1}\right]-\left[\sum_{i=1}^{N}\omega_{i0} \right]^{-1}\left[\sum_{i=1}^{N} D_i\omega_{i0}\right]} \nonumber\\
			&=& \frac{a^{\alpha_1}\left(\left[\sum_{i=1}^{N}\omega_{i1} \right]^{-1}\left[\sum_{i=1}^{N} \alpha_2 Y_i^{\alpha_1}\omega_{i1}\right]-\left[\sum_{i=1}^{N}\omega_{i0} \right]^{-1}\left[\sum_{i=1}^{N} \alpha_2 Y_i^{\alpha_1}\omega_{i0}\right]\right)}{\left[\sum_{i=1}^{N}\omega_{i1} \right]^{-1}\left[\sum_{i=1}^{N} D_i\omega_{i1}\right]-\left[\sum_{i=1}^{N}\omega_{i0} \right]^{-1}\left[\sum_{i=1}^{N} D_i\omega_{i0}\right]} \nonumber\\
	&&\pm \; \; \frac{a^{\alpha_1}\alpha_3\left(  \left[\sum_{i=1}^{N}\omega_{i1} \right]^{-1}\left[\sum_{i=1}^{N} \omega_{i1}\right]-\left[\sum_{i=1}^{N}\omega_{i0} \right]^{-1}\left[\sum_{i=1}^{N} \omega_{i0}\right]\right) } {\left[\sum_{i=1}^{N}\omega_{i1} \right]^{-1}\left[\sum_{i=1}^{N} D_i\omega_{i1}\right]-\left[\sum_{i=1}^{N}\omega_{i0} \right]^{-1}\left[\sum_{i=1}^{N} D_i\omega_{i0}\right]} \nonumber\\
				&=& \frac{a^{\alpha_1}\left(\left[\sum_{i=1}^{N}\omega_{i1} \right]^{-1}\left[\sum_{i=1}^{N} (\alpha_2 Y_i^{\alpha_1}-\alpha_3)\omega_{i1}\right]-\left[\sum_{i=1}^{N}\omega_{i0} \right]^{-1}\left[\sum_{i=1}^{N} (\alpha_2 Y_i^{\alpha_1}-\alpha_3)\omega_{i0}\right]\right)}{\left[\sum_{i=1}^{N}\omega_{i1} \right]^{-1}\left[\sum_{i=1}^{N} D_i\omega_{i1}\right]-\left[\sum_{i=1}^{N}\omega_{i0} \right]^{-1}\left[\sum_{i=1}^{N} D_i\omega_{i0}\right]} \nonumber\\
	&&+ \; \; \frac{a^{\alpha_1}\alpha_3\left(  \left[\sum_{i=1}^{N}\omega_{i1} \right]^{-1}\left[\sum_{i=1}^{N} \omega_{i1}\right]-\left[\sum_{i=1}^{N}\omega_{i0} \right]^{-1}\left[\sum_{i=1}^{N} \omega_{i0}\right]\right) } {\left[\sum_{i=1}^{N}\omega_{i1} \right]^{-1}\left[\sum_{i=1}^{N} D_i\omega_{i1}\right]-\left[\sum_{i=1}^{N}\omega_{i0} \right]^{-1}\left[\sum_{i=1}^{N} D_i\omega_{i0}\right]} \nonumber\\
	&=&a^{\alpha_1}	\hat{\tau}_{u} \left( f(\mathbf{Y}), \mathbf{W}\right),
\nonumber
\end{eqnarray}
\endgroup
which means that $\hat{\tau}_{u}$ is indeed scale equivariant. Similarly,
\begingroup
\allowdisplaybreaks
\begin{eqnarray}
	\hat{\tau}_{a,10} \left( f(a\mathbf{Y}), \mathbf{W}\right) &=&\left[\sum_{i=1}^{N}\kappa_{i1} \right]^{-1}\left[\sum_{i=1}^{N} \kappa_{i1}(\alpha_2(aY_i)^{\alpha_1}-\alpha_3)\right]-\left[\sum_{i=1}^{N}\kappa_{i0} \right]^{-1}\left[\sum_{i=1}^{N} \kappa_{i0}(\alpha_2(aY_i)^{\alpha_1}-\alpha_3)\right] \nonumber \\
	&=&\left[\sum_{i=1}^{N}\kappa_{i1} \right]^{-1}\left[\sum_{i=1}^{N} \kappa_{i1}\alpha_2(aY_i)^{\alpha_1}\right]-\left[\sum_{i=1}^{N}\kappa_{i0} \right]^{-1}\left[\sum_{i=1}^{N} \kappa_{i0}\alpha_2(aY_i)^{\alpha_1}\right]\nonumber \\
	&&- \; \; \alpha_3\left(\left[\sum_{i=1}^{N}\kappa_{i1} \right]^{-1}\left[\sum_{i=1}^{N} \kappa_{i1} \right]-\left[\sum_{i=1}^{N}\kappa_{i0} \right]^{-1}\left[\sum_{i=1}^{N} \kappa_{i0}\right] \right) \nonumber \\
	&=&a^{\alpha_1}\left(\left[\sum_{i=1}^{N}\kappa_{i1} \right]^{-1}\left[\sum_{i=1}^{N} \kappa_{i1}\alpha_2 Y_i^{\alpha_1}\right]-\left[\sum_{i=1}^{N}\kappa_{i0} \right]^{-1}\left[\sum_{i=1}^{N} \kappa_{i0}\alpha_2 Y_i^{\alpha_1}\right] \right)\nonumber \\
&&\pm \; \;  a^{\alpha_1}\alpha_3\left(\left[\sum_{i=1}^{N}\kappa_{i1} \right]^{-1}\left[\sum_{i=1}^{N} \kappa_{i1} \right]-\left[\sum_{i=1}^{N}\kappa_{i0} \right]^{-1}\left[\sum_{i=1}^{N} \kappa_{i0}\right] \right) \nonumber \\
&=&a^{\alpha_1}\left(\left[\sum_{i=1}^{N}\kappa_{i1} \right]^{-1}\left[\sum_{i=1}^{N} \kappa_{i1}(\alpha_2 Y_i^{\alpha_1}-\alpha_3)\right]-\left[\sum_{i=1}^{N}\kappa_{i0} \right]^{-1}\left[\sum_{i=1}^{N} \kappa_{i0}(\alpha_2 Y_i^{\alpha_1}-\alpha_3)\right] \right)\nonumber \\
&&+ \; \; a^{\alpha_1}\alpha_3 \left(\left[\sum_{i=1}^{N}\kappa_{i1} \right]^{-1}\left[\sum_{i=1}^{N} \kappa_{i1} \right]-\left[\sum_{i=1}^{N}\kappa_{i0} \right]^{-1}\left[\sum_{i=1}^{N} \kappa_{i0}\right] \right) \nonumber\\
&=&a^{\alpha_1}\hat{\tau}_{a,10} \left( f(\mathbf{Y}), \mathbf{W}\right),
\nonumber
\end{eqnarray}
\endgroup
which means that $\hat{\tau}_{a,10}$ is scale equivariant, too. On the other hand, we can write
\begingroup
\allowdisplaybreaks
\begin{eqnarray}
		\hat{\tau}_{a} \left( f(a\mathbf{Y}), \mathbf{W}\right) &=&\left[\sum_{i=1}^{N}\kappa_{i} \right]^{-1}\left[\sum_{i=1}^{N} \kappa_{i1}(\alpha_2(aY_i)^{\alpha_1}-\alpha_3)\right]-\left[\sum_{i=1}^{N}\kappa_{i} \right]^{-1}\left[\sum_{i=1}^{N} \kappa_{i0}(\alpha_2(aY_i)^{\alpha_1}-\alpha_3)\right] \nonumber \\
		&=&\left[\sum_{i=1}^{N}\kappa_{i} \right]^{-1}\left[\sum_{i=1}^{N} \kappa_{i1}\alpha_2(aY_i)^{\alpha_1}\right]-\left[\sum_{i=1}^{N}\kappa_{i} \right]^{-1}\left[\sum_{i=1}^{N} \kappa_{i0}\alpha_2(aY_i)^{\alpha_1}\right]\nonumber \\
		&&- \; \; \alpha_3\left(\left[\sum_{i=1}^{N}\kappa_{i} \right]^{-1}\left[\sum_{i=1}^{N} \kappa_{i1} \right]-\left[\sum_{i=1}^{N}\kappa_{i} \right]^{-1}\left[\sum_{i=1}^{N} \kappa_{i0}\right] \right) \nonumber \\
		&=&a^{\alpha_1}\left(\left[\sum_{i=1}^{N}\kappa_{i} \right]^{-1}\left[\sum_{i=1}^{N} \kappa_{i1}\alpha_2 Y_i^{\alpha_1}\right]-\left[\sum_{i=1}^{N}\kappa_{i} \right]^{-1}\left[\sum_{i=1}^{N} \kappa_{i0}\alpha_2 Y_i^{\alpha_1}\right] \right)\nonumber \\
		&&- \; \; \alpha_3\left(\left[\sum_{i=1}^{N}\kappa_{i} \right]^{-1}\left[\sum_{i=1}^{N} \kappa_{i1} \right]-\left[\sum_{i=1}^{N}\kappa_{i} \right]^{-1}\left[\sum_{i=1}^{N} \kappa_{i0}\right] \right) \nonumber \\
		&&\pm \; \;  a^{\alpha_1}\alpha_3\left(\left[\sum_{i=1}^{N}\kappa_{i} \right]^{-1}\left[\sum_{i=1}^{N} \kappa_{i1} \right]-\left[\sum_{i=1}^{N}\kappa_{i} \right]^{-1}\left[\sum_{i=1}^{N} \kappa_{i0}\right] \right) \nonumber \\
		&=&a^{\alpha_1}\left(\left[\sum_{i=1}^{N}\kappa_{i} \right]^{-1}\left[\sum_{i=1}^{N} \kappa_{i1}(\alpha_2 Y_i^{\alpha_1}-\alpha_3)\right]-\left[\sum_{i=1}^{N}\kappa_{i} \right]^{-1}\left[\sum_{i=1}^{N} \kappa_{i0}(\alpha_2 Y_i^{\alpha_1}-\alpha_3)\right] \right)\nonumber \\
		&&- \; \; (\alpha_3-a^{\alpha_1}\alpha_3)\left(\left[\sum_{i=1}^{N}\kappa_{i} \right]^{-1}\left[\sum_{i=1}^{N} \kappa_{i1} \right]-\left[\sum_{i=1}^{N}\kappa_{i} \right]^{-1}\left[\sum_{i=1}^{N} \kappa_{i0}\right] \right) \nonumber \\
		&=&a^{\alpha_1}\hat{\tau}_{a} \left( f(\mathbf{Y}), \mathbf{W}\right) \; - \; (\alpha_3-a^{\alpha_1}\alpha_3)\left(\left[\sum_{i=1}^{N}\kappa_{i} \right]^{-1}\left[\sum_{i=1}^{N} \kappa_{i1} \right]-\left[\sum_{i=1}^{N}\kappa_{i} \right]^{-1}\left[\sum_{i=1}^{N} \kappa_{i0}\right] \right) \nonumber 
\end{eqnarray}
\endgroup
and
\begingroup
\allowdisplaybreaks
\begin{eqnarray}
		\hat{\tau}_{a.1} \left( f(a\mathbf{Y}), \mathbf{W}\right) &=&\left[\sum_{i=1}^{N}\kappa_{i1} \right]^{-1}\left[\sum_{i=1}^{N} \kappa_{i1}(\alpha_2(aY_i)^{\alpha_1}-\alpha_3)\right]-\left[\sum_{i=1}^{N}\kappa_{i1} \right]^{-1}\left[\sum_{i=1}^{N} \kappa_{i0}(\alpha_2(aY_i)^{\alpha_1}-\alpha_3)\right] \nonumber \\
		&=&\left[\sum_{i=1}^{N}\kappa_{i1} \right]^{-1}\left[\sum_{i=1}^{N} \kappa_{i1}\alpha_2(aY_i)^{\alpha_1}\right]-\left[\sum_{i=1}^{N}\kappa_{i1} \right]^{-1}\left[\sum_{i=1}^{N} \kappa_{i0}\alpha_2(aY_i)^{\alpha_1}\right]\nonumber \\
		&&- \; \; \alpha_3\left(\left[\sum_{i=1}^{N}\kappa_{i1} \right]^{-1}\left[\sum_{i=1}^{N} \kappa_{i1} \right]-\left[\sum_{i=1}^{N}\kappa_{i1} \right]^{-1}\left[\sum_{i=1}^{N} \kappa_{i0}\right] \right) \nonumber \\
		&=&a^{\alpha_1}\left(\left[\sum_{i=1}^{N}\kappa_{i1} \right]^{-1}\left[\sum_{i=1}^{N} \kappa_{i1}\alpha_2 Y_i^{\alpha_1}\right]-\left[\sum_{i=1}^{N}\kappa_{i1} \right]^{-1}\left[\sum_{i=1}^{N} \kappa_{i0}\alpha_2 Y_i^{\alpha_1}\right] \right)\nonumber \\
		&&- \; \; \alpha_3\left(\left[\sum_{i=1}^{N}\kappa_{i1} \right]^{-1}\left[\sum_{i=1}^{N} \kappa_{i1} \right]-\left[\sum_{i=1}^{N}\kappa_{i1} \right]^{-1}\left[\sum_{i=1}^{N} \kappa_{i0}\right] \right) \nonumber \\
		&&\pm \; \;  a^{\alpha_1}\alpha_3\left(\left[\sum_{i=1}^{N}\kappa_{i1} \right]^{-1}\left[\sum_{i=1}^{N} \kappa_{i1} \right]-\left[\sum_{i=1}^{N}\kappa_{i1} \right]^{-1}\left[\sum_{i=1}^{N} \kappa_{i0}\right] \right) \nonumber \\
		&=&a^{\alpha_1}\left(\left[\sum_{i=1}^{N}\kappa_{i1} \right]^{-1}\left[\sum_{i=1}^{N} \kappa_{i1}(\alpha_2 Y_i^{\alpha_1}-\alpha_3)\right]-\left[\sum_{i=1}^{N}\kappa_{i1} \right]^{-1}\left[\sum_{i=1}^{N} \kappa_{i0}(\alpha_2 Y_i^{\alpha_1}-\alpha_3)\right] \right)\nonumber \\
		&&- \; \; (\alpha_3-a^{\alpha_1}\alpha_3)\left(\left[\sum_{i=1}^{N}\kappa_{i1} \right]^{-1}\left[\sum_{i=1}^{N} \kappa_{i1} \right]-\left[\sum_{i=1}^{N}\kappa_{i1} \right]^{-1}\left[\sum_{i=1}^{N} \kappa_{i0}\right] \right) \nonumber \\
		&=&a^{\alpha_1}\hat{\tau}_{a,1} \left( f(\mathbf{Y}), \mathbf{W}\right) \; - \; (\alpha_3-a^{\alpha_1}\alpha_3)\left(1-\left[\sum_{i=1}^{N}\kappa_{i1} \right]^{-1}\left[\sum_{i=1}^{N} \kappa_{i0}\right] \right) \nonumber 
\end{eqnarray}
\endgroup
and also
\begingroup
\allowdisplaybreaks
\begin{eqnarray}
		\hat{\tau}_{a,0} \left( f(a\mathbf{Y}), \mathbf{W}\right) &=&\left[\sum_{i=1}^{N}\kappa_{i0} \right]^{-1}\left[\sum_{i=1}^{N} \kappa_{i1}(\alpha_2(aY_i)^{\alpha_1}-\alpha_3)\right]-\left[\sum_{i=1}^{N}\kappa_{i0} \right]^{-1}\left[\sum_{i=1}^{N} \kappa_{i0}(\alpha_2(aY_i)^{\alpha_1}-\alpha_3)\right] \nonumber \\
		&=&\left[\sum_{i=1}^{N}\kappa_{i0} \right]^{-1}\left[\sum_{i=1}^{N} \kappa_{i1}\alpha_2(aY_i)^{\alpha_1}\right]-\left[\sum_{i=1}^{N}\kappa_{i0} \right]^{-1}\left[\sum_{i=1}^{N} \kappa_{i0}\alpha_2(aY_i)^{\alpha_1}\right]\nonumber \\
		&&- \; \; \alpha_3\left(\left[\sum_{i=1}^{N}\kappa_{i0} \right]^{-1}\left[\sum_{i=1}^{N} \kappa_{i1} \right]-\left[\sum_{i=1}^{N}\kappa_{i0} \right]^{-1}\left[\sum_{i=1}^{N} \kappa_{i0}\right] \right) \nonumber \\
		&=&a^{\alpha_1}\left(\left[\sum_{i=1}^{N}\kappa_{i0} \right]^{-1}\left[\sum_{i=1}^{N} \kappa_{i1}\alpha_2 Y_i^{\alpha_1}\right]-\left[\sum_{i=1}^{N}\kappa_{i0} \right]^{-1}\left[\sum_{i=1}^{N} \kappa_{i0}\alpha_2 Y_i^{\alpha_1}\right] \right)\nonumber \\
		&&- \; \; \alpha_3\left(\left[\sum_{i=1}^{N}\kappa_{i0} \right]^{-1}\left[\sum_{i=1}^{N} \kappa_{i1} \right]-\left[\sum_{i=1}^{N}\kappa_{i0} \right]^{-1}\left[\sum_{i=1}^{N} \kappa_{i0}\right] \right) \nonumber \\
		&&\pm \; \;  a^{\alpha_1}\alpha_3\left(\left[\sum_{i=1}^{N}\kappa_{i0} \right]^{-1}\left[\sum_{i=1}^{N} \kappa_{i1} \right]-\left[\sum_{i=1}^{N}\kappa_{i0} \right]^{-1}\left[\sum_{i=1}^{N} \kappa_{i0}\right] \right) \nonumber \\
		&=&a^{\alpha_1}\left(\left[\sum_{i=1}^{N}\kappa_{i0} \right]^{-1}\left[\sum_{i=1}^{N} \kappa_{i1}(\alpha_2 Y_i^{\alpha_1}-\alpha_3)\right]-\left[\sum_{i=1}^{N}\kappa_{i0} \right]^{-1}\left[\sum_{i=1}^{N} \kappa_{i0}(\alpha_2 Y_i^{\alpha_1}-\alpha_3)\right] \right)\nonumber \\
		&&- \; \; (\alpha_3-a^{\alpha_1}\alpha_3)\left(\left[\sum_{i=1}^{N}\kappa_{i0} \right]^{-1}\left[\sum_{i=1}^{N} \kappa_{i1} \right]-\left[\sum_{i=1}^{N}\kappa_{i0} \right]^{-1}\left[\sum_{i=1}^{N} \kappa_{i0}\right] \right) \nonumber \\
		&=&a^{\alpha_1}\hat{\tau}_{a,0} \left( f(\mathbf{Y}), \mathbf{W}\right) \; - \; (\alpha_3-a^{\alpha_1}\alpha_3)\left(\left[\sum_{i=1}^{N}\kappa_{i0} \right]^{-1}\left[\sum_{i=1}^{N} \kappa_{i1} \right]-1 \right). \nonumber 
\end{eqnarray}
\endgroup
Even though $(\alpha_3 \; - \; a^{\alpha_1}\alpha_3)\left( \left[ \sum_{i=1}^{N} \kappa_{i0} \right]^{-1} \left[ \sum_{i=1}^{N} \kappa_{i1} \right] \; - \; 1 \right)$, $(\alpha_3 \; - \; a^{\alpha_1}\alpha_3) \left( 1 \; - \; \left[ \sum_{i=1}^{N} \kappa_{i1} \right]^{-1} \left[ \sum_{i=1}^{N} \kappa_{i0} \right] \right)$, and $(\alpha_3 \; - \; a^{\alpha_1}\alpha_3) \left( \left[ \sum_{i=1}^{N} \kappa_{i} \right]^{-1} \left[ \sum_{i=1}^{N} \kappa_{i1} \right] \; - \; \left[ \sum_{i=1}^{N} \kappa_{i} \right]^{-1} \left[ \sum_{i=1}^{N} \kappa_{i0} \right] \right)$ are all $o_p(1)$, none of these objects is generally equal to zero in finite samples, which means that $\hat{\tau}_{a,0}$, $\hat{\tau}_{a,1}$, and $\hat{\tau}_{a}$, respectively, are not scale equivariant.

\paragraph{Proof of Proposition \ref{prop:cb}.}

The sample moment conditions in equation (\ref{eq:samplemoments}) can be written as \begin{equation*}
N^{-1}\sum_{i=1}^{N} X_{i} \frac{Z_{i} - \hat{p}_{cb}(X_{i})}{\hat{p}_{cb}(X_{i}) \left( 1 - \hat{p}_{cb}(X_{i}) \right)} = 0.
\end{equation*}If $X$ includes a constant, then one of these moment conditions is $N^{-1}\sum_{i=1}^{N} \frac{Z_{i} - \hat{p}_{cb}(X_{i})}{\hat{p}_{cb}(X_{i}) \left( 1 - \hat{p}_{cb}(X_{i}) \right)} = 0$, and this, together with Remark \ref{remark:kappaeq}, guarantees that $N^{-1}\sum_{i=1}^{N} \hat{\kappa}_{i1} = N^{-1}\sum_{i=1}^{N} \hat{\kappa}_{i0}$, where $\hat{\kappa}_{1}$ and $\hat{\kappa}_{0}$ use the covariate-balancing instrument propensity score, $\hat{p}_{cb}(X)$\@. If $N^{-1}\sum_{i=1}^{N} \hat{\kappa}_{i1} = N^{-1}\sum_{i=1}^{N} \hat{\kappa}_{i0}$, then it is also the case that $\hat{\tau}_{t}^{cb}$ ($= \hat{\tau}_{a,1}^{cb}$), $\hat{\tau}_{a,0}^{cb}$, and $\hat{\tau}_{a,10}^{cb}$ are numerically identical to each other. They are also identical to $\hat{\tau}_{u}^{cb}$ following the result in \cite{Heiler2022}, which says that $\hat{\tau}_{u}^{cb}$ is identical to $\hat{\tau}_{t}^{cb}$.

\paragraph{Asymptotic Derivations.}

As stated in Section \ref{sec:asymptotic}, all the weighting estimators considered in this paper can be represented as an M-estimator. Thus, for the asymptotic distributions of each estimator, we can rely on the results regarding the asymptotics of the M-estimator. The M-estimator, denoted as $\hat{\theta}$, for $\theta$, a $K\times1$ unknown parameter vector, can be derived as the solution to the sample moment equation
\begin{equation*}\label{m-est-gen}
	N^{-1}\sum_{i=1}^N \psi(O_i,\hat{\theta})=0,
\end{equation*}
where $O_i$ is the observed data. Thus, $\hat{\theta}$ is the estimator of $\theta$ that satisfies the population relation $\e \left[\psi(O,\theta)\right]=0$.\footnote{See, for example, \cite{Wooldridge2010} and \cite{BoosStefanski2013} for more on M-estimation.} Under standard regularity conditions\footnote{Theorem 7.2 in \cite{BoosStefanski2013} states the conditions for the asymptotic normality of M-estimators. A more general treatment of these regularity conditions can be found in \cite{NeweyMcFadden1994}.} and assuming that the relevant moments exist, i.e.~$\e \left[\pd{\psi(O,\theta)}{\theta'}\right]$ exists and is nonsingular, and $\e \left[\psi(O,\theta)\psi(O,\theta)'\right]$ exists and is finite, the asymptotic distribution of an M-estimator is given by
\begin{equation}\label{eq:mestgenvar}
	\sqrt{N}(\hat{\theta}-\theta) \stackrel{d}{\longrightarrow} \mathcal{N}(0, A^{-1}VA^{-1}{'})
\end{equation}
with
\begin{eqnarray*}
	\label{A-gen} A&=&\e \left[\pd{\psi(O,\theta)}{\theta'}\right], \\
	\label{V-gen} V&=&\e \left[\psi(O,\theta)\psi(O,\theta)'\right].
\end{eqnarray*}

We use different combinations of moment functions listed in Table \ref{tab:moments} for each of the weighting estimators.  For example, if $\late$ is estimated by $\hat{\tau}_{a}^{ml}$, then
\begin{eqnarray*}
	\psi_a^{ml}=\left(\begin{array}{c}
		\psi_{\alpha}^{ml} \\
		\psi_{\Gamma} \\
		\psi_{\Delta} \\
		\psi_{\tau_a}
	\end{array}\right)
\end{eqnarray*}
is used as the vector of moment functions. Under standard regularity conditions for M-estimation, all of the LATE estimators discussed above will be asymptotically normal with different asymptotic variances. A joint estimation of $\alpha$ and $\late$ allows us to conduct inference based on the asymptotic variance-covariance matrix of an M-estimator given in \eqref{eq:mestgenvar} without explicitly deriving the asymptotic distribution of $\late$\@. At the same time, the M-estimation framework also facilitates the derivations of the asymptotic variance terms for each of the LATE estimators. In what follows, we provide asymptotic distributions of all the estimators discussed in the body of the paper.

We first introduce some additional notation in order to simplify the representation of the asymptotic variances. Let us denote the population counterpart of the  numerator of the estimators $\hat{\tau}_{a}$, $\hat{\tau}_{a,1}$ ($= \hat{\tau}_{t}$), $\hat{\tau}_{a,0}$, and $\hat{\tau}_{u}$ by $\Delta$, i.e.,
\begin{eqnarray} \label{eq:delta}
	\Delta & \equiv & \e \left[Y  \frac{Z  - p(X)}{p(X) \left( 1 - p(X)\right)}\right].
\end{eqnarray}
Recall that the expectation on the right hand side is equal to $\e \left[ \left( \kappa_{1} - \kappa_{0} \right) Y \right]$; see equation \eqref{eq:late2}. Next, denote $\e ( \kappa_{1} Y )$  and $\e ( \kappa_{0} Y )$ by $\Delta_1$ and $\Delta_0$, respectively. Alternatively, we can write the expectation in equation (\ref{eq:delta}) as follows:
\begin{eqnarray*}
	\e \left[Y  \frac{Z  - p(X)}{p(X) \left( 1 - p(X)\right)}\right] &=& \e \left[  \frac{Y Z}{p(X)}\right]-\e \left[\frac{Y(1-Z)} {1-p(X)}\right].
\end{eqnarray*}
We denote $\e \left[  \frac{Y Z}{p(X)}\right]$  by $\mu_1$ and $\e \left[\frac{Y(1-Z)} {1-p(X)}\right]$ by $\mu_0$. Symmetrically, we denote $\e \left[  \frac{D Z}{p(X)}\right]$  and $\e \left[\frac{D(1-Z)} {1-p(X)}\right]$ by $m_1$ and $m_0$.  Additionally, the population proportion of compliers is denoted by $\Gamma$, $\Gamma_1$, or $\Gamma_0$, depending on which sample mean is used to estimate the population parameter, i.e., $\Gamma \equiv  \e ( \kappa )$, $\Gamma_1 \equiv \e(\kappa_{1})$, and $\Gamma_0 \equiv \e(\kappa_{0})$. Note that $\late = \frac{\Delta}{\Gamma} = \frac{\Delta}{\Gamma_1} =\frac{\Delta}{\Gamma_0} =\frac{\Delta_1}{\Gamma_1}-\frac{\Delta_0}{\Gamma_0} = \frac{\mu_1-\mu_0}{m_1-m_0}$. When the population parameters are replaced by their sample counterparts, we obtain the estimators $\hat{\tau}_{a}$, $\hat{\tau}_{a,1}$, $\hat{\tau}_{a,0}$, $\hat{\tau}_{a,10}$, and $\hat{\tau}_{t}$, respectively. If normalized weights are used to estimate $\mu_z$ and $m_z$ for $z=0,1$, the resulting ratio estimator corresponds to  $\hat{\tau}_{u}$. This is of course without taking into account how the propensity score is estimated. 

In what follows, we first consider ML-based estimation of the instrument propensity score. For the estimator $\hat{\tau}_{a}^{ml}$, we use the moment functions $\psi_{\alpha}^{ml}$, $\psi_{\Delta}$, and $\psi_{\Gamma}$. Based on the result given in equation (\ref{eq:mestgenvar}), the asymptotic distribution of $\hat{\tau}_{a}^{ml}$ can be derived as follows:
\begin{eqnarray*}\label{asytaua}
	&\sqrt{N}\left(\hat{\tau}_{a}^{ml}-\late\right) \stackrel{d}{\longrightarrow} \mathcal{N}(0,V_{{\tau}_{a}^{ml}}),
\end{eqnarray*}
where
\begingroup
\allowdisplaybreaks
\begin{eqnarray*}\label{vtaug}
	V_{\tau_a^{ml}} &=& -\left(\frac{1}{\Gamma}E_{\Delta,\alpha}-\frac{\late}{\Gamma}E_{\Gamma,\alpha}\right)\left(-E_{H_{\alpha}^{ml}}\right)^{-1}\left(\frac{1}{\Gamma}E_{\Delta,\alpha}-\frac{\late}{\Gamma}E_{\Gamma,\alpha}\right)' \; + \; \e \left[\left(\frac{1}{\Gamma}\psi_{\Delta}-\frac{\late}{
		\Gamma}\psi_{\Gamma}\right)^2\right]
\end{eqnarray*}
\endgroup
with
\begingroup
\allowdisplaybreaks
\begin{eqnarray*}
	\psi_{\Delta}&=&\frac{Z_i
		Y_i}{F(X_i,\alpha)}-\frac{(1-Z_i)Y_i}{1-F(X_i,\alpha)}-\Delta, \\
	\psi_{\Gamma}&=&1-\frac{(1-Z_i)D_i}{1-F(X_i,\alpha)} - \frac{Z_i
		(1-D_i)}{F(X_i,\alpha)}-\Gamma, \\
	E_{\Delta,\alpha} &=& \e \left[\frac{\partial \psi_{\Delta}}{\partial \alpha}\right] =   \e \left[-\left(\frac{Y Z}{F(X,\alpha)^2}+\frac{Y (1-Z)}{(1-F(X,\alpha))^2}\right)\nabla_{\alpha}F(X,\alpha)\right], \\
	E_{\Gamma,\alpha} &=& \e \left[\frac{\partial \psi_{\Gamma}}{\partial \alpha}\right] = \e \left[\left(\frac{(1-D)Z}{F(X,\alpha)^2}-\frac{D(1-Z)}{(1-F(X,\alpha))^2}\right)\nabla_{\alpha}F(X,\alpha)\right], \\
	E_{H_{\alpha}^{ml}}&=& \e \left[\pd{\psi_{\alpha}^{ml}}{\alpha'}\right] = \e \left[H(X,\alpha)\right],
\end{eqnarray*}
\endgroup
and $H(X,\alpha)$ denotes the Hessian of the log-likelihood of $\alpha$.

The estimators $\hat{\tau}_{a,1}^{ml}$ ($= \hat{\tau}_{t}^{ml}$) and  $\hat{\tau}_{a,0}^{ml}$ use the same moment functions for $\alpha$ and $\Delta$ as $\hat{\tau}_{a}^{ml}$. However, they estimate the population proportion of compliers using the moment functions derived from the population relations  $\Gamma_1$ and $\Gamma_0$, respectively. The variances of $\hat{\tau}_{a,1}^{ml}$ and  $\hat{\tau}_{a,0}^{ml}$ have the same form as  $\hat{\tau}_{a}^{ml}$, where $\Gamma$ is replaced with $\Gamma_1$ and $\Gamma_0$. Thus, the asymptotic distributions of $\hat{\tau}_{a,1}^{ml}$  and $\hat{\tau}_{a,0}^{ml}$ can be summarized as follows:
\begin{eqnarray*}\label{asytaua1}
	&\sqrt{N}\left(\hat{\tau}_{a,1}^{ml}-\late\right) \stackrel{d}{\longrightarrow} \mathcal{N}(0,V_{{\tau}_{a,1}^{ml}}),
\end{eqnarray*}
where
\begingroup
\allowdisplaybreaks
\begin{eqnarray*}\label{vtauga1}
	V_{{\tau}_{a,1}^{ml}} &=& -\left(\frac{1}{\Gamma_1}E_{\Delta,\alpha}-\frac{\late}{\Gamma_1}E_{\Gamma_1,\alpha}\right)\left(-E_{H_{\alpha}^{ml}}\right)^{-1}\left(\frac{1}{\Gamma_1}E_{\Delta,\alpha}-\frac{\late}{\Gamma_1}E_{\Gamma_1,\alpha}\right)' \; + \; \e \left[\left(\frac{1}{\Gamma_1}\psi_{\Delta}-\frac{\late}{
		\Gamma_1}\psi_{\Gamma_1}\right)^2\right]
\end{eqnarray*}
\endgroup
with
\begingroup
\allowdisplaybreaks
\begin{eqnarray*}
	\psi_{\Gamma_1}&=&\frac{Z_i
		Y_i}{F(X_i,\alpha)}-\frac{(1-Z_i)Y_i}{1-F(X_i,\alpha)}-\Gamma_1, \\
	E_{\Gamma_1,\alpha} &=& \e \left[-\left(\frac{D Z}{F(X,\alpha)^2}+\frac{D (1-Z)}{(1-F(X,\alpha))^2}\right)\nabla_{\alpha}F(X,\alpha)\right],
\end{eqnarray*}
\endgroup
and
\begin{eqnarray*}\label{asytaua0}
	&\sqrt{N}\left(\hat{\tau}_{a,0}^{ml}-\late\right) \stackrel{d}{\longrightarrow} \mathcal{N}(0,V_{{\tau}_{a,0}^{ml}}),
\end{eqnarray*}
where
\begingroup
\allowdisplaybreaks
\begin{eqnarray*}\label{vtauga0}
	V_{{\tau}_{a,0}^{ml}} &=& -\left(\frac{1}{\Gamma_0}E_{\Delta,\alpha}-\frac{\late}{\Gamma_0}E_{\Gamma_0,\alpha}\right)\left(-E_{H_{\alpha}^{ml}}\right)^{-1}\left(\frac{1}{\Gamma_0}E_{\Delta,\alpha}-\frac{\late}{\Gamma_0}E_{\Gamma_0,\alpha}\right)' \; + \; \e \left[\left(\frac{1}{\Gamma_0}\psi_{\Delta}-\frac{\late}{
		\Gamma_0}\psi_{\Gamma_0}\right)^2\right]
\end{eqnarray*}
\endgroup
with
\begingroup
\allowdisplaybreaks
\begin{eqnarray*}
	\psi_{\Gamma_0}&=&  \frac{Z_i
		(D_i-1)}{F(X_i,\alpha)}-\frac{(1-Z_i)(D_i-1)}{1-F(X_i,\alpha)}-\Gamma_0, \\
	E_{\Gamma_0,\alpha} &=& \e \left[\frac{\partial \psi_{\Gamma_0}}{\partial \alpha}\right]  = \e \left[-\left(\frac{(D-1)Z}{F(X,\alpha)^2}+\frac{(D-1)(1-Z)}{(1-F(X,\alpha))^2}\right)\nabla_{\alpha}F(X,\alpha)\right].
\end{eqnarray*}
\endgroup
The estimator $\hat{\tau}_{a,10}^{ml}$ is essentially the difference of two ratio estimators whose covariance is zero. Thus, the variance of the difference is the sum of variances of the two estimators. It follows that
\begin{eqnarray*}\label{asytaua10}
	&\sqrt{N}\left(\hat{\tau}_{a,10}^{ml}-\late\right) \stackrel{d}{\longrightarrow} \mathcal{N}(0,V_{{\tau}_{a,10}^{ml}}),
\end{eqnarray*}
where
\begingroup
\allowdisplaybreaks
\begin{eqnarray*}\label{vtauga10}
	V_{{\tau}_{a,10}^{ml}} &=& -\left(\frac{E_{\Delta_1,\alpha}}{\Gamma_1}-\frac{E_{\Delta_0,\alpha}}{\Gamma_0}-\frac{\Delta_1E_{\Gamma_1,\alpha}}{\Gamma_1^2}+\frac{\Delta_0E_{\Gamma_0,\alpha}}{\Gamma_0^2}\right)\left(-E_{H_{\alpha}^{ml}}\right)^{-1}\left(\frac{E_{\Delta_1,\alpha}}{\Gamma_1}-\frac{E_{\Delta_0,\alpha}}{\Gamma_0}-\frac{\Delta_1E_{\Gamma_1,\alpha}}{\Gamma_1^2}+\frac{\Delta_0E_{\Gamma_0,\alpha}}{\Gamma_0^2}\right)'	\nonumber\\
	&+&\e{\left(\frac{1}{\Gamma_1}\psi_{\Delta_1}-\frac{\Delta_1}{\Gamma_1^2}\psi_{\Gamma_1}\right)^2}+\e{\left(\frac{1}{\Gamma_0}\psi_{\Delta_0}-\frac{\Delta_0}{\Gamma_0^2}\psi_{\Gamma_0}\right)^2}
\end{eqnarray*}
\endgroup
with
\begingroup
\allowdisplaybreaks
\begin{eqnarray*}
	\psi_{\Delta_1}&=&D_i\frac{Z_i-F(X_i,\alpha)}{F(X_i,\alpha)(1-F(X_i,\alpha))}Y_i-\Delta_1, \\
	\psi_{\Delta_0}&=&(1-D_i)\frac{(1-Z_i)-(1-F(X_i,\alpha))}{F(X_i,\alpha)(1-F(X_i,\alpha))}Y_i-\Delta_0, \\
	E_{\Delta_1,\alpha} &=& \e \left[\frac{\partial \psi_{\Delta_1}}{\partial \alpha}\right] = \e \left[-\left(\frac{DYZ}{F(X,\alpha)^2}+\frac{DY(1-Z)}{(1-F(X,\alpha))^2}\right)\nabla_{\alpha}F(X,\alpha)\right], \\
	E_{\Delta_0,\alpha} &=&  \e \left[\frac{\partial \psi_{\Delta_0}}{\partial \alpha}\right] = \e \left[-\left(\frac{(D-1)YZ}{F(X,\alpha)^2}+\frac{(D-1)Y(1-Z)}{(1-F(X,\alpha))^2}\right)\nabla_{\alpha}F(X,\alpha)\right].
\end{eqnarray*}
\endgroup
Finally, we examine the estimators $\hat{\tau}_{u}^{ml}$ and  $\hat{\tau}_{u}^{cb}$. The key distinction between them is the method used to estimate the instrument propensity score. The instrument propensity score is estimated using maximum likelihood for   $\hat{\tau}_{u}^{ml}$, while it is estimated using covariate balancing for  $\hat{\tau}_{u}^{cb}$. As a result, the former employs $\psi_{\alpha}^{ml}$  whereas the latter uses $\psi_{\alpha}^{cb} $ within the M-estimation framework. Thus, the moment function related to the estimation of $\alpha$ and the appropriate moment functions that take normalization into account can be used to obtain the asymptotic distribution:
\begin{eqnarray*}\label{asytaunorm}
	&\sqrt{N}\left(\hat{\tau}_{u}^{ml}-\late\right) \stackrel{d}{\longrightarrow} \mathcal{N}(0,V_{{\tau}_{u}^{ml}}),
\end{eqnarray*}
where
\begingroup
\allowdisplaybreaks
\begin{eqnarray*}\label{vtaunorm}
	V_{{\tau}_{u}^{ml}} &=& -\left(\frac{1}{\Gamma}(E_{\mu_1, \alpha}-E_{\mu_0, \alpha})-\frac{\Delta}{\Gamma^2}(E_{m_1, \alpha}-E_{m_0, \alpha})\right)\left(-E_{H_{\alpha}^{ml}}\right)^{-1}\left(\frac{1}{\Gamma}(E_{\mu_1, \alpha}-E_{\mu_0, \alpha})-\frac{\Delta}{\Gamma^2}(E_{m_1, \alpha}-E_{m_0, \alpha})\right)' \nonumber\\
	&+& \e{\left(\frac{1}{\Gamma}\psi_{\mu_1}-\frac{\Delta}{\Gamma^2}\psi_{m_1}\right)^2}+\e{\left(\frac{1}{\Gamma}\psi_{\mu_0}-\frac{\Delta}{\Gamma^2}\psi_{m_0}\right)^2}
\end{eqnarray*}
\endgroup
with
\begingroup
\allowdisplaybreaks
\begin{eqnarray*}
	\psi_{\mu_1}&=&\frac{Z_i(Y_i-\mu_1)}{F(X_i,\alpha)}, \quad
	\psi_{\mu_0}=\frac{(1-Z_i)(Y_i-\mu_0)}{1-F(X_i,\alpha)}, \\
	\psi_{m_1}&=&\frac{Z_i(D_i-m_1)}{F(X_i,\alpha)}, \quad
	\psi_{m_0}=\frac{(1-Z_i)(D_i-m_0)}{1-F(X_i,\alpha)}, \\
	E_{\mu_1,\alpha} &=& \e \left[ \frac{\partial \psi_{\mu_1}}{\partial \alpha} \right] = \e\left[-  \frac{Z(Y-\mu_1)}{F(X,\alpha)^2} \nabla_{\alpha}F(X,\alpha) \right], \\
	E_{\mu_0,\alpha} &=& \e \left[ \frac{\partial \psi_{\mu_0}}{\partial \alpha} \right] = \e\left[-  \frac{(1-Z)(Y-\mu_1)}{(1-F(X,\alpha))^2} \nabla_{\alpha}F(X,\alpha) \right], \\
	E_{m_1,\alpha} &=& \e \left[ \frac{\partial \psi_{m_1}}{\partial \alpha} \right] = \e\left[-  \frac{Z(D-m_1)}{F(X,\alpha)^2} \nabla_{\alpha}F(X,\alpha) \right], \\
	E_{m_0,\alpha} &=& \e \left[ \frac{\partial \psi_{m_0}}{\partial \alpha} \right] = \e\left[-  \frac{(1-Z)(D-m_1)}{(1-F(X,\alpha))^2} \nabla_{\alpha}F(X,\alpha) \right],
\end{eqnarray*}
\endgroup
and 
\begin{eqnarray*}\label{asytaucb}
	&\sqrt{N}\left(\hat{\tau}_{u}^{cb}-\late\right) \stackrel{d}{\longrightarrow} \mathcal{N}(0,V_{{\tau}_{u}^{cb}}),
\end{eqnarray*}
where
\begingroup
\allowdisplaybreaks
\begin{eqnarray*}\label{vtaucb}
	V_{{\tau}_{u}^{cb}} &=& \left(\frac{1}{\Gamma}(E_{\mu_1, \alpha}-E_{\mu_0, \alpha})-\frac{\Delta}{\Gamma^2}(E_{m_1, \alpha}-E_{m_0, \alpha})\right)\left(-E_{H_{\alpha}^{cb}}\right)^{-1}V_{\alpha,cb}\left(-E_{H_{\alpha}^{cb}}\right)^{-1}\left(\frac{1}{\Gamma}(E_{\mu_1, \alpha}-E_{\mu_0, \alpha})-\frac{\Delta}{\Gamma^2}(E_{m_1, \alpha}-E_{m_0, \alpha})\right)' \nonumber\\
	&-&2\left(\frac{1}{\Gamma}(V_{\mu_1, \alpha}-V_{\mu_0, \alpha})-\frac{\Delta}{\Gamma^2}(V_{m_1, \alpha}-V_{m_0, \alpha})\right)\left(-E_{H_{\alpha}^{cb}}\right)^{-1}\left(\frac{1}{\Gamma}(E_{\mu_1, \alpha}-E_{\mu_0, \alpha})-\frac{\Delta}{\Gamma^2}(E_{m_1, \alpha}-E_{m_0, \alpha})\right)' \nonumber\\
	&+& \e{\left(\frac{1}{\Gamma}\psi_{\mu_1}-\frac{\Delta}{\Gamma^2}\psi_{m_1}\right)^2} \; + \; \e{\left(\frac{1}{\Gamma}\psi_{\mu_0}-\frac{\Delta}{\Gamma^2}\psi_{m_0}\right)^2}
\end{eqnarray*}
\endgroup
with
\begingroup
\allowdisplaybreaks
\begin{eqnarray*}
	E_{H_{\alpha}^{cb}}&=&\e  \left[ \frac{\partial \psi_{\alpha}^{cb}}{\partial \alpha} \right], \\
	V_{\alpha}^{cb} &=&\e \left[\psi_{\alpha}^{cb} (\cdot)\psi_{\alpha}^{cb} (\cdot)'\right], \\
	V_{\mu_1,\alpha} &=& \e \left[ \psi_{\mu_1}\psi_{\alpha}^{cb}  \right] = \e\left[\frac{Z_i(Y_i-\mu_1)}{F(X_i,\alpha)^2} \frac{(Z_i-F(X_i,\alpha))}{(1-F(X_i,\alpha))}X_i\right], \\
	V_{\mu_0,\alpha} &=& \e \left[ \psi_{\mu_0}\psi_{\alpha}^{cb}   \right] = \e\left[\frac{(1-Z_i)(Y_i-\mu_0)}{(1-F(X_i,\alpha))^2} \frac{(Z_i-F(X_i,\alpha))}{F(X_i,\alpha)}X_i\right], \\
	V_{m_1,\alpha} &=& \e \left[ \psi_{m_1}\psi_{\alpha}^{cb}  \right] = \e\left[\frac{Z_i(D_i-m_1)}{F(X_i,\alpha)} \frac{Z_i-F(X_i,\alpha)}{F(X_i,\alpha)(1-F(X_i,\alpha))}X_i\right], \\
	V_{m_0,\alpha} &=& \e \left[ \psi_{m_0}\psi_{\alpha}^{cb}   \right] = \e\left[\frac{(1-Z_i)(D_i-m_0)}{1-F(X_i,\alpha)} \frac{Z_i-F(X_i,\alpha)}{F(X_i,\alpha)(1-F(X_i,\alpha))}X_i\right].
\end{eqnarray*}
\endgroup
In fact, $V_{{\tau}_{u}^{ml}}$ has the same structure as $V_{\tau_{u}^{cb}}$, but it enjoys some additional simplifications when the ML-based moment condition is used to estimate $p(X)$. Namely, $ \e \left[\pd{\psi_{\alpha}^{cb} (\cdot) }{\alpha'}\right] = -\e \left[\psi_{\alpha}^{cb} (\cdot)\psi_{\alpha}^{cb} (\cdot)'\right]$, $\e \left[\frac{\partial \psi_{\mu_z}}{\partial \alpha}\right] = -\e \left[\psi_{\mu_z}(\cdot)\psi_{\alpha}^{cb} (\cdot)'\right]$, and $\e \left[\frac{\partial \psi_{m_z}}{\partial \alpha}\right] = -\e \left[\psi_{m_z}(\cdot)\psi_{\alpha}^{cb} (\cdot)'\right]$ for $z=0,1$.

\pagebreak

\setcounter{table}{0}
\renewcommand{\thetable}{A.\arabic{table}}
\setcounter{figure}{0}
\renewcommand{\thefigure}{A.\arabic{figure}}

\begin{table}[!t]
	\begin{adjustwidth}{-1in}{-1in}
		\centering
		\renewcommand*{\arraystretch}{1.75}
		\begin{threeparttable}
			\caption{Parameters and Moment Functions\label{tab:moments}}
			\begin{tabular}{lll}
				\hline\hline
				\multicolumn{1}{l}{Parameter} & Population Relation & Related Moment Condition \\
				\hline
				$\alpha$ & $\pr (Z=1 \mid X)=F(X,\alpha)$&$\psi_{\alpha}^{ml}=                 \frac{Z_i-F(X_i,\alpha)}{F(X_i,\alpha)(1-F(X_i,\alpha))}\nabla_{\alpha}F(X_i,\alpha)$ \\
				&&$\psi_{\alpha}^{cb}=\frac{Z_i-F(X_i,\alpha)}{F(X_i,\alpha)(1-F(X_i,\alpha))}X_i$\\
				$\Delta$ & $\Delta=\e \left[Y  \frac{Z  - p(X)}{p(X) \left( 1 - p(X)\right)}\right]$&$\psi_{\Delta}=\frac{Z_i
					Y_i}{F(X_i,\alpha)}-\frac{(1-Z_i)Y_i}{1-F(X_i,\alpha)}-\Delta$\\
				$\Gamma$ &$\Gamma=\e \left[1 - \frac{D \left( 1 - Z \right)}{1 - p(X)} - \frac{\left( 1 - D \right) Z}{p(X)}\right]$ &$\psi_{\Gamma}=1-\frac{(1-Z_i)D_i}{1-F(X_i,\alpha)} - \frac{Z_i
					(1-D_i)}{F(X_i,\alpha)}-\Gamma$ \\
				$\Gamma_1$ &$\Gamma_1=\e \left[D \frac{Z - p(X)}{p(X) \left( 1 - p(X) \right)}\right]$ &$\psi_{\Gamma_1}=\frac{Z_i
					D_i}{F(X_i,\alpha)}-\frac{(1-Z_i)D_i}{1-F(X_i,\alpha)}-\Gamma_1$ \\
				$\Gamma_0$ &$\Gamma_0=\e \left[\left( 1 - D \right) \frac{\left( 1 - Z \right) - \left( 1 - p(X) \right)}{p(X) \left( 1 - p(X) \right)}\right]$ &$\psi_{\Gamma_0}=\frac{Z_i
					(D_i-1)}{F(X_i,\alpha)}-\frac{(1-Z_i)(D_i-1)}{1-F(X_i,\alpha)}-\Gamma_0$ \\
				$\Delta_1$ & $\Delta_1=\e(\kappa_{1}Y)$&$\psi_{\Delta_1}=D_i\frac{Z_i-F(X_i,\alpha)}{F(X_i,\alpha)(1-F(X_i,\alpha))}Y_i-\Delta_1$\\
				$\Delta_0$ & $\Delta_0=\e(\kappa_{0}Y)$&$\psi_{\Delta_0}=(1-D_i)\frac{(1-Z_i)-(1-F(X_i,\alpha))}{F(X_i,\alpha)(1-F(X_i,\alpha))}Y_i-\Delta_0$\\
				$\mu_1$ & $\mu_1=\e ( Y \mid Z=1 )$&$\psi_{\mu_1}=\frac{Z_i(Y_i-\mu_1)}{F(X_i,\alpha)}$\\
				$\mu_0$ &$\mu_0=\e ( Y \mid Z=0 )$ &$\psi_{\mu_0}=\frac{(1-Z_i)(Y_i-\mu_0)}{1-F(X_i,\alpha)}$ \\
				$m_1$ &$m_1=\e ( D \mid Z=1 )$&$\psi_{m_1}=\frac{Z_i(D_i-m_1)}{F(X_i,\alpha)}$\\
				$m_0$ &$m_0=\e ( D \mid Z=0 )$&$\psi_{m_0}=\frac{(1-Z_i)(D_i-m_0)}{1-F(X_i,\alpha)}$\\
				$\late $ &$\late = \frac{\Delta}{\Gamma} = \frac{\Delta}{\Gamma_1} =\frac{\Delta}{\Gamma_0} =\frac{\Delta_1}{\Gamma_1}-\frac{\Delta_0}{\Gamma_0} = \frac{\mu_1-\mu_0}{m_1-m_0}$&$\psi_{\tau_a}=\frac{\Delta}{\Gamma}-\tau_a$\\
				& &$\psi_{\tau_{a,1}}=\frac{\Delta}{\Gamma_1}-\tau_{a,1}$\\
				& &$\psi_{\tau_{a,0}}=\frac{\Delta}{\Gamma_0}-\tau_{a,0}$\\
				& &$\psi_{\tau_{a,10}}=\frac{\Delta_1}{\Gamma_1}-\frac{\Delta_0}{\Gamma_0}-\tau_{a,10}$\\
				& &$\psi_{{\tau}_{u}}=\frac{\mu_1-\mu_0}{m_1-m_0}-{\tau}_{u}$\\
				\hline
			\end{tabular}
			\begin{footnotesize}
				\begin{tablenotes}[flushleft]
					\item
				\end{tablenotes}
			\end{footnotesize}
		\end{threeparttable}
	\end{adjustwidth}
\end{table}

\setcounter{table}{0}
\renewcommand{\thetable}{B.\arabic{table}}
\setcounter{figure}{0}
\renewcommand{\thefigure}{B.\arabic{figure}}

\begin{table}[!p]
\begin{adjustwidth}{-1in}{-1in}
\centering
\begin{threeparttable}
\caption{Simulation Results for Design A.1\label{tab:simA1}}
\begin{tabular}{cc>{\centering\arraybackslash}m{1.75cm} >{\centering\arraybackslash}m{1.75cm} >{\centering\arraybackslash}m{1.75cm} >{\centering\arraybackslash}m{1.75cm} >{\centering\arraybackslash}m{1.75cm} >{\centering\arraybackslash}m{1.75cm} >{\centering\arraybackslash}m{1.75cm}}
\hline\hline
     \multicolumn{3}{c}{} & \multicolumn{3}{c}{Normalized estimators} & \multicolumn{3}{c}{Unnormalized estimators} \\
    \cline{4-6}
    \cline{7-9}
          &       & 2SLS    & $\hat{\tau}_{u}^{cb}$ & $\hat{\tau}_{u}^{ml}$ & $\hat{\tau}_{a,10}^{ml}$ & $\hat{\tau}_{a}^{ml}$ & $\hat{\tau}_{t}^{ml} = \hat{\tau}_{a,1}^{ml}$ & $\hat{\tau}_{a,0}^{ml}$ \\
\hline
    \multicolumn{2}{c}{$\delta=0.01$} &       &       &       &       &       &  \\
    \cline{1-2}
     \multicolumn{1}{l}{$N=500$} & MSE   & 1     & 2.70 & 2.63  & 1093.84 & 14.16 & 1304.62 & 3.12 \\
          & |B|   & 0.0095 & 0.0215 & 0.0216 & 0.1852 & 0.0365 & 0.1813 & 0.0333 \\
          & Coverage rate & 0.96  & 0.88  & 0.92  & 0.93  & 0.94  & 0.94  & 0.93 \\
          &       &       &       &       &       &       &  \\
     \multicolumn{1}{l}{$N=1{,}000$} & MSE   & 1     & 2.75 & 2.72  & 4.11  & 3.45  & 4.36  & 3.07 \\
          & |B|   & 0.0052 & 0.0090 & 0.0080 & 0.0359 & 0.0096 & 0.0357 & 0.0130 \\
          & Coverage rate & 0.95  & 0.91  & 0.93  & 0.94  & 0.94  & 0.95  & 0.93 \\
          &       &       &       &       &       &       &  \\
     \multicolumn{1}{l}{$N=5{,}000$} & MSE   & 1     & 2.71  & 2.69  & 3.00  & 2.84  & 3.02  & 2.98 \\
          & |B|   & 0.0003 & 0.0023 & 0.0023 & 0.0058 & 0.0018 & 0.0057 & 0.0035 \\
          & Coverage rate & 0.95  & 0.94  & 0.95  & 0.95  & 0.95  & 0.95  & 0.95 \\
          &       &       &       &       &       &       &  \\
     \multicolumn{2}{c}{$\delta=0.02$} &       &       &       &       &       &  \\
     \cline{1-2}
     \multicolumn{1}{l}{$N=500$} & MSE   & 1     & 1.93 & 1.91  & 20.87 & 2.94  & 20.67 & 2.11 \\
          & |B|   & 0.0097 & 0.0154 & 0.0153 & 0.0492 & 0.0211 & 0.0495 & 0.0215 \\
          & Coverage rate & 0.96  & 0.91  & 0.93  & 0.94  & 0.94  & 0.94  & 0.93 \\
          &       &       &       &       &       &       &  \\
     \multicolumn{1}{l}{$N=1{,}000$} & MSE   & 1     & 1.89 & 1.88  & 2.14  & 2.00  & 2.18  & 2.03 \\
          & |B|   & 0.0027 & 0.0057 & 0.0056 & 0.0148 & 0.0058 & 0.0149 & 0.0082 \\
          & Coverage rate & 0.95  & 0.93  & 0.94  & 0.95  & 0.95  & 0.95  & 0.94 \\
          &       &       &       &       &       &       &  \\
     \multicolumn{1}{l}{$N=5{,}000$} & MSE   & 1     & 1.86  & 1.85  & 2.00  & 1.90  & 2.01  & 1.98 \\
          & |B|   & 0.0026 & 0.0032 & 0.0032 & 0.0048 & 0.0030 & 0.0048 & 0.0037 \\
          & Coverage rate & 0.95  & 0.95  & 0.95  & 0.95  & 0.95  & 0.95  & 0.95 \\
          &       &       &       &       &       &       &  \\
     \multicolumn{2}{c}{$\delta=0.05$} &       &       &       &       &       &  \\
     \cline{1-2}
     \multicolumn{1}{l}{$N=500$} & MSE   & 1     & 1.33 & 1.32  & 1.43  & 1.36  & 1.46  & 1.37 \\
          & |B|   & 0.0016 & 0.0026 & 0.0024 & 0.0089 & 0.0025 & 0.0088 & 0.0036 \\
          & Coverage rate & 0.95  & 0.94  & 0.94  & 0.95  & 0.94  & 0.95  & 0.94 \\
          &       &       &       &       &       &       &  \\
     \multicolumn{1}{l}{$N=1{,}000$} & MSE   & 1     & 1.32 & 1.31  & 1.38  & 1.33  & 1.39  & 1.36 \\
          & |B|   & 0.0022 & 0.0001 & 0.0001 & 0.0024 & 0.0001 & 0.0024 & 0.0009 \\
          & Coverage rate & 0.95  & 0.94  & 0.95  & 0.95  & 0.95  & 0.95  & 0.95 \\
          &       &       &       &       &       &       &  \\
     \multicolumn{1}{l}{$N=5{,}000$} & MSE   & 1     & 1.31  & 1.31  & 1.35  & 1.32  & 1.35  & 1.36 \\
          & |B|   & 0.0000 & 0.0000 & 0.0000 & 0.0005 & 0.0000 & 0.0005 & 0.0001 \\
          & Coverage rate & 0.95  & 0.95  & 0.95  & 0.95  & 0.95  & 0.95  & 0.95 \\
\hline
\end{tabular}
\begin{footnotesize}
\begin{tablenotes}[flushleft]
\item \textit{Notes:} The details of this simulation design are provided in Section \ref{sec:simulation}. ``MSE'' is the mean squared error of an estimator, normalized by the mean squared error of 2SLS\@. ``|B|'' is the absolute bias. ``Coverage rate'' is the coverage rate for a nominal 95\% confidence interval. ``2SLS'' is the 2SLS estimator that additively controls for $X$\@. The weighting estimators are defined in Section \ref{sec:estandinf}. All weighting estimators also control for $X$\@. Results are based on 10,000 replications.
\end{tablenotes}
\end{footnotesize}
\end{threeparttable}
\end{adjustwidth}
\end{table}

\begin{table}[!p]
\begin{adjustwidth}{-1in}{-1in}
\centering
\begin{threeparttable}
\caption{Simulation Results for Design A.2\label{tab:simA2}}
\begin{tabular}{cc>{\centering\arraybackslash}m{1.75cm} >{\centering\arraybackslash}m{1.75cm} >{\centering\arraybackslash}m{1.75cm} >{\centering\arraybackslash}m{1.75cm} >{\centering\arraybackslash}m{1.75cm} >{\centering\arraybackslash}m{1.75cm} >{\centering\arraybackslash}m{1.75cm}}
\hline\hline
     \multicolumn{3}{c}{} & \multicolumn{3}{c}{Normalized estimators} & \multicolumn{3}{c}{Unnormalized estimators} \\
    \cline{4-6}
    \cline{7-9}
          &       & 2SLS    & $\hat{\tau}_{u}^{cb}$ & $\hat{\tau}_{u}^{ml}$ & $\hat{\tau}_{a,10}^{ml}$ & $\hat{\tau}_{a}^{ml}$ & $\hat{\tau}_{t}^{ml} = \hat{\tau}_{a,1}^{ml}$ & $\hat{\tau}_{a,0}^{ml}$ \\
\hline
    \multicolumn{2}{c}{$\delta=0.01$} &       &       &       &       &       &  \\
    \cline{1-2}
     \multicolumn{1}{l}{$N=500$} & MSE   & 1     & 2.75 & 2.78  & 2.30e+04 & 6.83  & 3.09  & 2.52e+04 \\
          & |B|   & 0.0023 & 0.0033 & 0.0028 & 0.4066 & 0.0046 & 0.0025 & 0.4334 \\
          & Coverage rate & 0.96  & 0.88  & 0.93  & 0.93  & 0.96  & 0.93  & 0.94 \\
          &       &       &       &       &       &       &  \\
     \multicolumn{1}{l}{$N=1{,}000$} & MSE   & 1     & 2.63 & 2.60  & 3.03  & 2.92  & 2.72  & 3.26 \\
          & |B|   & 0.0017 & 0.0013 & 0.0010 & 0.0008 & 0.0006 & 0.0011 & 0.0008 \\
          & Coverage rate & 0.95  & 0.91  & 0.94  & 0.94  & 0.96  & 0.94  & 0.95 \\
          &       &       &       &       &       &       &  \\
     \multicolumn{1}{l}{$N=5{,}000$} & MSE   & 1     & 2.72  & 2.71  & 2.76  & 2.76  & 2.73  & 2.79 \\
          & |B|   & 0.0008 & 0.0018 & 0.0018 & 0.0018 & 0.0017 & 0.0017 & 0.0017 \\
          & Coverage rate & 0.95  & 0.94  & 0.95  & 0.95  & 0.95  & 0.95  & 0.95 \\
          &       &       &       &       &       &       &  \\
     \multicolumn{2}{c}{$\delta=0.02$} &       &       &       &       &       &  \\
     \cline{1-2}
     \multicolumn{1}{l}{$N=500$} & MSE   & 1     & 1.93 & 1.91  & 2.31  & 2.16  & 2.00  & 2.44 \\
          & |B|   & 0.0029 & 0.0027 & 0.0025 & 0.0026 & 0.0034 & 0.0028 & 0.0031 \\
          & Coverage rate & 0.95  & 0.91  & 0.93  & 0.94  & 0.95  & 0.94  & 0.95 \\
          &       &       &       &       &       &       &  \\
     \multicolumn{1}{l}{$N=1{,}000$} & MSE   & 1     & 1.86 & 1.84  & 1.92  & 1.90  & 1.88  & 1.96 \\
          & |B|   & 0.0019 & 0.0028 & 0.0032 & 0.0035 & 0.0034 & 0.0034 & 0.0035 \\
          & Coverage rate & 0.95  & 0.93  & 0.94  & 0.95  & 0.95  & 0.95  & 0.95 \\
          &       &       &       &       &       &       &  \\
     \multicolumn{1}{l}{$N=5{,}000$} & MSE   & 1     & 1.91  & 1.90  & 1.92  & 1.91  & 1.91  & 1.93 \\
          & |B|   & 0.0006 & 0.0007 & 0.0008 & 0.0008 & 0.0008 & 0.0008 & 0.0008 \\
          & Coverage rate & 0.95  & 0.94  & 0.95  & 0.95  & 0.95  & 0.95  & 0.95 \\
          &       &       &       &       &       &       &  \\
     \multicolumn{2}{c}{$\delta=0.05$} &       &       &       &       &       &  \\
     \cline{1-2}
     \multicolumn{1}{l}{$N=500$} & MSE   & 1     & 1.32 & 1.31  & 1.36  & 1.34  & 1.32  & 1.39 \\
          & |B|   & 0.0008 & 0.0012 & 0.0013 & 0.0018 & 0.0016 & 0.0015 & 0.0017 \\
          & Coverage rate & 0.95  & 0.94  & 0.94  & 0.94  & 0.94  & 0.94  & 0.95 \\
          &       &       &       &       &       &       &  \\
     \multicolumn{1}{l}{$N=1{,}000$} & MSE   & 1     & 1.30 & 1.30  & 1.31  & 1.31  & 1.31  & 1.32 \\
          & |B|   & 0.0003 & 0.0008 & 0.0008 & 0.0007 & 0.0007 & 0.0010 & 0.0005 \\
          & Coverage rate & 0.95  & 0.95  & 0.95  & 0.95  & 0.95  & 0.95  & 0.95 \\
          &       &       &       &       &       &       &  \\
     \multicolumn{1}{l}{$N=5{,}000$} & MSE   & 1     & 1.30  & 1.30  & 1.30  & 1.30  & 1.30  & 1.30 \\
          & |B|   & 0.0005 & 0.0008 & 0.0008 & 0.0008 & 0.0008 & 0.0008 & 0.0008 \\
          & Coverage rate & 0.95  & 0.95  & 0.95  & 0.95  & 0.95  & 0.95  & 0.95 \\
\hline
\end{tabular}
\begin{footnotesize}
\begin{tablenotes}[flushleft]
\item \textit{Notes:} The details of this simulation design are provided in Section \ref{sec:simulation}. ``MSE'' is the mean squared error of an estimator, normalized by the mean squared error of 2SLS\@. ``|B|'' is the absolute bias. ``Coverage rate'' is the coverage rate for a nominal 95\% confidence interval. ``2SLS'' is the 2SLS estimator that additively controls for $X$\@. The weighting estimators are defined in Section \ref{sec:estandinf}. All weighting estimators also control for $X$\@. Results are based on 10,000 replications.
\end{tablenotes}
\end{footnotesize}
\end{threeparttable}
\end{adjustwidth}
\end{table}

\begin{table}[!p]
\begin{adjustwidth}{-1in}{-1in}
\centering
\begin{threeparttable}
\caption{Simulation Results for Design B\label{tab:simB}}
\begin{tabular}{cc>{\centering\arraybackslash}m{1.75cm} >{\centering\arraybackslash}m{1.75cm} >{\centering\arraybackslash}m{1.75cm} >{\centering\arraybackslash}m{1.75cm} >{\centering\arraybackslash}m{1.75cm} >{\centering\arraybackslash}m{1.75cm} >{\centering\arraybackslash}m{1.75cm}}
\hline\hline
     \multicolumn{3}{c}{} & \multicolumn{3}{c}{Normalized estimators} & \multicolumn{3}{c}{Unnormalized estimators} \\
    \cline{4-6}
    \cline{7-9}
          &       & 2SLS    & $\hat{\tau}_{u}^{cb}$ & $\hat{\tau}_{u}^{ml}$ & $\hat{\tau}_{a,10}^{ml}$ & $\hat{\tau}_{a}^{ml}$ & $\hat{\tau}_{t}^{ml} = \hat{\tau}_{a,1}^{ml}$ & $\hat{\tau}_{a,0}^{ml}$ \\
\hline
    \multicolumn{2}{c}{$\delta=0.01$} &       &       &       &       &       &  \\
    \cline{1-2}
     \multicolumn{1}{l}{$N=500$} & MSE   & 1     & 2.57 & 2.74  & 189.22 & 210.94 & 761.97 & 4.02 \\
          & |B|   & 0.0614 & 0.0140 & 0.0103 & 0.0490 & 0.0927 & 0.0059 & 0.0197 \\
          & Coverage rate & 0.96  & 0.88  & 0.94  & 0.95  & 0.95  & 0.94  & 0.94 \\
          &       &       &       &       &       &       &  \\
     \multicolumn{1}{l}{$N=1{,}000$} & MSE   & 1     & 2.50 & 2.51  & 6.59  & 3.20  & 7.00  & 2.82 \\
          & |B|   & 0.0551 & 0.0035 & 0.0024 & 0.0323 & 0.0094 & 0.0340 & 0.0065 \\
          & Coverage rate & 0.95  & 0.91  & 0.94  & 0.95  & 0.95  & 0.95  & 0.94 \\
          &       &       &       &       &       &       &  \\
     \multicolumn{1}{l}{$N=5{,}000$} & MSE   & 1     & 1.96  & 1.95  & 2.19  & 2.06  & 2.20  & 2.10 \\
          & |B|   & 0.0531 & 0.0009 & 0.0006 & 0.0046 & 0.0009 & 0.0045 & 0.0014 \\
          & Coverage rate & 0.92  & 0.94  & 0.95  & 0.95  & 0.95  & 0.95  & 0.95 \\
          &       &       &       &       &       &       &  \\
     \multicolumn{2}{c}{$\delta=0.02$} &       &       &       &       &       &  \\
     \cline{1-2}
     \multicolumn{1}{l}{$N=500$} & MSE   & 1     & 1.92 & 1.93  & 11.76 & 2.61  & 16.46 & 2.09 \\
          & |B|   & 0.0498 & 0.0129 & 0.0117 & 0.0534 & 0.0186 & 0.0568 & 0.0142 \\
          & Coverage rate & 0.95  & 0.91  & 0.93  & 0.95  & 0.95  & 0.95  & 0.94 \\
          &       &       &       &       &       &       &  \\
     \multicolumn{1}{l}{$N=1{,}000$} & MSE   & 1     & 1.81 & 1.80  & 2.20  & 1.96  & 2.23  & 1.92 \\
          & |B|   & 0.0473 & 0.0063 & 0.0058 & 0.0182 & 0.0075 & 0.0180 & 0.0069 \\
          & Coverage rate & 0.95  & 0.93  & 0.95  & 0.95  & 0.96  & 0.96  & 0.95 \\
          &       &       &       &       &       &       &  \\
     \multicolumn{1}{l}{$N=5{,}000$} & MSE   & 1     & 1.46  & 1.45  & 1.58  & 1.50  & 1.58  & 1.53 \\
          & |B|   & 0.0436 & 0.0003 & 0.0003 & 0.0021 & 0.0004 & 0.0021 & 0.0006 \\
          & Coverage rate & 0.93  & 0.95  & 0.95  & 0.95  & 0.95  & 0.95  & 0.95 \\
          &       &       &       &       &       &       &  \\
     \multicolumn{2}{c}{$\delta=0.05$} &       &       &       &       &       &  \\
     \cline{1-2}
     \multicolumn{1}{l}{$N=500$} & MSE   & 1     & 1.30 & 1.30  & 5.79  & 1.35  & 5.22  & 1.34 \\
          & |B|   & 0.0334 & 0.0018 & 0.0014 & 0.0141 & 0.0022 & 0.0137 & 0.0016 \\
          & Coverage rate & 0.96  & 0.94  & 0.95  & 0.95  & 0.95  & 0.96  & 0.95 \\
          &       &       &       &       &       &       &  \\
     \multicolumn{1}{l}{$N=1{,}000$} & MSE   & 1     & 1.29 & 1.29  & 1.36  & 1.31  & 1.37  & 1.33 \\
          & |B|   & 0.0335 & 0.0042 & 0.0040 & 0.0073 & 0.0041 & 0.0073 & 0.0041 \\
          & Coverage rate & 0.95  & 0.94  & 0.95  & 0.95  & 0.95  & 0.95  & 0.94 \\
          &       &       &       &       &       &       &  \\
     \multicolumn{1}{l}{$N=5{,}000$} & MSE   & 1     & 1.12  & 1.12  & 1.16  & 1.13  & 1.16  & 1.15 \\
          & |B|   & 0.0309 & 0.0008 & 0.0007 & 0.0012 & 0.0007 & 0.0013 & 0.0008 \\
          & Coverage rate & 0.94  & 0.95  & 0.95  & 0.95  & 0.95  & 0.95  & 0.95 \\
\hline
\end{tabular}
\begin{footnotesize}
\begin{tablenotes}[flushleft]
\item \textit{Notes:} The details of this simulation design are provided in Section \ref{sec:simulation}. ``MSE'' is the mean squared error of an estimator, normalized by the mean squared error of 2SLS\@. ``|B|'' is the absolute bias. ``Coverage rate'' is the coverage rate for a nominal 95\% confidence interval. ``2SLS'' is the 2SLS estimator that additively controls for $X$\@. The weighting estimators are defined in Section \ref{sec:estandinf}. All weighting estimators also control for $X$\@. Results are based on 10,000 replications.
\end{tablenotes}
\end{footnotesize}
\end{threeparttable}
\end{adjustwidth}
\end{table}

\begin{table}[!p]
\begin{adjustwidth}{-1in}{-1in}
\centering
\begin{threeparttable}
\caption{Simulation Results for Design C\label{tab:simC}}
\begin{tabular}{cc>{\centering\arraybackslash}m{1.75cm} >{\centering\arraybackslash}m{1.75cm} >{\centering\arraybackslash}m{1.75cm} >{\centering\arraybackslash}m{1.75cm} >{\centering\arraybackslash}m{1.75cm} >{\centering\arraybackslash}m{1.75cm} >{\centering\arraybackslash}m{1.75cm}}
\hline\hline
     \multicolumn{3}{c}{} & \multicolumn{3}{c}{Normalized estimators} & \multicolumn{3}{c}{Unnormalized estimators} \\
    \cline{4-6}
    \cline{7-9}
          &       & 2SLS    & $\hat{\tau}_{u}^{cb}$ & $\hat{\tau}_{u}^{ml}$ & $\hat{\tau}_{a,10}^{ml}$ & $\hat{\tau}_{a}^{ml}$ & $\hat{\tau}_{t}^{ml} = \hat{\tau}_{a,1}^{ml}$ & $\hat{\tau}_{a,0}^{ml}$ \\
\hline
    \multicolumn{2}{c}{$\delta=0.01$} &       &       &       &       &       &  \\
    \cline{1-2}
     \multicolumn{1}{l}{$N=500$} & MSE   & 1     & 0.75 & 3.82  & 4.95e+04 & 2010.01 & 4.92e+04 & 219.69 \\
          & |B|   & 4.6994 & 0.1184 & 0.7953 & 7.2631 & 2.5598 & 7.2230 & 2.4048 \\
          & Coverage rate & 0.33  & 0.78  & 0.82  & 0.83  & 0.96  & 0.83  & 0.93 \\
          &       &       &       &       &       &       &  \\
     \multicolumn{1}{l}{$N=1{,}000$} & MSE   & 1     & 0.42 & 1.47  & 95.93 & 23.83 & 96.38 & 38.68 \\
          & |B|   & 4.7053 & 0.0938 & 0.3867 & 0.8364 & 1.4320 & 0.8401 & 1.1898 \\
          & Coverage rate & 0.07  & 0.84  & 0.87  & 0.88  & 0.97  & 0.88  & 0.94 \\
          &       &       &       &       &       &       &  \\
     \multicolumn{1}{l}{$N=5{,}000$} & MSE   & 1     & 0.09  & 0.30  & 0.34  & 2.24  & 0.34  & 7.35 \\
          & |B|   & 4.6729 & 0.0415 & 0.0568 & 0.0848 & 0.2707 & 0.0849 & 0.2319 \\
          & Coverage rate & 0.00  & 0.92  & 0.94  & 0.94  & 0.96  & 0.94  & 0.95 \\
          &       &       &       &       &       &       &  \\
     \multicolumn{2}{c}{$\delta=0.02$} &       &       &       &       &       &  \\
     \cline{1-2}
     \multicolumn{1}{l}{$N=500$} & MSE   & 1     & 0.64 & 1.82  & 20.02 & 52.38 & 20.36 & 53.85 \\
          & |B|   & 3.9155 & 0.0580 & 0.4457 & 0.4927 & 1.8422 & 0.4896 & 1.5703 \\
          & Coverage rate & 0.44  & 0.84  & 0.87  & 0.89  & 0.97  & 0.89  & 0.94 \\
          &       &       &       &       &       &       &  \\
     \multicolumn{1}{l}{$N=1{,}000$} & MSE   & 1     & 0.36 & 0.97  & 1.29  & 7.64  & 1.29  & 24.19 \\
          & |B|   & 3.8732 & 0.0521 & 0.1726 & 0.2334 & 0.7182 & 0.2335 & 0.5280 \\
          & Coverage rate & 0.15  & 0.89  & 0.91  & 0.92  & 0.96  & 0.92  & 0.95 \\
          &       &       &       &       &       &       &  \\
     \multicolumn{1}{l}{$N=5{,}000$} & MSE   & 1     & 0.08  & 0.20  & 0.23  & 1.52  & 0.23  & 5.09 \\
          & |B|   & 3.8464 & 0.0124 & 0.0109 & 0.0589 & 0.1196 & 0.0589 & 0.0763 \\
          & Coverage rate & 0.00  & 0.93  & 0.94  & 0.95  & 0.95  & 0.95  & 0.95 \\
          &       &       &       &       &       &       &  \\
     \multicolumn{2}{c}{$\delta=0.05$} &       &       &       &       &       &  \\
     \cline{1-2}
     \multicolumn{1}{l}{$N=500$} & MSE   & 1     & 0.62 & 1.13  & 1.44  & 7.88  & 1.44  & 24.77 \\
          & |B|   & 2.6174 & 0.0767 & 0.1027 & 0.1660 & 0.5604 & 0.1661 & 0.2451 \\
          & Coverage rate & 0.66  & 0.91  & 0.93  & 0.94  & 0.97  & 0.94  & 0.95 \\
          &       &       &       &       &       &       &  \\
     \multicolumn{1}{l}{$N=1{,}000$} & MSE   & 1     & 0.37 & 0.65  & 0.74  & 4.29  & 0.74  & 13.98 \\
          & |B|   & 2.6376 & 0.0319 & 0.0268 & 0.0894 & 0.2009 & 0.0894 & 0.1782 \\
          & Coverage rate & 0.40  & 0.93  & 0.94  & 0.95  & 0.95  & 0.95  & 0.95 \\
          &       &       &       &       &       &       &  \\
     \multicolumn{1}{l}{$N=5{,}000$} & MSE   & 1     & 0.09  & 0.15  & 0.16  & 0.93  & 0.16  & 3.10 \\
          & |B|   & 2.6232 & 0.0029 & 0.0161 & 0.0035 & 0.0294 & 0.0035 & 0.0586 \\
          & Coverage rate & 0.00  & 0.95  & 0.95  & 0.95  & 0.95  & 0.95  & 0.95 \\
\hline
\end{tabular}
\begin{footnotesize}
\begin{tablenotes}[flushleft]
\item \textit{Notes:} The details of this simulation design are provided in Section \ref{sec:simulation}. ``MSE'' is the mean squared error of an estimator, normalized by the mean squared error of 2SLS\@. ``|B|'' is the absolute bias. ``Coverage rate'' is the coverage rate for a nominal 95\% confidence interval. ``2SLS'' is the 2SLS estimator that additively controls for $X$\@. The weighting estimators are defined in Section \ref{sec:estandinf}. All weighting estimators also control for $X$\@. Results are based on 10,000 replications.
\end{tablenotes}
\end{footnotesize}
\end{threeparttable}
\end{adjustwidth}
\end{table}

\begin{table}[!p]
\begin{adjustwidth}{-1in}{-1in}
\centering
\begin{threeparttable}
\caption{Simulation Results for Design D\label{tab:simD}}
\begin{tabular}{cc>{\centering\arraybackslash}m{1.75cm} >{\centering\arraybackslash}m{1.75cm} >{\centering\arraybackslash}m{1.75cm} >{\centering\arraybackslash}m{1.75cm} >{\centering\arraybackslash}m{1.75cm} >{\centering\arraybackslash}m{1.75cm} >{\centering\arraybackslash}m{1.75cm}}
\hline\hline
     \multicolumn{3}{c}{} & \multicolumn{3}{c}{Normalized estimators} & \multicolumn{3}{c}{Unnormalized estimators} \\
    \cline{4-6}
    \cline{7-9}
          &       & 2SLS    & $\hat{\tau}_{u}^{cb}$ & $\hat{\tau}_{u}^{ml}$ & $\hat{\tau}_{a,10}^{ml}$ & $\hat{\tau}_{a}^{ml}$ & $\hat{\tau}_{t}^{ml} = \hat{\tau}_{a,1}^{ml}$ & $\hat{\tau}_{a,0}^{ml}$ \\
\hline
    \multicolumn{2}{c}{$\delta=0.01$} &       &       &       &       &       &  \\
    \cline{1-2}
     \multicolumn{1}{l}{$N=500$} & MSE   & 1     & 0.08 & 7.06  & 0.56  & 2.69e+05 & 0.32  & 1.75e+04 \\
          & |B|   & 17.6766 & 0.6047 & 4.2535 & 0.6326 & 102.1028 & 0.7343 & 82.6894 \\
          & Coverage rate & 0.00  & 0.85  & 0.77  & 0.75  & 0.93  & 0.74  & 0.91 \\
          &       &       &       &       &       &       &  \\
     \multicolumn{1}{l}{$N=1{,}000$} & MSE   & 1     & 0.04 & 3.98  & 2.64  & 1.44e+04 & 0.12  & 1.91e+05 \\
          & |B|   & 17.5275 & 0.4052 & 6.1212 & 1.9580 & 46.4242 & 2.4467 & 46.6583 \\
          & Coverage rate & 0.00  & 0.88  & 0.80  & 0.79  & 0.86  & 0.79  & 0.82 \\
          &       &       &       &       &       &       &  \\
     \multicolumn{1}{l}{$N=5{,}000$} & MSE   & 1     & 0.01  & 0.26  & 0.07  & 11.68 & 0.07  & 23.12 \\
          & |B|   & 17.4073 & 0.3154 & 7.9930 & 3.7953 & 55.3392 & 3.7955 & 78.2082 \\
          & Coverage rate & 0.00  & 0.93  & 0.42  & 0.58  & 0.13  & 0.58  & 0.09 \\
          &       &       &       &       &       &       &  \\
     \multicolumn{2}{c}{$\delta=0.02$} &       &       &       &       &       &  \\
     \cline{1-2}
     \multicolumn{1}{l}{$N=500$} & MSE   & 1     & 0.06 & 0.40  & 0.21  & 7978.30 & 0.16  & 1.12e+04 \\
          & |B|   & 14.1078 & 0.3874 & 4.0705 & 1.3717 & 17.2726 & 1.3658 & 40.6495 \\
          & Coverage rate & 0.00  & 0.89  & 0.84  & 0.84  & 0.89  & 0.83  & 0.86 \\
          &       &       &       &       &       &       &  \\
     \multicolumn{1}{l}{$N=1{,}000$} & MSE   & 1     & 0.03 & 0.27  & 0.09  & 10.24 & 0.09  & 25.76 \\
          & |B|   & 13.9940 & 0.3326 & 4.7909 & 2.0492 & 35.2328 & 2.0474 & 51.9926 \\
          & Coverage rate & 0.00  & 0.91  & 0.83  & 0.84  & 0.75  & 0.84  & 0.70 \\
          &       &       &       &       &       &       &  \\
     \multicolumn{1}{l}{$N=5{,}000$} & MSE   & 1     & 0.01  & 0.18  & 0.05  & 6.64  & 0.05  & 13.56 \\
          & |B|   & 13.9115 & 0.2707 & 5.3737 & 2.5524 & 34.3929 & 2.5523 & 49.3305 \\
          & Coverage rate & 0.00  & 0.95  & 0.36  & 0.61  & 0.02  & 0.61  & 0.01 \\
          &       &       &       &       &       &       &  \\
     \multicolumn{2}{c}{$\delta=0.05$} &       &       &       &       &       &  \\
     \cline{1-2}
     \multicolumn{1}{l}{$N=500$} & MSE   & 1     & 0.06 & 0.24  & 0.12  & 5.29  & 0.12  & 11.84 \\
          & |B|   & 9.1248 & 0.2697 & 2.2155 & 0.8326 & 16.2049 & 0.8327 & 24.8322 \\
          & Coverage rate & 0.01  & 0.93  & 0.90  & 0.91  & 0.82  & 0.91  & 0.80 \\
          &       &       &       &       &       &       &  \\
     \multicolumn{1}{l}{$N=1{,}000$} & MSE   & 1     & 0.03 & 0.15  & 0.06  & 4.01  & 0.06  & 8.93 \\
          & |B|   & 9.0882 & 0.2770 & 2.3381 & 0.9487 & 15.9970 & 0.9487 & 24.1235 \\
          & Coverage rate & 0.00  & 0.94  & 0.87  & 0.91  & 0.57  & 0.91  & 0.54 \\
          &       &       &       &       &       &       &  \\
     \multicolumn{1}{l}{$N=5{,}000$} & MSE   & 1     & 0.01  & 0.09  & 0.02  & 3.28  & 0.02  & 7.27 \\
          & |B|   & 9.0474 & 0.2702 & 2.4706 & 1.0592 & 15.9694 & 1.0591 & 23.7925 \\
          & Coverage rate & 0.00  & 0.95  & 0.46  & 0.79  & 0.01  & 0.79  & 0.00 \\
\hline
\end{tabular}
\begin{footnotesize}
\begin{tablenotes}[flushleft]
\item \textit{Notes:} The details of this simulation design are provided in Section \ref{sec:simulation}. ``MSE'' is the mean squared error of an estimator, normalized by the mean squared error of 2SLS\@. ``|B|'' is the absolute bias. ``Coverage rate'' is the coverage rate for a nominal 95\% confidence interval. ``2SLS'' is the 2SLS estimator that additively controls for $X$\@. The weighting estimators are defined in Section \ref{sec:estandinf}. All weighting estimators also control for $X$\@. Results are based on 10,000 replications.
\end{tablenotes}
\end{footnotesize}
\end{threeparttable}
\end{adjustwidth}
\end{table}

\begin{figure}[!p]
\begin{adjustwidth}{-1in}{-1in}
\centering
\caption{Simulation Results for the Proportion of Compliers in Design A.1\label{fig:simA1}}
\includegraphics[width=21cm]{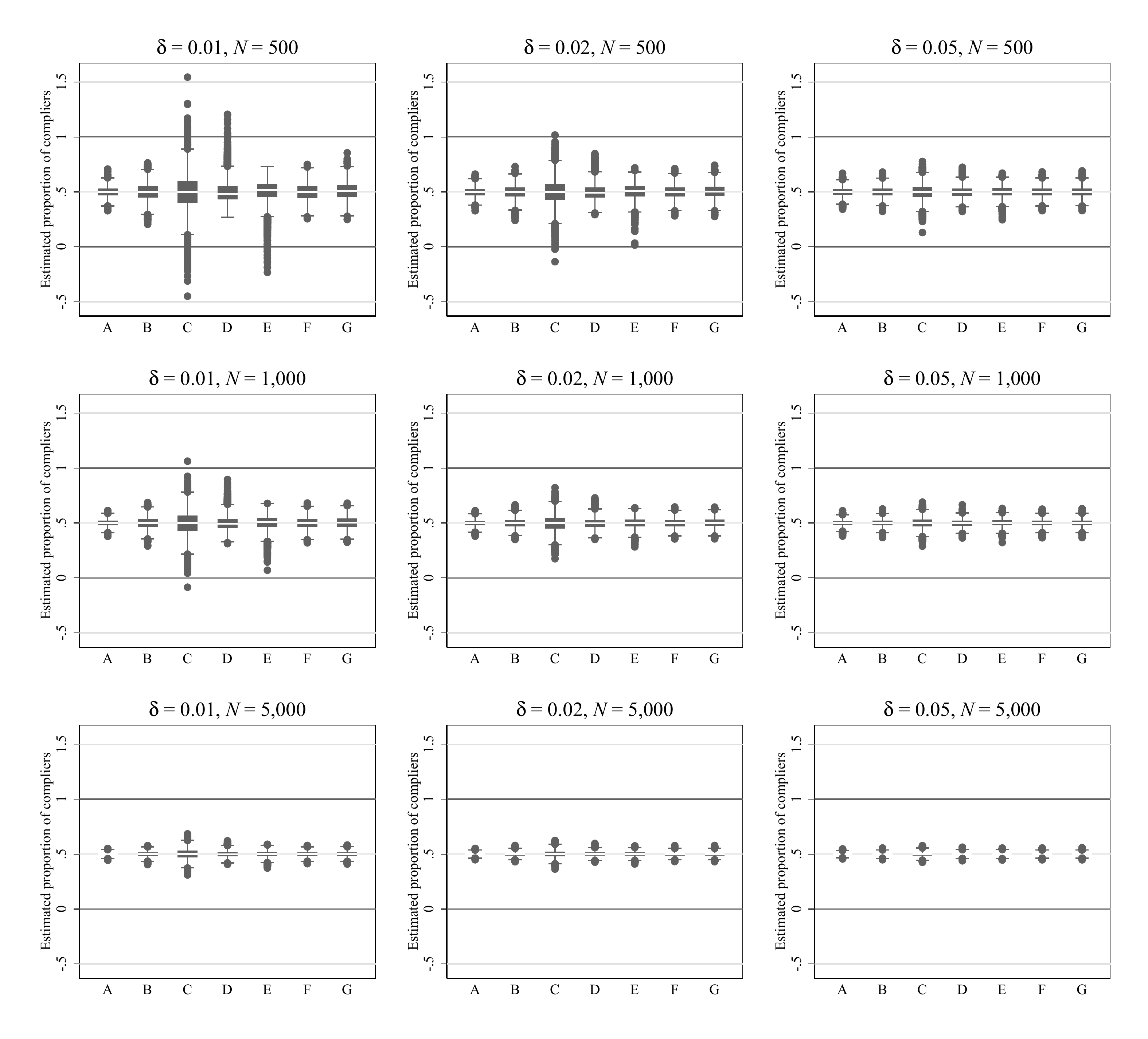}
\begin{footnotesize}
\vspace{0.075cm}
\begin{tabular}{p{19.5cm}}
\textit{Notes:} The details of this simulation design are provided in Section \ref{sec:simulation}. ``A'' corresponds to the first-stage coefficient on $Z$ in 2SLS, controlling additively for $X$\@. ``B'' corresponds to the denominator of $\hat{\tau}_{u}^{ml}$. ``C,'' ``D,'' and ``E'' correspond to $N^{-1}\sum_{i=1}^{N} \hat{\kappa}_{i1}$, $N^{-1}\sum_{i=1}^{N} \hat{\kappa}_{i0}$, and $N^{-1}\sum_{i=1}^{N} \hat{\kappa}_{i}$, respectively. These estimators, as well as the denominator of $\hat{\tau}_{u}^{ml}$, are based on an instrument propensity score, which is estimated using logit ML, also controlling for $X$\@. ``F'' corresponds to the denominator of $\hat{\tau}_{u}^{cb}$. ``G'' corresponds to $N^{-1}\sum_{i=1}^{N} \hat{\kappa}_{i1} = N^{-1}\sum_{i=1}^{N} \hat{\kappa}_{i0}$, where the instrument propensity score is estimated using the logit model and the moment conditions in equation (\ref{eq:samplemoments}), also controlling for $X$, as in the case of the denominator of $\hat{\tau}_{u}^{cb}$. Results are based on 10,000 replications.
\end{tabular}
\end{footnotesize}
\end{adjustwidth}
\end{figure}

\begin{figure}[!p]
\begin{adjustwidth}{-1in}{-1in}
\centering
\caption{Simulation Results for the Proportion of Compliers in Design A.2\label{fig:simA2}}
\includegraphics[width=21cm]{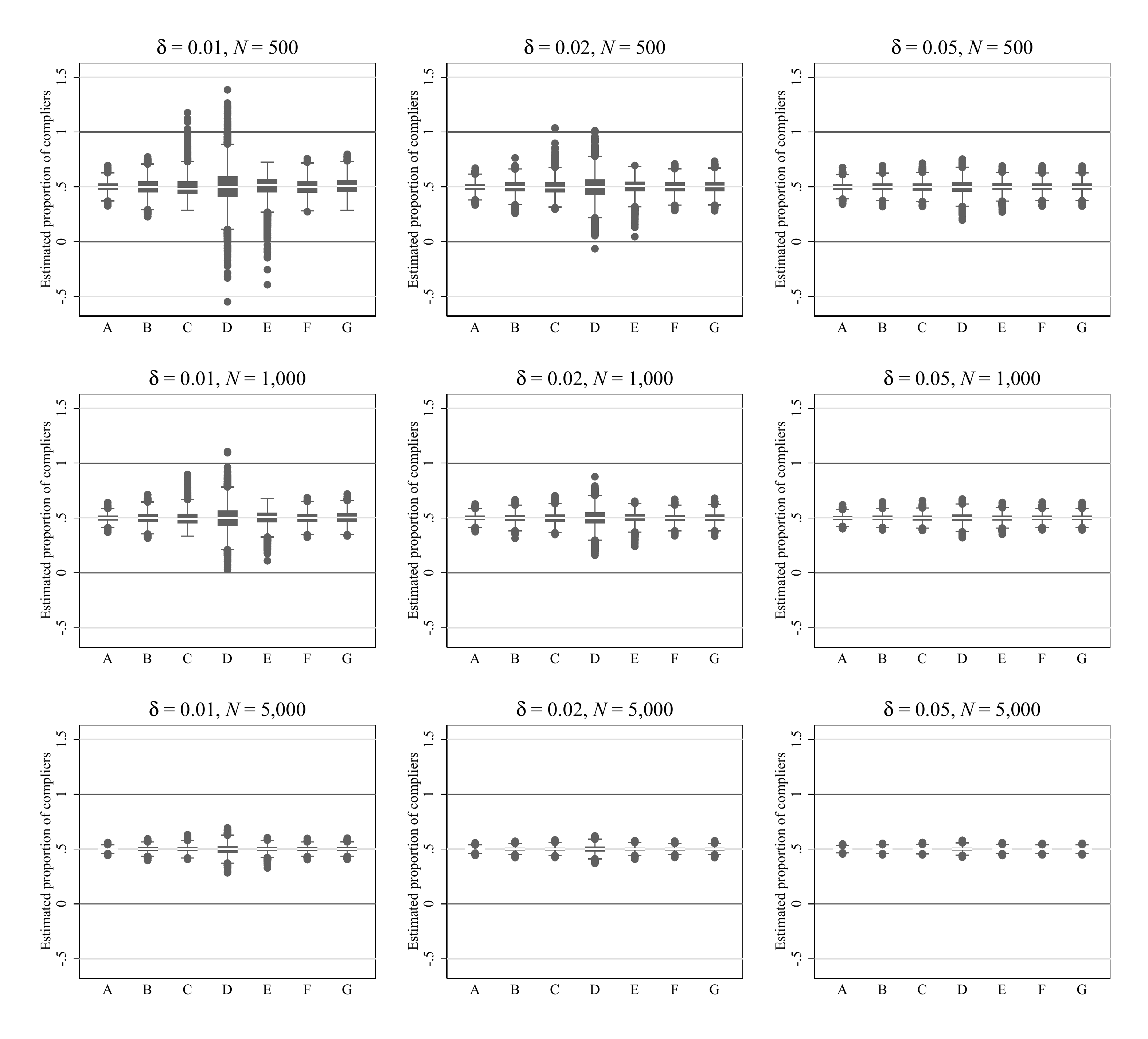}
\begin{footnotesize}
\vspace{0.075cm}
\begin{tabular}{p{19.5cm}}
\textit{Notes:} The details of this simulation design are provided in Section \ref{sec:simulation}. ``A'' corresponds to the first-stage coefficient on $Z$ in 2SLS, controlling additively for $X$\@. ``B'' corresponds to the denominator of $\hat{\tau}_{u}^{ml}$. ``C,'' ``D,'' and ``E'' correspond to $N^{-1}\sum_{i=1}^{N} \hat{\kappa}_{i1}$, $N^{-1}\sum_{i=1}^{N} \hat{\kappa}_{i0}$, and $N^{-1}\sum_{i=1}^{N} \hat{\kappa}_{i}$, respectively. These estimators, as well as the denominator of $\hat{\tau}_{u}^{ml}$, are based on an instrument propensity score, which is estimated using logit ML, also controlling for $X$\@. ``F'' corresponds to the denominator of $\hat{\tau}_{u}^{cb}$. ``G'' corresponds to $N^{-1}\sum_{i=1}^{N} \hat{\kappa}_{i1} = N^{-1}\sum_{i=1}^{N} \hat{\kappa}_{i0}$, where the instrument propensity score is estimated using the logit model and the moment conditions in equation (\ref{eq:samplemoments}), also controlling for $X$, as in the case of the denominator of $\hat{\tau}_{u}^{cb}$. Results are based on 10,000 replications.
\end{tabular}
\end{footnotesize}
\end{adjustwidth}
\end{figure}

\begin{figure}[!p]
\begin{adjustwidth}{-1in}{-1in}
\centering
\caption{Simulation Results for the Proportion of Compliers in Design B\label{fig:simB}}
\includegraphics[width=21cm]{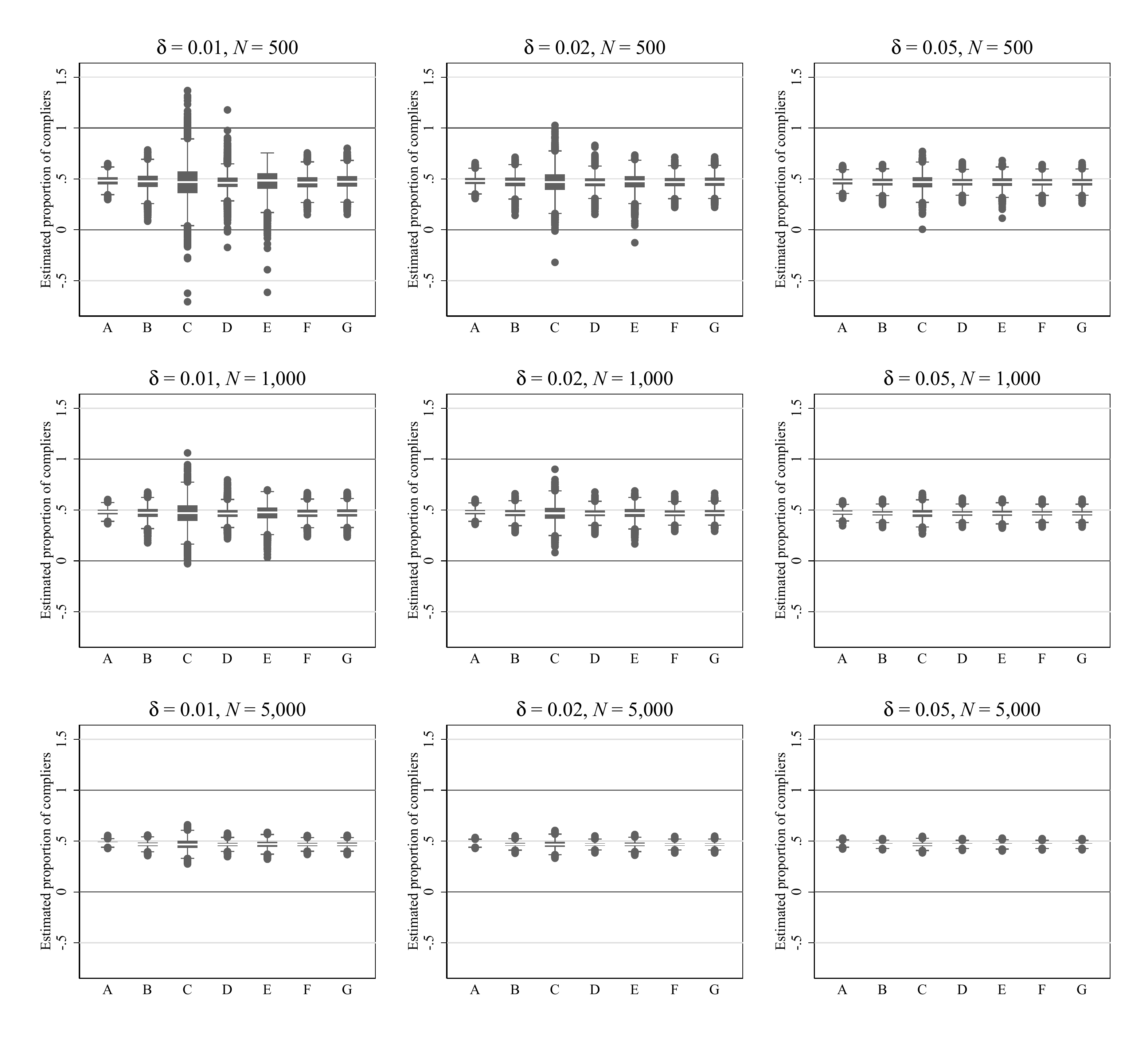}
\begin{footnotesize}
\vspace{0.075cm}
\begin{tabular}{p{19.5cm}}
\textit{Notes:} The details of this simulation design are provided in Section \ref{sec:simulation}. ``A'' corresponds to the first-stage coefficient on $Z$ in 2SLS, controlling additively for $X$\@. ``B'' corresponds to the denominator of $\hat{\tau}_{u}^{ml}$. ``C,'' ``D,'' and ``E'' correspond to $N^{-1}\sum_{i=1}^{N} \hat{\kappa}_{i1}$, $N^{-1}\sum_{i=1}^{N} \hat{\kappa}_{i0}$, and $N^{-1}\sum_{i=1}^{N} \hat{\kappa}_{i}$, respectively. These estimators, as well as the denominator of $\hat{\tau}_{u}^{ml}$, are based on an instrument propensity score, which is estimated using logit ML, also controlling for $X$\@. ``F'' corresponds to the denominator of $\hat{\tau}_{u}^{cb}$. ``G'' corresponds to $N^{-1}\sum_{i=1}^{N} \hat{\kappa}_{i1} = N^{-1}\sum_{i=1}^{N} \hat{\kappa}_{i0}$, where the instrument propensity score is estimated using the logit model and the moment conditions in equation (\ref{eq:samplemoments}), also controlling for $X$, as in the case of the denominator of $\hat{\tau}_{u}^{cb}$. Results are based on 10,000 replications.
\end{tabular}
\end{footnotesize}
\end{adjustwidth}
\end{figure}

\begin{figure}[!p]
\begin{adjustwidth}{-1in}{-1in}
\centering
\caption{Simulation Results for the Proportion of Compliers in Design C\label{fig:simC}}
\includegraphics[width=21cm]{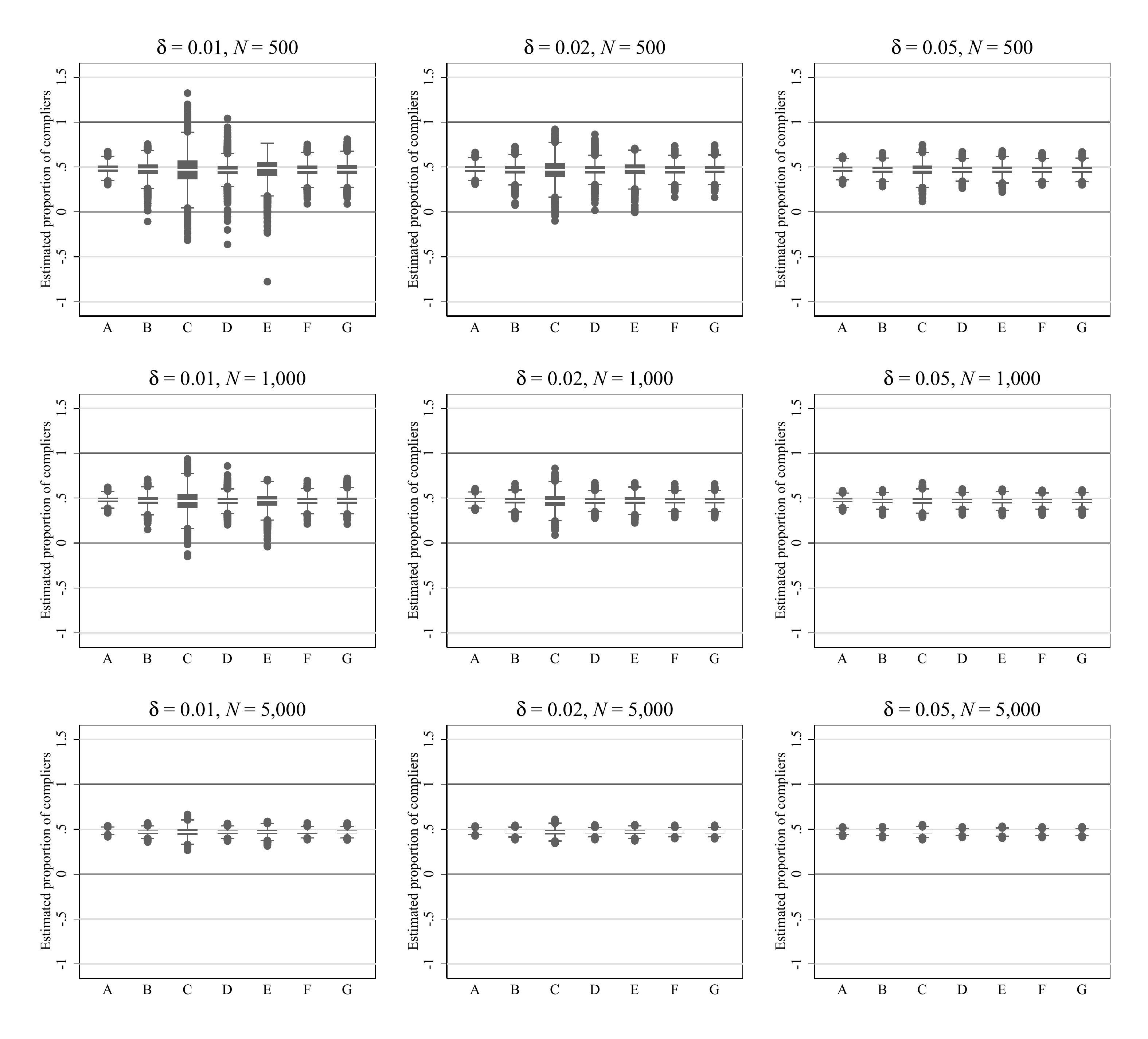}
\begin{footnotesize}
\vspace{0.075cm}
\begin{tabular}{p{19.5cm}}
\textit{Notes:} The details of this simulation design are provided in Section \ref{sec:simulation}. ``A'' corresponds to the first-stage coefficient on $Z$ in 2SLS, controlling additively for $X$\@. ``B'' corresponds to the denominator of $\hat{\tau}_{u}^{ml}$. ``C,'' ``D,'' and ``E'' correspond to $N^{-1}\sum_{i=1}^{N} \hat{\kappa}_{i1}$, $N^{-1}\sum_{i=1}^{N} \hat{\kappa}_{i0}$, and $N^{-1}\sum_{i=1}^{N} \hat{\kappa}_{i}$, respectively. These estimators, as well as the denominator of $\hat{\tau}_{u}^{ml}$, are based on an instrument propensity score, which is estimated using logit ML, also controlling for $X$\@. ``F'' corresponds to the denominator of $\hat{\tau}_{u}^{cb}$. ``G'' corresponds to $N^{-1}\sum_{i=1}^{N} \hat{\kappa}_{i1} = N^{-1}\sum_{i=1}^{N} \hat{\kappa}_{i0}$, where the instrument propensity score is estimated using the logit model and the moment conditions in equation (\ref{eq:samplemoments}), also controlling for $X$, as in the case of the denominator of $\hat{\tau}_{u}^{cb}$. Results are based on 10,000 replications.
\end{tabular}
\end{footnotesize}
\end{adjustwidth}
\end{figure}

\begin{figure}[!p]
\begin{adjustwidth}{-1in}{-1in}
\centering
\caption{Simulation Results for the Proportion of Compliers in Design D\label{fig:simD}}
\includegraphics[width=21cm]{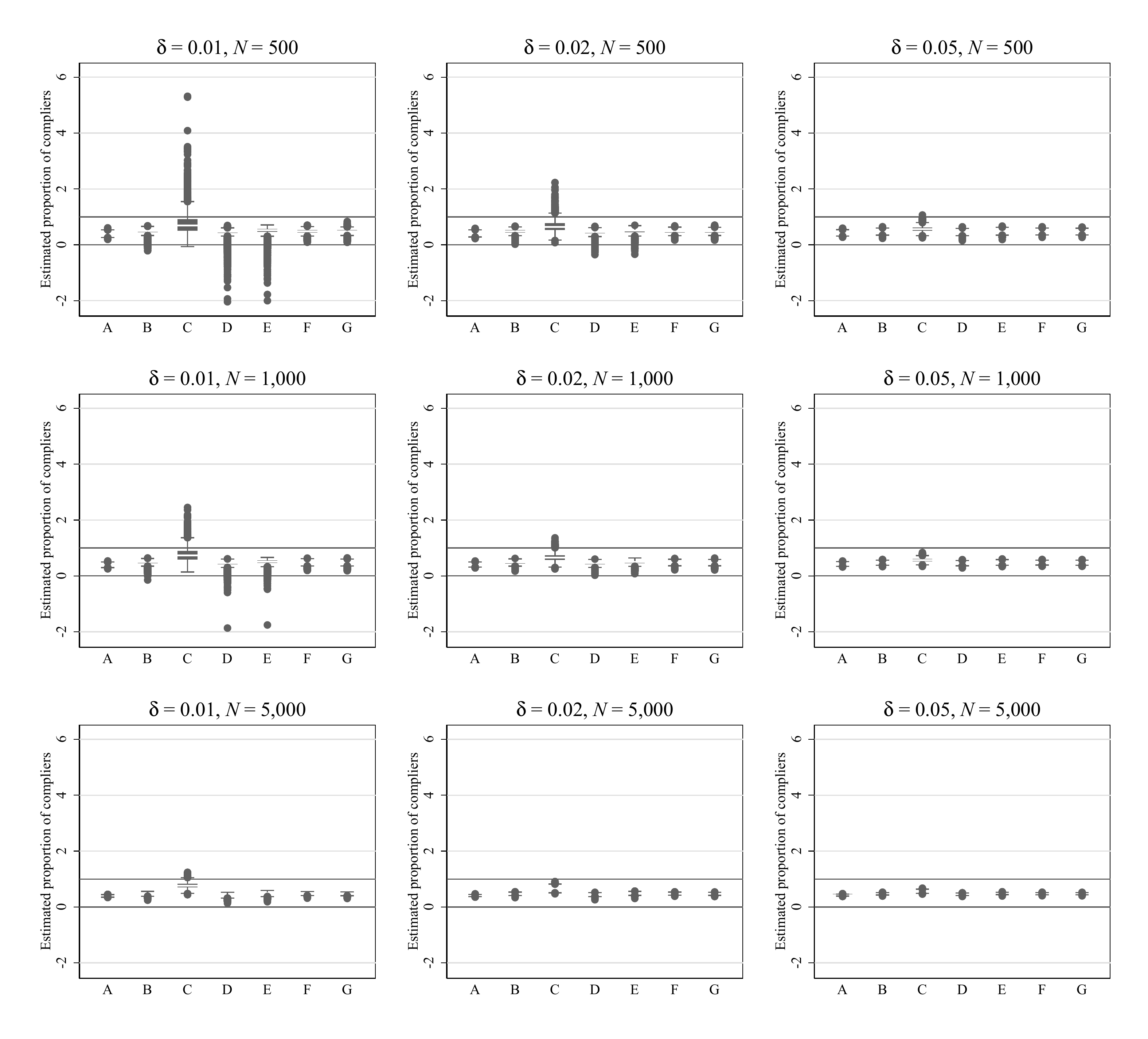}
\begin{footnotesize}
\vspace{0.075cm}
\begin{tabular}{p{19.5cm}}
\textit{Notes:} The details of this simulation design are provided in Section \ref{sec:simulation}. ``A'' corresponds to the first-stage coefficient on $Z$ in 2SLS, controlling additively for $X$\@. ``B'' corresponds to the denominator of $\hat{\tau}_{u}^{ml}$. ``C,'' ``D,'' and ``E'' correspond to $N^{-1}\sum_{i=1}^{N} \hat{\kappa}_{i1}$, $N^{-1}\sum_{i=1}^{N} \hat{\kappa}_{i0}$, and $N^{-1}\sum_{i=1}^{N} \hat{\kappa}_{i}$, respectively. These estimators, as well as the denominator of $\hat{\tau}_{u}^{ml}$, are based on an instrument propensity score, which is estimated using logit ML, also controlling for $X$\@. ``F'' corresponds to the denominator of $\hat{\tau}_{u}^{cb}$. ``G'' corresponds to $N^{-1}\sum_{i=1}^{N} \hat{\kappa}_{i1} = N^{-1}\sum_{i=1}^{N} \hat{\kappa}_{i0}$, where the instrument propensity score is estimated using the logit model and the moment conditions in equation (\ref{eq:samplemoments}), also controlling for $X$, as in the case of the denominator of $\hat{\tau}_{u}^{cb}$. Results are based on 10,000 replications.
\end{tabular}
\end{footnotesize}
\end{adjustwidth}
\end{figure}

\setcounter{table}{0}
\renewcommand{\thetable}{C.\arabic{table}}
\setcounter{figure}{0}
\renewcommand{\thefigure}{C.\arabic{figure}}

\clearpage
\begin{figure}[!p]
\begin{adjustwidth}{-1in}{-1in}
\centering
\caption{Simulation Results for Design A.1, $\delta=0.01$, $N=500$\label{fig:hist_first}}
\includegraphics[width=21cm]{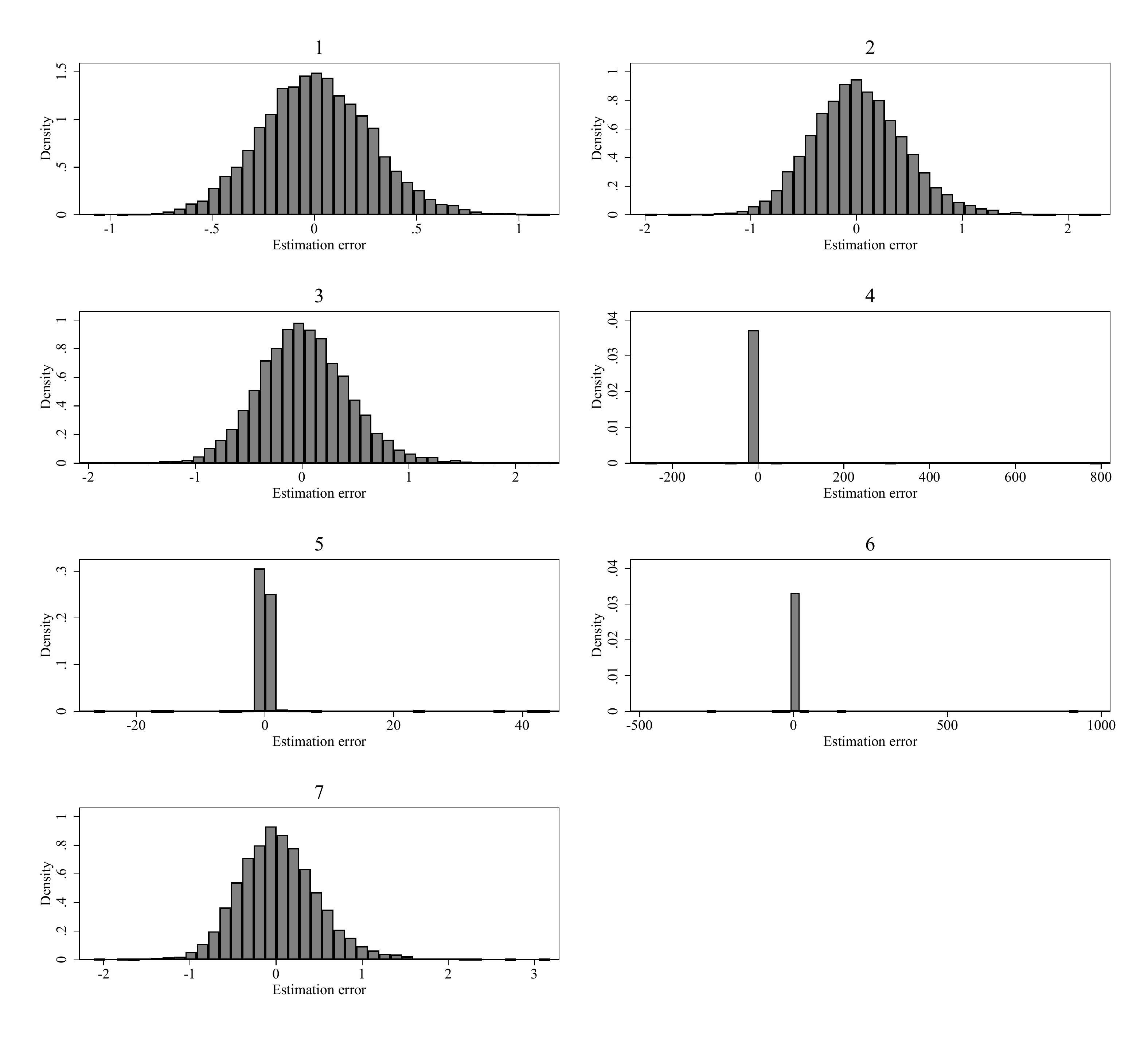}
\begin{footnotesize}
\vspace{0.075cm}
\begin{tabular}{p{19.5cm}}
\textit{Notes:} The details of this simulation design are provided in Section \ref{sec:simulation}. ``1'' corresponds to the 2SLS estimator that additively controls for $X$\@. ``2'' corresponds to $\hat{\tau}_{u}^{cb}$. ``3'' corresponds to $\hat{\tau}_{u}^{ml}$. ``4'' corresponds to $\hat{\tau}_{a,10}^{ml}$. ``5'' corresponds to $\hat{\tau}_{a}^{ml}$. ``6'' corresponds to $\hat{\tau}_{t}^{ml}$ ($= \hat{\tau}_{a,1}^{ml}$). ``7'' corresponds to $\hat{\tau}_{a,0}^{ml}$. All weighting estimators also control for $X$\@. Results are based on 10,000 replications.
\end{tabular}
\end{footnotesize}
\end{adjustwidth}
\end{figure}

\begin{figure}[!p]
\begin{adjustwidth}{-1in}{-1in}
\centering
\caption{Simulation Results for Design A.1, $\delta=0.01$, $N=1{,}000$}
\includegraphics[width=21cm]{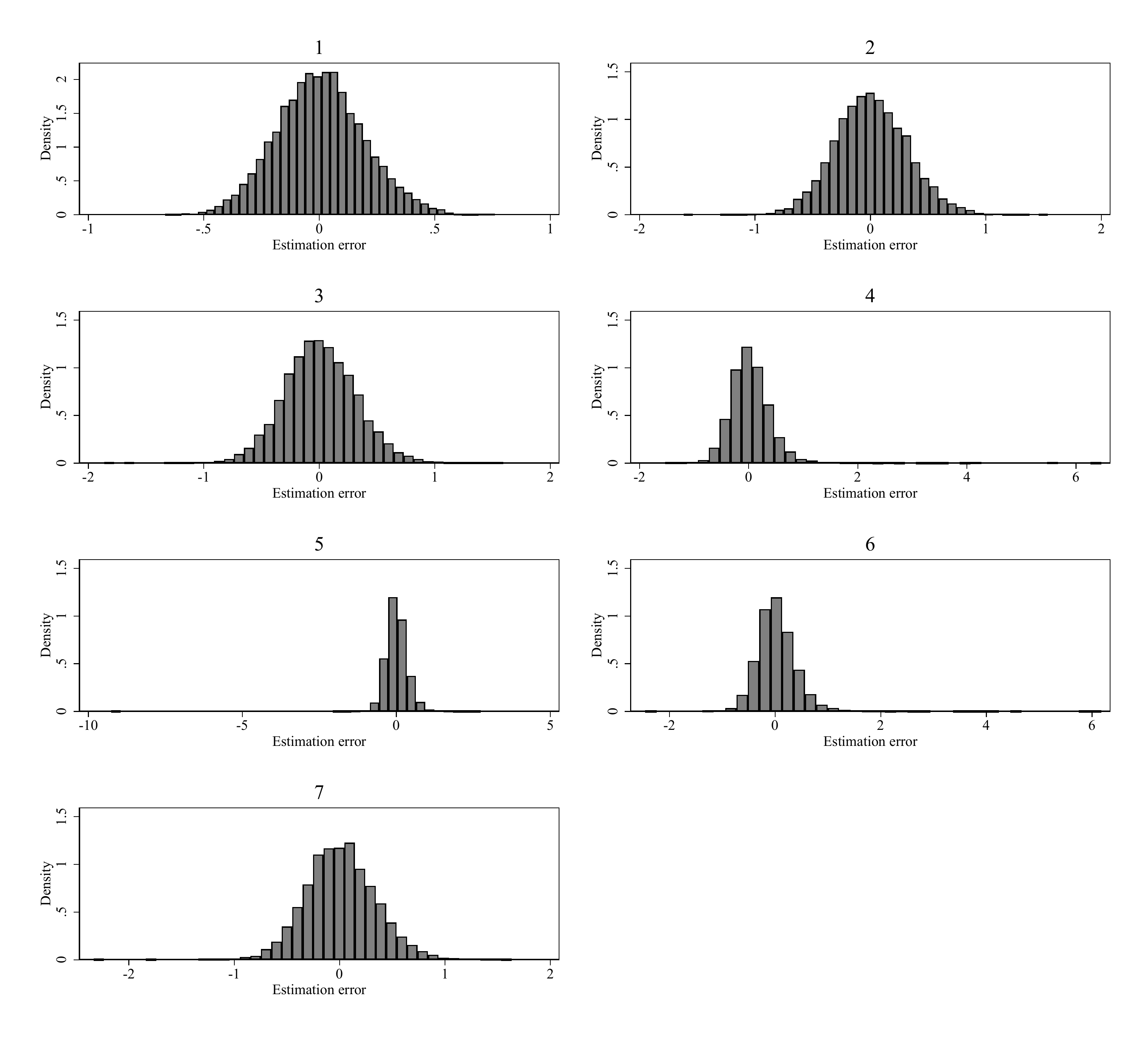}
\begin{footnotesize}
\vspace{0.075cm}
\begin{tabular}{p{19.5cm}}
\textit{Notes:} The details of this simulation design are provided in Section \ref{sec:simulation}. ``1'' corresponds to the 2SLS estimator that additively controls for $X$\@. ``2'' corresponds to $\hat{\tau}_{u}^{cb}$. ``3'' corresponds to $\hat{\tau}_{u}^{ml}$. ``4'' corresponds to $\hat{\tau}_{a,10}^{ml}$. ``5'' corresponds to $\hat{\tau}_{a}^{ml}$. ``6'' corresponds to $\hat{\tau}_{t}^{ml}$ ($= \hat{\tau}_{a,1}^{ml}$). ``7'' corresponds to $\hat{\tau}_{a,0}^{ml}$. All weighting estimators also control for $X$\@. Results are based on 10,000 replications.
\end{tabular}
\end{footnotesize}
\end{adjustwidth}
\end{figure}

\begin{figure}[!p]
\begin{adjustwidth}{-1in}{-1in}
\centering
\caption{Simulation Results for Design A.1, $\delta=0.01$, $N=5{,}000$}
\includegraphics[width=21cm]{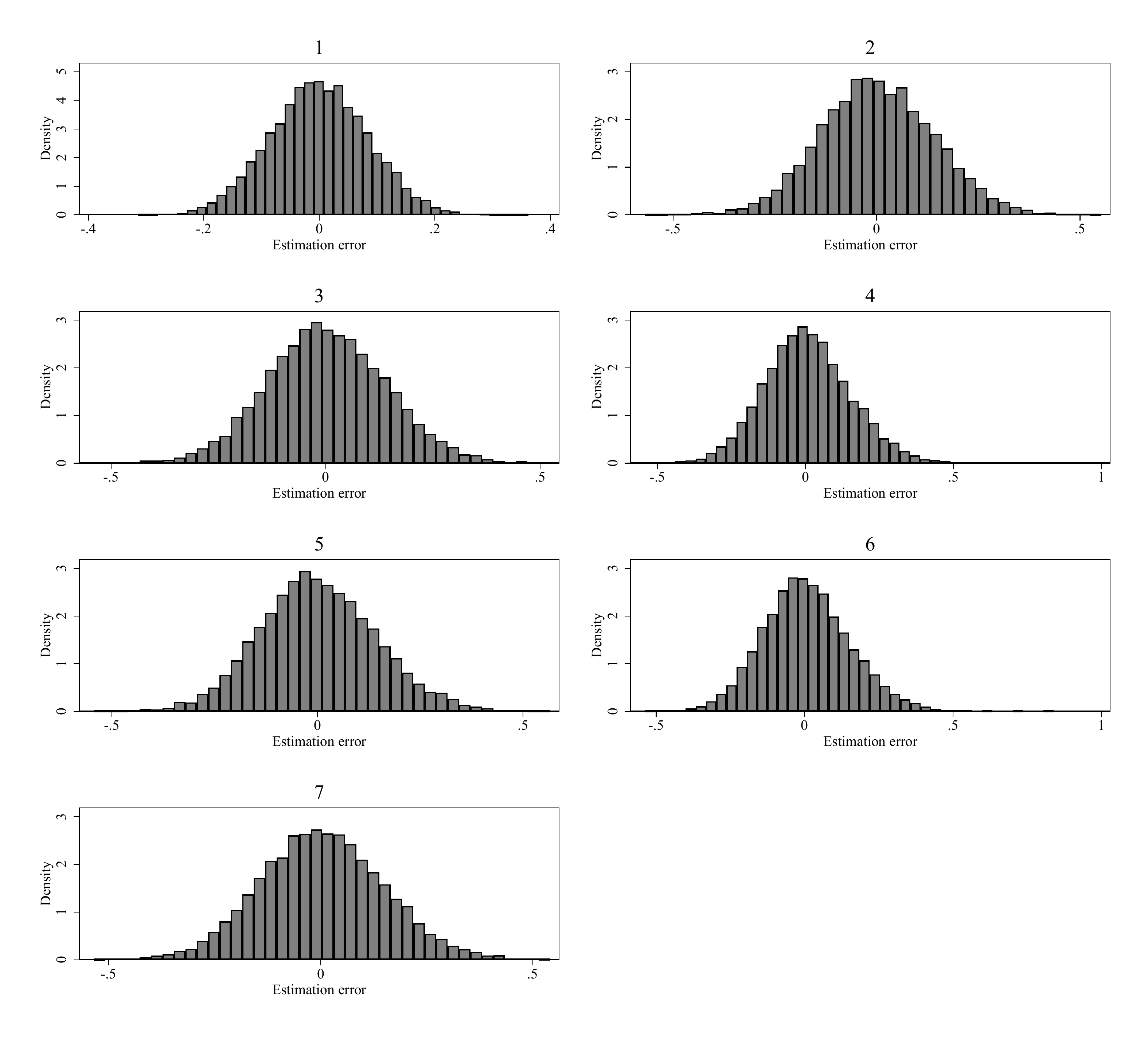}
\begin{footnotesize}
\vspace{0.075cm}
\begin{tabular}{p{19.5cm}}
\textit{Notes:} The details of this simulation design are provided in Section \ref{sec:simulation}. ``1'' corresponds to the 2SLS estimator that additively controls for $X$\@. ``2'' corresponds to $\hat{\tau}_{u}^{cb}$. ``3'' corresponds to $\hat{\tau}_{u}^{ml}$. ``4'' corresponds to $\hat{\tau}_{a,10}^{ml}$. ``5'' corresponds to $\hat{\tau}_{a}^{ml}$. ``6'' corresponds to $\hat{\tau}_{t}^{ml}$ ($= \hat{\tau}_{a,1}^{ml}$). ``7'' corresponds to $\hat{\tau}_{a,0}^{ml}$. All weighting estimators also control for $X$\@. Results are based on 10,000 replications.
\end{tabular}
\end{footnotesize}
\end{adjustwidth}
\end{figure}

\begin{figure}[!p]
\begin{adjustwidth}{-1in}{-1in}
\centering
\caption{Simulation Results for Design A.1, $\delta=0.02$, $N=500$}
\includegraphics[width=21cm]{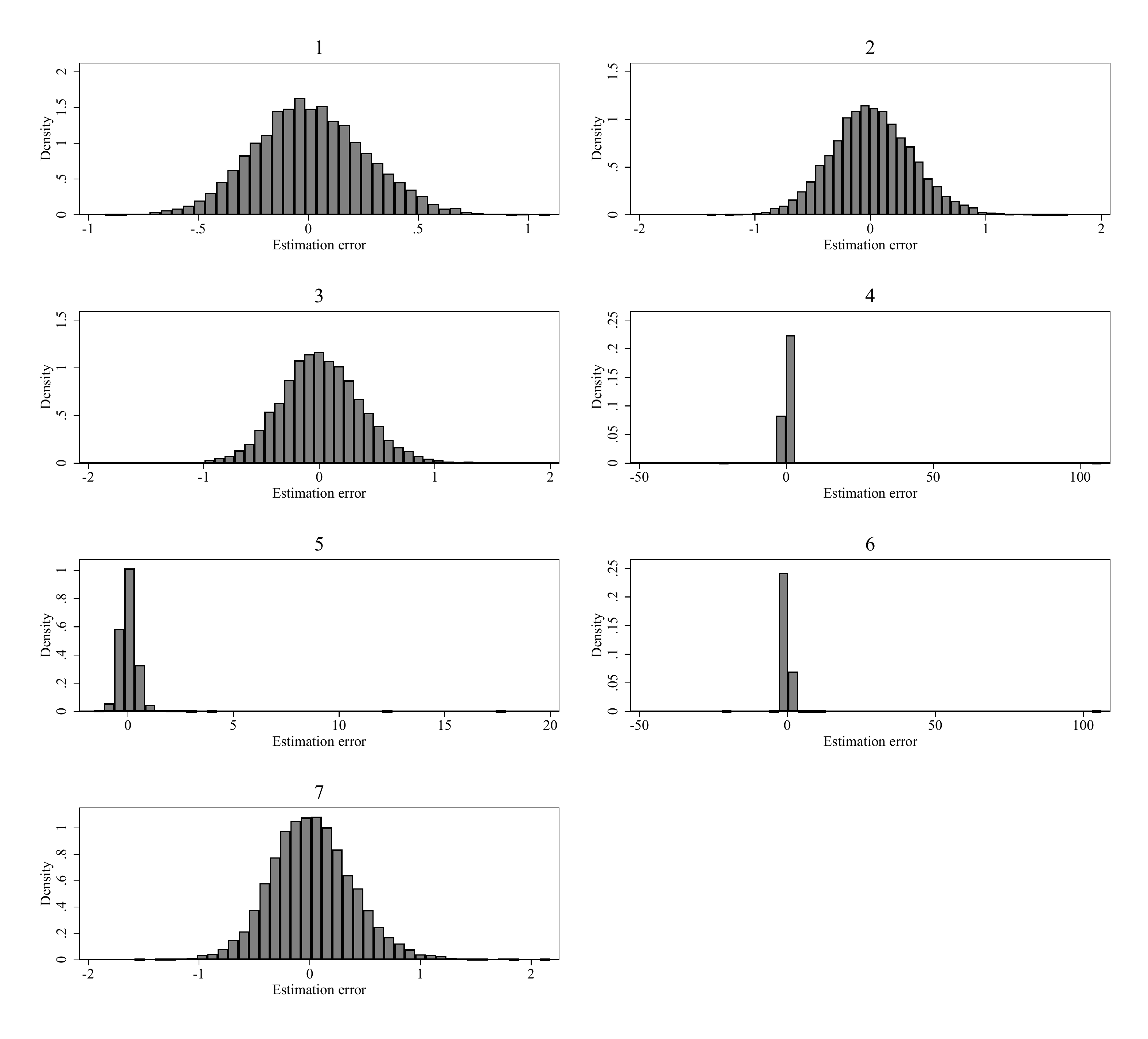}
\begin{footnotesize}
\vspace{0.075cm}
\begin{tabular}{p{19.5cm}}
\textit{Notes:} The details of this simulation design are provided in Section \ref{sec:simulation}. ``1'' corresponds to the 2SLS estimator that additively controls for $X$\@. ``2'' corresponds to $\hat{\tau}_{u}^{cb}$. ``3'' corresponds to $\hat{\tau}_{u}^{ml}$. ``4'' corresponds to $\hat{\tau}_{a,10}^{ml}$. ``5'' corresponds to $\hat{\tau}_{a}^{ml}$. ``6'' corresponds to $\hat{\tau}_{t}^{ml}$ ($= \hat{\tau}_{a,1}^{ml}$). ``7'' corresponds to $\hat{\tau}_{a,0}^{ml}$. All weighting estimators also control for $X$\@. Results are based on 10,000 replications.
\end{tabular}
\end{footnotesize}
\end{adjustwidth}
\end{figure}

\begin{figure}[!p]
\begin{adjustwidth}{-1in}{-1in}
\centering
\caption{Simulation Results for Design A.1, $\delta=0.02$, $N=1{,}000$}
\includegraphics[width=21cm]{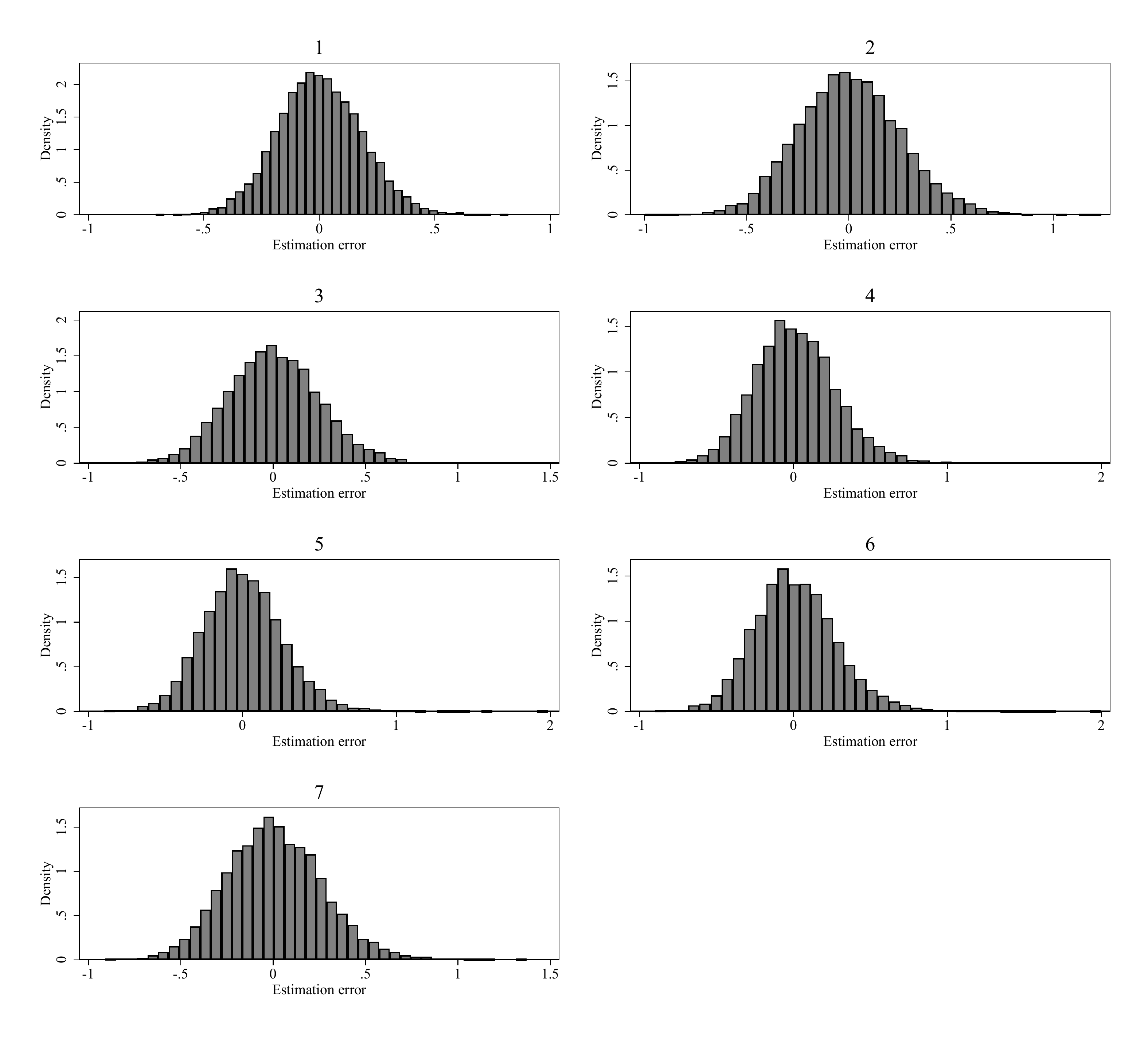}
\begin{footnotesize}
\vspace{0.075cm}
\begin{tabular}{p{19.5cm}}
\textit{Notes:} The details of this simulation design are provided in Section \ref{sec:simulation}. ``1'' corresponds to the 2SLS estimator that additively controls for $X$\@. ``2'' corresponds to $\hat{\tau}_{u}^{cb}$. ``3'' corresponds to $\hat{\tau}_{u}^{ml}$. ``4'' corresponds to $\hat{\tau}_{a,10}^{ml}$. ``5'' corresponds to $\hat{\tau}_{a}^{ml}$. ``6'' corresponds to $\hat{\tau}_{t}^{ml}$ ($= \hat{\tau}_{a,1}^{ml}$). ``7'' corresponds to $\hat{\tau}_{a,0}^{ml}$. All weighting estimators also control for $X$\@. Results are based on 10,000 replications.
\end{tabular}
\end{footnotesize}
\end{adjustwidth}
\end{figure}

\begin{figure}[!p]
\begin{adjustwidth}{-1in}{-1in}
\centering
\caption{Simulation Results for Design A.1, $\delta=0.02$, $N=5{,}000$}
\includegraphics[width=21cm]{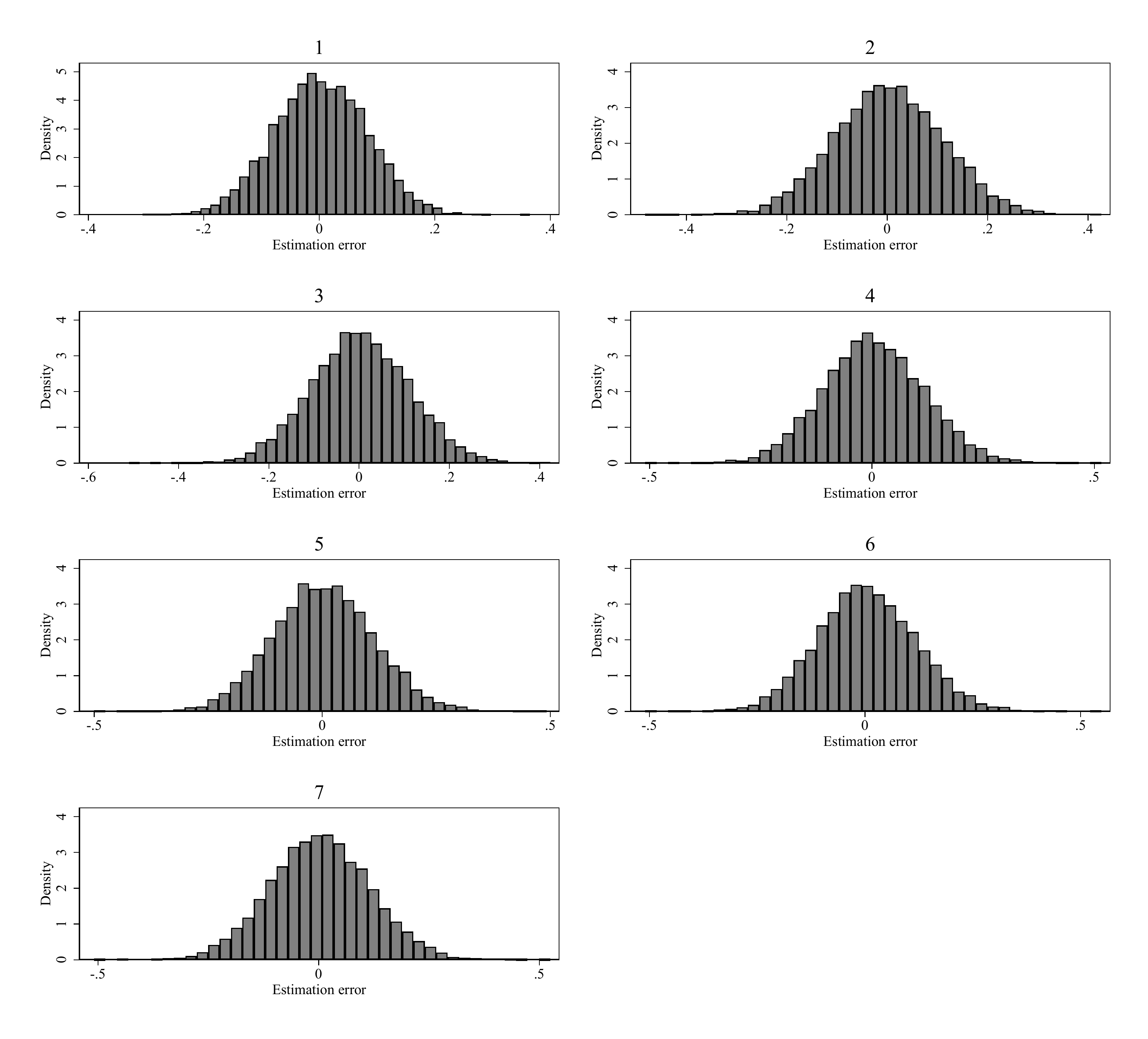}
\begin{footnotesize}
\vspace{0.075cm}
\begin{tabular}{p{19.5cm}}
\textit{Notes:} The details of this simulation design are provided in Section \ref{sec:simulation}. ``1'' corresponds to the 2SLS estimator that additively controls for $X$\@. ``2'' corresponds to $\hat{\tau}_{u}^{cb}$. ``3'' corresponds to $\hat{\tau}_{u}^{ml}$. ``4'' corresponds to $\hat{\tau}_{a,10}^{ml}$. ``5'' corresponds to $\hat{\tau}_{a}^{ml}$. ``6'' corresponds to $\hat{\tau}_{t}^{ml}$ ($= \hat{\tau}_{a,1}^{ml}$). ``7'' corresponds to $\hat{\tau}_{a,0}^{ml}$. All weighting estimators also control for $X$\@. Results are based on 10,000 replications.
\end{tabular}
\end{footnotesize}
\end{adjustwidth}
\end{figure}

\begin{figure}[!p]
\begin{adjustwidth}{-1in}{-1in}
\centering
\caption{Simulation Results for Design A.1, $\delta=0.05$, $N=500$}
\includegraphics[width=21cm]{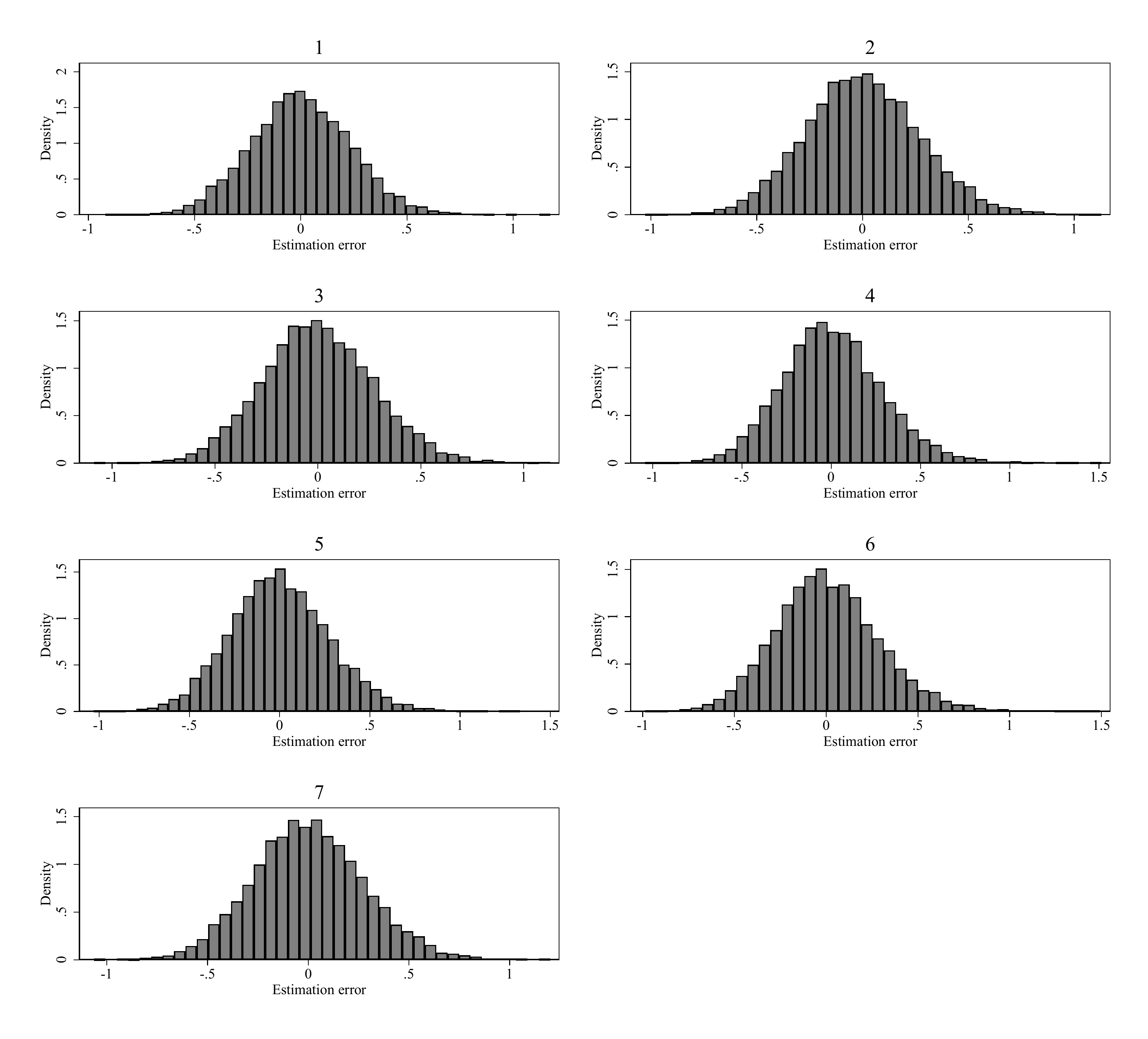}
\begin{footnotesize}
\vspace{0.075cm}
\begin{tabular}{p{19.5cm}}
\textit{Notes:} The details of this simulation design are provided in Section \ref{sec:simulation}. ``1'' corresponds to the 2SLS estimator that additively controls for $X$\@. ``2'' corresponds to $\hat{\tau}_{u}^{cb}$. ``3'' corresponds to $\hat{\tau}_{u}^{ml}$. ``4'' corresponds to $\hat{\tau}_{a,10}^{ml}$. ``5'' corresponds to $\hat{\tau}_{a}^{ml}$. ``6'' corresponds to $\hat{\tau}_{t}^{ml}$ ($= \hat{\tau}_{a,1}^{ml}$). ``7'' corresponds to $\hat{\tau}_{a,0}^{ml}$. All weighting estimators also control for $X$\@. Results are based on 10,000 replications.
\end{tabular}
\end{footnotesize}
\end{adjustwidth}
\end{figure}

\begin{figure}[!p]
\begin{adjustwidth}{-1in}{-1in}
\centering
\caption{Simulation Results for Design A.1, $\delta=0.05$, $N=1{,}000$}
\includegraphics[width=21cm]{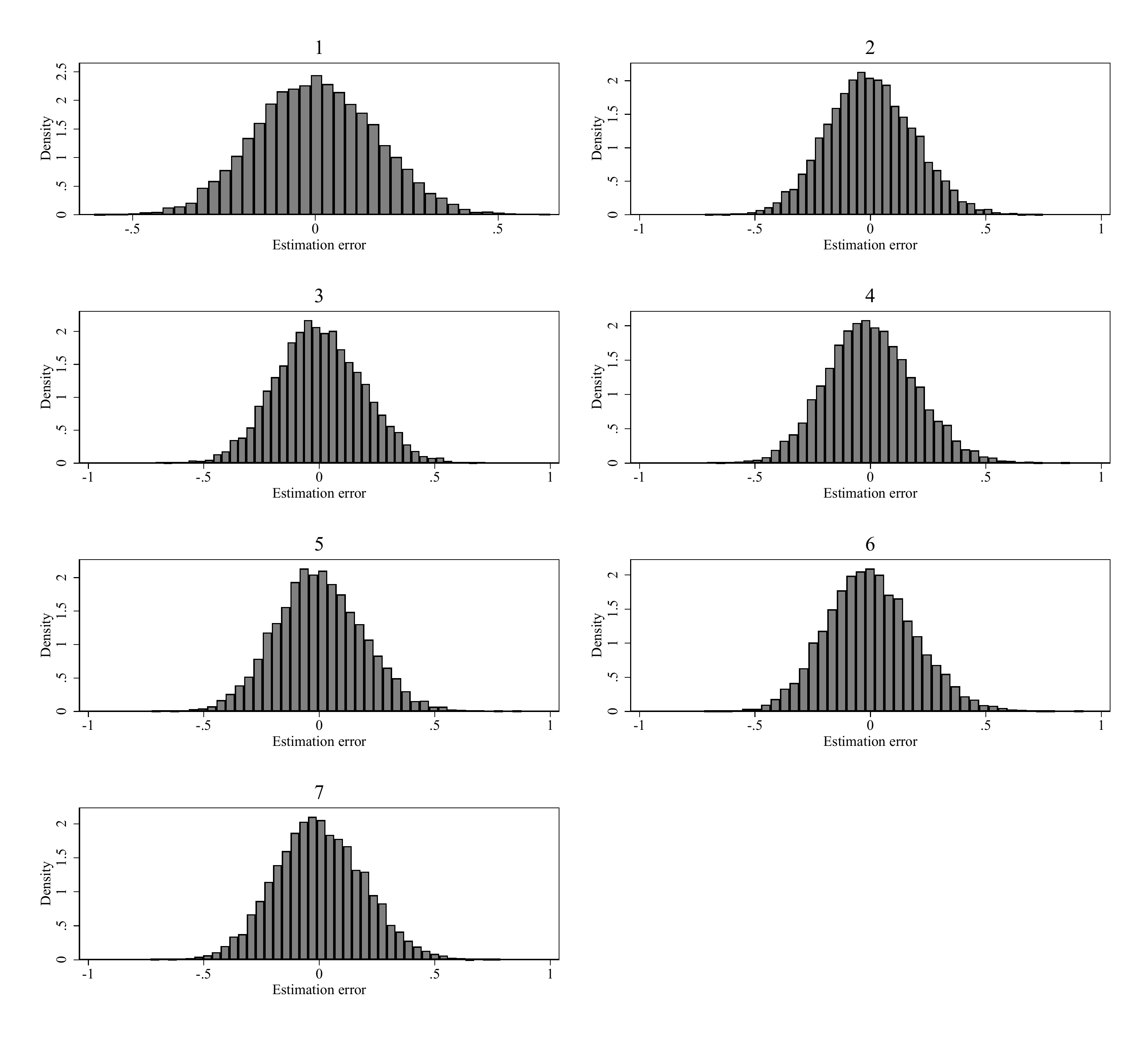}
\begin{footnotesize}
\vspace{0.075cm}
\begin{tabular}{p{19.5cm}}
\textit{Notes:} The details of this simulation design are provided in Section \ref{sec:simulation}. ``1'' corresponds to the 2SLS estimator that additively controls for $X$\@. ``2'' corresponds to $\hat{\tau}_{u}^{cb}$. ``3'' corresponds to $\hat{\tau}_{u}^{ml}$. ``4'' corresponds to $\hat{\tau}_{a,10}^{ml}$. ``5'' corresponds to $\hat{\tau}_{a}^{ml}$. ``6'' corresponds to $\hat{\tau}_{t}^{ml}$ ($= \hat{\tau}_{a,1}^{ml}$). ``7'' corresponds to $\hat{\tau}_{a,0}^{ml}$. All weighting estimators also control for $X$\@. Results are based on 10,000 replications.
\end{tabular}
\end{footnotesize}
\end{adjustwidth}
\end{figure}

\begin{figure}[!p]
\begin{adjustwidth}{-1in}{-1in}
\centering
\caption{Simulation Results for Design A.1, $\delta=0.05$, $N=5{,}000$}
\includegraphics[width=21cm]{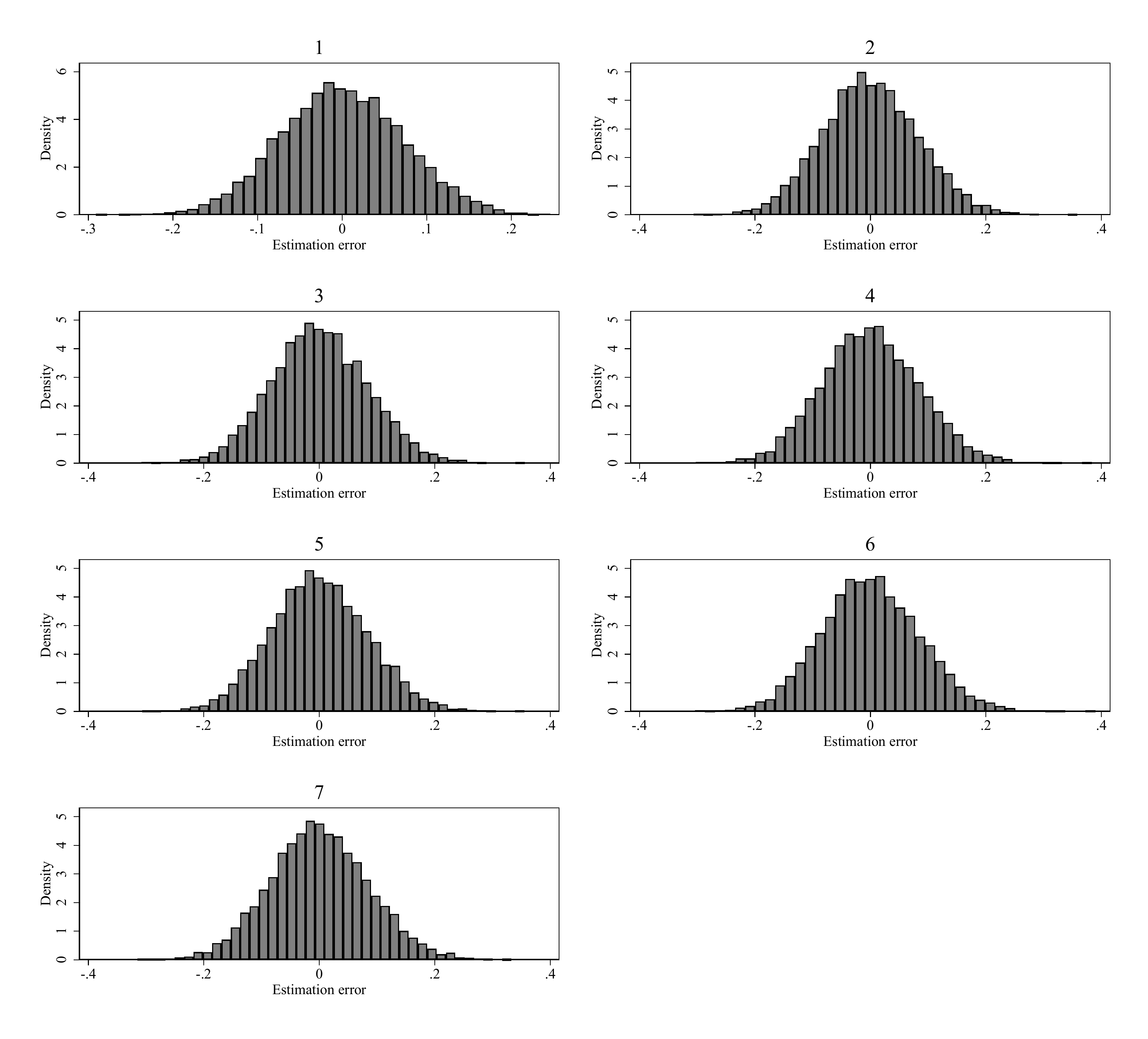}
\begin{footnotesize}
\vspace{0.075cm}
\begin{tabular}{p{19.5cm}}
\textit{Notes:} The details of this simulation design are provided in Section \ref{sec:simulation}. ``1'' corresponds to the 2SLS estimator that additively controls for $X$\@. ``2'' corresponds to $\hat{\tau}_{u}^{cb}$. ``3'' corresponds to $\hat{\tau}_{u}^{ml}$. ``4'' corresponds to $\hat{\tau}_{a,10}^{ml}$. ``5'' corresponds to $\hat{\tau}_{a}^{ml}$. ``6'' corresponds to $\hat{\tau}_{t}^{ml}$ ($= \hat{\tau}_{a,1}^{ml}$). ``7'' corresponds to $\hat{\tau}_{a,0}^{ml}$. All weighting estimators also control for $X$\@. Results are based on 10,000 replications.
\end{tabular}
\end{footnotesize}
\end{adjustwidth}
\end{figure}

\begin{figure}[!p]
\begin{adjustwidth}{-1in}{-1in}
\centering
\caption{Simulation Results for Design A.2, $\delta=0.01$, $N=500$}
\includegraphics[width=21cm]{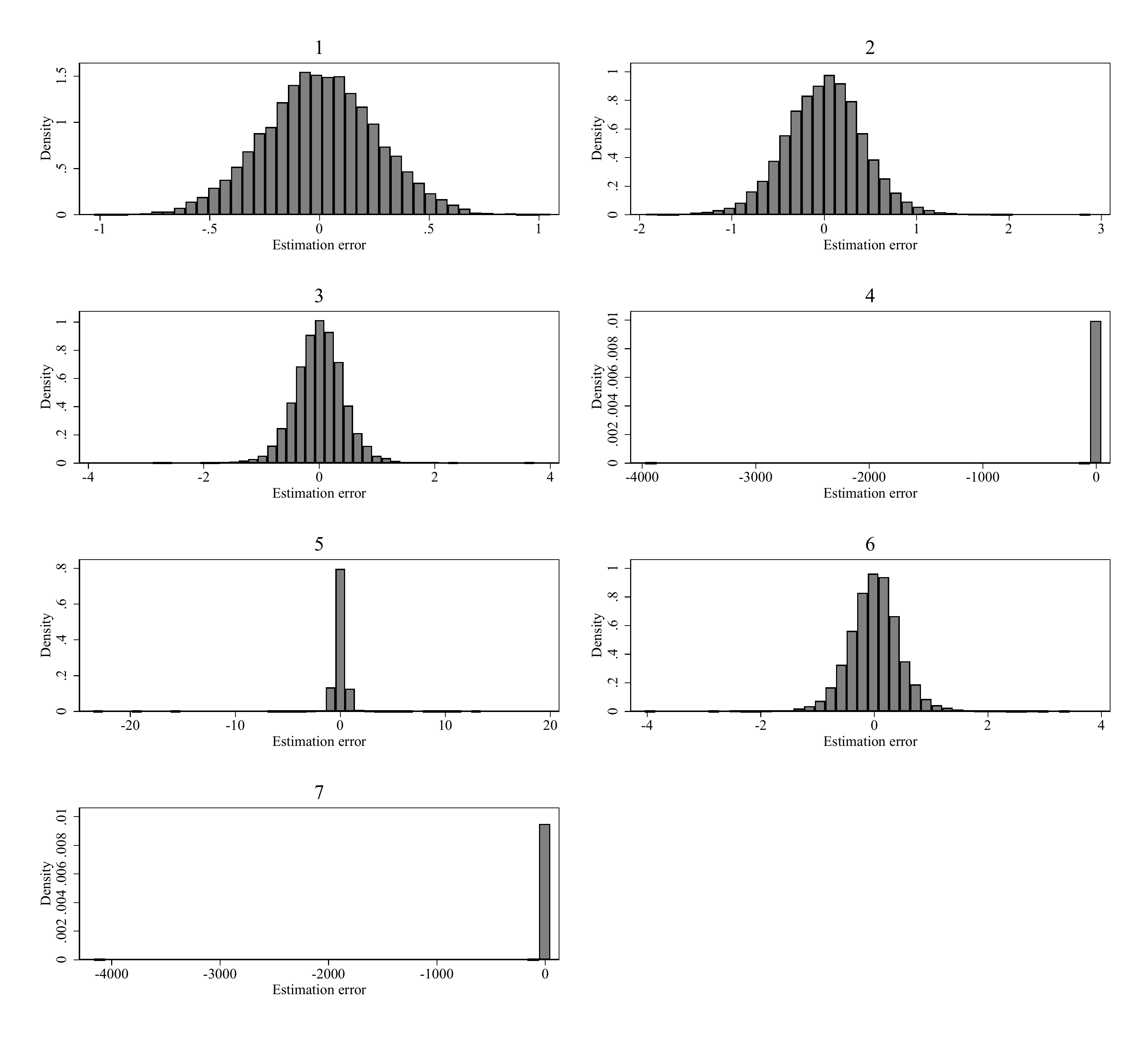}
\begin{footnotesize}
\vspace{0.075cm}
\begin{tabular}{p{19.5cm}}
\textit{Notes:} The details of this simulation design are provided in Section \ref{sec:simulation}. ``1'' corresponds to the 2SLS estimator that additively controls for $X$\@. ``2'' corresponds to $\hat{\tau}_{u}^{cb}$. ``3'' corresponds to $\hat{\tau}_{u}^{ml}$. ``4'' corresponds to $\hat{\tau}_{a,10}^{ml}$. ``5'' corresponds to $\hat{\tau}_{a}^{ml}$. ``6'' corresponds to $\hat{\tau}_{t}^{ml}$ ($= \hat{\tau}_{a,1}^{ml}$). ``7'' corresponds to $\hat{\tau}_{a,0}^{ml}$. All weighting estimators also control for $X$\@. Results are based on 10,000 replications.
\end{tabular}
\end{footnotesize}
\end{adjustwidth}
\end{figure}

\begin{figure}[!p]
\begin{adjustwidth}{-1in}{-1in}
\centering
\caption{Simulation Results for Design A.2, $\delta=0.01$, $N=1{,}000$}
\includegraphics[width=21cm]{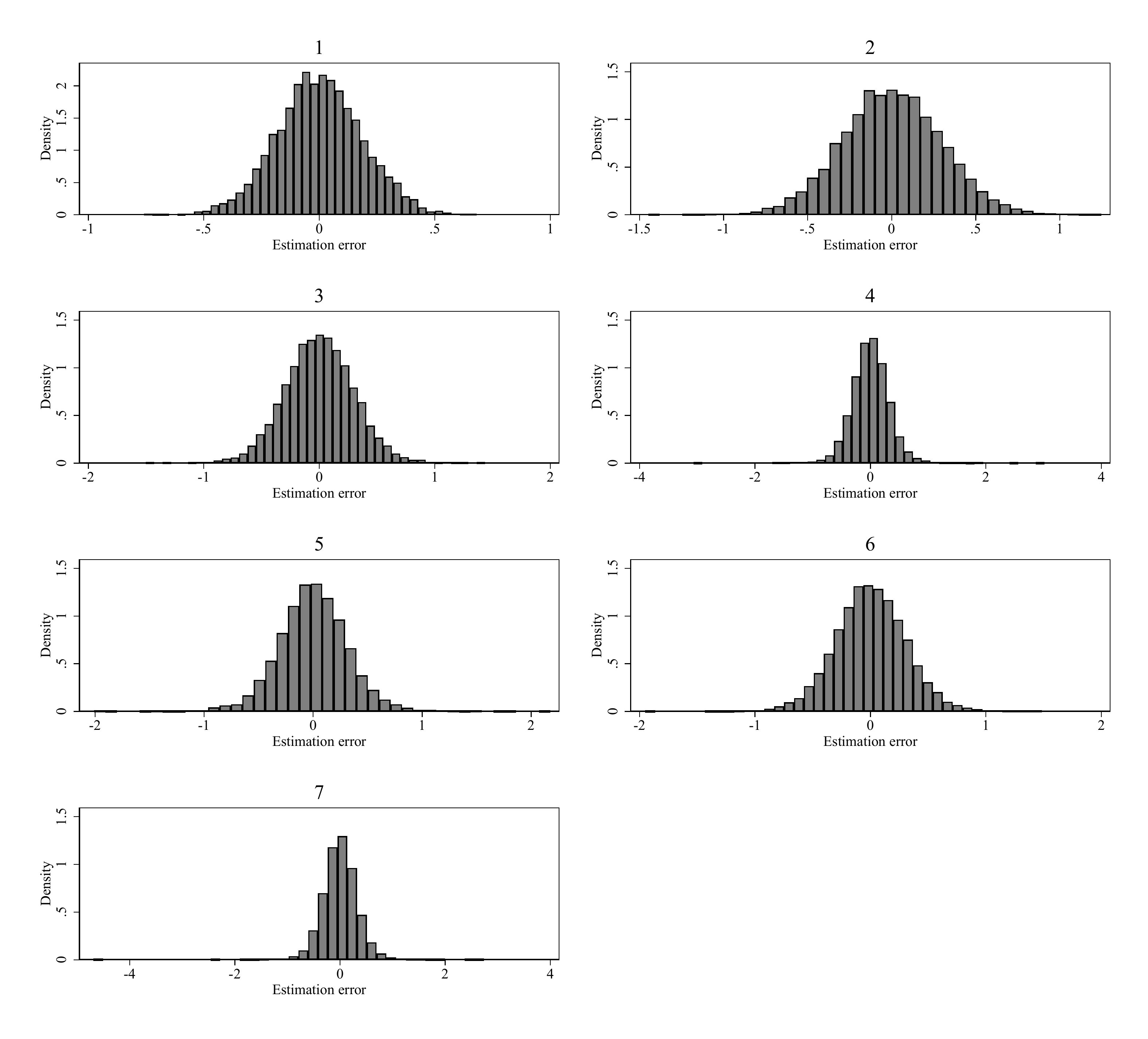}
\begin{footnotesize}
\vspace{0.075cm}
\begin{tabular}{p{19.5cm}}
\textit{Notes:} The details of this simulation design are provided in Section \ref{sec:simulation}. ``1'' corresponds to the 2SLS estimator that additively controls for $X$\@. ``2'' corresponds to $\hat{\tau}_{u}^{cb}$. ``3'' corresponds to $\hat{\tau}_{u}^{ml}$. ``4'' corresponds to $\hat{\tau}_{a,10}^{ml}$. ``5'' corresponds to $\hat{\tau}_{a}^{ml}$. ``6'' corresponds to $\hat{\tau}_{t}^{ml}$ ($= \hat{\tau}_{a,1}^{ml}$). ``7'' corresponds to $\hat{\tau}_{a,0}^{ml}$. All weighting estimators also control for $X$\@. Results are based on 10,000 replications.
\end{tabular}
\end{footnotesize}
\end{adjustwidth}
\end{figure}

\begin{figure}[!p]
\begin{adjustwidth}{-1in}{-1in}
\centering
\caption{Simulation Results for Design A.2, $\delta=0.01$, $N=5{,}000$}
\includegraphics[width=21cm]{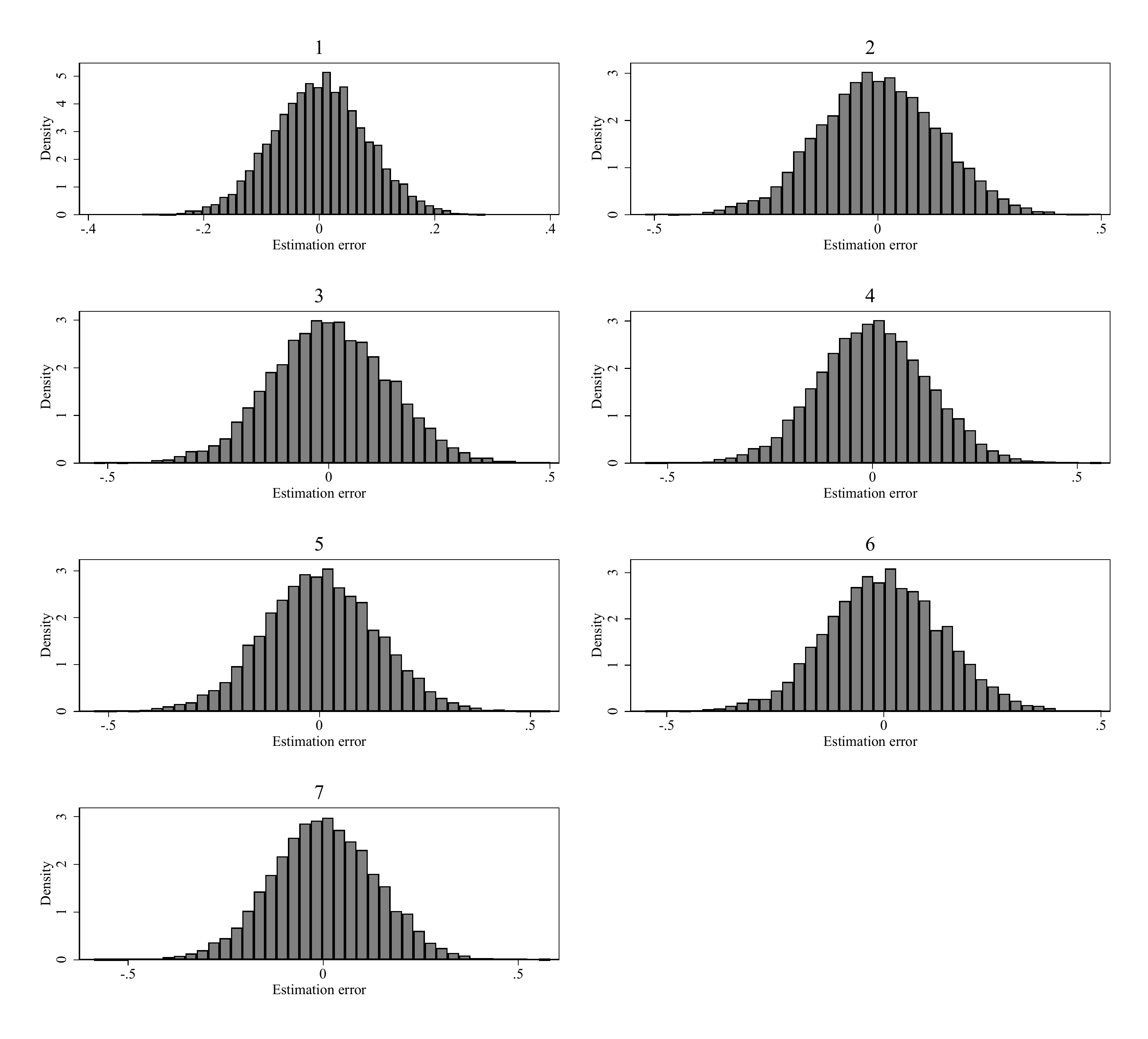}
\begin{footnotesize}
\vspace{0.075cm}
\begin{tabular}{p{19.5cm}}
\textit{Notes:} The details of this simulation design are provided in Section \ref{sec:simulation}. ``1'' corresponds to the 2SLS estimator that additively controls for $X$\@. ``2'' corresponds to $\hat{\tau}_{u}^{cb}$. ``3'' corresponds to $\hat{\tau}_{u}^{ml}$. ``4'' corresponds to $\hat{\tau}_{a,10}^{ml}$. ``5'' corresponds to $\hat{\tau}_{a}^{ml}$. ``6'' corresponds to $\hat{\tau}_{t}^{ml}$ ($= \hat{\tau}_{a,1}^{ml}$). ``7'' corresponds to $\hat{\tau}_{a,0}^{ml}$. All weighting estimators also control for $X$\@. Results are based on 10,000 replications.
\end{tabular}
\end{footnotesize}
\end{adjustwidth}
\end{figure}

\begin{figure}[!p]
\begin{adjustwidth}{-1in}{-1in}
\centering
\caption{Simulation Results for Design A.2, $\delta=0.02$, $N=500$}
\includegraphics[width=21cm]{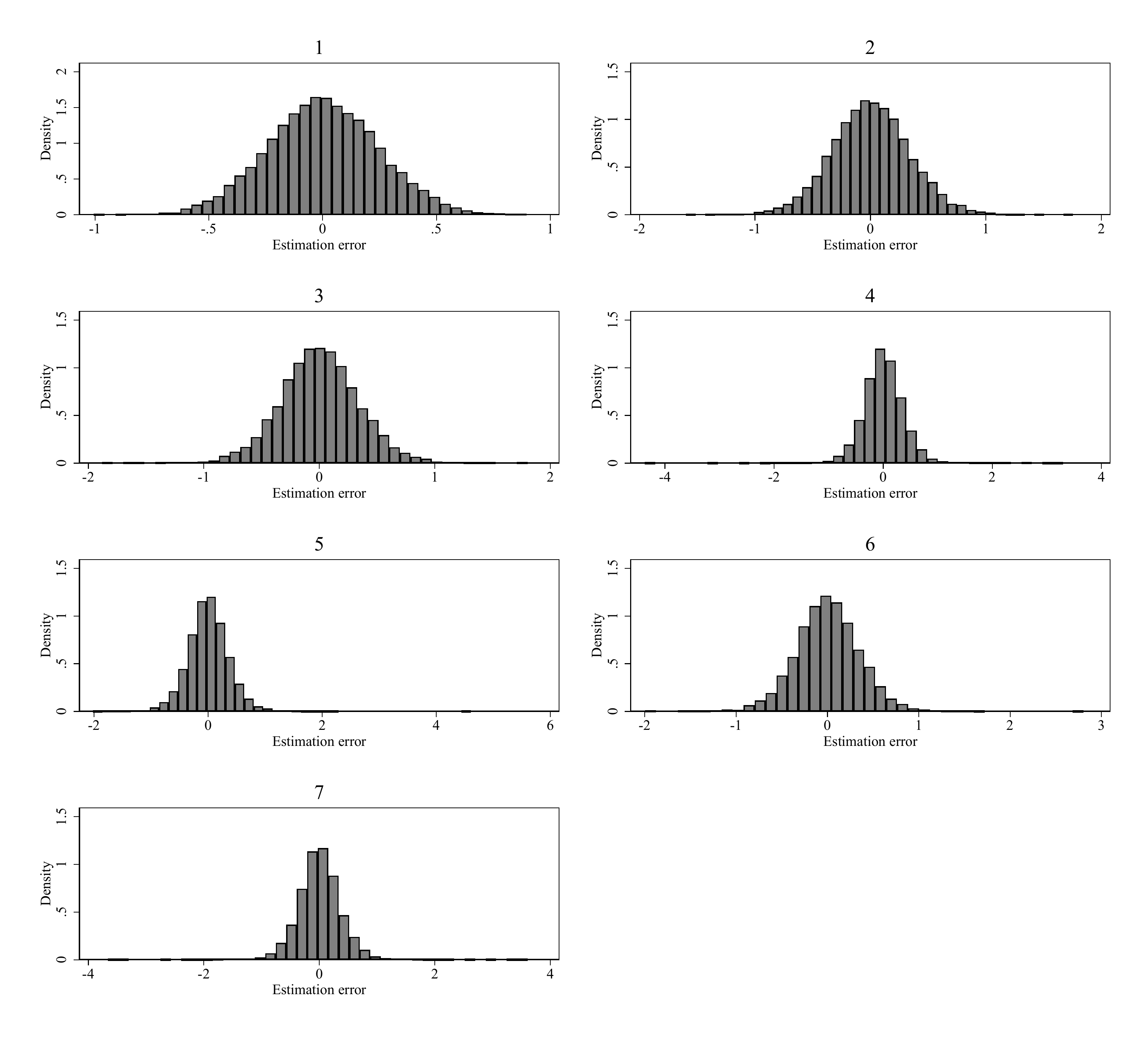}
\begin{footnotesize}
\vspace{0.075cm}
\begin{tabular}{p{19.5cm}}
\textit{Notes:} The details of this simulation design are provided in Section \ref{sec:simulation}. ``1'' corresponds to the 2SLS estimator that additively controls for $X$\@. ``2'' corresponds to $\hat{\tau}_{u}^{cb}$. ``3'' corresponds to $\hat{\tau}_{u}^{ml}$. ``4'' corresponds to $\hat{\tau}_{a,10}^{ml}$. ``5'' corresponds to $\hat{\tau}_{a}^{ml}$. ``6'' corresponds to $\hat{\tau}_{t}^{ml}$ ($= \hat{\tau}_{a,1}^{ml}$). ``7'' corresponds to $\hat{\tau}_{a,0}^{ml}$. All weighting estimators also control for $X$\@. Results are based on 10,000 replications.
\end{tabular}
\end{footnotesize}
\end{adjustwidth}
\end{figure}

\begin{figure}[!p]
\begin{adjustwidth}{-1in}{-1in}
\centering
\caption{Simulation Results for Design A.2, $\delta=0.02$, $N=1{,}000$}
\includegraphics[width=21cm]{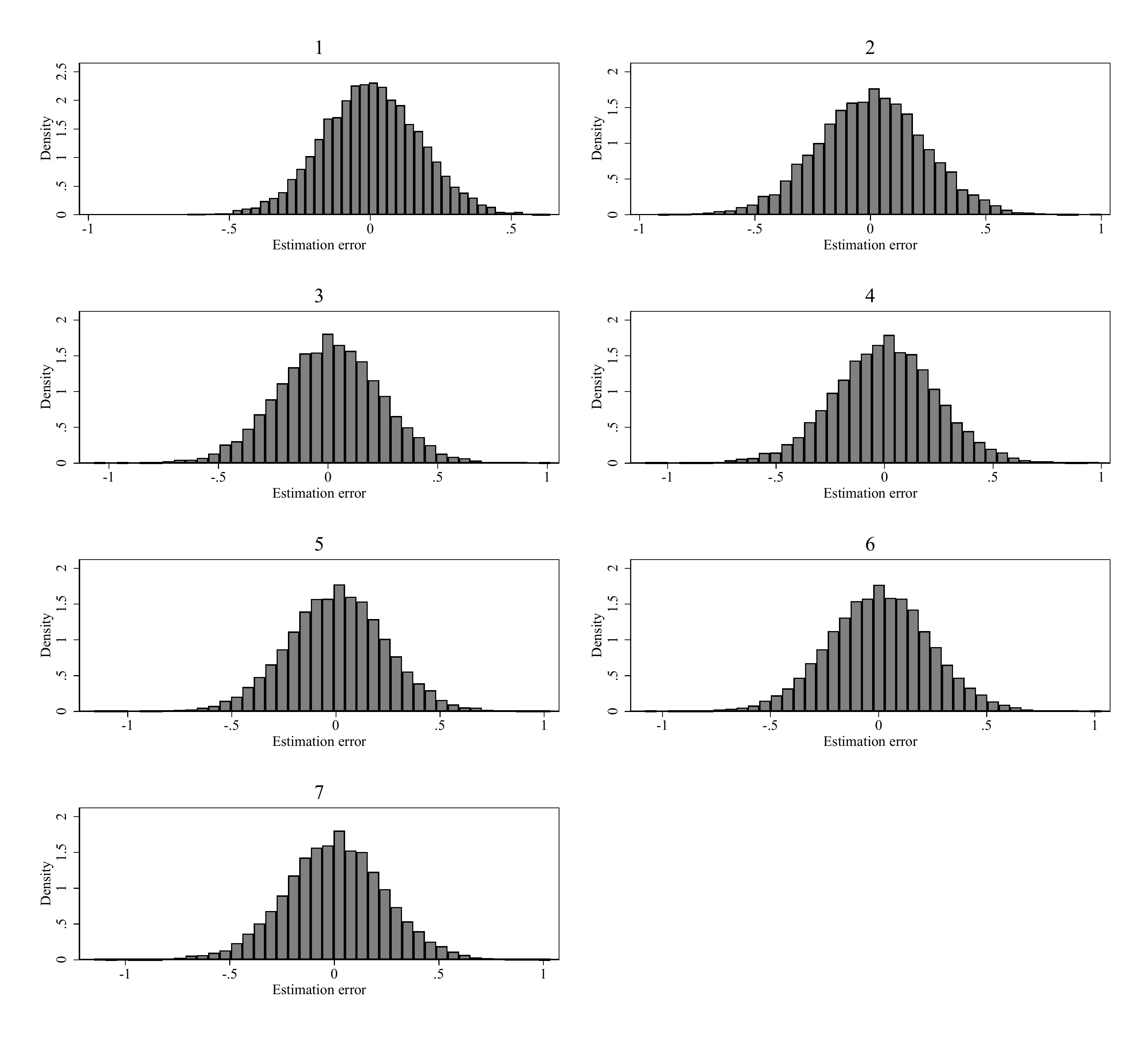}
\begin{footnotesize}
\vspace{0.075cm}
\begin{tabular}{p{19.5cm}}
\textit{Notes:} The details of this simulation design are provided in Section \ref{sec:simulation}. ``1'' corresponds to the 2SLS estimator that additively controls for $X$\@. ``2'' corresponds to $\hat{\tau}_{u}^{cb}$. ``3'' corresponds to $\hat{\tau}_{u}^{ml}$. ``4'' corresponds to $\hat{\tau}_{a,10}^{ml}$. ``5'' corresponds to $\hat{\tau}_{a}^{ml}$. ``6'' corresponds to $\hat{\tau}_{t}^{ml}$ ($= \hat{\tau}_{a,1}^{ml}$). ``7'' corresponds to $\hat{\tau}_{a,0}^{ml}$. All weighting estimators also control for $X$\@. Results are based on 10,000 replications.
\end{tabular}
\end{footnotesize}
\end{adjustwidth}
\end{figure}

\begin{figure}[!p]
\begin{adjustwidth}{-1in}{-1in}
\centering
\caption{Simulation Results for Design A.2, $\delta=0.02$, $N=5{,}000$}
\includegraphics[width=21cm]{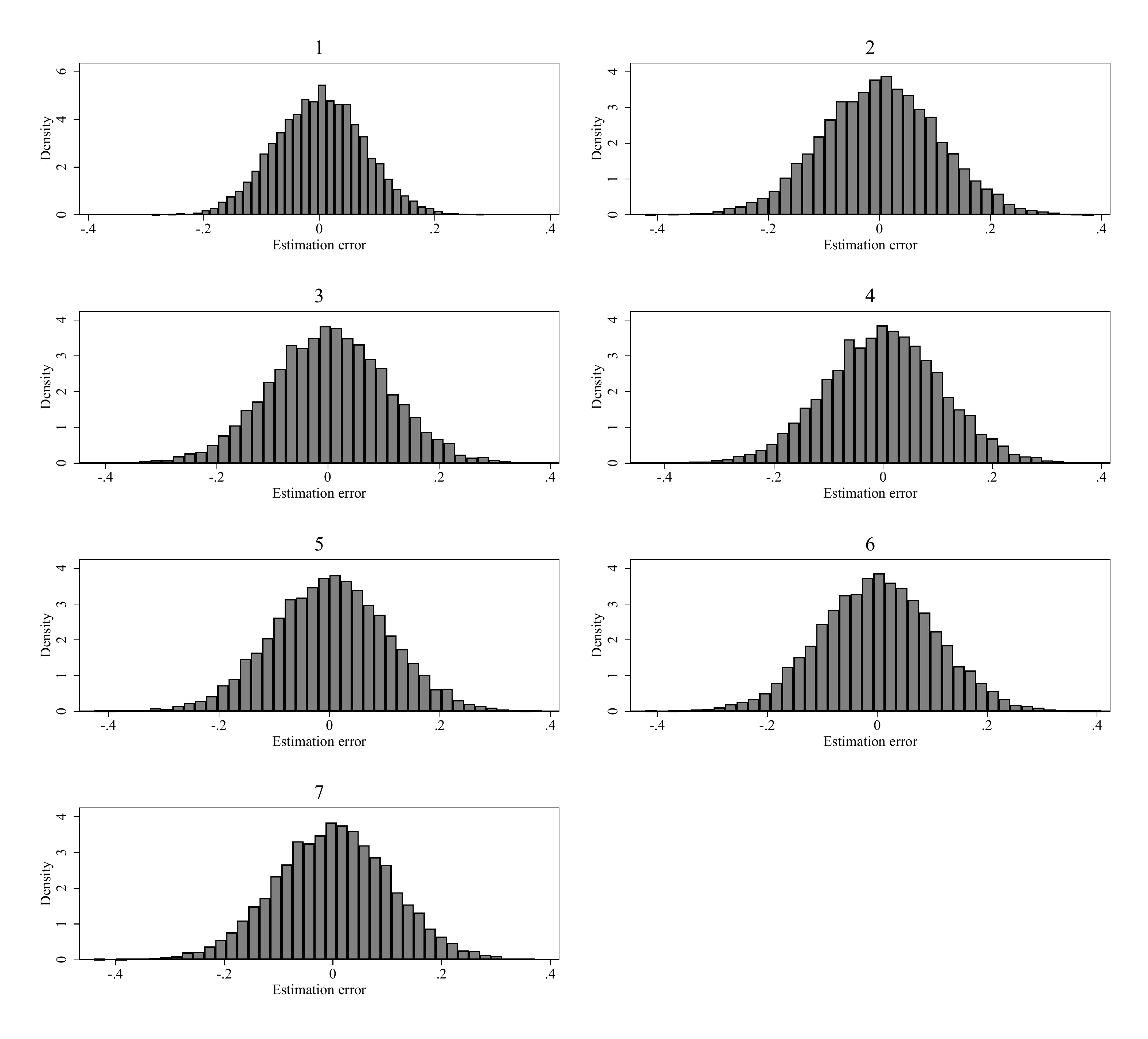}
\begin{footnotesize}
\vspace{0.075cm}
\begin{tabular}{p{19.5cm}}
\textit{Notes:} The details of this simulation design are provided in Section \ref{sec:simulation}. ``1'' corresponds to the 2SLS estimator that additively controls for $X$\@. ``2'' corresponds to $\hat{\tau}_{u}^{cb}$. ``3'' corresponds to $\hat{\tau}_{u}^{ml}$. ``4'' corresponds to $\hat{\tau}_{a,10}^{ml}$. ``5'' corresponds to $\hat{\tau}_{a}^{ml}$. ``6'' corresponds to $\hat{\tau}_{t}^{ml}$ ($= \hat{\tau}_{a,1}^{ml}$). ``7'' corresponds to $\hat{\tau}_{a,0}^{ml}$. All weighting estimators also control for $X$\@. Results are based on 10,000 replications.
\end{tabular}
\end{footnotesize}
\end{adjustwidth}
\end{figure}

\begin{figure}[!p]
\begin{adjustwidth}{-1in}{-1in}
\centering
\caption{Simulation Results for Design A.2, $\delta=0.05$, $N=500$}
\includegraphics[width=21cm]{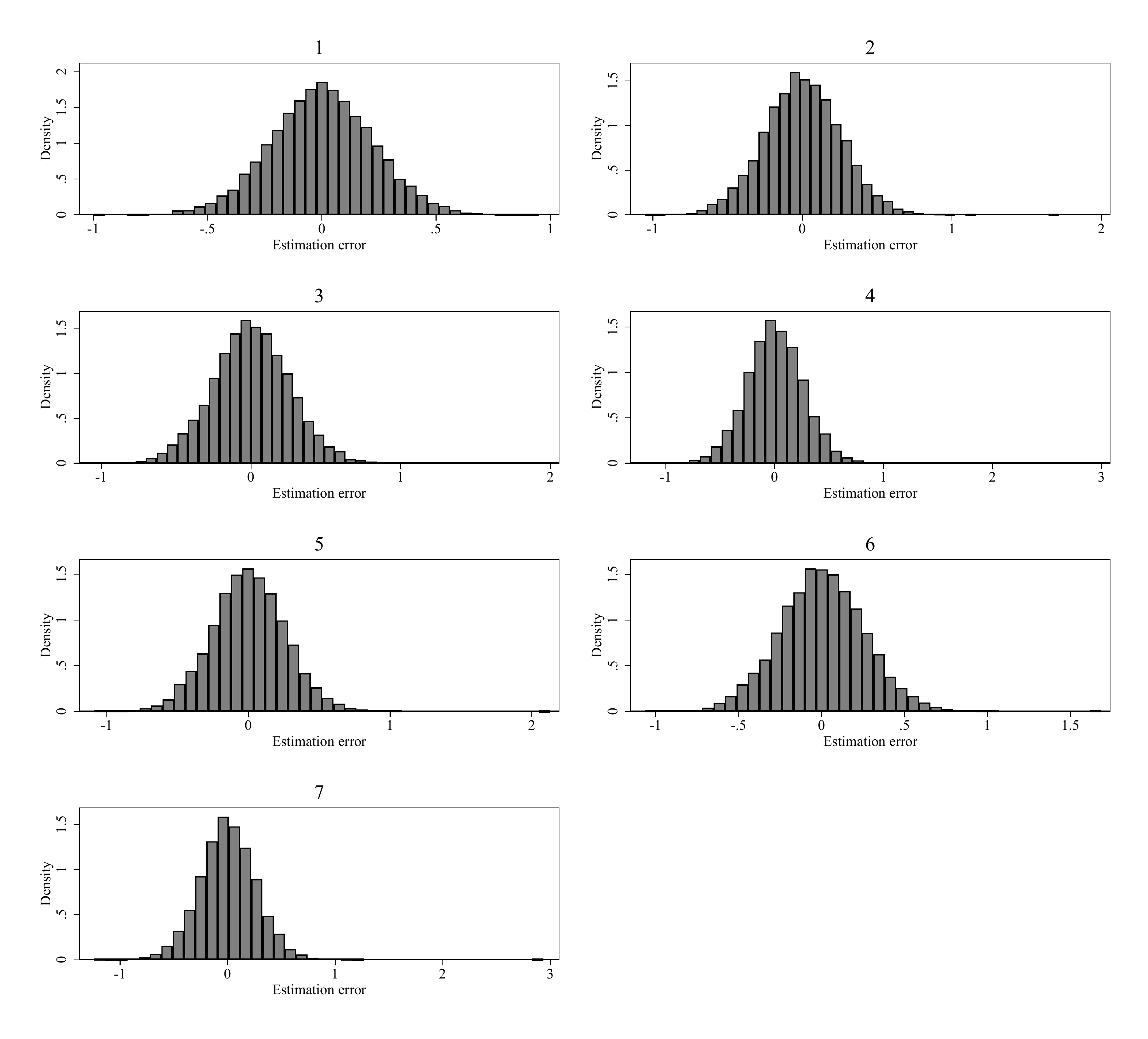}
\begin{footnotesize}
\vspace{0.075cm}
\begin{tabular}{p{19.5cm}}
\textit{Notes:} The details of this simulation design are provided in Section \ref{sec:simulation}. ``1'' corresponds to the 2SLS estimator that additively controls for $X$\@. ``2'' corresponds to $\hat{\tau}_{u}^{cb}$. ``3'' corresponds to $\hat{\tau}_{u}^{ml}$. ``4'' corresponds to $\hat{\tau}_{a,10}^{ml}$. ``5'' corresponds to $\hat{\tau}_{a}^{ml}$. ``6'' corresponds to $\hat{\tau}_{t}^{ml}$ ($= \hat{\tau}_{a,1}^{ml}$). ``7'' corresponds to $\hat{\tau}_{a,0}^{ml}$. All weighting estimators also control for $X$\@. Results are based on 10,000 replications.
\end{tabular}
\end{footnotesize}
\end{adjustwidth}
\end{figure}

\begin{figure}[!p]
\begin{adjustwidth}{-1in}{-1in}
\centering
\caption{Simulation Results for Design A.2, $\delta=0.05$, $N=1{,}000$}
\includegraphics[width=21cm]{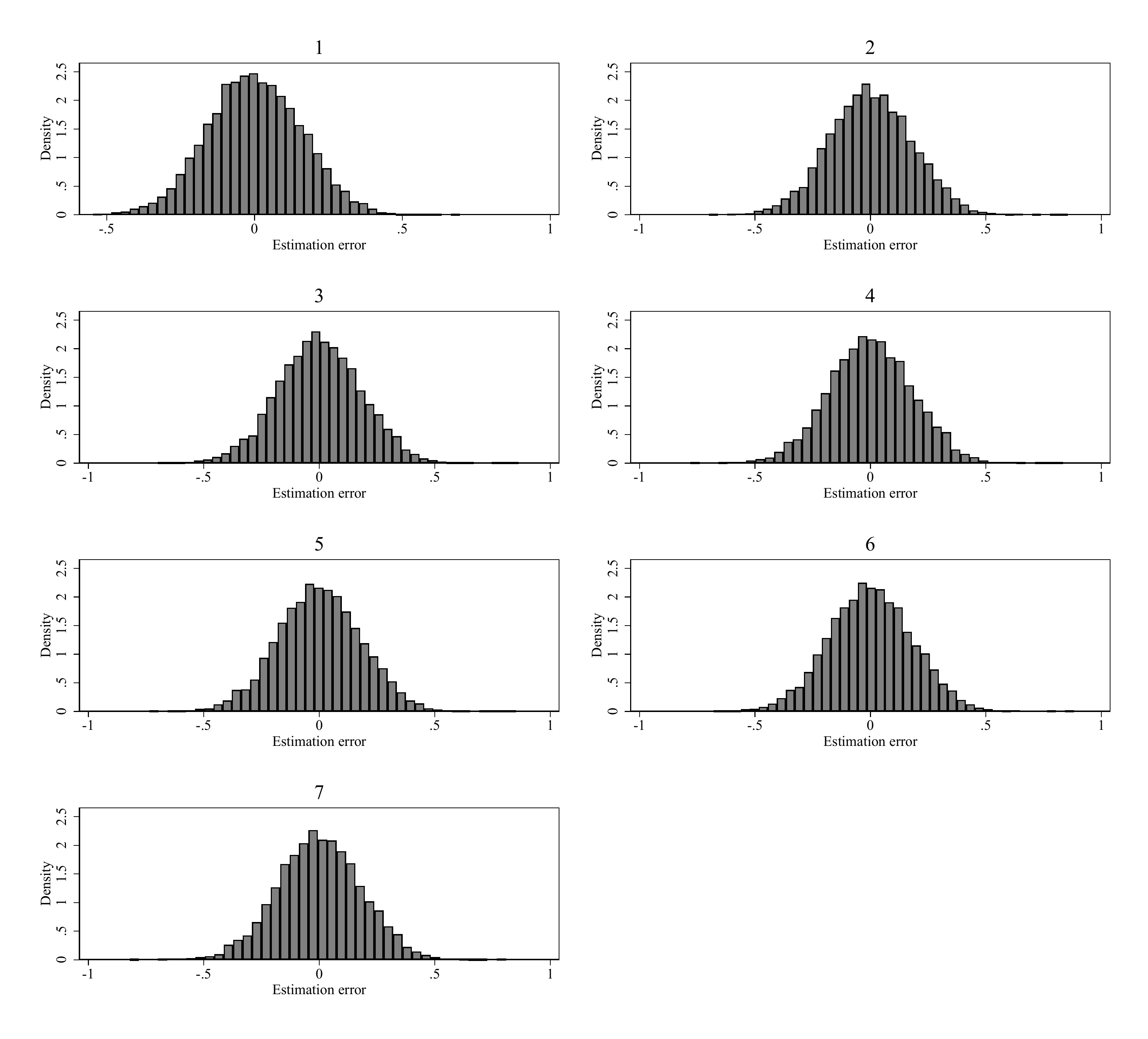}
\begin{footnotesize}
\vspace{0.075cm}
\begin{tabular}{p{19.5cm}}
\textit{Notes:} The details of this simulation design are provided in Section \ref{sec:simulation}. ``1'' corresponds to the 2SLS estimator that additively controls for $X$\@. ``2'' corresponds to $\hat{\tau}_{u}^{cb}$. ``3'' corresponds to $\hat{\tau}_{u}^{ml}$. ``4'' corresponds to $\hat{\tau}_{a,10}^{ml}$. ``5'' corresponds to $\hat{\tau}_{a}^{ml}$. ``6'' corresponds to $\hat{\tau}_{t}^{ml}$ ($= \hat{\tau}_{a,1}^{ml}$). ``7'' corresponds to $\hat{\tau}_{a,0}^{ml}$. All weighting estimators also control for $X$\@. Results are based on 10,000 replications.
\end{tabular}
\end{footnotesize}
\end{adjustwidth}
\end{figure}

\begin{figure}[!p]
\begin{adjustwidth}{-1in}{-1in}
\centering
\caption{Simulation Results for Design A.2, $\delta=0.05$, $N=5{,}000$}
\includegraphics[width=21cm]{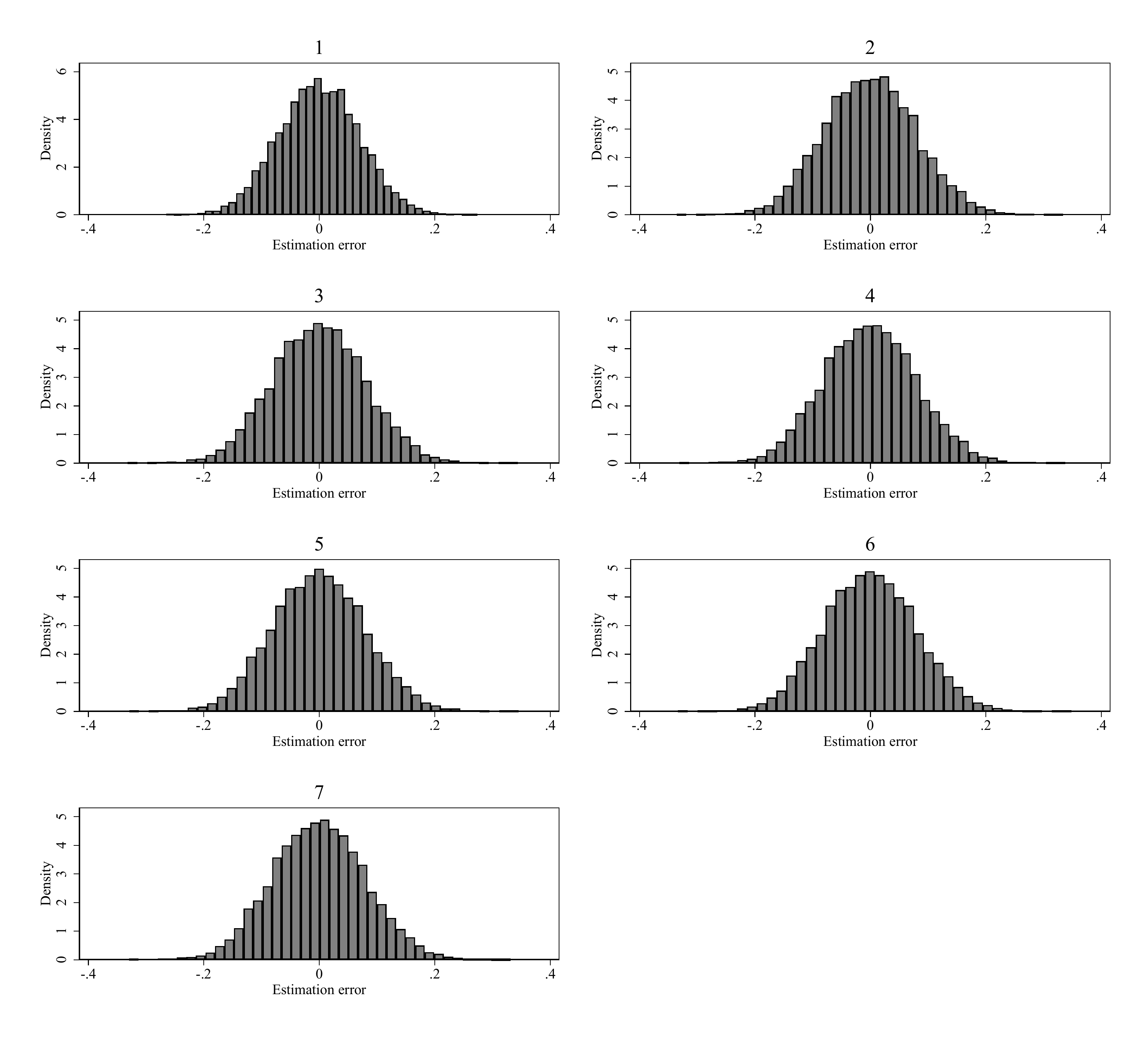}
\begin{footnotesize}
\vspace{0.075cm}
\begin{tabular}{p{19.5cm}}
\textit{Notes:} The details of this simulation design are provided in Section \ref{sec:simulation}. ``1'' corresponds to the 2SLS estimator that additively controls for $X$\@. ``2'' corresponds to $\hat{\tau}_{u}^{cb}$. ``3'' corresponds to $\hat{\tau}_{u}^{ml}$. ``4'' corresponds to $\hat{\tau}_{a,10}^{ml}$. ``5'' corresponds to $\hat{\tau}_{a}^{ml}$. ``6'' corresponds to $\hat{\tau}_{t}^{ml}$ ($= \hat{\tau}_{a,1}^{ml}$). ``7'' corresponds to $\hat{\tau}_{a,0}^{ml}$. All weighting estimators also control for $X$\@. Results are based on 10,000 replications.
\end{tabular}
\end{footnotesize}
\end{adjustwidth}
\end{figure}

\clearpage
\begin{figure}[!p]
\begin{adjustwidth}{-1in}{-1in}
\centering
\caption{Simulation Results for Design B, $\delta=0.01$, $N=500$}
\includegraphics[width=21cm]{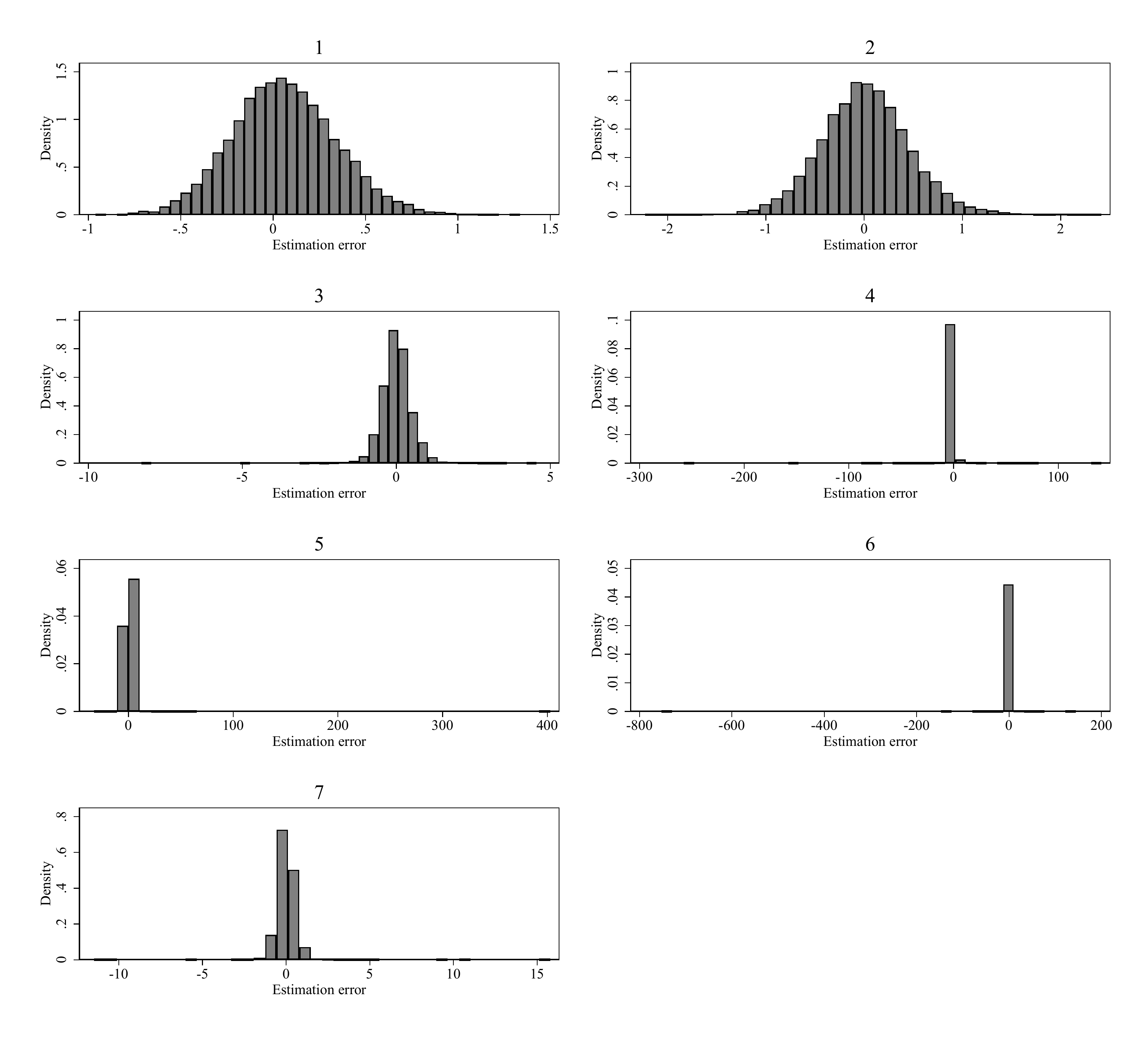}
\begin{footnotesize}
\vspace{0.075cm}
\begin{tabular}{p{19.5cm}}
\textit{Notes:} The details of this simulation design are provided in Section \ref{sec:simulation}. ``1'' corresponds to the 2SLS estimator that additively controls for $X$\@. ``2'' corresponds to $\hat{\tau}_{u}^{cb}$. ``3'' corresponds to $\hat{\tau}_{u}^{ml}$. ``4'' corresponds to $\hat{\tau}_{a,10}^{ml}$. ``5'' corresponds to $\hat{\tau}_{a}^{ml}$. ``6'' corresponds to $\hat{\tau}_{t}^{ml}$ ($= \hat{\tau}_{a,1}^{ml}$). ``7'' corresponds to $\hat{\tau}_{a,0}^{ml}$. All weighting estimators also control for $X$\@. Results are based on 10,000 replications.
\end{tabular}
\end{footnotesize}
\end{adjustwidth}
\end{figure}

\begin{figure}[!p]
\begin{adjustwidth}{-1in}{-1in}
\centering
\caption{Simulation Results for Design B, $\delta=0.01$, $N=1{,}000$}
\includegraphics[width=21cm]{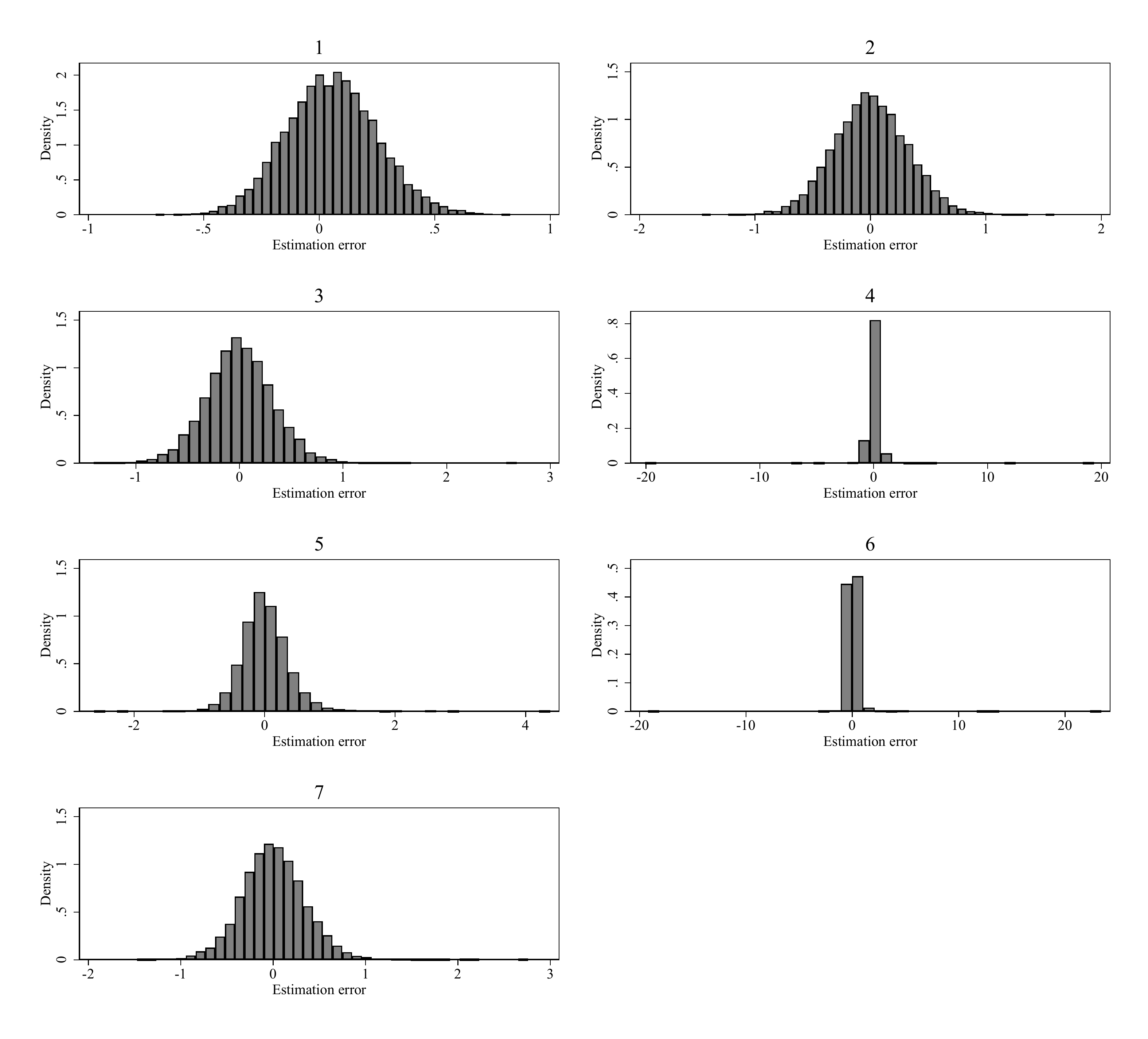}
\begin{footnotesize}
\vspace{0.075cm}
\begin{tabular}{p{19.5cm}}
\textit{Notes:} The details of this simulation design are provided in Section \ref{sec:simulation}. ``1'' corresponds to the 2SLS estimator that additively controls for $X$\@. ``2'' corresponds to $\hat{\tau}_{u}^{cb}$. ``3'' corresponds to $\hat{\tau}_{u}^{ml}$. ``4'' corresponds to $\hat{\tau}_{a,10}^{ml}$. ``5'' corresponds to $\hat{\tau}_{a}^{ml}$. ``6'' corresponds to $\hat{\tau}_{t}^{ml}$ ($= \hat{\tau}_{a,1}^{ml}$). ``7'' corresponds to $\hat{\tau}_{a,0}^{ml}$. All weighting estimators also control for $X$\@. Results are based on 10,000 replications.
\end{tabular}
\end{footnotesize}
\end{adjustwidth}
\end{figure}

\begin{figure}[!p]
\begin{adjustwidth}{-1in}{-1in}
\centering
\caption{Simulation Results for Design B, $\delta=0.01$, $N=5{,}000$}
\includegraphics[width=21cm]{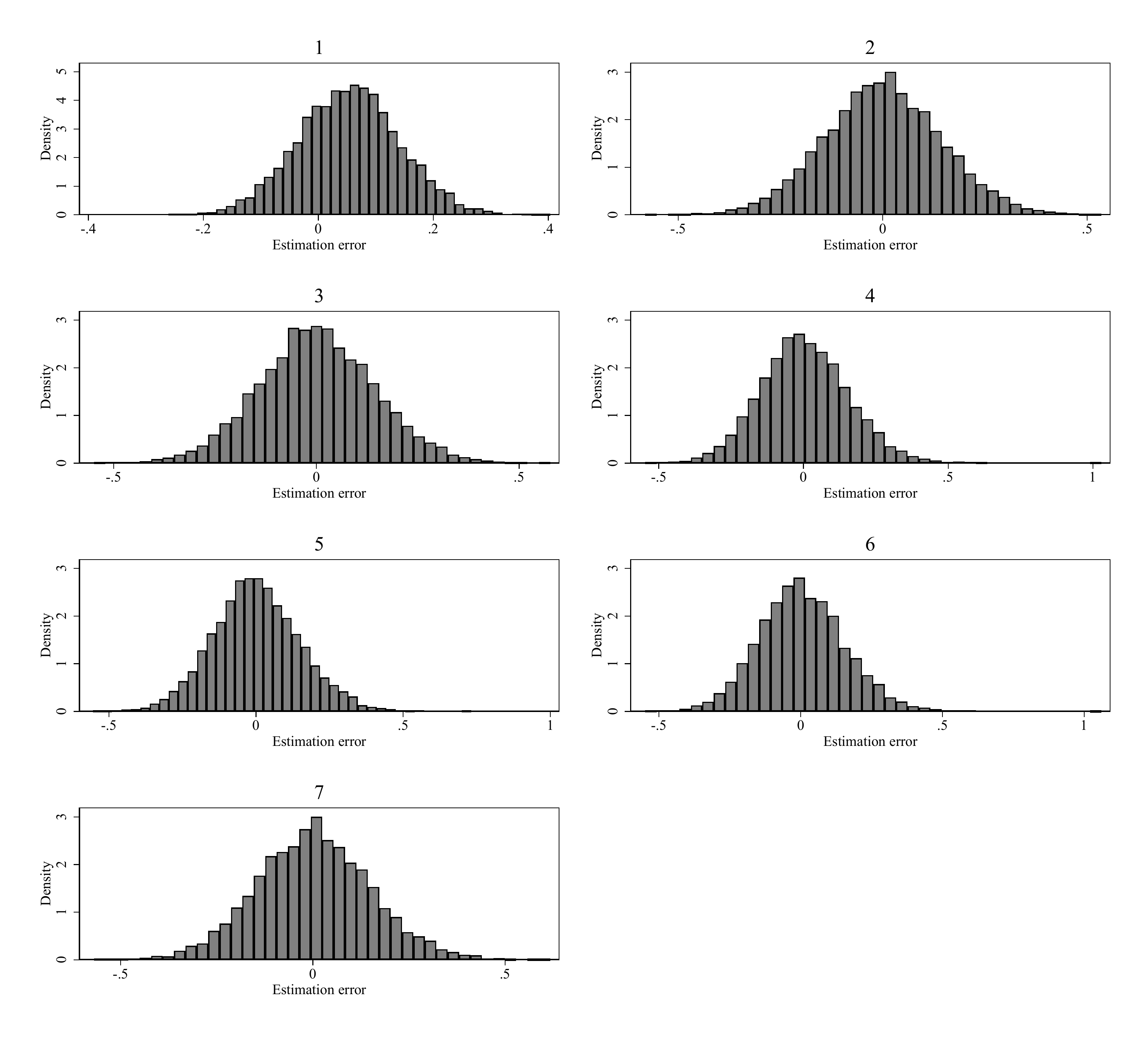}
\begin{footnotesize}
\vspace{0.075cm}
\begin{tabular}{p{19.5cm}}
\textit{Notes:} The details of this simulation design are provided in Section \ref{sec:simulation}. ``1'' corresponds to the 2SLS estimator that additively controls for $X$\@. ``2'' corresponds to $\hat{\tau}_{u}^{cb}$. ``3'' corresponds to $\hat{\tau}_{u}^{ml}$. ``4'' corresponds to $\hat{\tau}_{a,10}^{ml}$. ``5'' corresponds to $\hat{\tau}_{a}^{ml}$. ``6'' corresponds to $\hat{\tau}_{t}^{ml}$ ($= \hat{\tau}_{a,1}^{ml}$). ``7'' corresponds to $\hat{\tau}_{a,0}^{ml}$. All weighting estimators also control for $X$\@. Results are based on 10,000 replications.
\end{tabular}
\end{footnotesize}
\end{adjustwidth}
\end{figure}

\begin{figure}[!p]
\begin{adjustwidth}{-1in}{-1in}
\centering
\caption{Simulation Results for Design B, $\delta=0.02$, $N=500$}
\includegraphics[width=21cm]{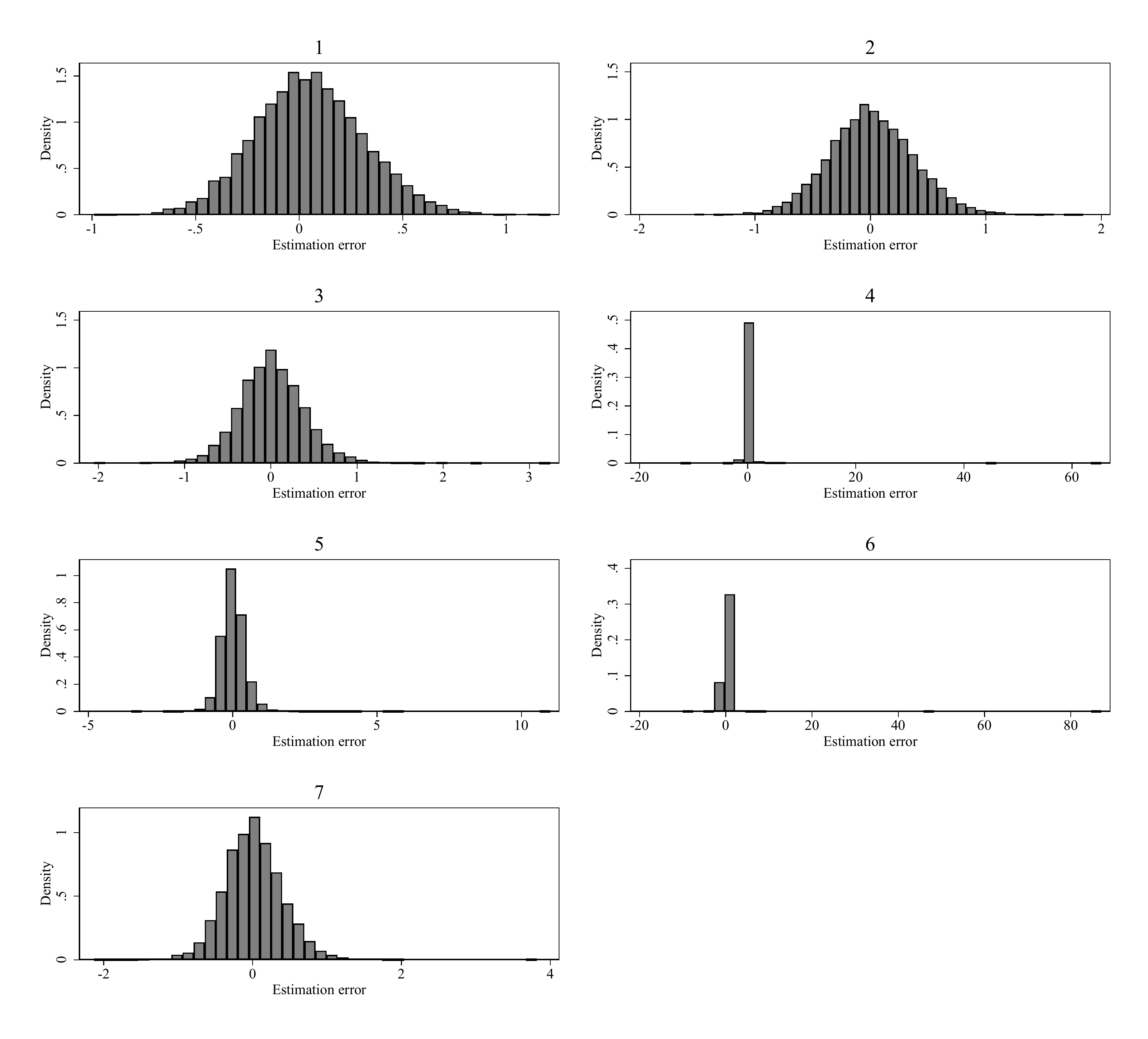}
\begin{footnotesize}
\vspace{0.075cm}
\begin{tabular}{p{19.5cm}}
\textit{Notes:} The details of this simulation design are provided in Section \ref{sec:simulation}. ``1'' corresponds to the 2SLS estimator that additively controls for $X$\@. ``2'' corresponds to $\hat{\tau}_{u}^{cb}$. ``3'' corresponds to $\hat{\tau}_{u}^{ml}$. ``4'' corresponds to $\hat{\tau}_{a,10}^{ml}$. ``5'' corresponds to $\hat{\tau}_{a}^{ml}$. ``6'' corresponds to $\hat{\tau}_{t}^{ml}$ ($= \hat{\tau}_{a,1}^{ml}$). ``7'' corresponds to $\hat{\tau}_{a,0}^{ml}$. All weighting estimators also control for $X$\@. Results are based on 10,000 replications.
\end{tabular}
\end{footnotesize}
\end{adjustwidth}
\end{figure}

\begin{figure}[!p]
\begin{adjustwidth}{-1in}{-1in}
\centering
\caption{Simulation Results for Design B, $\delta=0.02$, $N=1{,}000$}
\includegraphics[width=21cm]{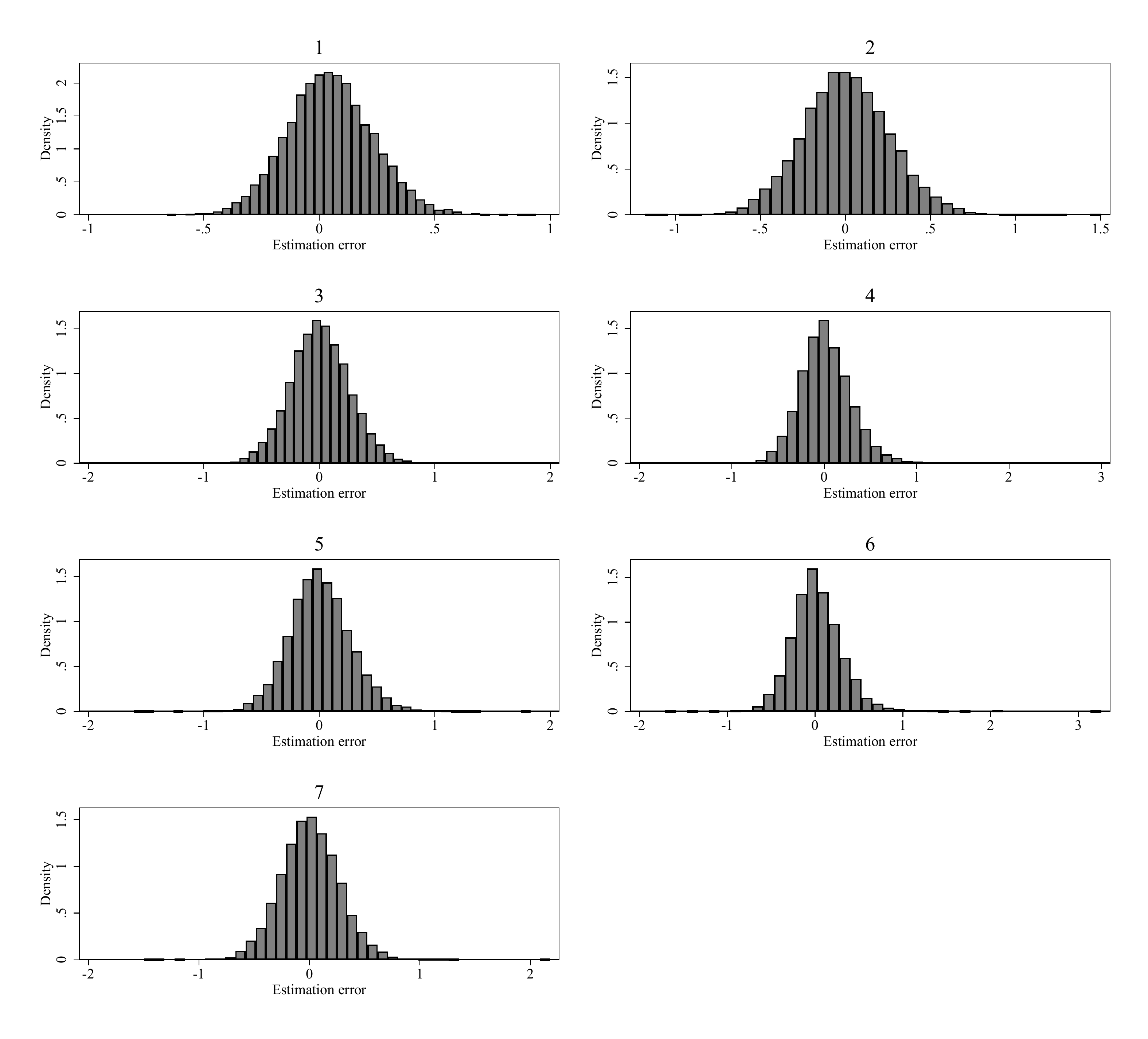}
\begin{footnotesize}
\vspace{0.075cm}
\begin{tabular}{p{19.5cm}}
\textit{Notes:} The details of this simulation design are provided in Section \ref{sec:simulation}. ``1'' corresponds to the 2SLS estimator that additively controls for $X$\@. ``2'' corresponds to $\hat{\tau}_{u}^{cb}$. ``3'' corresponds to $\hat{\tau}_{u}^{ml}$. ``4'' corresponds to $\hat{\tau}_{a,10}^{ml}$. ``5'' corresponds to $\hat{\tau}_{a}^{ml}$. ``6'' corresponds to $\hat{\tau}_{t}^{ml}$ ($= \hat{\tau}_{a,1}^{ml}$). ``7'' corresponds to $\hat{\tau}_{a,0}^{ml}$. All weighting estimators also control for $X$\@. Results are based on 10,000 replications.
\end{tabular}
\end{footnotesize}
\end{adjustwidth}
\end{figure}

\begin{figure}[!p]
\begin{adjustwidth}{-1in}{-1in}
\centering
\caption{Simulation Results for Design B, $\delta=0.02$, $N=5{,}000$}
\includegraphics[width=21cm]{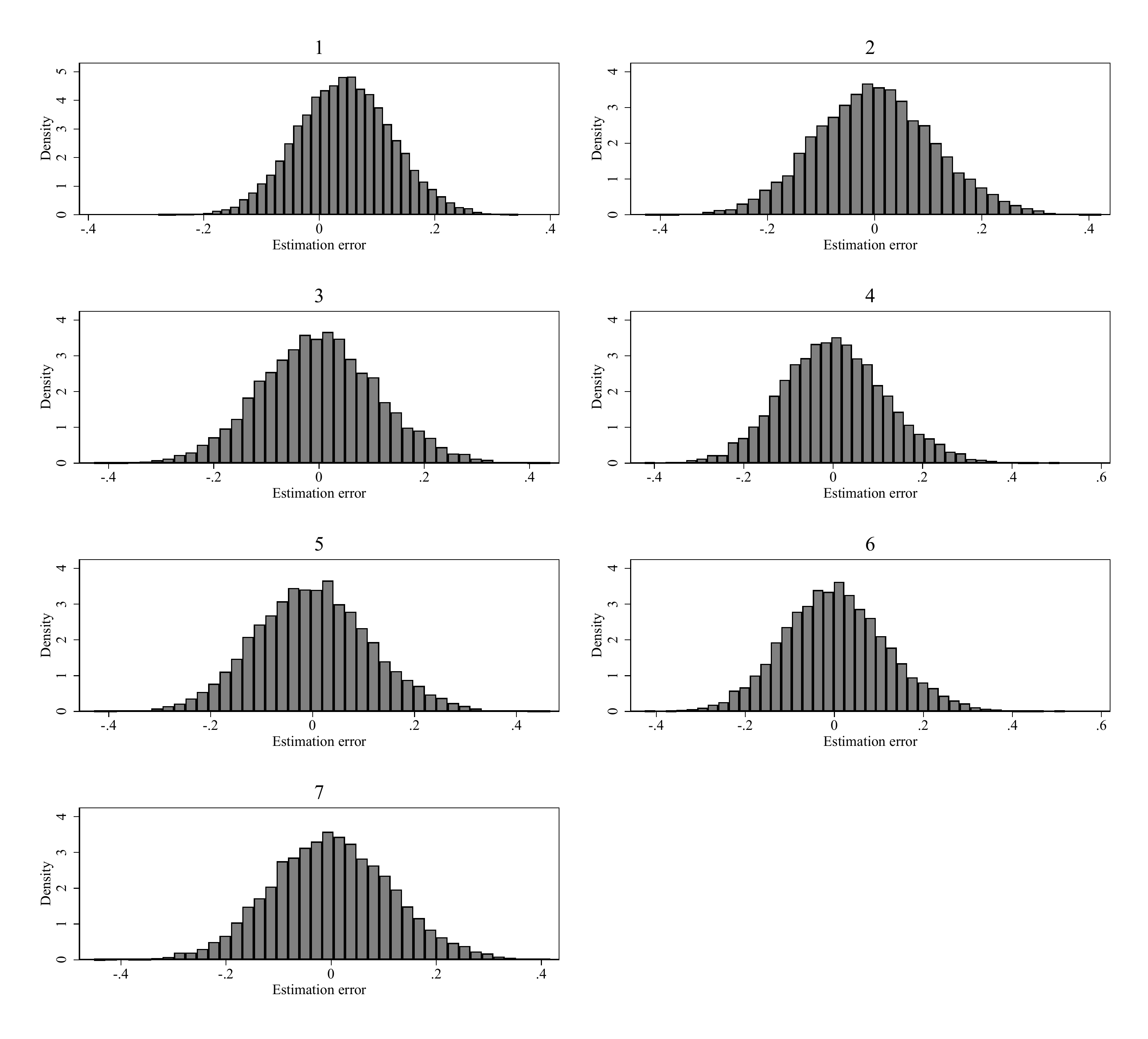}
\begin{footnotesize}
\vspace{0.075cm}
\begin{tabular}{p{19.5cm}}
\textit{Notes:} The details of this simulation design are provided in Section \ref{sec:simulation}. ``1'' corresponds to the 2SLS estimator that additively controls for $X$\@. ``2'' corresponds to $\hat{\tau}_{u}^{cb}$. ``3'' corresponds to $\hat{\tau}_{u}^{ml}$. ``4'' corresponds to $\hat{\tau}_{a,10}^{ml}$. ``5'' corresponds to $\hat{\tau}_{a}^{ml}$. ``6'' corresponds to $\hat{\tau}_{t}^{ml}$ ($= \hat{\tau}_{a,1}^{ml}$). ``7'' corresponds to $\hat{\tau}_{a,0}^{ml}$. All weighting estimators also control for $X$\@. Results are based on 10,000 replications.
\end{tabular}
\end{footnotesize}
\end{adjustwidth}
\end{figure}

\begin{figure}[!p]
\begin{adjustwidth}{-1in}{-1in}
\centering
\caption{Simulation Results for Design B, $\delta=0.05$, $N=500$}
\includegraphics[width=21cm]{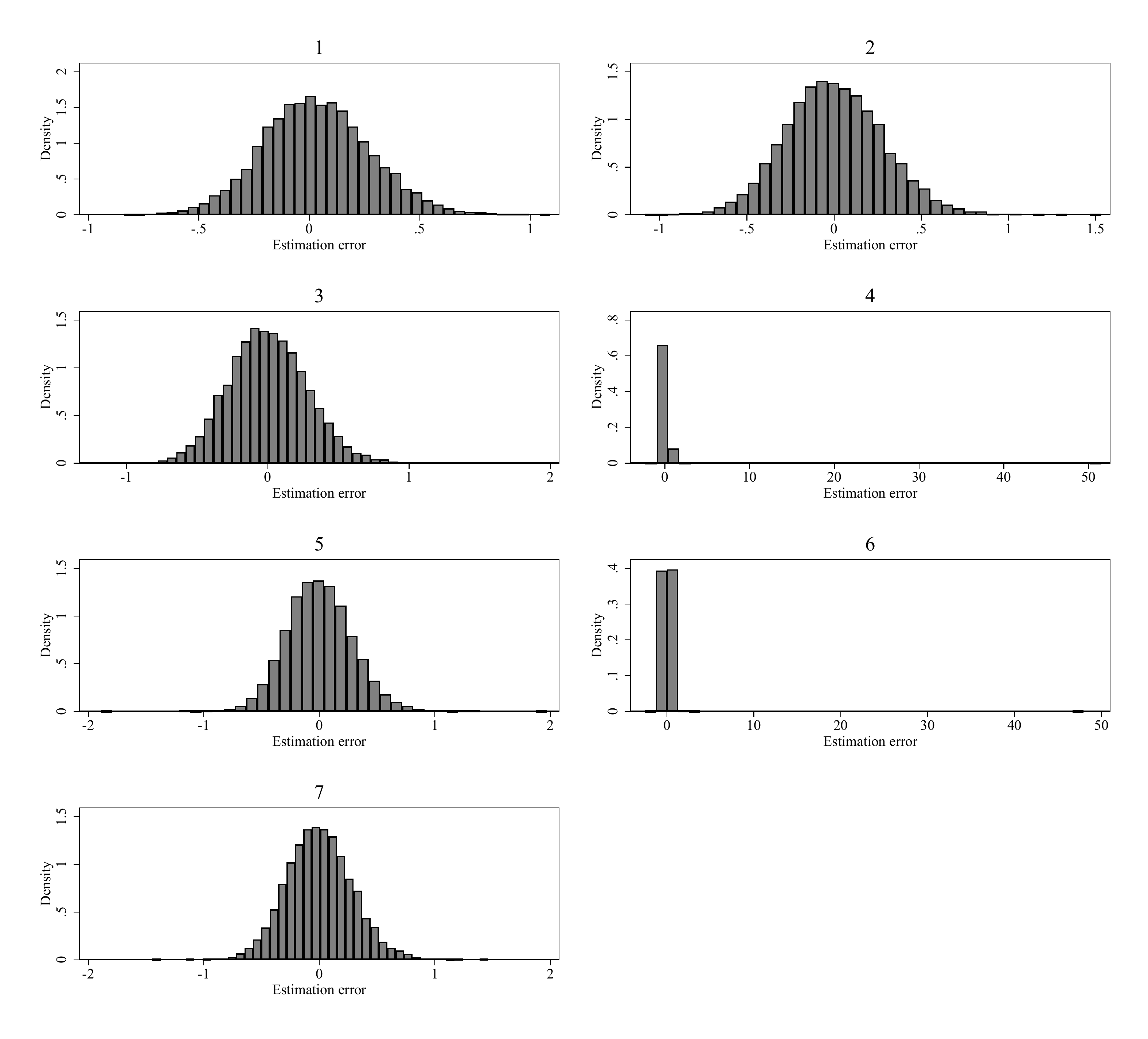}
\begin{footnotesize}
\vspace{0.075cm}
\begin{tabular}{p{19.5cm}}
\textit{Notes:} The details of this simulation design are provided in Section \ref{sec:simulation}. ``1'' corresponds to the 2SLS estimator that additively controls for $X$\@. ``2'' corresponds to $\hat{\tau}_{u}^{cb}$. ``3'' corresponds to $\hat{\tau}_{u}^{ml}$. ``4'' corresponds to $\hat{\tau}_{a,10}^{ml}$. ``5'' corresponds to $\hat{\tau}_{a}^{ml}$. ``6'' corresponds to $\hat{\tau}_{t}^{ml}$ ($= \hat{\tau}_{a,1}^{ml}$). ``7'' corresponds to $\hat{\tau}_{a,0}^{ml}$. All weighting estimators also control for $X$\@. Results are based on 10,000 replications.
\end{tabular}
\end{footnotesize}
\end{adjustwidth}
\end{figure}

\begin{figure}[!p]
\begin{adjustwidth}{-1in}{-1in}
\centering
\caption{Simulation Results for Design B, $\delta=0.05$, $N=1{,}000$}
\includegraphics[width=21cm]{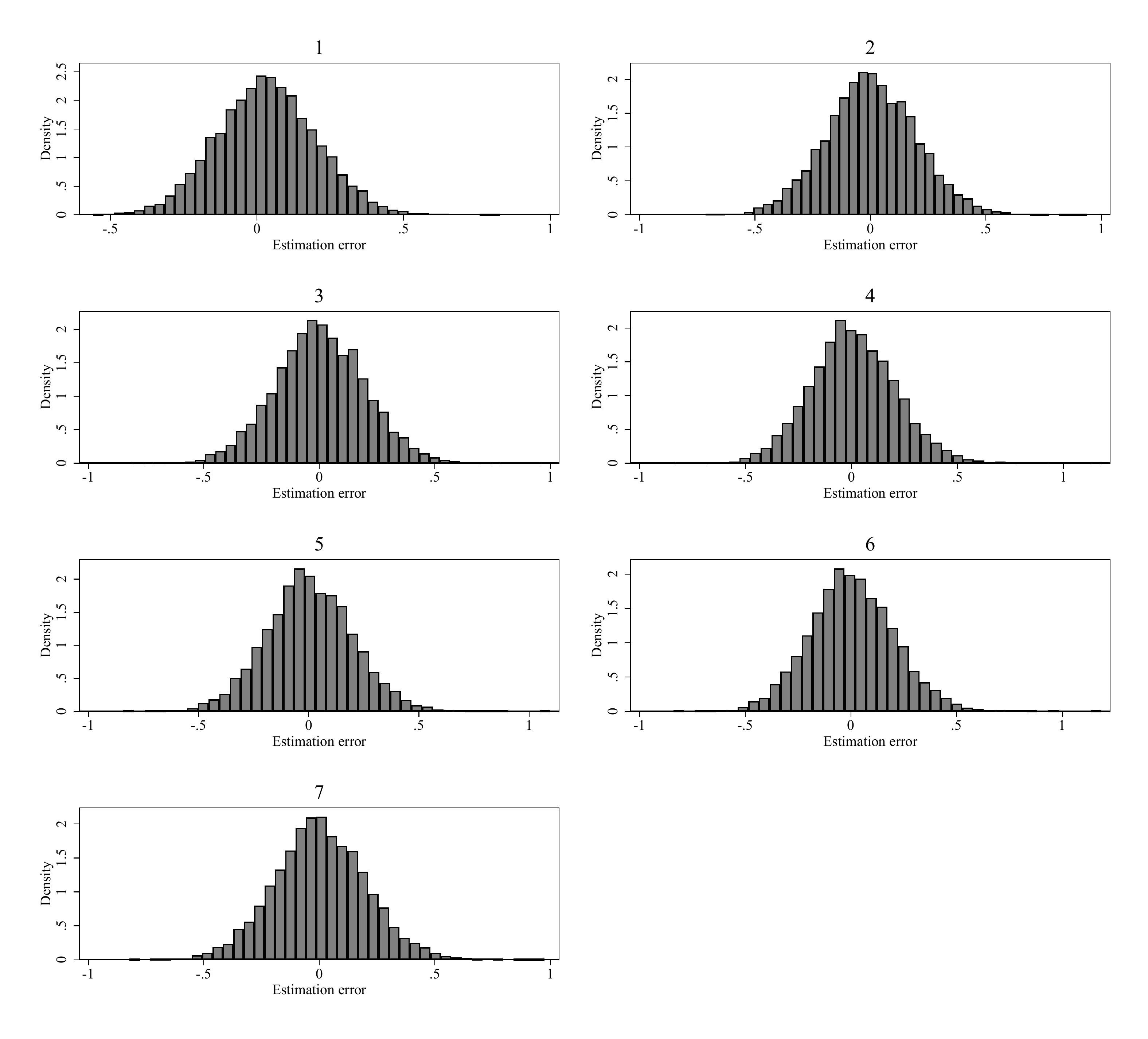}
\begin{footnotesize}
\vspace{0.075cm}
\begin{tabular}{p{19.5cm}}
\textit{Notes:} The details of this simulation design are provided in Section \ref{sec:simulation}. ``1'' corresponds to the 2SLS estimator that additively controls for $X$\@. ``2'' corresponds to $\hat{\tau}_{u}^{cb}$. ``3'' corresponds to $\hat{\tau}_{u}^{ml}$. ``4'' corresponds to $\hat{\tau}_{a,10}^{ml}$. ``5'' corresponds to $\hat{\tau}_{a}^{ml}$. ``6'' corresponds to $\hat{\tau}_{t}^{ml}$ ($= \hat{\tau}_{a,1}^{ml}$). ``7'' corresponds to $\hat{\tau}_{a,0}^{ml}$. All weighting estimators also control for $X$\@. Results are based on 10,000 replications.
\end{tabular}
\end{footnotesize}
\end{adjustwidth}
\end{figure}

\begin{figure}[!p]
\begin{adjustwidth}{-1in}{-1in}
\centering
\caption{Simulation Results for Design B, $\delta=0.05$, $N=5{,}000$}
\includegraphics[width=21cm]{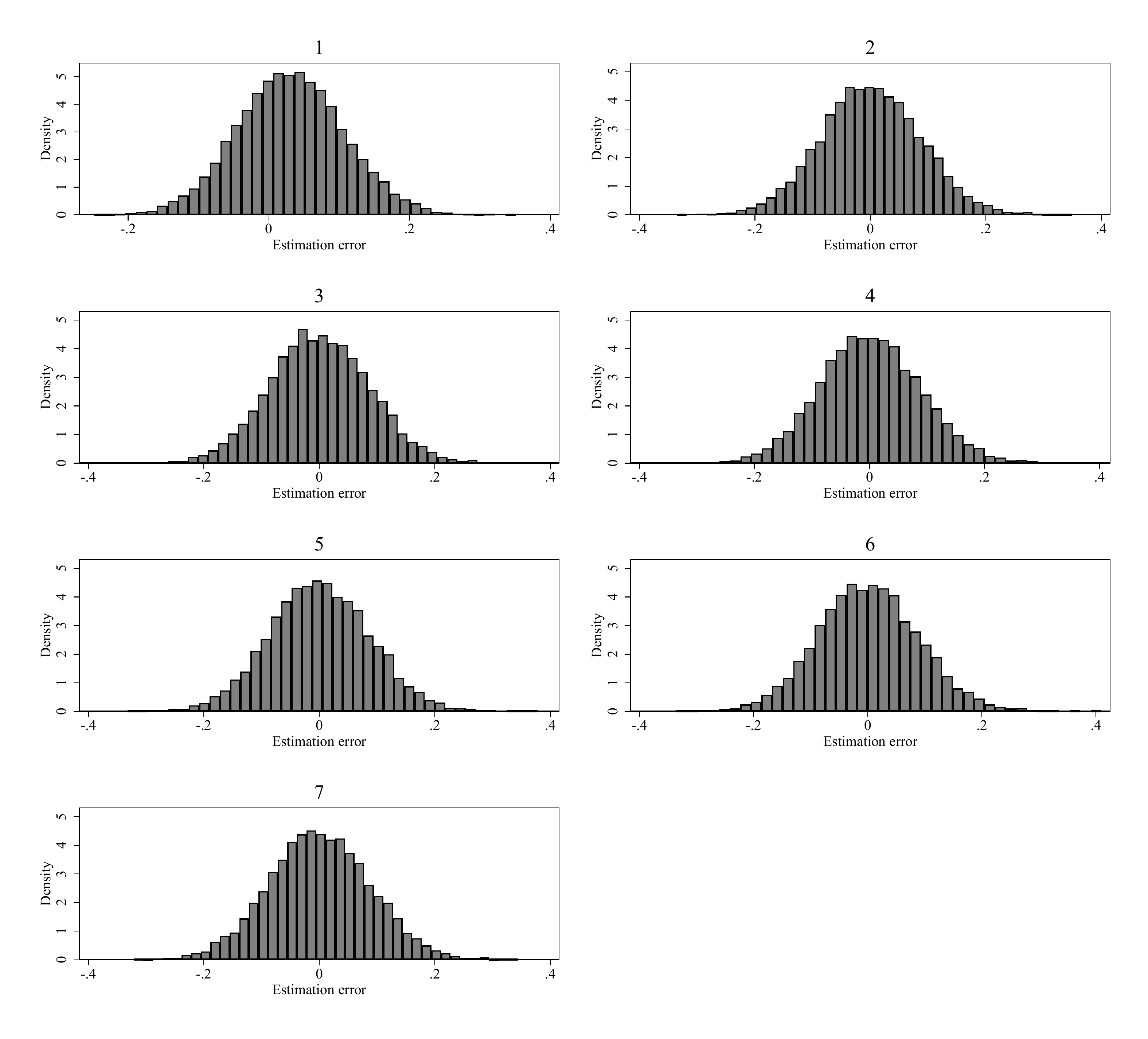}
\begin{footnotesize}
\vspace{0.075cm}
\begin{tabular}{p{19.5cm}}
\textit{Notes:} The details of this simulation design are provided in Section \ref{sec:simulation}. ``1'' corresponds to the 2SLS estimator that additively controls for $X$\@. ``2'' corresponds to $\hat{\tau}_{u}^{cb}$. ``3'' corresponds to $\hat{\tau}_{u}^{ml}$. ``4'' corresponds to $\hat{\tau}_{a,10}^{ml}$. ``5'' corresponds to $\hat{\tau}_{a}^{ml}$. ``6'' corresponds to $\hat{\tau}_{t}^{ml}$ ($= \hat{\tau}_{a,1}^{ml}$). ``7'' corresponds to $\hat{\tau}_{a,0}^{ml}$. All weighting estimators also control for $X$\@. Results are based on 10,000 replications.
\end{tabular}
\end{footnotesize}
\end{adjustwidth}
\end{figure}

\begin{figure}[!p]
\begin{adjustwidth}{-1in}{-1in}
\centering
\caption{Simulation Results for Design C, $\delta=0.01$, $N=500$}
\includegraphics[width=21cm]{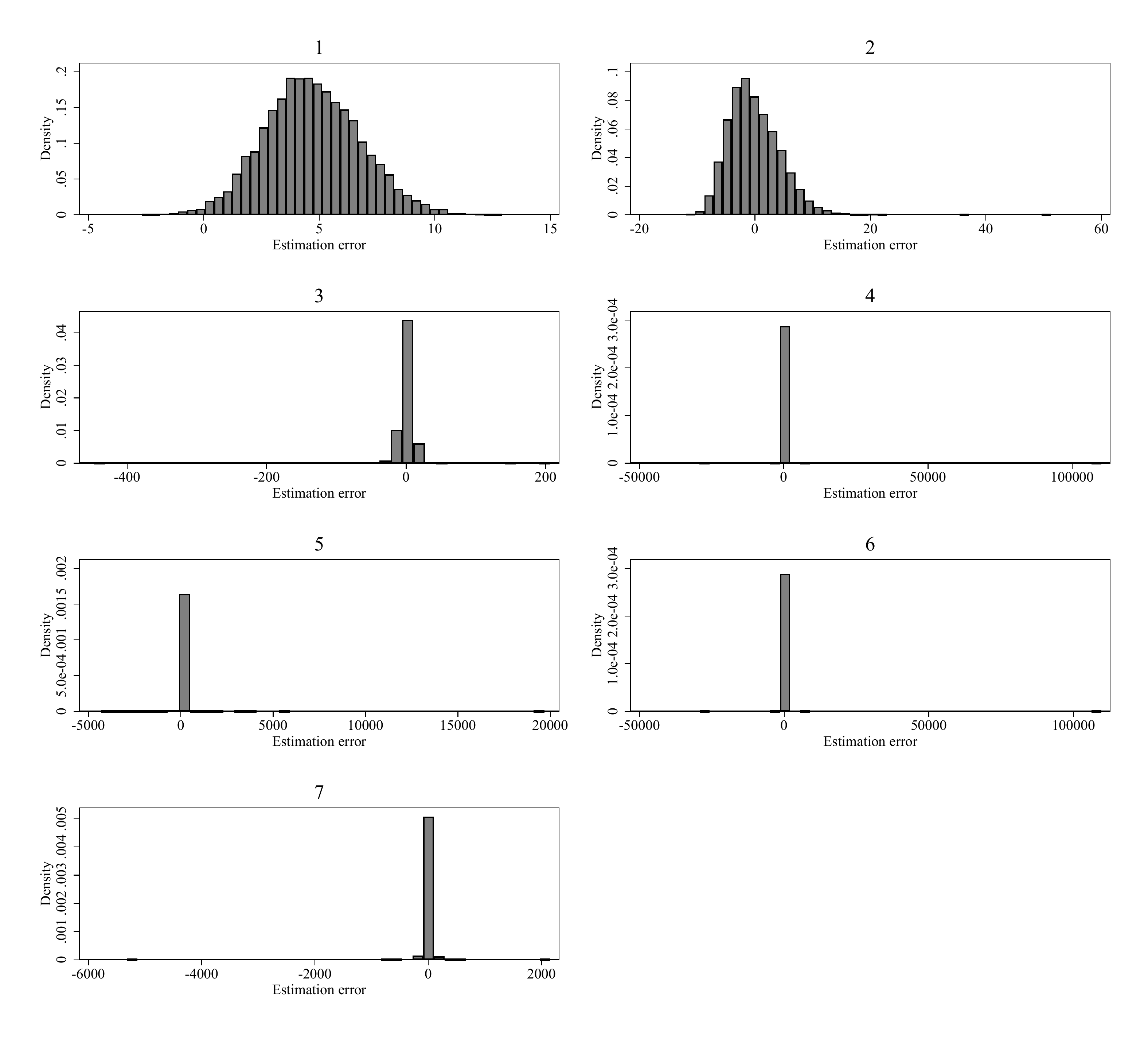}
\begin{footnotesize}
\vspace{0.075cm}
\begin{tabular}{p{19.5cm}}
\textit{Notes:} The details of this simulation design are provided in Section \ref{sec:simulation}. ``1'' corresponds to the 2SLS estimator that additively controls for $X$\@. ``2'' corresponds to $\hat{\tau}_{u}^{cb}$. ``3'' corresponds to $\hat{\tau}_{u}^{ml}$. ``4'' corresponds to $\hat{\tau}_{a,10}^{ml}$. ``5'' corresponds to $\hat{\tau}_{a}^{ml}$. ``6'' corresponds to $\hat{\tau}_{t}^{ml}$ ($= \hat{\tau}_{a,1}^{ml}$). ``7'' corresponds to $\hat{\tau}_{a,0}^{ml}$. All weighting estimators also control for $X$\@. Results are based on 10,000 replications.
\end{tabular}
\end{footnotesize}
\end{adjustwidth}
\end{figure}

\begin{figure}[!p]
\begin{adjustwidth}{-1in}{-1in}
\centering
\caption{Simulation Results for Design C, $\delta=0.01$, $N=1{,}000$}
\includegraphics[width=21cm]{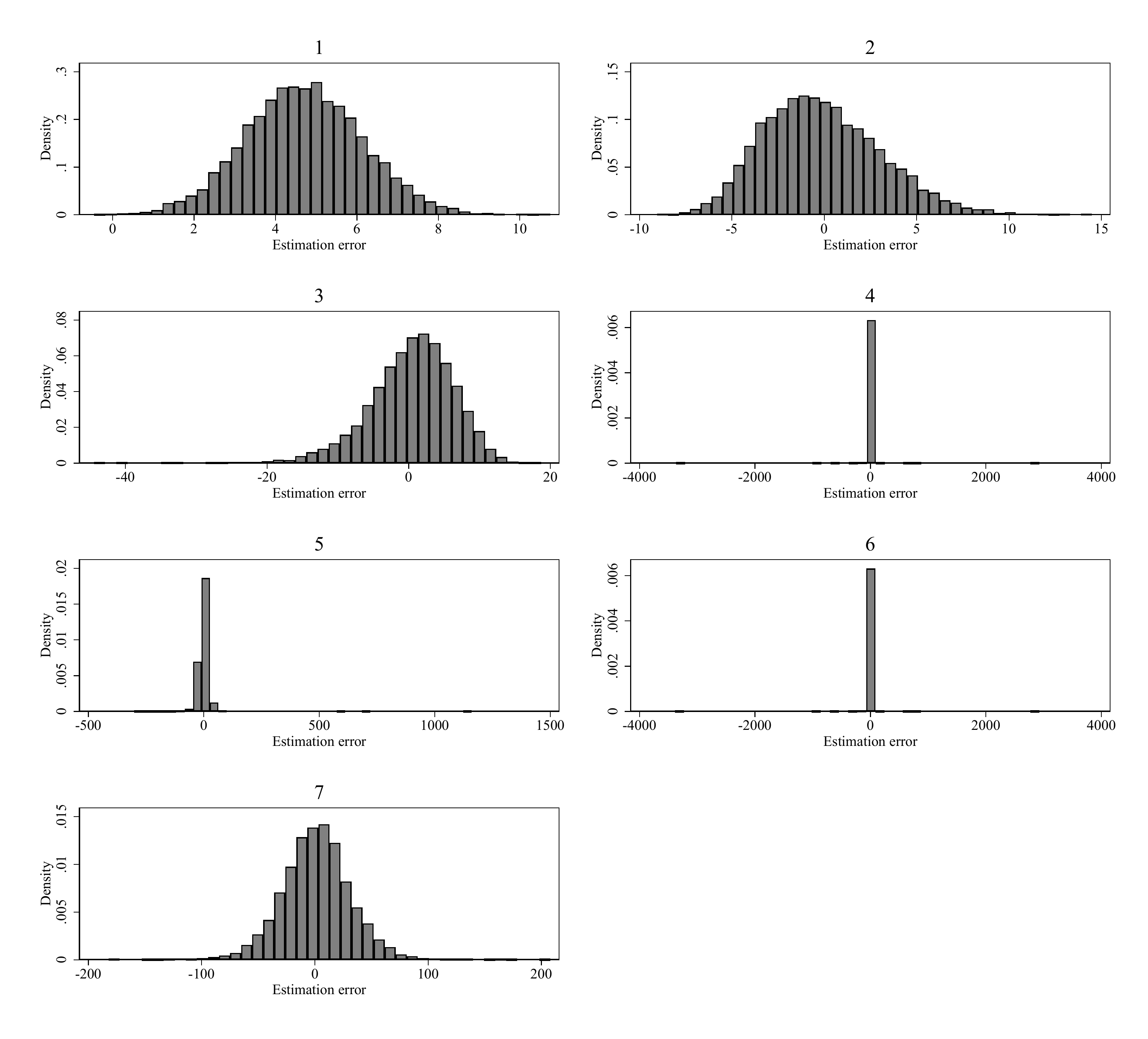}
\begin{footnotesize}
\vspace{0.075cm}
\begin{tabular}{p{19.5cm}}
\textit{Notes:} The details of this simulation design are provided in Section \ref{sec:simulation}. ``1'' corresponds to the 2SLS estimator that additively controls for $X$\@. ``2'' corresponds to $\hat{\tau}_{u}^{cb}$. ``3'' corresponds to $\hat{\tau}_{u}^{ml}$. ``4'' corresponds to $\hat{\tau}_{a,10}^{ml}$. ``5'' corresponds to $\hat{\tau}_{a}^{ml}$. ``6'' corresponds to $\hat{\tau}_{t}^{ml}$ ($= \hat{\tau}_{a,1}^{ml}$). ``7'' corresponds to $\hat{\tau}_{a,0}^{ml}$. All weighting estimators also control for $X$\@. Results are based on 10,000 replications.
\end{tabular}
\end{footnotesize}
\end{adjustwidth}
\end{figure}

\begin{figure}[!p]
\begin{adjustwidth}{-1in}{-1in}
\centering
\caption{Simulation Results for Design C, $\delta=0.01$, $N=5{,}000$}
\includegraphics[width=21cm]{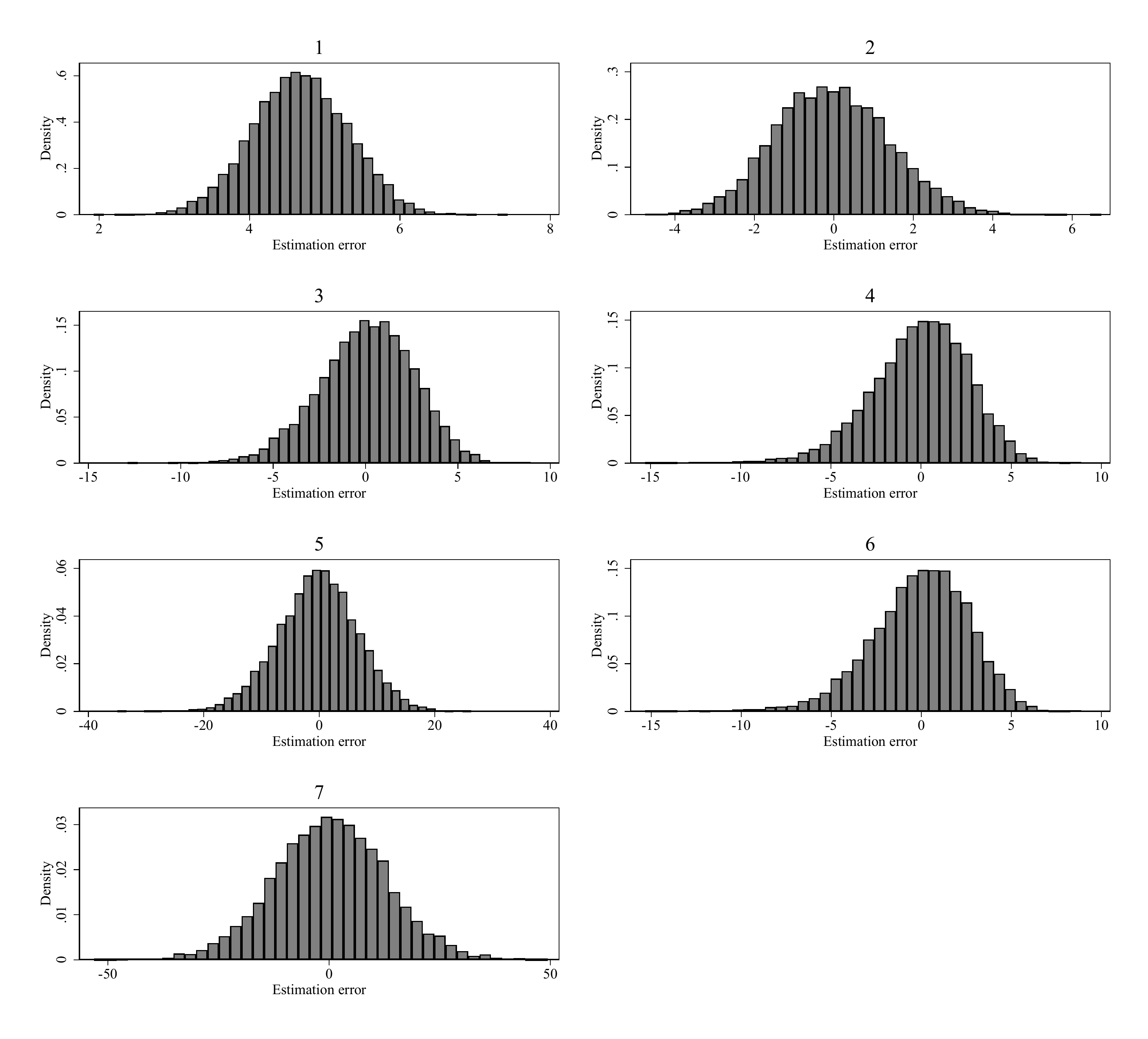}
\begin{footnotesize}
\vspace{0.075cm}
\begin{tabular}{p{19.5cm}}
\textit{Notes:} The details of this simulation design are provided in Section \ref{sec:simulation}. ``1'' corresponds to the 2SLS estimator that additively controls for $X$\@. ``2'' corresponds to $\hat{\tau}_{u}^{cb}$. ``3'' corresponds to $\hat{\tau}_{u}^{ml}$. ``4'' corresponds to $\hat{\tau}_{a,10}^{ml}$. ``5'' corresponds to $\hat{\tau}_{a}^{ml}$. ``6'' corresponds to $\hat{\tau}_{t}^{ml}$ ($= \hat{\tau}_{a,1}^{ml}$). ``7'' corresponds to $\hat{\tau}_{a,0}^{ml}$. All weighting estimators also control for $X$\@. Results are based on 10,000 replications.
\end{tabular}
\end{footnotesize}
\end{adjustwidth}
\end{figure}

\begin{figure}[!p]
\begin{adjustwidth}{-1in}{-1in}
\centering
\caption{Simulation Results for Design C, $\delta=0.02$, $N=500$}
\includegraphics[width=21cm]{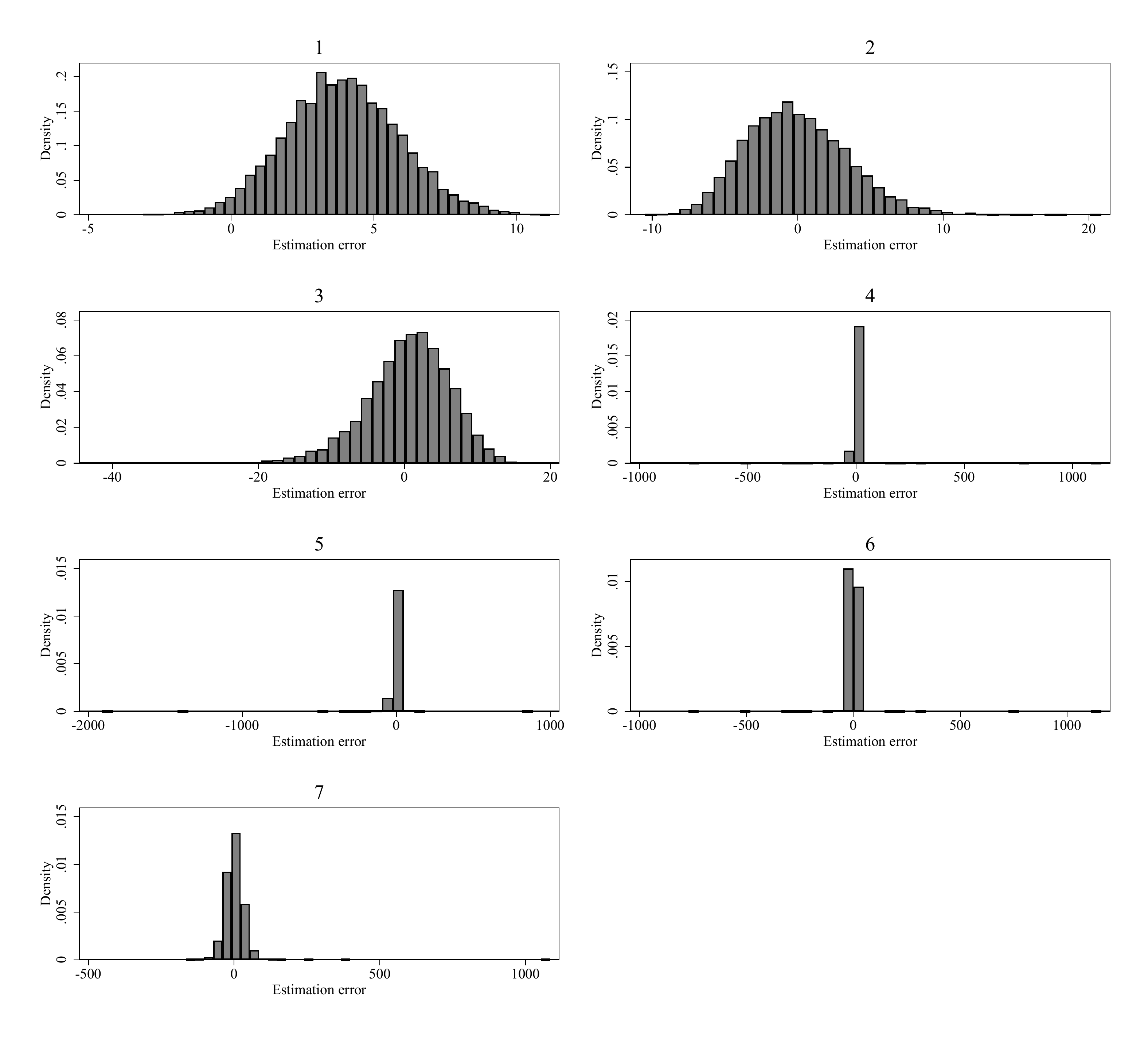}
\begin{footnotesize}
\vspace{0.075cm}
\begin{tabular}{p{19.5cm}}
\textit{Notes:} The details of this simulation design are provided in Section \ref{sec:simulation}. ``1'' corresponds to the 2SLS estimator that additively controls for $X$\@. ``2'' corresponds to $\hat{\tau}_{u}^{cb}$. ``3'' corresponds to $\hat{\tau}_{u}^{ml}$. ``4'' corresponds to $\hat{\tau}_{a,10}^{ml}$. ``5'' corresponds to $\hat{\tau}_{a}^{ml}$. ``6'' corresponds to $\hat{\tau}_{t}^{ml}$ ($= \hat{\tau}_{a,1}^{ml}$). ``7'' corresponds to $\hat{\tau}_{a,0}^{ml}$. All weighting estimators also control for $X$\@. Results are based on 10,000 replications.
\end{tabular}
\end{footnotesize}
\end{adjustwidth}
\end{figure}

\begin{figure}[!p]
\begin{adjustwidth}{-1in}{-1in}
\centering
\caption{Simulation Results for Design C, $\delta=0.02$, $N=1{,}000$}
\includegraphics[width=21cm]{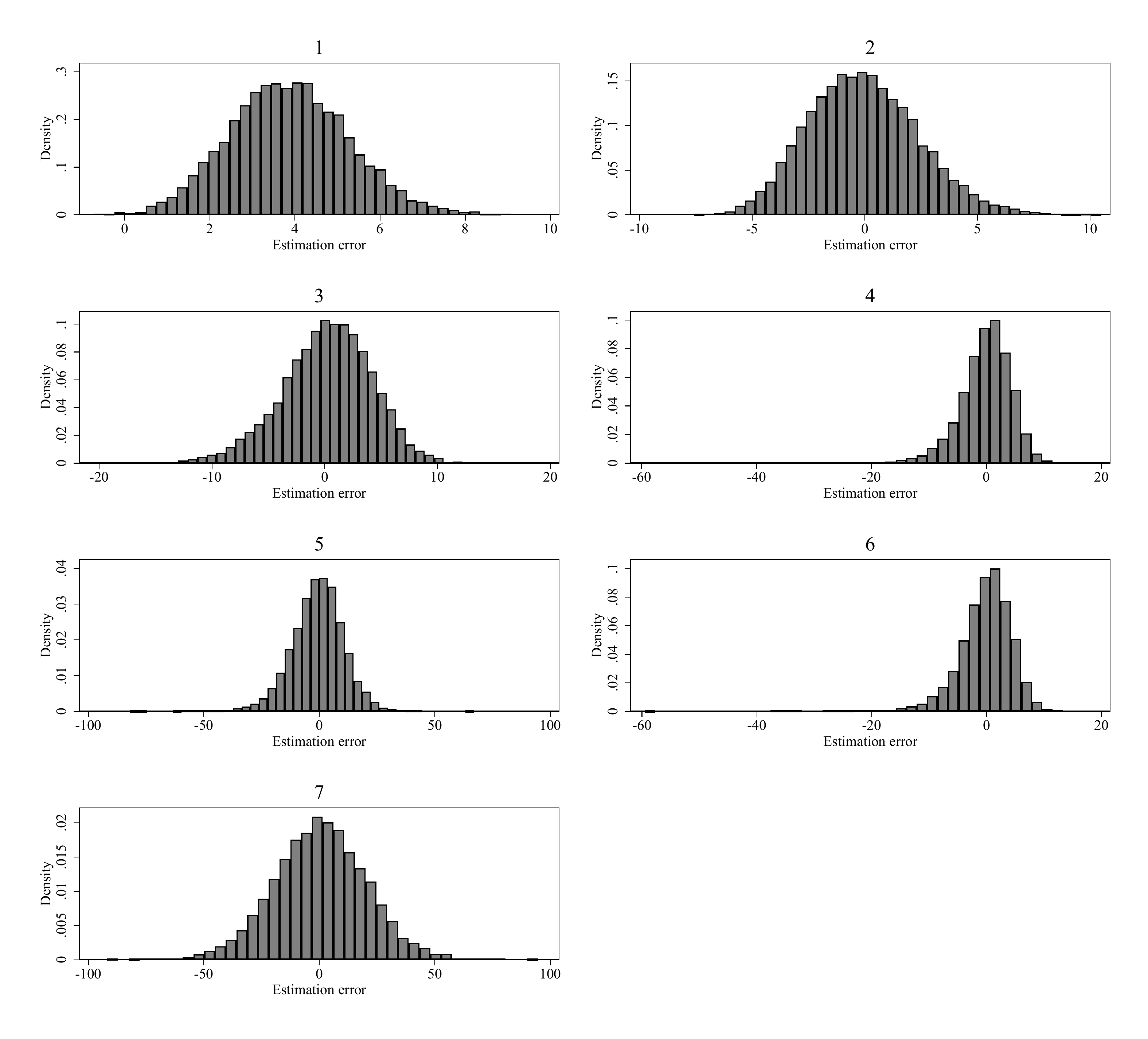}
\begin{footnotesize}
\vspace{0.075cm}
\begin{tabular}{p{19.5cm}}
\textit{Notes:} The details of this simulation design are provided in Section \ref{sec:simulation}. ``1'' corresponds to the 2SLS estimator that additively controls for $X$\@. ``2'' corresponds to $\hat{\tau}_{u}^{cb}$. ``3'' corresponds to $\hat{\tau}_{u}^{ml}$. ``4'' corresponds to $\hat{\tau}_{a,10}^{ml}$. ``5'' corresponds to $\hat{\tau}_{a}^{ml}$. ``6'' corresponds to $\hat{\tau}_{t}^{ml}$ ($= \hat{\tau}_{a,1}^{ml}$). ``7'' corresponds to $\hat{\tau}_{a,0}^{ml}$. All weighting estimators also control for $X$\@. Results are based on 10,000 replications.
\end{tabular}
\end{footnotesize}
\end{adjustwidth}
\end{figure}

\begin{figure}[!p]
\begin{adjustwidth}{-1in}{-1in}
\centering
\caption{Simulation Results for Design C, $\delta=0.02$, $N=5{,}000$}
\includegraphics[width=21cm]{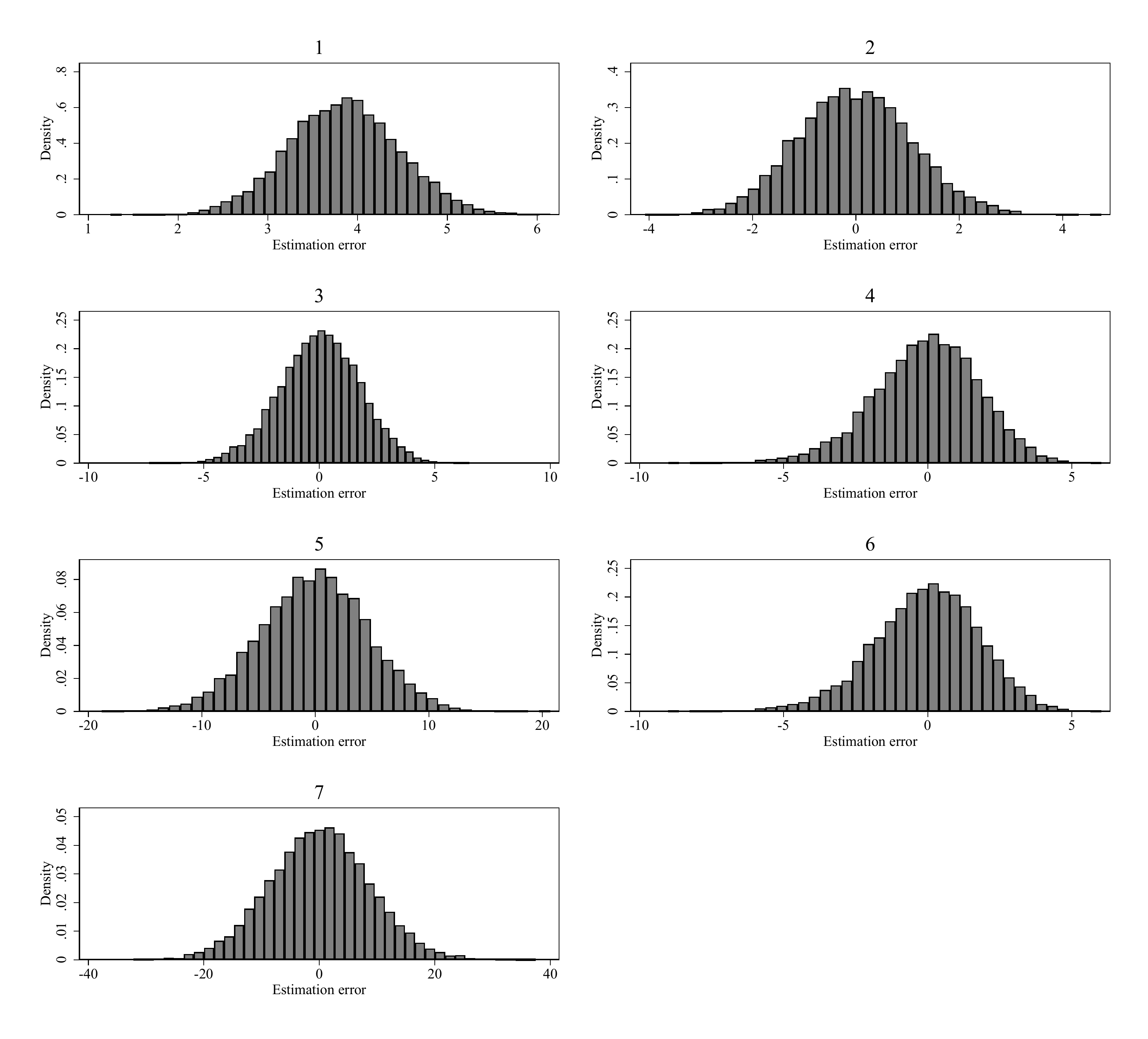}
\begin{footnotesize}
\vspace{0.075cm}
\begin{tabular}{p{19.5cm}}
\textit{Notes:} The details of this simulation design are provided in Section \ref{sec:simulation}. ``1'' corresponds to the 2SLS estimator that additively controls for $X$\@. ``2'' corresponds to $\hat{\tau}_{u}^{cb}$. ``3'' corresponds to $\hat{\tau}_{u}^{ml}$. ``4'' corresponds to $\hat{\tau}_{a,10}^{ml}$. ``5'' corresponds to $\hat{\tau}_{a}^{ml}$. ``6'' corresponds to $\hat{\tau}_{t}^{ml}$ ($= \hat{\tau}_{a,1}^{ml}$). ``7'' corresponds to $\hat{\tau}_{a,0}^{ml}$. All weighting estimators also control for $X$\@. Results are based on 10,000 replications.
\end{tabular}
\end{footnotesize}
\end{adjustwidth}
\end{figure}

\begin{figure}[!p]
\begin{adjustwidth}{-1in}{-1in}
\centering
\caption{Simulation Results for Design C, $\delta=0.05$, $N=500$}
\includegraphics[width=21cm]{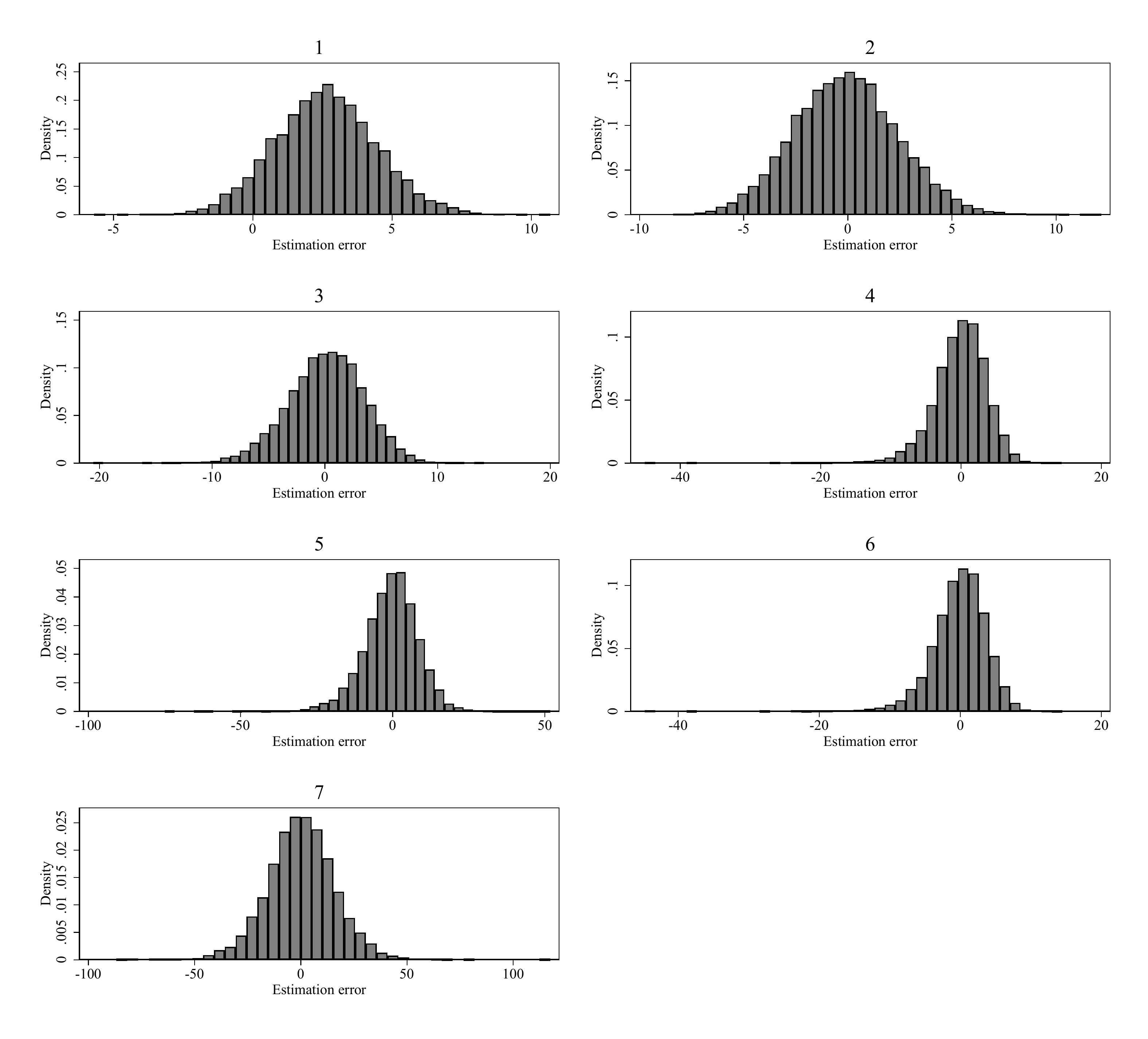}
\begin{footnotesize}
\vspace{0.075cm}
\begin{tabular}{p{19.5cm}}
\textit{Notes:} The details of this simulation design are provided in Section \ref{sec:simulation}. ``1'' corresponds to the 2SLS estimator that additively controls for $X$\@. ``2'' corresponds to $\hat{\tau}_{u}^{cb}$. ``3'' corresponds to $\hat{\tau}_{u}^{ml}$. ``4'' corresponds to $\hat{\tau}_{a,10}^{ml}$. ``5'' corresponds to $\hat{\tau}_{a}^{ml}$. ``6'' corresponds to $\hat{\tau}_{t}^{ml}$ ($= \hat{\tau}_{a,1}^{ml}$). ``7'' corresponds to $\hat{\tau}_{a,0}^{ml}$. All weighting estimators also control for $X$\@. Results are based on 10,000 replications.
\end{tabular}
\end{footnotesize}
\end{adjustwidth}
\end{figure}

\begin{figure}[!p]
\begin{adjustwidth}{-1in}{-1in}
\centering
\caption{Simulation Results for Design C, $\delta=0.05$, $N=1{,}000$}
\includegraphics[width=21cm]{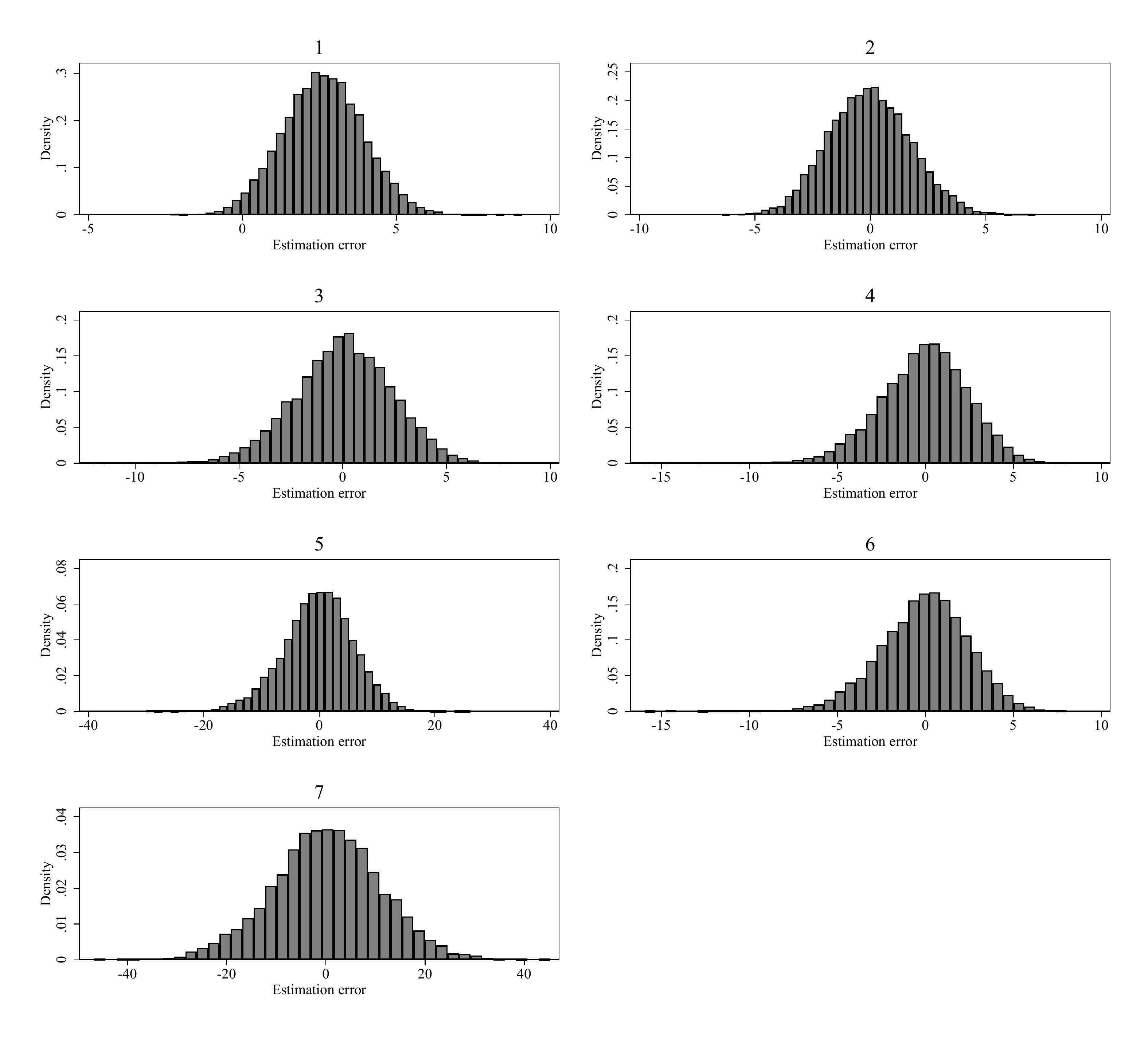}
\begin{footnotesize}
\vspace{0.075cm}
\begin{tabular}{p{19.5cm}}
\textit{Notes:} The details of this simulation design are provided in Section \ref{sec:simulation}. ``1'' corresponds to the 2SLS estimator that additively controls for $X$\@. ``2'' corresponds to $\hat{\tau}_{u}^{cb}$. ``3'' corresponds to $\hat{\tau}_{u}^{ml}$. ``4'' corresponds to $\hat{\tau}_{a,10}^{ml}$. ``5'' corresponds to $\hat{\tau}_{a}^{ml}$. ``6'' corresponds to $\hat{\tau}_{t}^{ml}$ ($= \hat{\tau}_{a,1}^{ml}$). ``7'' corresponds to $\hat{\tau}_{a,0}^{ml}$. All weighting estimators also control for $X$\@. Results are based on 10,000 replications.
\end{tabular}
\end{footnotesize}
\end{adjustwidth}
\end{figure}

\begin{figure}[!p]
\begin{adjustwidth}{-1in}{-1in}
\centering
\caption{Simulation Results for Design C, $\delta=0.05$, $N=5{,}000$}
\includegraphics[width=21cm]{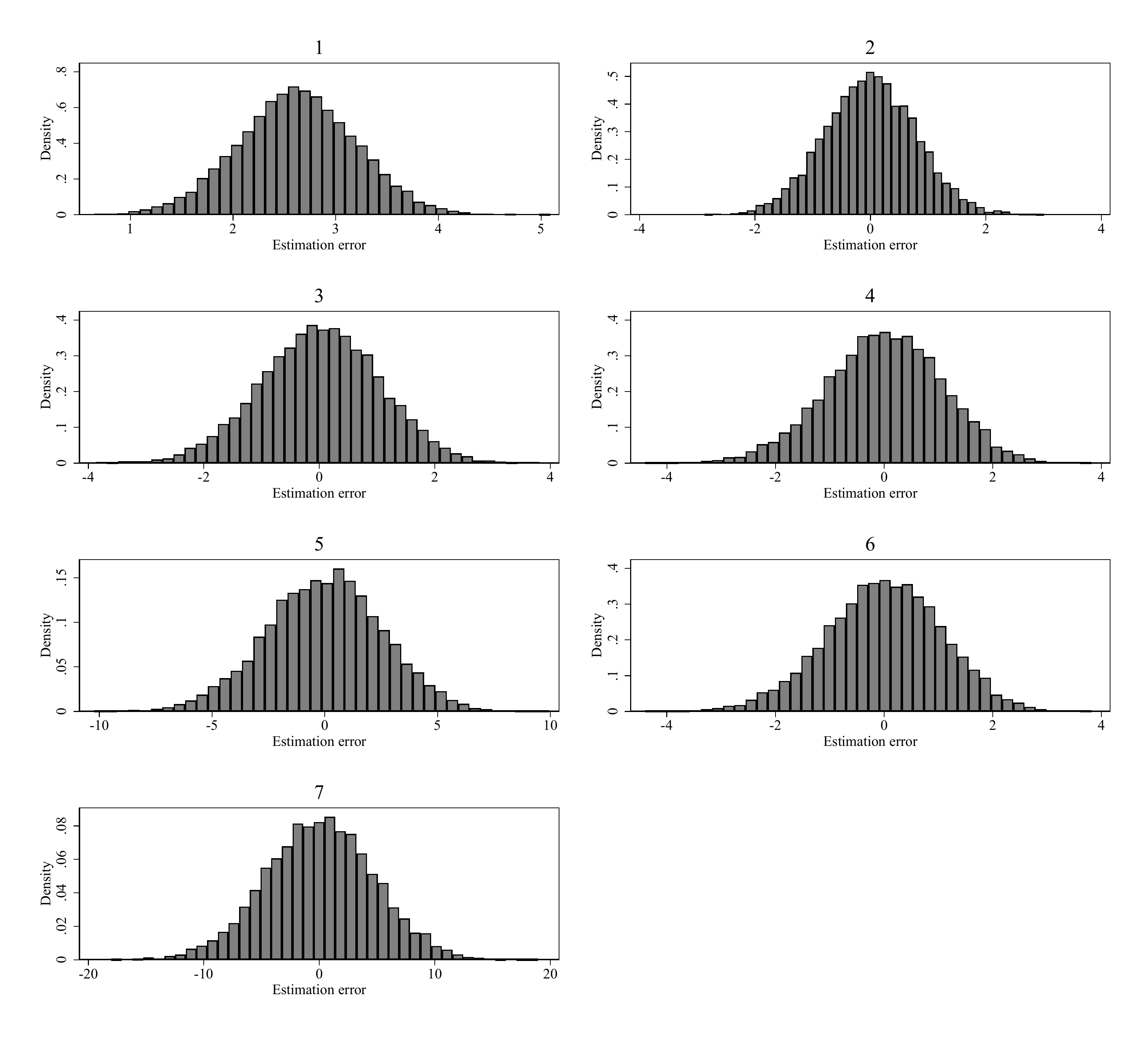}
\begin{footnotesize}
\vspace{0.075cm}
\begin{tabular}{p{19.5cm}}
\textit{Notes:} The details of this simulation design are provided in Section \ref{sec:simulation}. ``1'' corresponds to the 2SLS estimator that additively controls for $X$\@. ``2'' corresponds to $\hat{\tau}_{u}^{cb}$. ``3'' corresponds to $\hat{\tau}_{u}^{ml}$. ``4'' corresponds to $\hat{\tau}_{a,10}^{ml}$. ``5'' corresponds to $\hat{\tau}_{a}^{ml}$. ``6'' corresponds to $\hat{\tau}_{t}^{ml}$ ($= \hat{\tau}_{a,1}^{ml}$). ``7'' corresponds to $\hat{\tau}_{a,0}^{ml}$. All weighting estimators also control for $X$\@. Results are based on 10,000 replications.
\end{tabular}
\end{footnotesize}
\end{adjustwidth}
\end{figure}

\clearpage
\begin{figure}[!p]
\begin{adjustwidth}{-1in}{-1in}
\centering
\caption{Simulation Results for Design D, $\delta=0.01$, $N=500$}
\includegraphics[width=21cm]{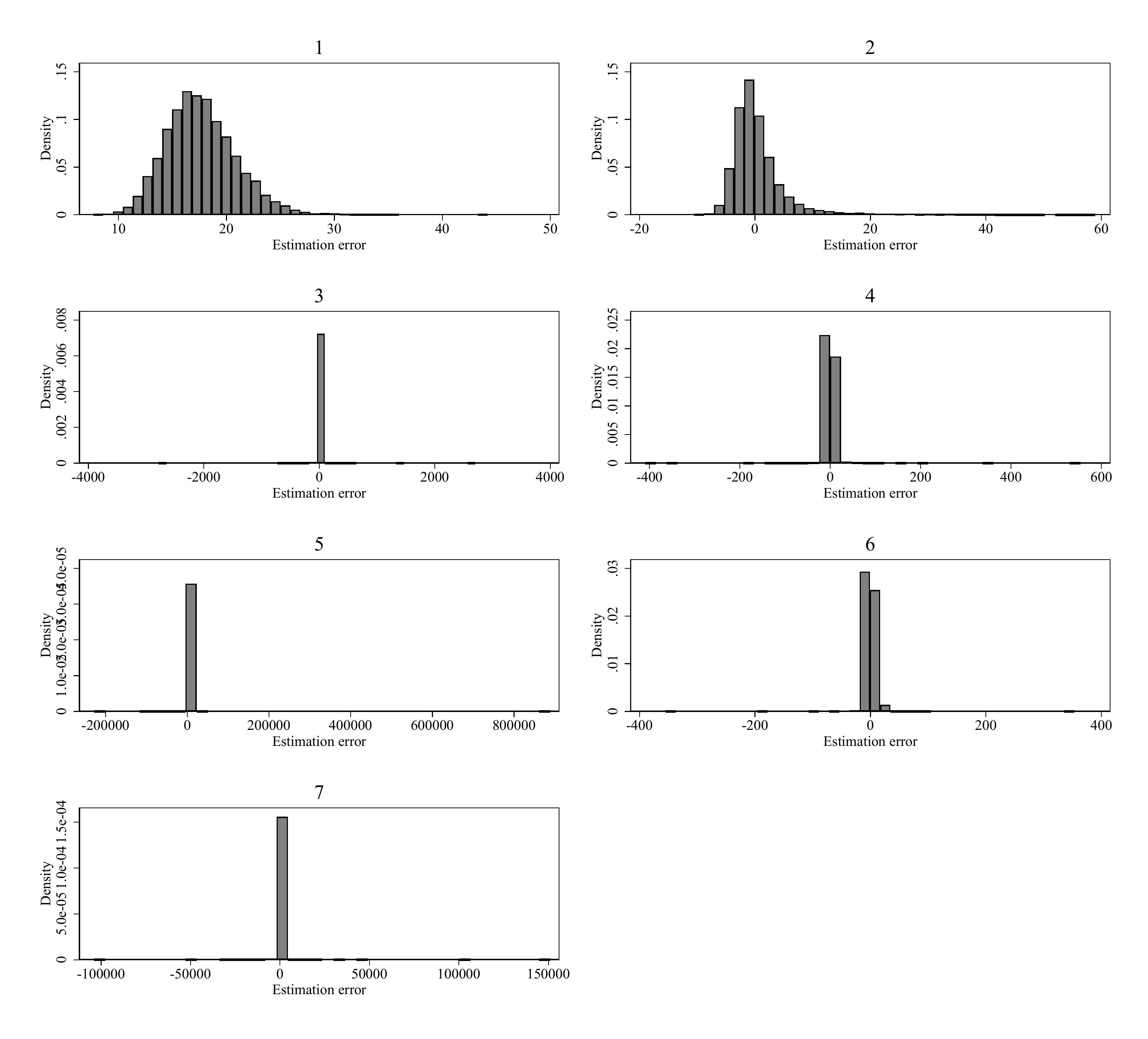}
\begin{footnotesize}
\vspace{0.075cm}
\begin{tabular}{p{19.5cm}}
\textit{Notes:} The details of this simulation design are provided in Section \ref{sec:simulation}. ``1'' corresponds to the 2SLS estimator that additively controls for $X$\@. ``2'' corresponds to $\hat{\tau}_{u}^{cb}$. ``3'' corresponds to $\hat{\tau}_{u}^{ml}$. ``4'' corresponds to $\hat{\tau}_{a,10}^{ml}$. ``5'' corresponds to $\hat{\tau}_{a}^{ml}$. ``6'' corresponds to $\hat{\tau}_{t}^{ml}$ ($= \hat{\tau}_{a,1}^{ml}$). ``7'' corresponds to $\hat{\tau}_{a,0}^{ml}$. All weighting estimators also control for $X$\@. Results are based on 10,000 replications.
\end{tabular}
\end{footnotesize}
\end{adjustwidth}
\end{figure}

\begin{figure}[!p]
\begin{adjustwidth}{-1in}{-1in}
\centering
\caption{Simulation Results for Design D, $\delta=0.01$, $N=1{,}000$}
\includegraphics[width=21cm]{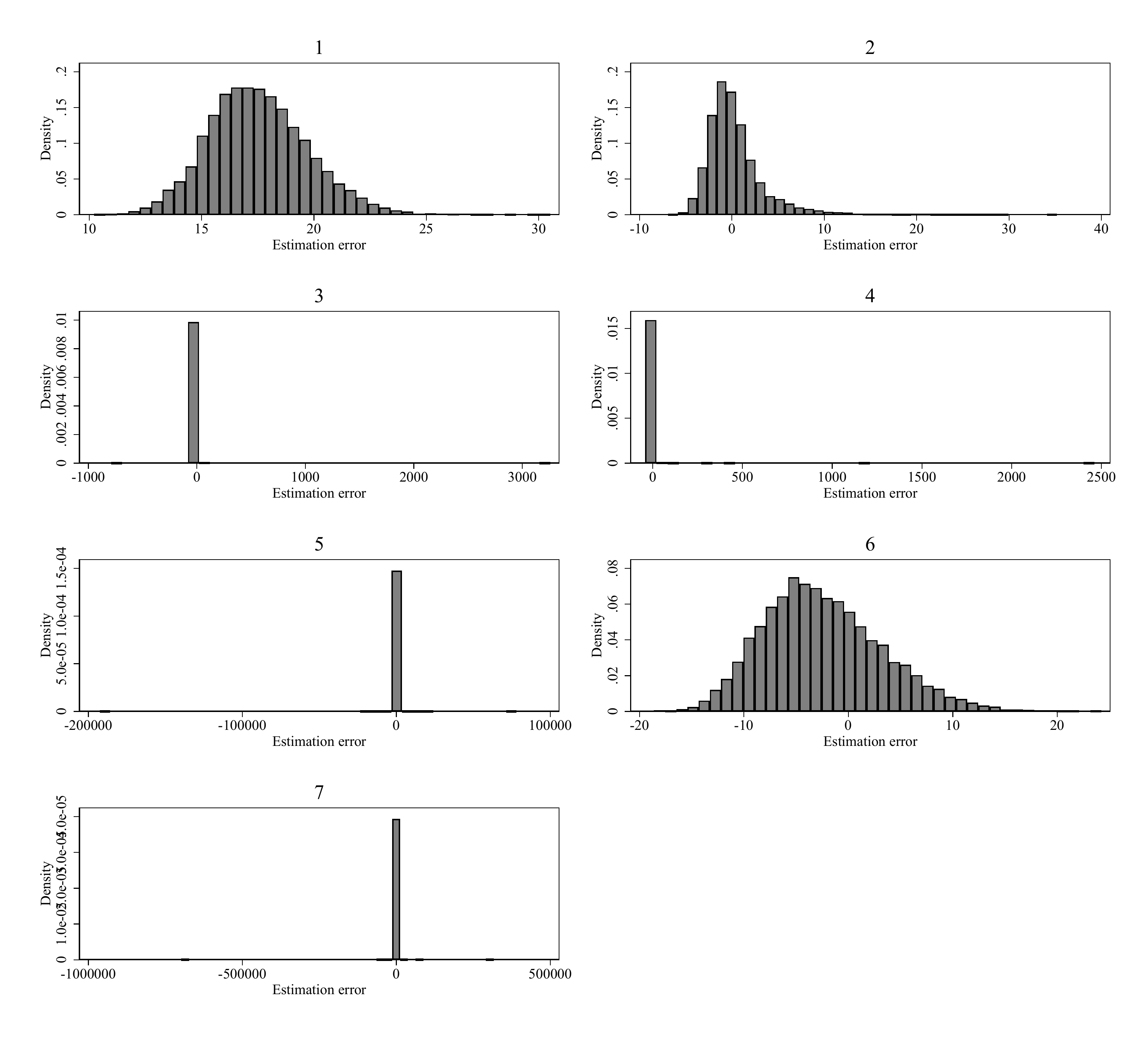}
\begin{footnotesize}
\vspace{0.075cm}
\begin{tabular}{p{19.5cm}}
\textit{Notes:} The details of this simulation design are provided in Section \ref{sec:simulation}. ``1'' corresponds to the 2SLS estimator that additively controls for $X$\@. ``2'' corresponds to $\hat{\tau}_{u}^{cb}$. ``3'' corresponds to $\hat{\tau}_{u}^{ml}$. ``4'' corresponds to $\hat{\tau}_{a,10}^{ml}$. ``5'' corresponds to $\hat{\tau}_{a}^{ml}$. ``6'' corresponds to $\hat{\tau}_{t}^{ml}$ ($= \hat{\tau}_{a,1}^{ml}$). ``7'' corresponds to $\hat{\tau}_{a,0}^{ml}$. All weighting estimators also control for $X$\@. Results are based on 10,000 replications.
\end{tabular}
\end{footnotesize}
\end{adjustwidth}
\end{figure}

\begin{figure}[!p]
\begin{adjustwidth}{-1in}{-1in}
\centering
\caption{Simulation Results for Design D, $\delta=0.01$, $N=5{,}000$}
\includegraphics[width=21cm]{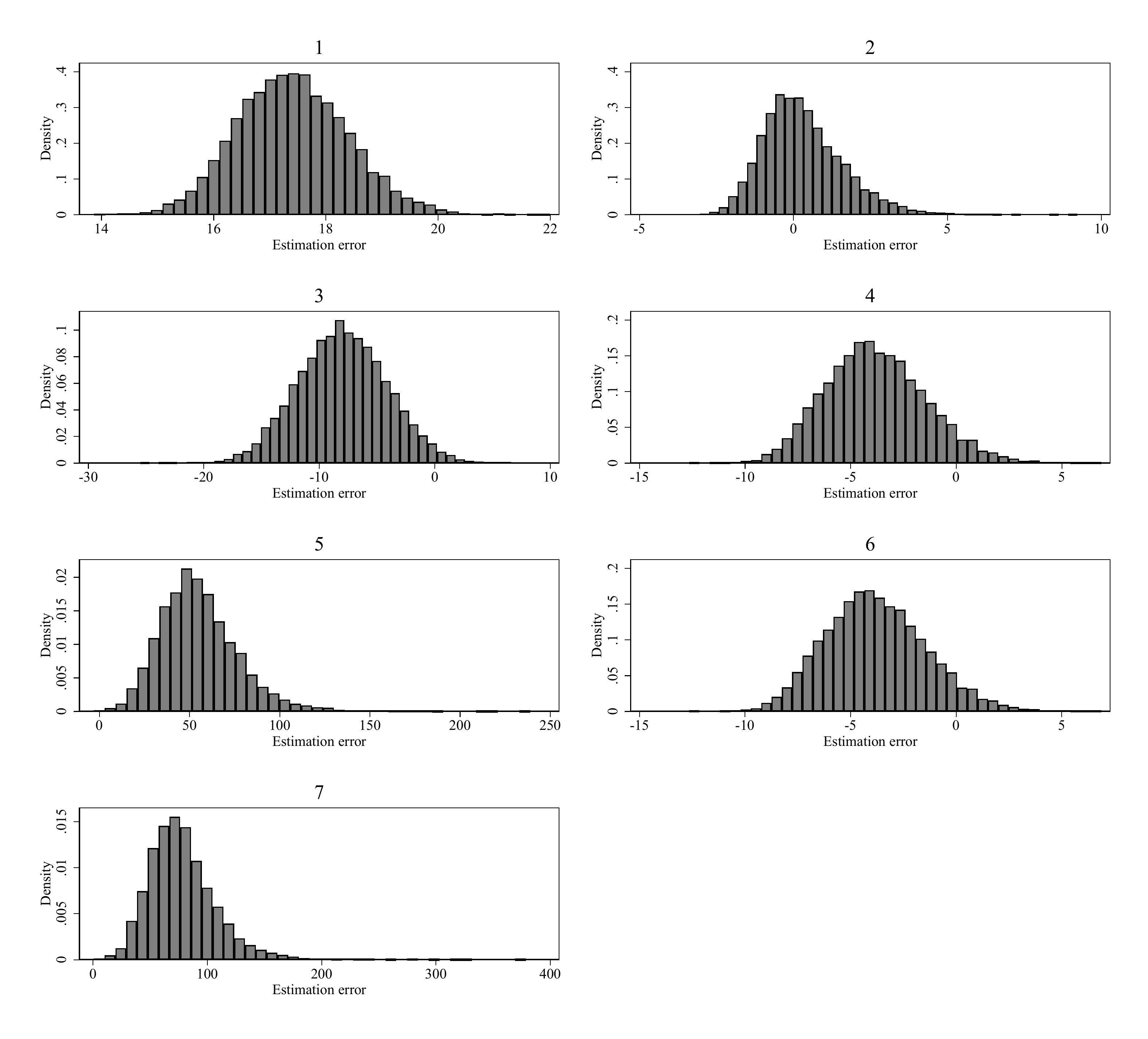}
\begin{footnotesize}
\vspace{0.075cm}
\begin{tabular}{p{19.5cm}}
\textit{Notes:} The details of this simulation design are provided in Section \ref{sec:simulation}. ``1'' corresponds to the 2SLS estimator that additively controls for $X$\@. ``2'' corresponds to $\hat{\tau}_{u}^{cb}$. ``3'' corresponds to $\hat{\tau}_{u}^{ml}$. ``4'' corresponds to $\hat{\tau}_{a,10}^{ml}$. ``5'' corresponds to $\hat{\tau}_{a}^{ml}$. ``6'' corresponds to $\hat{\tau}_{t}^{ml}$ ($= \hat{\tau}_{a,1}^{ml}$). ``7'' corresponds to $\hat{\tau}_{a,0}^{ml}$. All weighting estimators also control for $X$\@. Results are based on 10,000 replications.
\end{tabular}
\end{footnotesize}
\end{adjustwidth}
\end{figure}

\begin{figure}[!p]
\begin{adjustwidth}{-1in}{-1in}
\centering
\caption{Simulation Results for Design D, $\delta=0.02$, $N=500$}
\includegraphics[width=21cm]{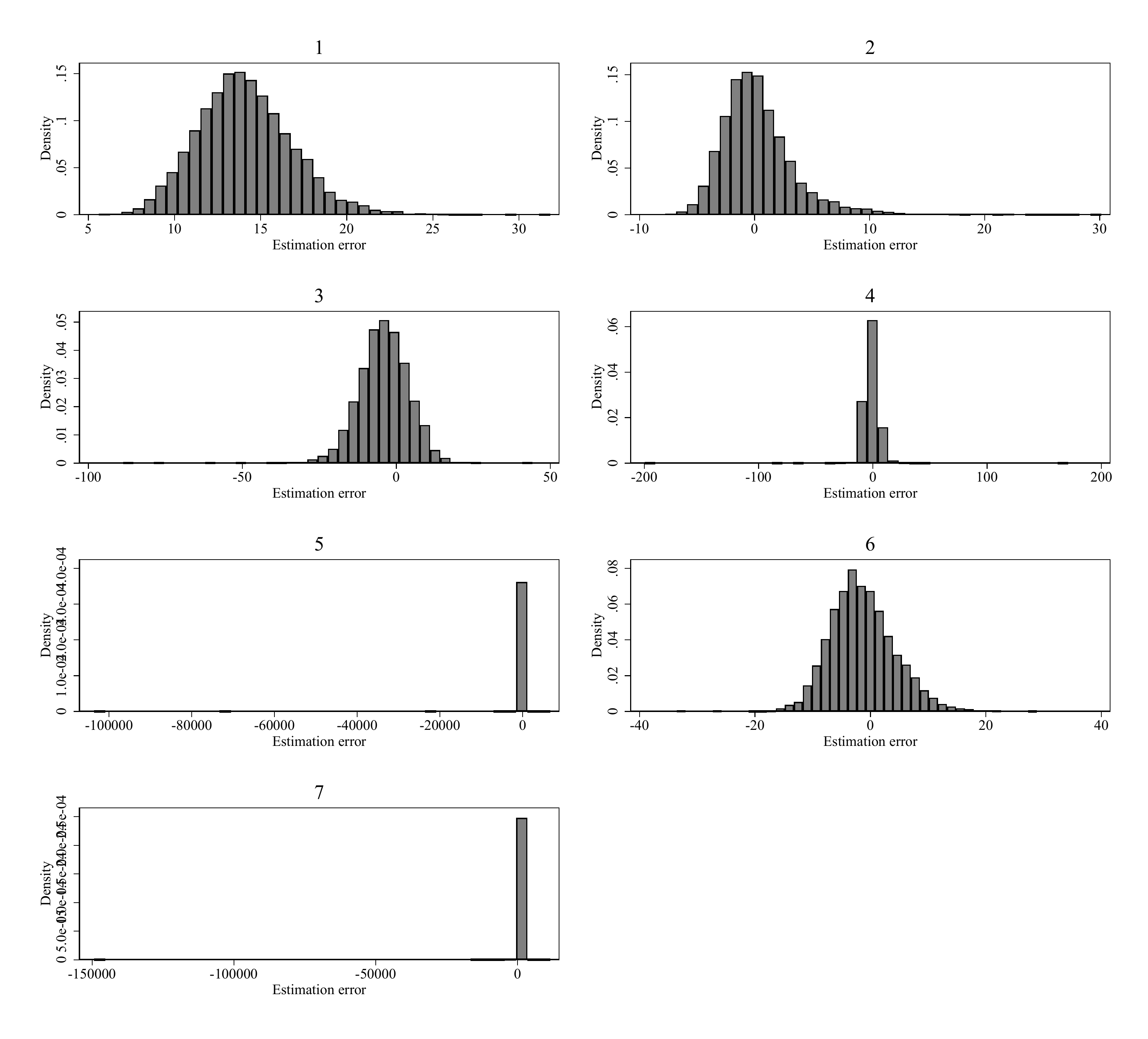}
\begin{footnotesize}
\vspace{0.075cm}
\begin{tabular}{p{19.5cm}}
\textit{Notes:} The details of this simulation design are provided in Section \ref{sec:simulation}. ``1'' corresponds to the 2SLS estimator that additively controls for $X$\@. ``2'' corresponds to $\hat{\tau}_{u}^{cb}$. ``3'' corresponds to $\hat{\tau}_{u}^{ml}$. ``4'' corresponds to $\hat{\tau}_{a,10}^{ml}$. ``5'' corresponds to $\hat{\tau}_{a}^{ml}$. ``6'' corresponds to $\hat{\tau}_{t}^{ml}$ ($= \hat{\tau}_{a,1}^{ml}$). ``7'' corresponds to $\hat{\tau}_{a,0}^{ml}$. All weighting estimators also control for $X$\@. Results are based on 10,000 replications.
\end{tabular}
\end{footnotesize}
\end{adjustwidth}
\end{figure}

\begin{figure}[!p]
\begin{adjustwidth}{-1in}{-1in}
\centering
\caption{Simulation Results for Design D, $\delta=0.02$, $N=1{,}000$}
\includegraphics[width=21cm]{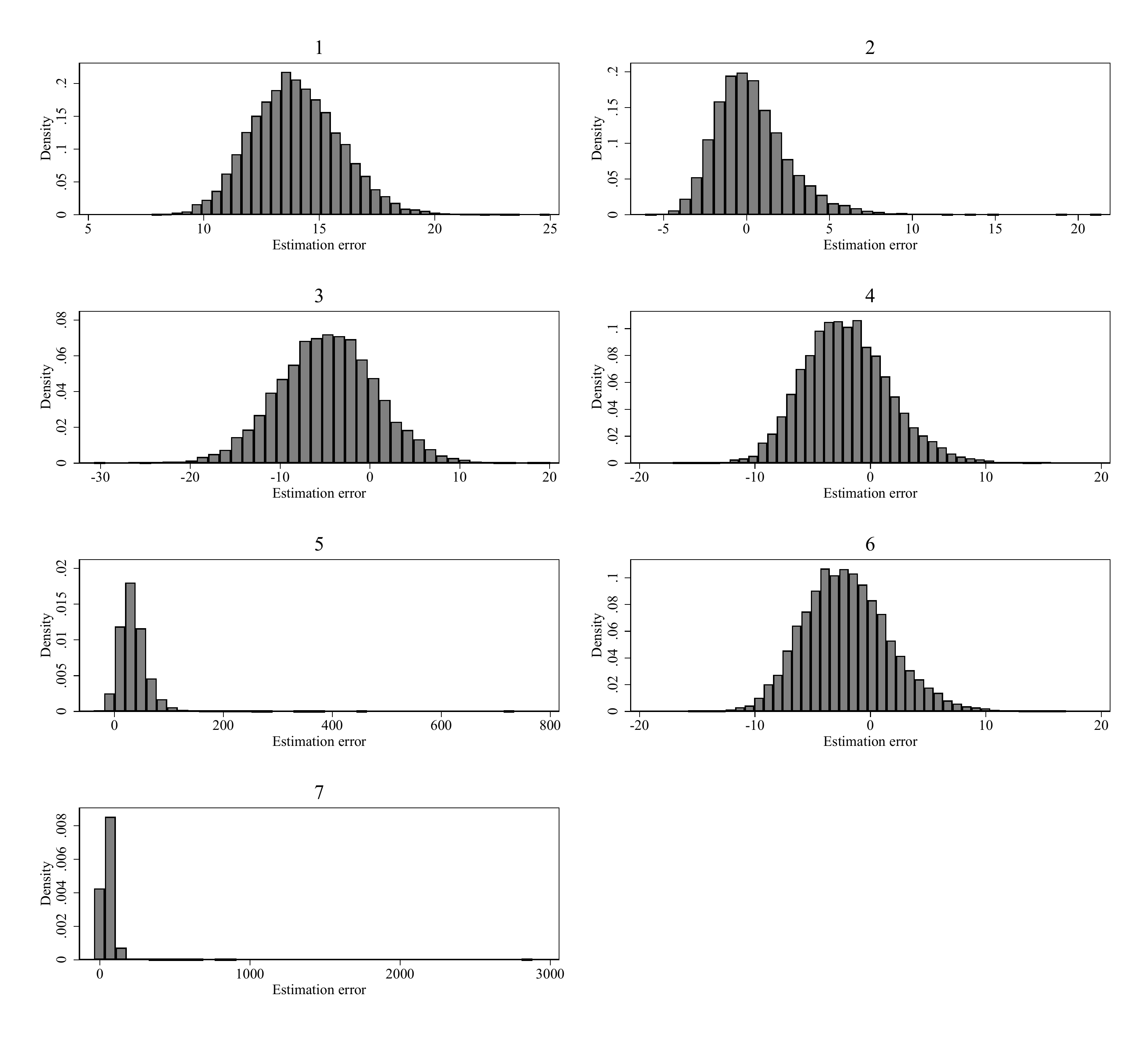}
\begin{footnotesize}
\vspace{0.075cm}
\begin{tabular}{p{19.5cm}}
\textit{Notes:} The details of this simulation design are provided in Section \ref{sec:simulation}. ``1'' corresponds to the 2SLS estimator that additively controls for $X$\@. ``2'' corresponds to $\hat{\tau}_{u}^{cb}$. ``3'' corresponds to $\hat{\tau}_{u}^{ml}$. ``4'' corresponds to $\hat{\tau}_{a,10}^{ml}$. ``5'' corresponds to $\hat{\tau}_{a}^{ml}$. ``6'' corresponds to $\hat{\tau}_{t}^{ml}$ ($= \hat{\tau}_{a,1}^{ml}$). ``7'' corresponds to $\hat{\tau}_{a,0}^{ml}$. All weighting estimators also control for $X$\@. Results are based on 10,000 replications.
\end{tabular}
\end{footnotesize}
\end{adjustwidth}
\end{figure}

\begin{figure}[!p]
\begin{adjustwidth}{-1in}{-1in}
\centering
\caption{Simulation Results for Design D, $\delta=0.02$, $N=5{,}000$}
\includegraphics[width=21cm]{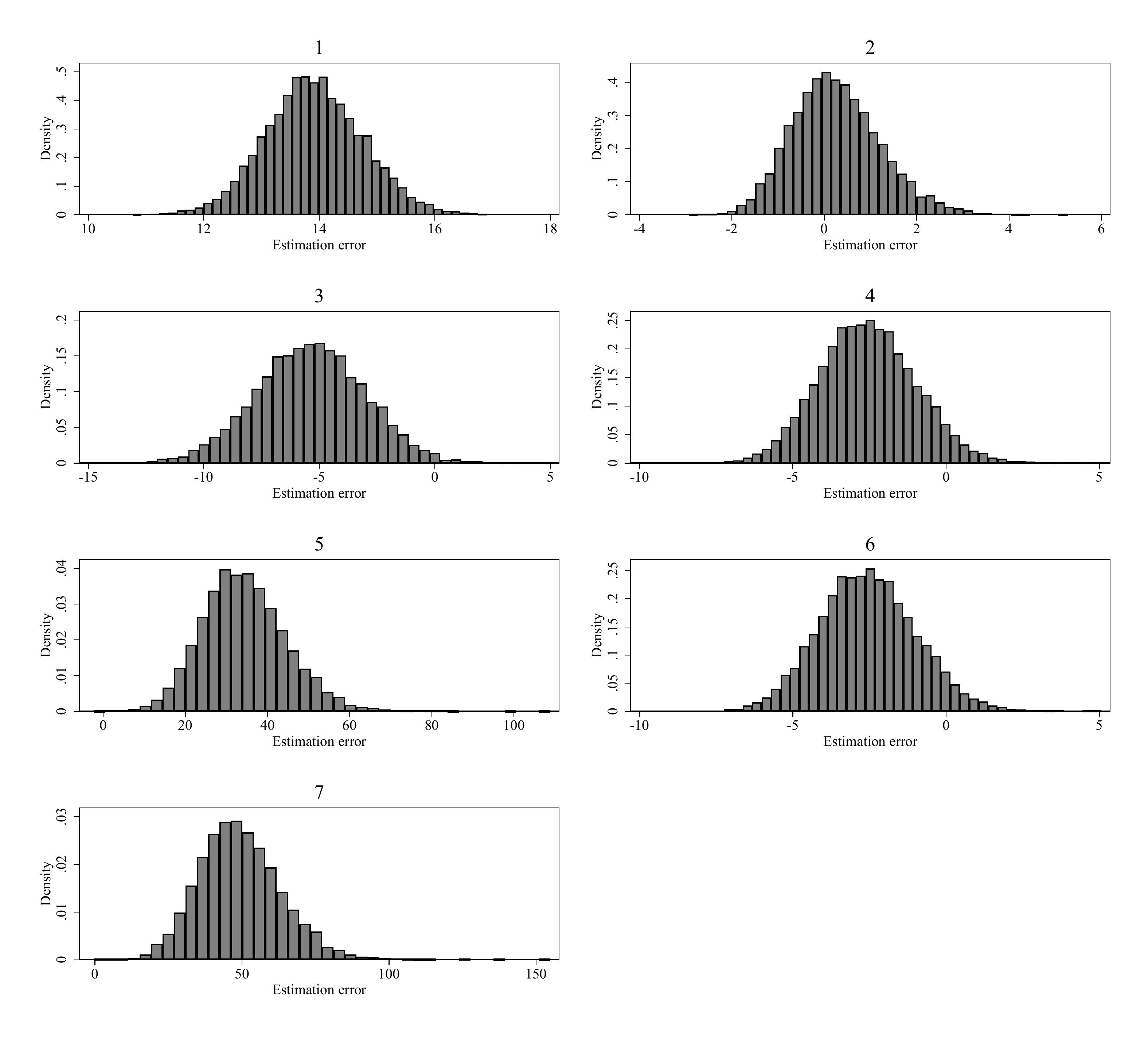}
\begin{footnotesize}
\vspace{0.075cm}
\begin{tabular}{p{19.5cm}}
\textit{Notes:} The details of this simulation design are provided in Section \ref{sec:simulation}. ``1'' corresponds to the 2SLS estimator that additively controls for $X$\@. ``2'' corresponds to $\hat{\tau}_{u}^{cb}$. ``3'' corresponds to $\hat{\tau}_{u}^{ml}$. ``4'' corresponds to $\hat{\tau}_{a,10}^{ml}$. ``5'' corresponds to $\hat{\tau}_{a}^{ml}$. ``6'' corresponds to $\hat{\tau}_{t}^{ml}$ ($= \hat{\tau}_{a,1}^{ml}$). ``7'' corresponds to $\hat{\tau}_{a,0}^{ml}$. All weighting estimators also control for $X$\@. Results are based on 10,000 replications.
\end{tabular}
\end{footnotesize}
\end{adjustwidth}
\end{figure}

\begin{figure}[!p]
\begin{adjustwidth}{-1in}{-1in}
\centering
\caption{Simulation Results for Design D, $\delta=0.05$, $N=500$}
\includegraphics[width=21cm]{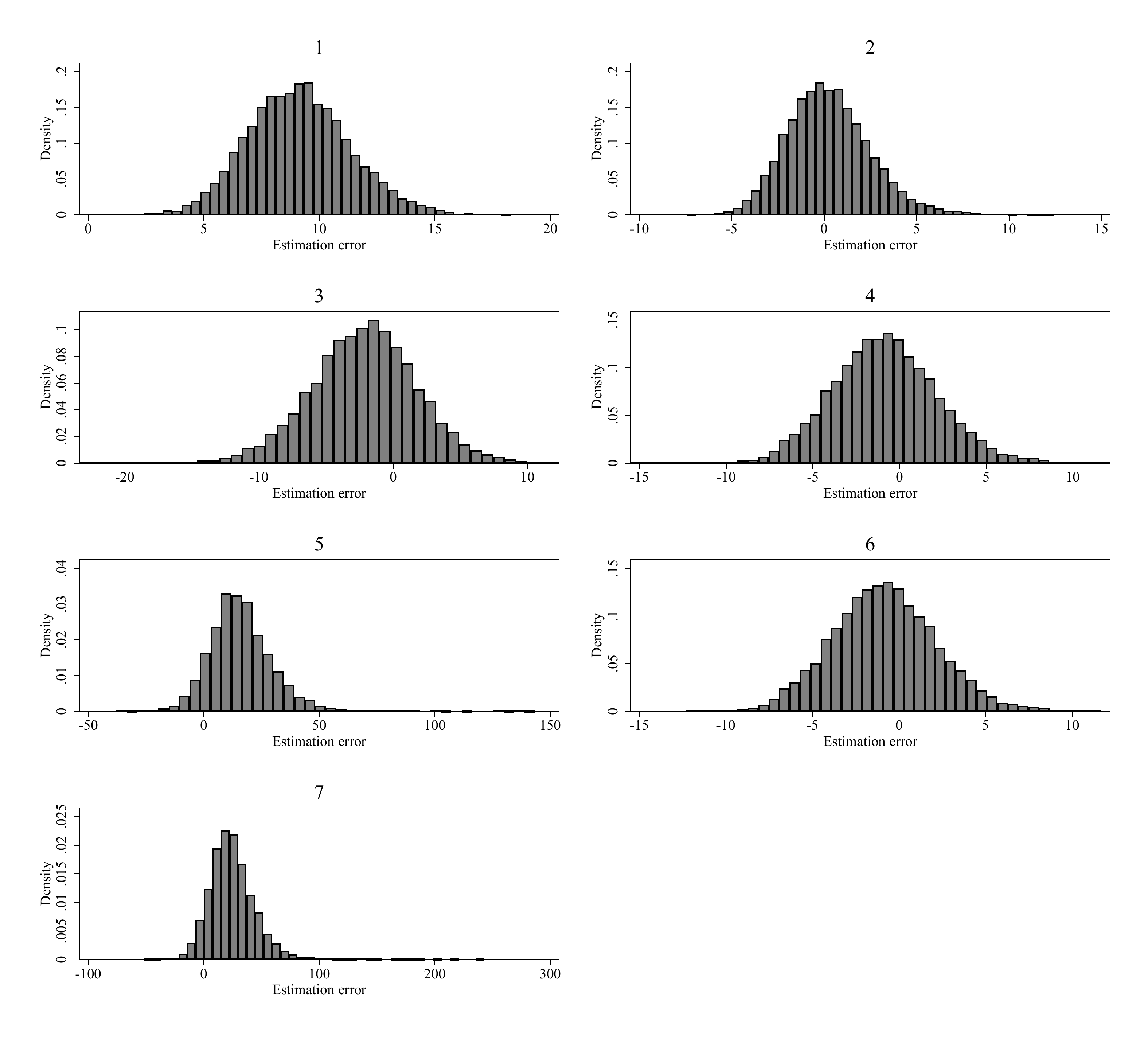}
\begin{footnotesize}
\vspace{0.075cm}
\begin{tabular}{p{19.5cm}}
\textit{Notes:} The details of this simulation design are provided in Section \ref{sec:simulation}. ``1'' corresponds to the 2SLS estimator that additively controls for $X$\@. ``2'' corresponds to $\hat{\tau}_{u}^{cb}$. ``3'' corresponds to $\hat{\tau}_{u}^{ml}$. ``4'' corresponds to $\hat{\tau}_{a,10}^{ml}$. ``5'' corresponds to $\hat{\tau}_{a}^{ml}$. ``6'' corresponds to $\hat{\tau}_{t}^{ml}$ ($= \hat{\tau}_{a,1}^{ml}$). ``7'' corresponds to $\hat{\tau}_{a,0}^{ml}$. All weighting estimators also control for $X$\@. Results are based on 10,000 replications.
\end{tabular}
\end{footnotesize}
\end{adjustwidth}
\end{figure}

\begin{figure}[!p]
\begin{adjustwidth}{-1in}{-1in}
\centering
\caption{Simulation Results for Design D, $\delta=0.05$, $N=1{,}000$}
\includegraphics[width=21cm]{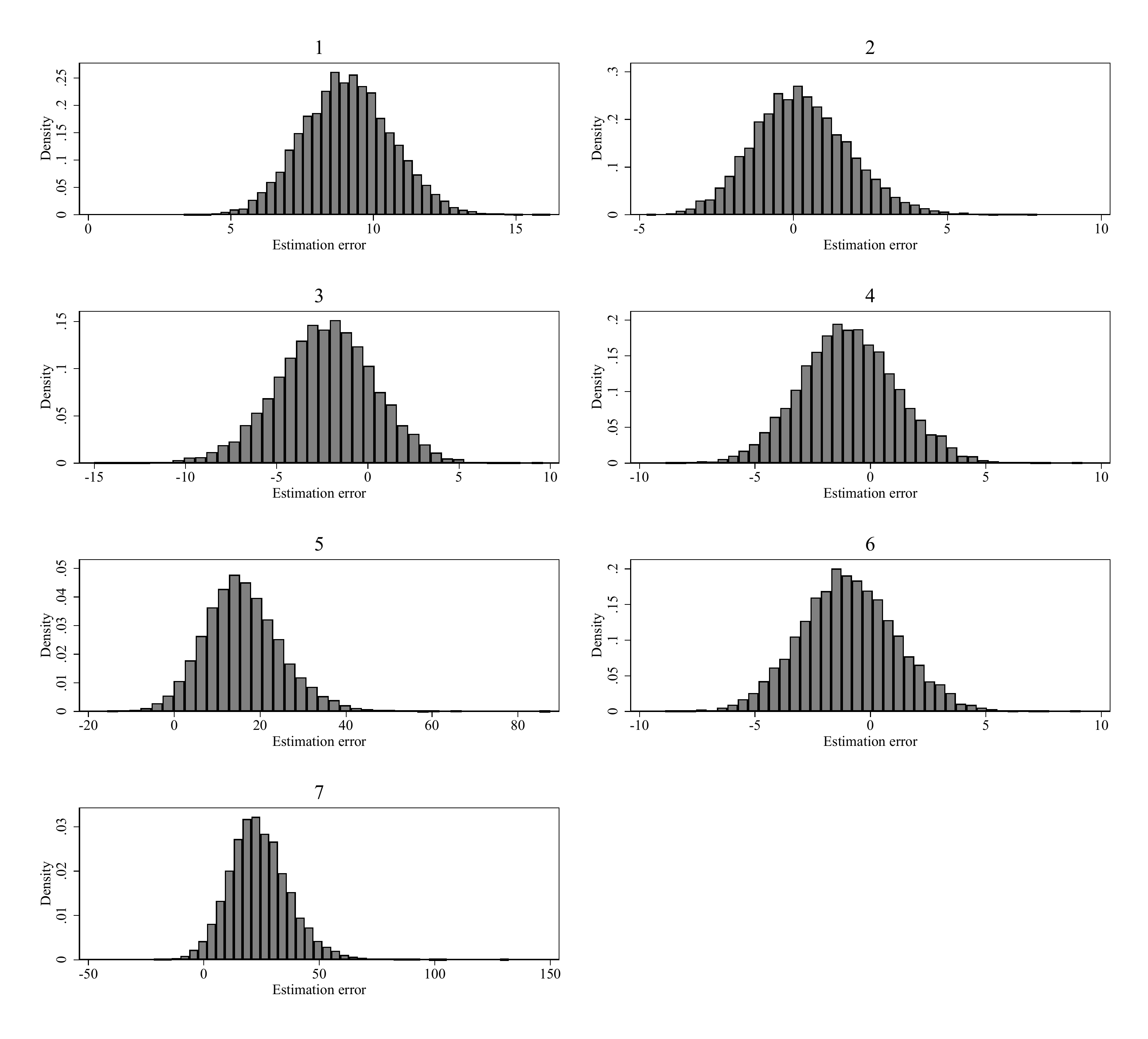}
\begin{footnotesize}
\vspace{0.075cm}
\begin{tabular}{p{19.5cm}}
\textit{Notes:} The details of this simulation design are provided in Section \ref{sec:simulation}. ``1'' corresponds to the 2SLS estimator that additively controls for $X$\@. ``2'' corresponds to $\hat{\tau}_{u}^{cb}$. ``3'' corresponds to $\hat{\tau}_{u}^{ml}$. ``4'' corresponds to $\hat{\tau}_{a,10}^{ml}$. ``5'' corresponds to $\hat{\tau}_{a}^{ml}$. ``6'' corresponds to $\hat{\tau}_{t}^{ml}$ ($= \hat{\tau}_{a,1}^{ml}$). ``7'' corresponds to $\hat{\tau}_{a,0}^{ml}$. All weighting estimators also control for $X$\@. Results are based on 10,000 replications.
\end{tabular}
\end{footnotesize}
\end{adjustwidth}
\end{figure}

\begin{figure}[!p]
\begin{adjustwidth}{-1in}{-1in}
\centering
\caption{Simulation Results for Design D, $\delta=0.05$, $N=5{,}000$\label{fig:hist_last}}
\includegraphics[width=21cm]{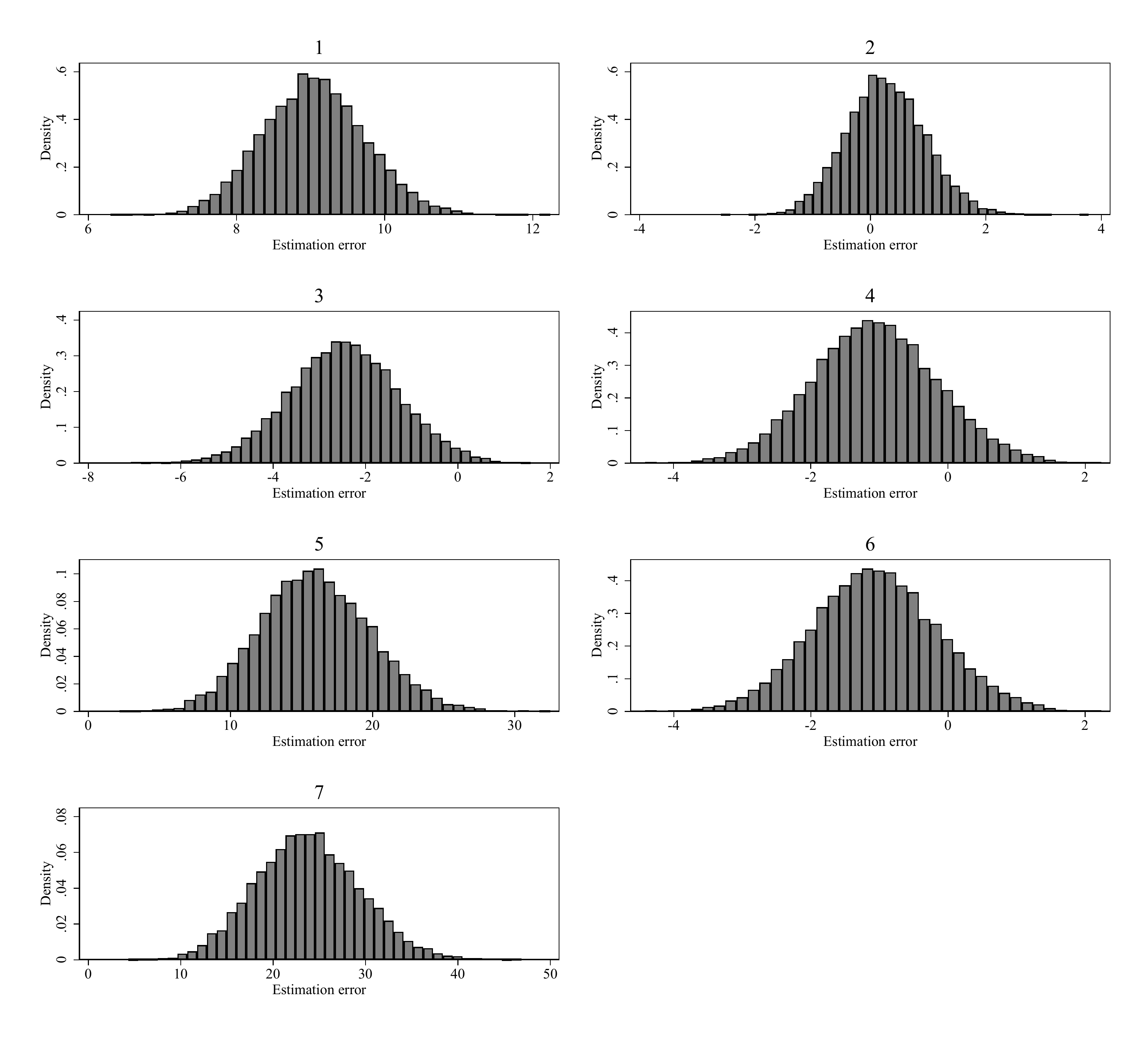}
\begin{footnotesize}
\vspace{0.075cm}
\begin{tabular}{p{19.5cm}}
\textit{Notes:} The details of this simulation design are provided in Section \ref{sec:simulation}. ``1'' corresponds to the 2SLS estimator that additively controls for $X$\@. ``2'' corresponds to $\hat{\tau}_{u}^{cb}$. ``3'' corresponds to $\hat{\tau}_{u}^{ml}$. ``4'' corresponds to $\hat{\tau}_{a,10}^{ml}$. ``5'' corresponds to $\hat{\tau}_{a}^{ml}$. ``6'' corresponds to $\hat{\tau}_{t}^{ml}$ ($= \hat{\tau}_{a,1}^{ml}$). ``7'' corresponds to $\hat{\tau}_{a,0}^{ml}$. All weighting estimators also control for $X$\@. Results are based on 10,000 replications.
\end{tabular}
\end{footnotesize}
\end{adjustwidth}
\end{figure}

\end{appendices}

\end{document}